\documentclass[12pt]{article}
\pdfoutput=1

\usepackage[pdftex]{graphicx}
\usepackage{amssymb}
\usepackage{amsmath}
\usepackage{cite}
\usepackage{multirow}

\usepackage{longtable}
\usepackage{booktabs}
\usepackage{array}

\usepackage{anyfontsize}

\usepackage{wrapfig}
\usepackage{color}
\definecolor{tit}{rgb}{0.1,0.2,0.4}

\renewcommand{\arraystretch}{1.2}

\newcommand{\eq}[1]{\begin{equation} #1 \end{equation}}

\newcommand{\av}[1]{\langle #1 \rangle}

\newcommand{\C}[1]{{\cal C}_{#1}}

\definecolor{DRed}{rgb}{0.8,0,0.1}
\definecolor{DBlue}{rgb}{0,0,0.8}

\begin{document}

\begin{flushright}
{\small
LPT-ORSAY/15-68\\
QFET-2015-29\\
SI-HEP-2015-19 
}
\end{flushright}
$\ $
\vspace{-2mm}
\begin{center}
\fontsize{19.2}{23}\selectfont
\bf 
\color{tit}
Global analysis of $b\to s\ell\ell$ anomalies
\end{center}

\vspace{1mm}
\begin{center}
{\sf S\'ebastien Descotes-Genon}\\[2mm]
{\em\small
Laboratoire de Physique Th\'eorique (UMR 8627), CNRS, Univ. Paris-Sud,\\
Universit\'e Paris-Saclay,
91405 Orsay, France
}
\end{center}

\begin{center}
{\sf  Lars Hofer, Joaquim Matias}\\[2mm]
{\em \small
Universitat Aut\`onoma de Barcelona, 08193 Bellaterra, Barcelona,\\[0.2cm]
Institut de Fisica d'Altes Energies (IFAE),\\ The Barcelona Institute of Science and Technology, Campus UAB, 08193 Bellaterra (Barcelona) Spain
}
\end{center}

\begin{center}
{\sf  Javier Virto}\\[2mm]
{\em \small Theoretische Physik 1, Naturwissenschaftlich-Technische Fakult\"at,\\
Universit\"at Siegen, 57068 Siegen, Germany
}
\end{center}

\vspace{1mm}
\begin{abstract}\noindent
\vspace{-5mm}

We present a detailed discussion of  the current theoretical and experimental situation of the anomaly in the angular distribution of $B \to K^*(\to K\pi)\mu^+\mu^-$, observed at LHCb in the  1 fb$^{-1}$ dataset and recently confirmed by the 3 fb$^{-1}$ dataset. The impact of this data and other recent measurements on $b\to s\ell^+\ell^-$ transitions ($\ell=e,\mu$) is considered. We review the observables of interest, focusing on their theoretical uncertainties and their sensitivity to New Physics, based on an analysis employing the QCD factorisation approach including several sources of hadronic uncertainties (form factors, power corrections, charm-loop effects). We perform fits to New Physics contributions including experimental and theoretical correlations.
The solution that we proposed in 2013 to solve the $B\to K^*\mu^+\mu^-$ anomaly, with a contribution $\C9^{\rm NP}\simeq -1$, is confirmed and reinforced. A wider range of New-Physics scenarios with high significances (between 4 and 5~$\sigma$) emerges from the fit, some of them being particularly relevant for model building. More data is needed to discriminate among them conclusively. The inclusion of $b\to s e^+ e^-$ observables increases the significance of the favoured scenarios under the hypothesis of New Physics breaking lepton flavour universality. Several tests illustrate the robustness of our conclusions.
\end{abstract}

\newpage

\setcounter{tocdepth}{2}
\tableofcontents

\newpage

\section{Introduction}
\label{sec:intro}

Flavour-Changing Neutral Currents (FCNC) have been prominent tools in high-energy physics in the search for new degrees of freedom, due to their quantum sensitivity to energies much higher than the external particles involved. In the current context where the LHC has discovered a scalar boson completing the Standard Model (SM) picture but no additional particles that would go beyond this framework, FCNC can be instrumental in order to determine where to look for New Physics (NP). One particularly interesting instance of FCNC is provided by $b\to s\ell\ell$ and $b\to s\gamma$ transitions, which can be probed through various decay channels, currently studied in detail at the LHCb, CMS and ATLAS  experiments. In addition, in some kinematic configurations it is possible to build observables with a very limited sensitivity to hadronic uncertainties, and thus enhancing the discovery potential of these decays for NP, based on the use of effective field theories adapted to the problem at hand. Finally, it is possible to analyse all these decays using a model-independent approach, namely the effective Hamiltonian~\cite{Grinstein:1987vj,Buchalla:1995vs} where heavy degrees of freedom have been integrated out in short-distance Wilson coefficients $\C{i}$, leaving only a set of operators $O_i$ describing the physics at long distances:
\begin{equation}
{\cal H}_{\rm eff}=-\frac{4G_F}{\sqrt{2}} V_{tb}V_{ts}^*\sum_i \C{i}  O_i
\end{equation}
(up to small corrections proportional to $V_{ub}V_{us}^*$ in the SM). In the following, the factorisation scale for the Wilson coefficients is $\mu_b=$ 4.8 GeV. We focus  our attention on the operators
\begin{align}
{\mathcal{O}}_{7} &= \frac{e}{16 \pi^2} m_b
(\bar{s} \sigma_{\mu \nu} P_R b) F^{\mu \nu} ,&
{\mathcal{O}}_{{7}^\prime} &= \frac{e}{16 \pi^2} m_b
(\bar{s} \sigma_{\mu \nu} P_L b) F^{\mu \nu} , \nonumber
\\
{\mathcal{O}}_{9} &= \frac{e^2}{16 \pi^2} 
(\bar{s} \gamma_{\mu} P_L b)(\bar{\ell} \gamma^\mu \ell) ,&
{\mathcal{O}}_{{9}^\prime} &= \frac{e^2}{16 \pi^2} 
(\bar{s} \gamma_{\mu} P_R b)(\bar{\ell} \gamma^\mu \ell) , \nonumber
\\
\label{eq:O10}
{\mathcal{O}}_{10} &=\frac{e^2}{16 \pi^2}
(\bar{s}  \gamma_{\mu} P_L b)(  \bar{\ell} \gamma^\mu \gamma_5 \ell) ,&
{\mathcal{O}}_{{10}^\prime} &=\frac{e^2}{16\pi^2}
(\bar{s}  \gamma_{\mu} P_R b)(  \bar{\ell} \gamma^\mu \gamma_5 \ell) ,
\end{align}
where $P_{L,R}=(1 \mp \gamma_5)/2$ and $m_b \equiv m_b(\mu_b)$ denotes the running $b$ quark mass in the $\overline{\mathrm{MS}}$ scheme. 
In the SM, three operators play a leading role in the discussion, namely the electromagnetic operator $O_7$ and the semileptonic operators $O_9$ and $O_{10}$, differing with respect to the chirality of the emitted charged leptons~(see Ref.~\cite{DescotesGenon:2011yn} for more detail). NP contributions could either modify the value of the short-distance Wilson coefficients $\C{7,9,10}$, or make other operators contribute in a significant manner (such as $O_{7',9',10'}$ defined above, or the scalar and pseudoscalar operators $O_{S,S',P,P'}$).

Recent experimental results have shown interesting deviations from the SM.
In 2013, the LHCb collaboration announced the measurement of angular observables describing the decay $B\to K^*\mu\mu$
in both regions of low- and large-$K^*$ recoil~\cite{Aaij:2013qta}.
Two observables, $P_2$ and $P_5^\prime$~\cite{Matias:2012xw,DescotesGenon:2012zf,Descotes-Genon:2013vna},
were in significant disagreement with the SM expectations in the large-$K^*$ recoil region~\cite{Descotes-Genon:2013wba}.
A few months later, an improved measurement of the branching ratio for $B\to K\mu\mu$ turned out to be slightly on the
low side compared to theoretical expectations~\cite{Aaij:2014pli}.
Both results were interpreted as indications for a large negative contribution to the Wilson coefficient of the
semileptonic operator $O_9$. Contributions to other Wilson coefficients could also occur, in particular to
$\C{9'}$~\cite{Altmannshofer:2013foa,Beaujean:2013soa,Hurth:2013ssa,Mahmoudi:2014mja,Hurth:2014vma,Altmannshofer:2014rta,Altmannshofer:2015sma}.
This triggered several theoretical studies reassessing the different long-distance effects that could contribute in
these decays, in particular charm resonances and loop contributions, form factors,
and power corrections~\cite{Khodjamirian:2010vf,Khodjamirian:2012rm,Lyon:2014hpa,Straub:2015ica,Jager:2012uw,Jager:2014rwa,Descotes-Genon:2014uoa}.

Another measurement has also raised a lot of attention recently, namely $R_K=Br(B\to K\mu\mu)/Br(B\to Kee)$, measured as $0.745^{+0.090}_{-0.074}\pm 0.036$ by LHCb in the dilepton mass range from 1 to 6 GeV$^2$~\cite{Aaij:2014ora} while predicted to be equal to 1 (to a very good accuracy) in the SM. This 2.6~$\sigma$ deviation can be naturally interpreted by the same negative shift to $\C9$, but applied only to the dimuon component of the operator $O_9$, whereas the dielectron 
component keeps the SM value~\cite{Hiller:2014yaa}. This could stem from heavy particles (typically a $Z'$ meson) coupling preferentially to muons in the lepton sector, with a flavour-changing $bs$ coupling~\cite{Buras:2013dea,Buras:2014yna,Gauld:2013qja,Gauld:2013qba,Boucenna:2016wpr}. On the other hand, hadronic effects should cancel in the ratio $R_K$ and thus are not able to explain this measurement.

Since the previous analysis of $B\to K^*\mu\mu$ data performed in Ref.~\cite{Descotes-Genon:2013wba},
several improvements have occurred on both theoretical and experimental sides. LHCb has recently released
new data on $B\to K^*\mu\mu$ with a finer binning~\cite{Aaij:2015oid}, confirming the pattern of deviations
observed in 2013, based on extended statistics (3 fb$^{-1}$). The same collaboration has also studied
$B_s\to\phi\mu\mu$~\cite{Aaij:2013aln} and $B\to K^*ee$ at very large recoil (the intermediate photon being
almost on shell)~\cite{Aaij:2015dea}. Concerning inclusive radiative decays, updated theoretical predictions
are available for $B\to X_s\gamma$ \cite{Misiak:2015xwa} and $B\to X_s \ell\ell$ \cite{Huber:2015sra}.
These various elements call for an update of the previous analysis, which can be compared to other recent global
analyses~\cite{Altmannshofer:2014rta,Altmannshofer:2015sma,Hurth:2016fbr}.

We start in Sec.~\ref{sec:BtoKstarmumu} by discussing salient features of $B\to K^*\mu\mu$ observables, detailing
the ingredients for their theoretical predictions, as well as their sensitivity to NP, before briefly considering
other $b\to s\gamma$ and $b\to s\mu\mu$ decays (both inclusive and exclusive) 
in Sec.~\ref{sec:otherobs}. In Sec.~\ref{sec:NPfits} we discuss a set of scenarios with large NP contributions
to one or two Wilson coefficients, confirming that a negative contribution to $\C{9}$ yields a significant
improvement compared to the SM. We  discuss which of these scenarios are able to reduce the anomalies observed in
$b\to s\ell\ell$ transitions. By performing a global fit to all six Wilson coefficients simultaneously, we show that
the most economic scenarios do indeed capture the main patterns suggested by the data.
In this case we provide, in addition, confidence-level regions for all Wilson coefficients when all of them are allowed
to deviate from their SM values simultaneously.
We also consider scenarios with violation of lepton-flavour universality,
and describe tests of the robustness of the fits presented. In Sec.~\ref{sec:SMfit},
we provide tests of the various sources of hadronic uncertainties that could affect our results (choice of form factors,
power corrections, long-distance charm corrections). We present our conclusions in Sec.~\ref{sec:Conclusions}.
Apps.~\ref{app:SMpred} and \ref{app:NPpred} are devoted to tables presenting our predictions for the SM as well as
the best-fit point for NP in $\C9$ only. In App.~\ref{app:ImpFit}, the confidence regions for less favoured,
but theoretically interesting, scenarios are shown. App.~\ref{app:C9} describes how various changes in the analysis
affect its outcome for the scenario with NP in $\C9$ only.
App.~\ref{app:PowCorr} contains further details on power corrections to $B_s\to\phi$ and $B\to K$ form factors.
App.~\ref{sec:Zcouplings} gathers basic features of $Z'$ models relevant for the $b\to s$ anomalies.

\section{\boldmath$B\to K^*\mu\mu$}
\label{sec:BtoKstarmumu}

\subsection{General approach}
\label{sec:BtoKstarmumu-gen}

In the effective Hamiltonian approach and in the SM (the extension to NP operators is straightforward), the $B\to K^*\mu\mu$ transversity amplitudes can be written in a compact way as
\begin{eqnarray}\label{eq:ampBtoKs}
A &\propto& \left[\C7 \frac{2im_b}{q^2} q_\rho \langle \bar{K}^*|\bar{s}\sigma^{\rho\mu}(1+\gamma_5)b|\bar{B}\rangle
 + \C9 \langle \bar{K}^*|\bar{s}\gamma^\mu(1-\gamma_5)b|\bar{B}\rangle
 + H^\mu
 \right]\bar{u}_\ell \gamma_\mu v_\ell\nonumber\\
 &&\qquad\qquad +\C{10}\langle \bar{K}^*|\bar{s}\gamma^\mu(1-\gamma_5)b|\bar{B}\rangle
 \bar{u}_\ell \gamma_\mu\gamma_5v_\ell\,,\\
 \text{with}&&H^\mu\,\propto\, i\int d^4x\ e^{iq\cdot x} 
 \langle \bar{K}^*|T[\bar{c}\gamma^\mu c]  {\cal H}_c |\bar{B}\rangle\,,
\end{eqnarray}
where ${\cal H}_c$ denotes the part of the weak effective Hamiltonian involving  four-quark operators with two charm fields. For simplicity, we have neglected contributions from CKM-suppressed terms here (they are included in our numerical evaluations). One can see from eq.~(\ref{eq:ampBtoKs}) the existence of two different kinds of contributions: local ones yielding form factors (seven for $B\to K^*$) and non-local ones (involving $c\bar{c}$ loops propagating). The former can be determined using non-perturbative methods (light-cone sum rules, lattice), whereas the latter must be estimated using $1/m_b$ expansion (QCD factorisation, OPE), with different tools depending on the kinematic regime considered (large- or low-$K^*$ recoil). We will illustrate these points in the large-recoil region where the strongest deviations have been observed between SM predictions and data. 

A first step in the evaluation of the amplitudes comes from the contributions due to $O_{7,9,10}$, involving seven form factors. In the large-recoil region there are basically two approaches:
\begin{itemize}
\item ``Improved QCD Factorisation (QCDF) approach": In this framework~\cite{Descotes-Genon:2013vna}  the large-recoil symmetries between form factors are used to implement the dominant correlations among them. This general approach is easy to cross-check and to implement for any form factor parametrisation (e.g. for the light-cone sum rules parametrisations \cite{Ball:2004rg,Khodjamirian:2010vf,Straub:2015ica}). The symmetries allow  the 7 form factors to be written in terms of only  two so-called  soft form factors $\xi_{\perp,\|}$ \cite{Charles:1998dr}:
\begin{eqnarray} \label{correlations} 
\frac{m_B}{m_B+m_{K^*}}{V(q^2)} = \frac{m_B+m_{K^*}}{2E}{A_1(q^2)}
={T_1(q^2)} = \frac{m_B}{2 E} {T_2(q^2)} =  {\xi_\bot(E)}, \,\,\,\,\,\,\,\,\,\,\, \\[2mm]
\frac{m_{K^*}}{E} {A_0(q^2)} =
\frac{m_B+m_{K^*}}{2E} {A_1(q^2)} - \frac{m_B-m_{K^*}}{m_B}{A_2(q^2)}
= \frac{m_B}{2E} {T_2(q^2)} - {T_3(q^2)} = {\xi_\|(E)}\,.  \nonumber
\end{eqnarray}
To this soft-form factor representation one should add (perturbatively computable) hard-gluon $O(\alpha_s)$ corrections as well as (non-perturbative) $O(\Lambda/m_b)$ corrections~\cite{Beneke:2000wa}. The soft form factors can be computed in a specific parametrisation. 
The basis of optimized observables $P_i$ is usually taken in this approach~\cite{Descotes-Genon:2013vna, Kruger:2005ep,Lunghi:2006hc, Egede:2010zc, Matias:2012xw, DescotesGenon:2012zf}.
We follow Ref.~\cite{Descotes-Genon:2014uoa} where we considered all symmetry-breaking corrections  to the relations in Eq.(\ref{correlations}). Our predictions take into account factorizable $\alpha_s$-corrections computed within QCDF \cite{Beneke:2000wa,Beneke:2001at,Beneke:2004dp}, as well as  factorizable power corrections. We will consider most of the time
the full form factors of Ref.~\cite{Khodjamirian:2010vf}, but  for completeness we will also compare some of our results with
the results using the form factors in Ref.~\cite{Straub:2015ica}.

\item ``Full Form Factor approach": Here  a specific set of full form factors determined from light-cone sum rules~\cite{Ball:2004rg,Straub:2015ica} is used. Factorizable $\alpha_s$ and factorizable power corrections are automatically included  with correlations associated to this particular parametrisation.  Other corrections
to the amplitudes (non-factorisable pieces, see below) have to be included and/or estimated exactly as in the previous approach. This approach has been employed in Refs.~\cite{Altmannshofer:2008dz,Altmannshofer:2013foa,Altmannshofer:2014rta}.
\end{itemize}
Both approaches are useful and complementary, should converge and give comparable results and error sizes,
as long as the correlations among the form factors are dominated by the large-recoil relations.
It is interesting to notice that the relevant form factors for the transversity amplitudes are not those defined in
the usual transversity basis ($V,A_i,T_i$) but rather the helicity form factors~\cite{Bharucha:2010im,Jager:2012uw}
being linear combinations of the usual transversity ones. It is therefore important to determine properly the correlations
among the usual form factors in order to determine correctly the transversity amplitudes. The first approach allows one to
restore correlations that are expected among the various form factors, even when these correlations were not given initially.
The second one requires one to compute  the complete set of form factors and to achieve a very good control of the
applied theoretical method in order to determine a meaningful correlation matrix.
Of course, both methods can be used to compute both types of observables $P_i$ and $S_i$, and they are expected to yield
similar results. We will discuss this point further in Sec.~\ref{sec:SMfit}.

Once the issue of the form factors has been settled, one can proceed with the determination of the amplitudes involving not only the form factors but also non-local $c\bar{c}$ loop contributions. QCD factorisation~\cite{Beneke:2000wa,Beneke:2001at,Beneke:2004dp} yields an expression of the amplitudes in terms of soft form factors, $\alpha_s$- and power corrections, which can be further split into factorisable and non-factorisable contributions (stemming or not from the expression of full form factors in terms of soft form factors). The factorisable power corrections have already been considered at the level of the form factors, whereas the non-factorisable ones still have to be addressed. First we
take the three hadronic form factors ${\cal T}_i(q^2)$ that parametrise the matrix element  $\langle K^* \gamma^*|H_{eff}|B\rangle$~\cite{Beneke:2000wa}, and we single out the hadronic contribution that is not related to the radiative Wilson coefficients (obtained taking the limit ${\cal T}_i^{\rm had} = {\cal T}_i|_{\C{7^{(\prime)}}\to0}$). We multiply each of these amplitudes serving as a normalisation with a complex $q^2$-dependent factor~\cite{Descotes-Genon:2014uoa}
\begin{equation}
{\cal T}_i^{\rm had}\to \big(1+r_i(q^2)\big) {\cal T}_i^{\rm had}\,,
\end{equation}
where
\begin{equation} r_i(s) = r_i^a e^{i\phi_i^a} + r_i^b e^{i\phi_i^b} (s/m_B^2) + r_i^c e^{i\phi_i^c} (s/m_B^2)^2\,. \end{equation}
We define our central values as the ones with $r_i(s)\equiv0$, and estimate the uncertainties from non-factorizable power corrections
by varying $r_i^{a,b,c}\in [0,0.1]$ and $\phi_i^{a,b,c} \in [-\pi,\pi]$ independently, corresponding to a
$\sim 10\%$ correction with an arbitrary phase. 

\begin{figure}[!t]
\begin{center}
\begin{tabular}{cc}
\includegraphics[width=7.5cm]{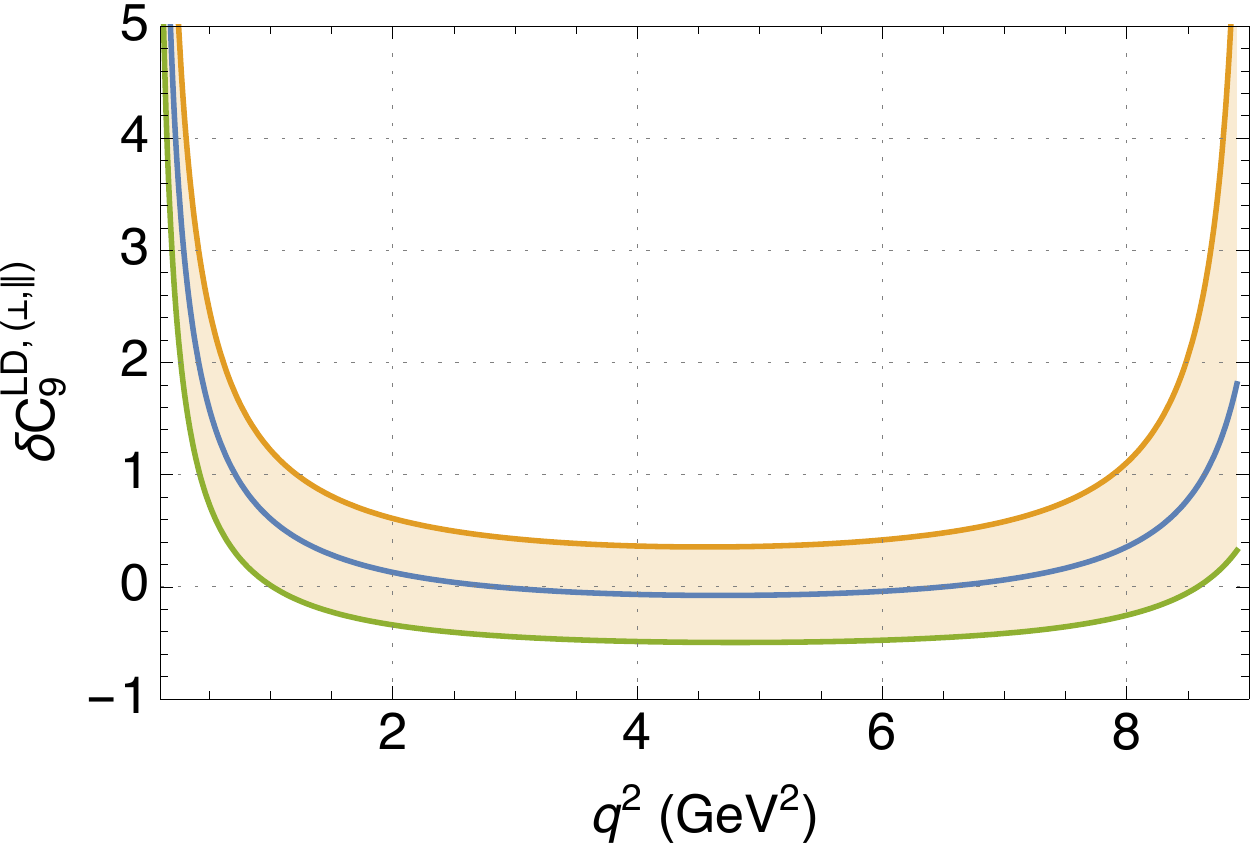} &  \includegraphics[width=7.5cm]{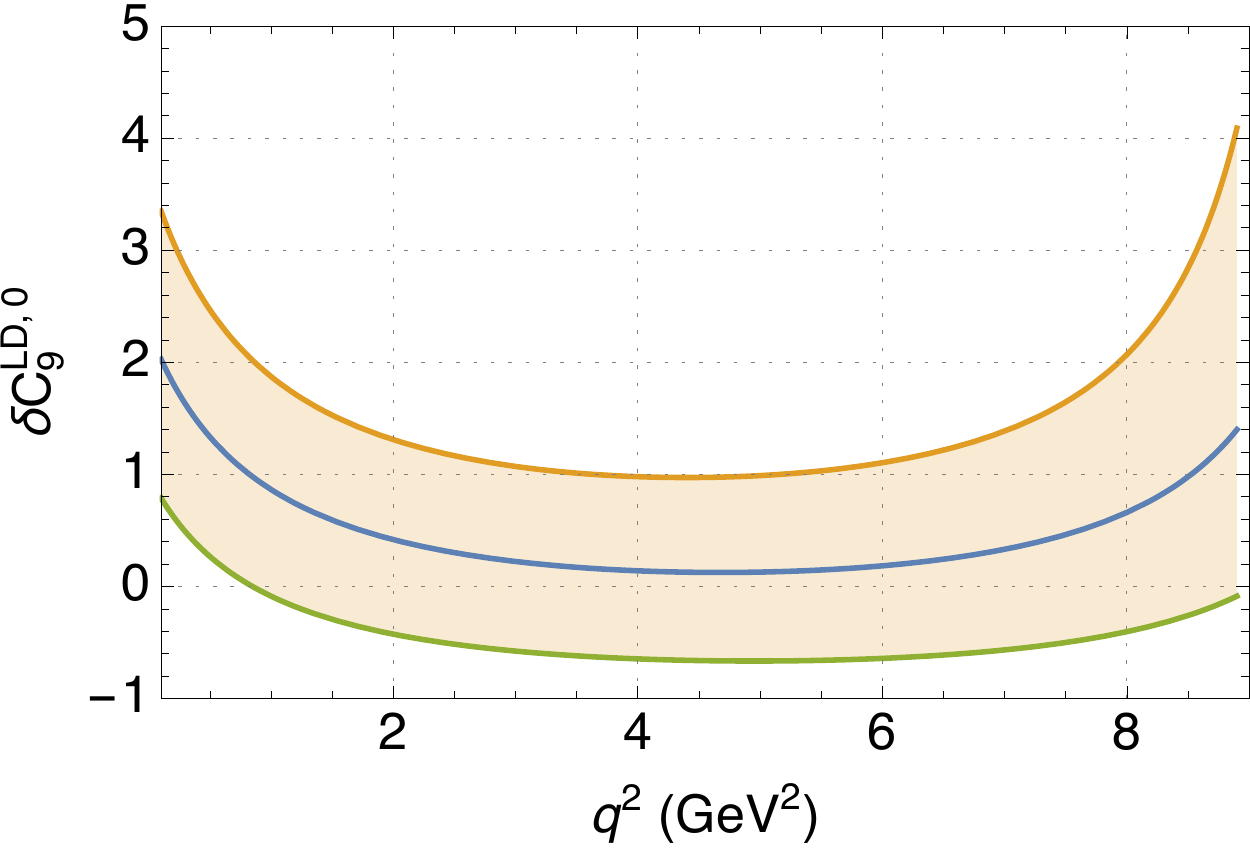}
\end{tabular}
\end{center}
\caption{\it Model used for the long-distance charm contribution for transverse (left) and longitudinal (right) $B\to K^*\ell\ell$ amplitudes.}\label{fig:longdistcharm}
\end{figure}

Part of the $c{\bar c}$-loop contributions have been already included in  the non-factorizable contributions
(hard-gluon exchange). The remaining long-distance contributions from  $c{\bar c}$ loops are still under debate.
For these contributions we will rely on the  partial computation in Ref.~\cite{Khodjamirian:2010vf}.
It is important to remark that the soft-gluon contribution of Ref.~\cite{Khodjamirian:2010vf} coming from 4-quark and
penguin operators induces a {\it positive} contribution to $\C9^{\rm eff}$ whose effect is to {\it enhance} the anomaly.
Since we are  interested only in the long-distance contribution $\delta \C9^{\rm LD}(q^2)$,
we subtract the  perturbative LO part  and include the shift due to a different reference value for $m_c$.
Ref.~\cite{Descotes-Genon:2014uoa} provides more details on this procedure. We introduce
two different parametrisations, corresponding to the contribution to transverse amplitudes
\begin{equation}
\delta \C9^{{\rm LD},\perp}(q^2)=\frac{a^\perp+b^\perp q^2(c^\perp -q^2)}{q^2(c^\perp -q^2)}\,,\qquad
\delta \C9^{{\rm LD},||}(q^2)=\frac{a^{||}+b^{||} q^2(c^{||}-q^2)}{q^2(c^{||}-q^2)}\,,
\end{equation}
and to the longitudinal amplitude (which does not exhibit a pole at $q^2=0$)
\begin{equation}
\delta \C9^{{\rm LD},0}(q^2)=\frac{a^0+b^0 (q^2+s_0)(c^0 -q^2)}{(q^2+s_0)(c^0-q^2)}\,,
\end{equation}
setting $s_0=1$ GeV$^2$.
We tune the parameters in order to cover the results obtained in Sec.~7 of Ref.~\cite{Khodjamirian:2010vf}
in the $q^2$-region between 1 and 9 GeV$^2$, where results for the three transversity amplitudes
(denoted ${\cal M}_1$, ${\cal M}_2$ and ${\cal M}_3$) have been derived~\footnote{
The transverse amplitudes $\delta \C9^{{\rm LD},(\|,\perp)}$ are a combination of ${\cal M}_1$ and ${\cal M}_2$,
while the longitudinal amplitude $\delta \C9^{{\rm LD},0}$ is a combination of ${\cal M}_2$ and ${\cal M}_3$.
Both ${\cal M}_2$ and ${\cal M}_3$ contain poles at $q^2=0$, but these cancel in the combination yielding the
longitudinal amplitude. Here we parametrise the regular part that remains after the cancellation of such poles.
}. We get
\begin{eqnarray}
&&a^\perp,a^{||}=9.25\pm 2.25\,, \qquad a^0=33\pm 7\,,\\
&&b^\perp,b^{||}=-0.5\pm 0.3\,, \qquad b^0=-0.9\pm 0.5\,,\\
&&c^\perp,c^{||}=9.35\pm 0.25\,, \qquad c^0=10.35\pm 0.55\,,
\end{eqnarray}
where all parameters will be taken as uncorrelated. The resulting functions $\delta \C9^{{\rm LD},(\perp,||)}(q^2)$ and $\delta \C9^{{\rm LD},0}(q^2)$
are shown in Fig.~\ref{fig:longdistcharm}.
In order to be  conservative, and in particular  given the discussion on the sign of this contribution, we use the result of Ref.~\cite{Khodjamirian:2010vf}
as an order of magnitude estimate, performing the following shift in each pair of transversity amplitudes
\begin{equation}
A_{i}^{L,R}: \qquad {\cal C}_{9}^{\rm eff}(q^2) \to {\cal C}_{9}^{\rm eff}(q^2) + s_i \,\delta \C9^{{\rm LD},i}(q^2)\,, \qquad i=0,\perp,||\,,
\end{equation}
with three independent parameters $s_i=0\pm 1$  (we recall that we include the perturbative $c\bar{c}$ contribution in ${\cal C}_{9}^{\rm eff}$ and that the direct inclusion of the result from Ref.~\cite{Khodjamirian:2010vf} would correspond to choosing $s_i=1$).

For the low-recoil region~\cite{Bobeth:2010wg,Bobeth:2011gi,Bobeth:2012vn}, one can perform a similar analysis based on Operator Product Expansion and Heavy-Quark Effective Theory, or using directly form factors provided by lattice QCD simulations. In the following, we will use the latter approach for the computation of the observables at low recoil. In this region, one has also to deal with resonances such as those observed by LHCb in the data of the partner channel $B^+ \to K^+ \mu^+\mu^-$.  This observation prevents one from taking small bins afflicted by the resonance structures. In Ref.~\cite{Beylich:2011aq} a quantitative estimate of duality violation is given. Unavoidably, one needs to use a model for this estimate, still the result is that the  low recoil bin, integrated over a large energy range, gets a duality-violation impact of a few percent at the level of the branching ratio (estimated to 5$\%$ in Ref.~\cite{Grinstein:2004vb} or 2$\%$ in Ref.~\cite{Beylich:2011aq}). It remains to be determined if this estimate also applies for angular observables in $B\to K^*\mu\mu$.  Moreover, the exact definition of the ends of the single large bin has 
some impact on the analysis in the framework of the effective Hamiltonian \cite{Horgan:2013pva}. In order to take into account such effect of duality violation for angular observables and the sensitivity to the position of the ends of the bin, we add a contribution of ${\mathcal O}(10\%)$ (with an arbitrary phase) to the term proportional to $\C{9}^{\rm eff}$ for each transversity amplitude. We notice that for all exclusive processes at low recoil, we include the NNLL corrections for $b\to s\ell\ell$ processes as described in Ref.~\cite{Greub:2008cy}.

\subsection{Optimised basis of observables: definition, properties and impact of data}

The  structure of the amplitudes at large recoil led to the construction of the optimised observables 
$P_i$ and $P_i^{\rm CP}$ \cite{Kruger:2005ep,Lunghi:2006hc,Egede:2010zc,Matias:2012xw,DescotesGenon:2012zf,Descotes-Genon:2013vna}
that exhibit a sensitivity to the soft form factors suppressed by $\alpha_s$ or $\Lambda/m_b$.
The observables that we consider can be found in App.~\ref{app:SMpred}, including the branching ratio,
its longitudinal fraction $F_L$ and the optimised observables $P_i$.
As discussed in Ref.~\cite{Matias:2012xw,DescotesGenon:2012zf}, the optimised observables $P_i$ together with two
additional (form factor dependent) observables exhaust the information provided by the angular coefficients\footnote{
For a discussion of additional observables including scalars or lepton masses see Ref.~\cite{Matias:2012xw},
for S-wave observables see Refs.~\cite{Matias:2012qz,Hofer:2015kka}.
}. These optimised observables have been measured by LHCb: the latest results incorporating the full 3 fb$^{-1}$ of data
collected during LHC run I can be found in Ref.~\cite{Aaij:2015oid},
which includes the results for the CP-averaged coefficients $S_i$ introduced in Ref.~\cite{Altmannshofer:2008dz}, as well as the corresponding correlation matrices. 

We should stress at this point that our definition of some optimised observables $P_i$ and CP-averaged angular coefficients $S_i$ differs from that adopted by the LHCb collaboration, due to two different issues. First, our convention for the angles to define the $\bar{B}\to \bar{K}^*\ell\ell$ kinematics  (identical to Ref.~\cite{Altmannshofer:2008dz}) differs from the LHCb choice. Refs.~\cite{Gratrex:2015hna,Becirevic:2016zri} provided the angular coefficients $J_i$ in terms of the transversity amplitudes using the LHCb convention. Comparing with the expressions in Ref.~\cite{Altmannshofer:2008dz}, one can confirm that the two conventions can be related using
\begin{equation}
 \theta_K^{\rm LHCb}= \theta_K,\quad  \theta_\ell^{\rm LHCb}= \pi-\theta_\ell, \qquad \phi^{\rm LHCb}= - \phi\,.
\end{equation}
This induces different signs in both conventions when the angular coefficients $J_i$ (and their CP-averaged versions $S_i$) are expressed in terms of transversity amplitudes, leading to the identification
\begin{equation}
S_{4,6c,6s,7,9}^{\rm LHCb} = -S_{4,6c,6s,7,9}\,,
\end{equation}
the other coefficients $S_i$ being identical in both conventions.

Second, our definition of the optimised observables $P_i$ in terms of the angular coefficients $J_i$ is different from the definition used by the LHCb collaboration~\cite{Aaij:2013iag}. This induces further sign and normalisation differences when expressing $P_i$ in terms of transversity amplitudes, finally leading to~\footnote{The updated dictionary eq.~(\ref{eq:dictionary}) differs from Ref.~\cite{Descotes-Genon:2013wba} through the sign of the optimised observables $P'_6$ and $P'_8$. These observables are predicted tiny in the case of real NP contributions and are measured compatible with zero, so that this update of dictionary has no actual consequences on the results of the fit in Ref.~\cite{Descotes-Genon:2013wba}.}
\begin{eqnarray} \label{eq:dictionary}
&&P_1^{\rm LHCb}  = P_1\,, \ 
P_2^{\rm LHCb}   = - P_2\,, \ 
P_3^{\rm LHCb}   = - P_3\,, \\
&&{P'_4}^{\rm LHCb} = -1/2 P'_4\,, \ 
{P'_5}^{\rm LHCb}  = P'_5\,, \ 
{P'_6}^{\rm LHCb}  = P'_6\,, \ 
{P'_8}^{\rm LHCb}  = -1/2 P'_8\,. \nonumber
\end{eqnarray}

The presence of discrepancies with respect to the SM in the LHCb measurements at
1 fb$^{-1}$ and 3 fb$^{-1}$
can be interpreted as a sign of additional contributions to some
of the Wilson coefficients. It is thus interesting to study the sensitivity of the $P_i$ observables to  such shifts,
see Table~\ref{tab:PiNPsens}. One can see interesting patterns, and in particular the global preference for a negative
contribution to $\C9$, as already observed with previous data~\cite{Descotes-Genon:2013wba} and in other
frameworks~\cite{Altmannshofer:2013foa,Altmannshofer:2014rta,Altmannshofer:2015sma}. We will now discuss the
features of each of the $P_i$ observables in more detail, as well as the status of LHCb data for these quantities.
The results given here are based on the final results provided in Ref.~\cite{Aaij:2015oid}.
We will focus on the results obtained using the maximum likelihood approach, and we will not consider the results
obtained using the amplitude method discussed recently in Ref.~\cite{Egede:2015kha}.

\begin{table}
\begin{center}
{\small 
\begin{tabular}{@{}lcccccccc@{}}
\toprule[1.6pt]
& & $|\delta \C7| =  0.1$ &  $|\delta \C9| =  1 $ &  $|\delta \C{10}| =  1$ & $|\delta \C{7'}| =  0.1$ &  $|\delta \C{9'}| =  1$ &  $|\delta \C{10'}| =  1$\\
\midrule
\multirow{2}{*}{$\av{P_1}_{[0.1,.98]}$} &  $+|\delta C_i|$ & $--$ & $--$ & $--$ & $-0.53$ & $-0.05$ & $--$ \\
& $-|\delta C_i|$  & $--$ & $--$ & $--$ & $+0.52$ & $ +0.05$ & $--$ \\
\midrule
\multirow{2}{*}{$\av{P_1}_{[6,8]}$} &  $+|\delta C_i|$ & $--$ & $--$ & $--$ & $+0.11$ & $+0.16$ & $\bf -0.37$ \\
& $-|\delta C_i|$  & $--$ & $--$ & $--$ & $\bf-0.12$ & $\bf -0.17$ & $+0.37$ \\
\midrule
\multirow{2}{*}{$\av{P_1}_{[15,19]}$}  & $+|\delta C_i|$  & $--$ & $--$ & $--$ & $\bf+0.03$ & $\bf+0.15$ & $ -0.14$ \\
& $-|\delta C_i|$ & $--$ & $--$ & $--$ & $-0.03$ & $ -0.11$ & $\bf +0.19$ \\
\midrule
\multirow{2}{*}{$\av{P_2}_{[2.5,4]}$}   & $+|\delta C_i|$ & $ -0.31$ & $-0.21$ & $\bf +0.05$ & $--$ & $--$ & $--$ \\
& $-|\delta C_i|$  & $\bf+0.19$ & $\bf+0.15$ & $-0.04$ & $-0.03$ & $--$ & $--$ \\
\midrule
\multirow{2}{*}{$\av{P_2}_{[6,8]}$}   &
$+|\delta C_i|$ & $ -0.07$ & $-0.09$ & $-0.06$ & $--$ & $--$ & $--$ \\
& $-|\delta C_i|$  & $\bf+0.11$ & $\bf+0.17$ & $\bf+0.05$ & $--$ & $--$ & $--$ \\
\midrule
\multirow{2}{*}{$\av{P_2}_{[15,19]}$}  &
$+|\delta C_i|$  & $--$ & $--$ & $--$ & $--$ & $-0.05$ & $ +0.06$ \\
& $-|\delta C_i|$ & $--$ & $+0.04$ & $--$ & $--$ & $ +0.05$ & $-0.06$ \\
\midrule
\multirow{2}{*}{$\av{P_4^\prime}_{[6,8]}$}   &
$+|\delta C_i|$ & $\bf +0.04$ & $--$ & $--$ & $-0.11$ & $-0.10$ & $\bf +0.17$ \\
& $-|\delta C_i|$  & $-0.05$ & $--$ & $--$ & $\bf+0.09$ & $\bf +0.10$ & $-0.20$ \\
\midrule
\multirow{2}{*}{$\av{P_4^\prime}_{[15,19]}$}  &
$+|\delta C_i|$  & $--$ & $--$ & $--$ & $--$ & $\bf-0.06$ & $ +0.05$ \\
& $-|\delta C_i|$ & $--$ & $--$ & $--$ & $--$ & $ +0.04$ & $\bf -0.08$ \\
\midrule
\multirow{2}{*}{$\av{P_5^\prime}_{[4,6]}$}   &
$+|\delta C_i|$ & $ -0.11$ & $-0.15$ & $-0.10$ & $-0.11$ & $-0.06$ & $\bf+0.21$ \\
& $-|\delta C_i|$  & $\bf+0.16$ & $\bf+0.28$ & $\bf+0.09$ & $\bf+0.15$ & $\bf+0.10$ & $-0.21$ \\
\midrule
\multirow{2}{*}{$\av{P_5^\prime}_{[6,8]}$}  &
$+|\delta C_i|$ & $ -0.04$ & $-0.07$ & $-0.07$ & $-0.08$ & $-0.08$ & $\bf+0.19$ \\
& $-|\delta C_i|$  & $\bf+0.07$ & $\bf+0.19$ & $\bf+0.09$ & $\bf+0.10$ & $\bf+0.11$ & $-0.18$ \\
\midrule
\multirow{2}{*}{$\av{P_5^\prime}_{[15,19]}$ } &
$+|\delta C_i|$  & $--$ & $--$ & $--$ & $\bf-0.03$ & $\bf-0.11$ & $ +0.12$ \\
& $-|\delta C_i|$ & $--$ & $+0.06$ & $+0.03$ & $+0.03$ & $ +0.10$ & $\bf-0.14$ \\
\bottomrule[1.6pt]
\end{tabular}}
\end{center}
\caption{\it Impact on a given observable of the shift of a single Wilson coefficient by an amount $\delta \C{i}$ (the other Wilson coefficients keeping their SM value). The first row corresponds to a variation of $+|\delta \C{i}|$ and the second row to $-|\delta \C{i}|$. The changes significantly improving the agreement with the 2015 LHCb data are highlighted in boldface. Notice that the dependence of the observables on the Wilson coefficients may exhibit non-linearities.} \label{tab:PiNPsens}\end{table}

\subsubsection{\boldmath$P_1$ or $A_T^{(2)}$}
Let us first consider the observable~\cite{Kruger:2005ep}\footnote{
In this definition and in the following ones in this section, it should be understood that each term is combined with the corresponding CP-conjugated term and the two leptonic chiralities are included (for instance, 
$|A_{i}|^2=|A_i^L|^2+|A_i^R|^2+|\bar A_i^L|^2+|\bar A_i^R|^2$).
In addition, we will ignore various factors of $\beta_\mu \equiv \sqrt{1-4m_\mu^2/q^2}$,
which are important for the observables at very low $q^2$.
For precise definitions see \cite{Matias:2012xw,DescotesGenon:2012zf,Descotes-Genon:2013vna},
where also the bin-integrated observables are given. Evidently, we use the exact expressions in all the numerical
results throughout the paper.
}
\begin{equation}
P_1=A_T^{(2)}=\frac{|A_{\perp}|^2-|A_{\|}|^2}{|A_{\perp}|^2+|A_{\|}|^2}\,.
\end{equation}
$P_1$ is particularly well suited to detect the presence of right-handed currents. The left-handed structure of the SM implies that a $b$ quark in the helicity state $-1/2$  would produce an $s$ quark in the same helicity state (neglecting the $s$ quark mass), combined with the spectator quark to generate a $K^*$ meson in an helicity state $-1$ or $0$, but not $+1$. The suppression of $H_{+1}=(A_{\|}+A_{\perp})/\sqrt{2} \simeq 0$ implies $A_{\perp}\simeq -A_{\|}$ and consequently $P_1^{\rm SM} \simeq 0$. In an completely analogous manner, a $b$ quark in the helicity state $+1/2$ leads to $H_{-1}=(A_{\|}-A_{\perp})/\sqrt{2} \simeq 0$ implying again $P_1^{\rm SM} \simeq 0$. Deviations from this prediction would signal contributions from a new right-handed structure.  

\begin{figure}[!t]
\begin{center}
\includegraphics[width=7.5cm]{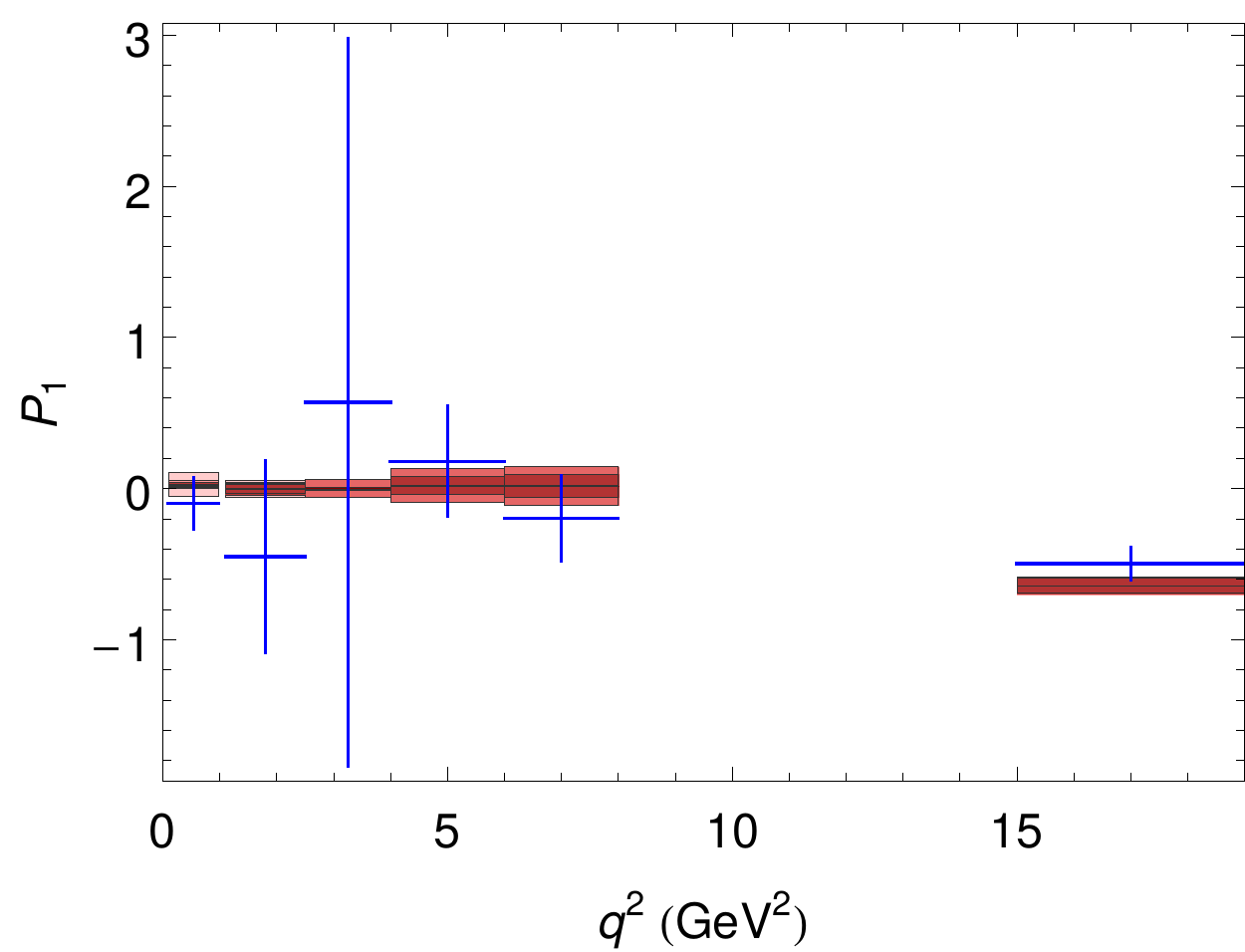}\hspace{4mm}
\includegraphics[width=7.5cm]{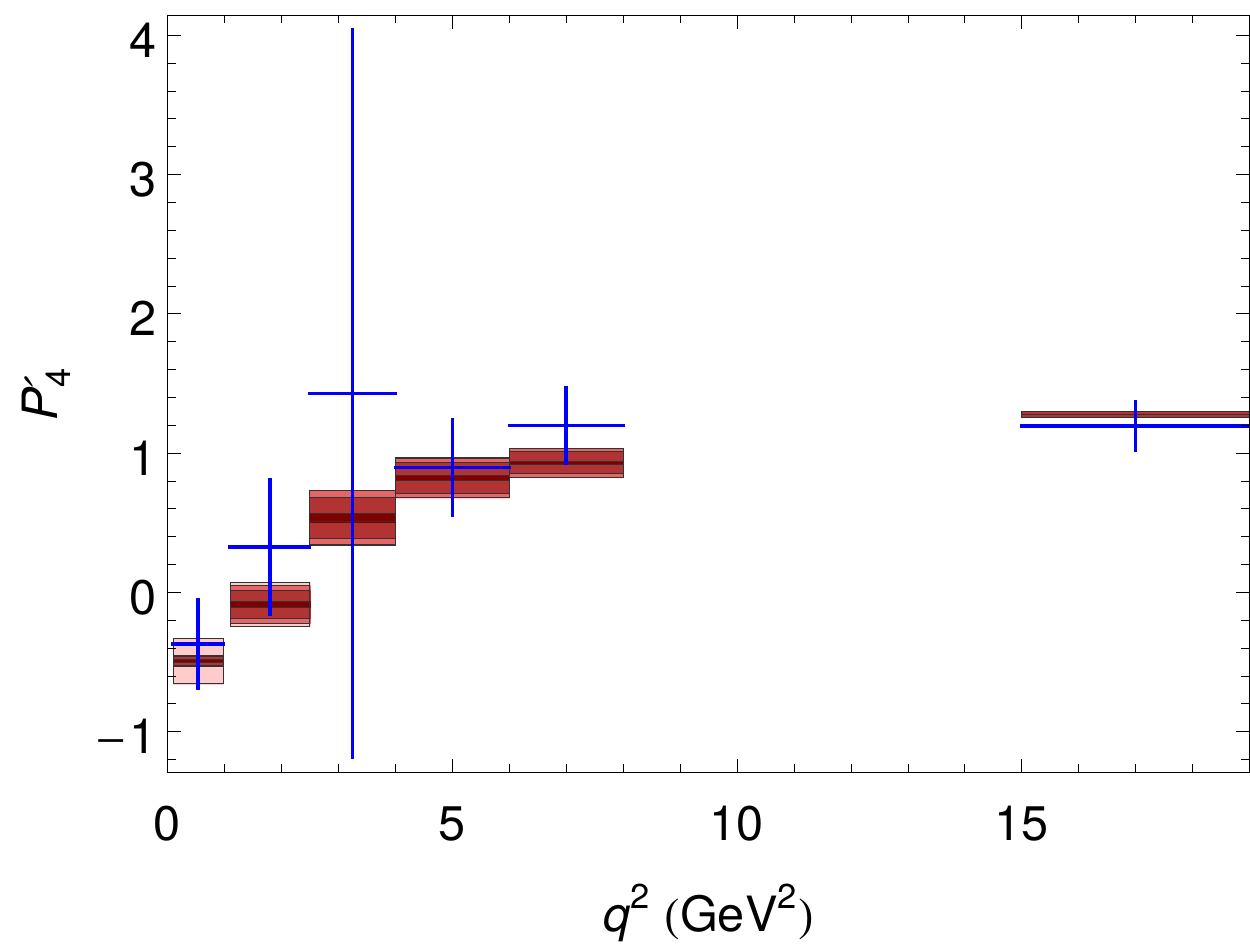}
\end{center}
\caption{\it Data (blue crosses) and SM prediction (red boxes) for $P_1,P_4'$.  The sources of uncertainties (added in quadrature) are shown as boxes in the following order from the center towards the outside: parametric, form factors, factorisable corrections, non-factorisable corrections, charm loop.}
\label{fig:P1P4}
\end{figure}

As seen in Fig.~\ref{fig:P1P4}, all bins are consistent with the SM, however  with very large error bars, so that no robust conclusion can be extracted from this observable with present data.. In Table \ref{tab:PiNPsens} we present the impact on $\av{P_1}_{ [0.1,0.98]}$, $\av{P_1}_{ [6,8]}$ and $\av{P_1}_{[15,19]}$ of shifting one 
of Wilson coefficients $\C7^{(\prime)},\C9^{(\prime)},\C{10}^{(\prime)}$ at a time. This is useful to see the relative size of the impact and if 
a corresponding NP contribution improves or not the agreement with data. Only significant improvements towards data are indicated. As expected, shifting Wilson coefficients for the SM operators does not induce any sizeable change. On the other hand, $P_1$ exhibits a relatively large sensitivity to right-handed operators. In particular should be noted a high sensitivity to contributions to $\C7^\prime$ in the first bin~\cite{Aaij:2015dea} as compared to other coefficients and also to other bins.

\subsubsection{\boldmath$P_4^\prime$}

The next observable that we would like to discuss is
\begin{equation} P_4^\prime=\sqrt{2}\frac{{\rm Re} (A_0^L A_{\|}^{L*}+ A_{0}^R A_{\|}^{R*})}{\sqrt{|A_0|^2 (|A_{\perp}|^2+|A_{\|}|^2)}}\,. \end{equation}
In conjunction with $P_5^\prime$, $P'_4$ establishes bounds on $P_1$ and enters consistency relations~\cite{Matias:2014jua}. In particular, the bound 
\begin{equation} P_5^{\prime 2}-1 \leq P_1 \leq 1-P_4^{\prime 2}\,. \label{bound}\end{equation}
is very efficient in two bins: [6,8] and low recoil. The preference of data for  $P_4^\prime \geq 1$ in the [6,8] bin requires $P_1 \leq 0$, in agreement with the 2015 LHCb data. Strictly speaking, this bound holds among the $q^2$ dependent observables, but it should also apply when the functions are only slowly varying (or almost constant)
for the binned observables. As an illustration of the usefulness as a test on data of the bounds provided by Eq.~(\ref{bound}) we have checked which value would imply for $P_1$ the measured values of $P_4^\prime$ and $P_5^\prime$ at low recoil. Taking central experimental values for this illustrative example we find that $P_1$ should be roughly in the range $-0.54 \leq P_1 \leq -0.44$ which is the right ball park as compared to the  central measured value $P_1\simeq -0.50$.
 A similar exercise using the SM central values for $P_{4,5}^{\prime\,\rm SM}$ gives $-0.67 \leq P_1^{\rm SM} \leq -0.64$ versus $P_1^{\rm SM} \simeq -0.64$. 

As can be seen in Fig.~\ref{fig:P1P4},  $P_4^\prime$  exhibits a perfect agreement with the SM in all bins, still with very large error bars. For completeness we provide also the bins [6,8] and [15,19] in Table~\ref{tab:PiNPsens} to make manifest  the lower sensitivity of this observable to  shifts of Wilson coefficients (particularly at low recoil) as compared to other observables, a fact that should not downgrade its status to a mere ``control" observable. 

\begin{figure}[!t]
\begin{center}
\includegraphics[width=7.5cm]{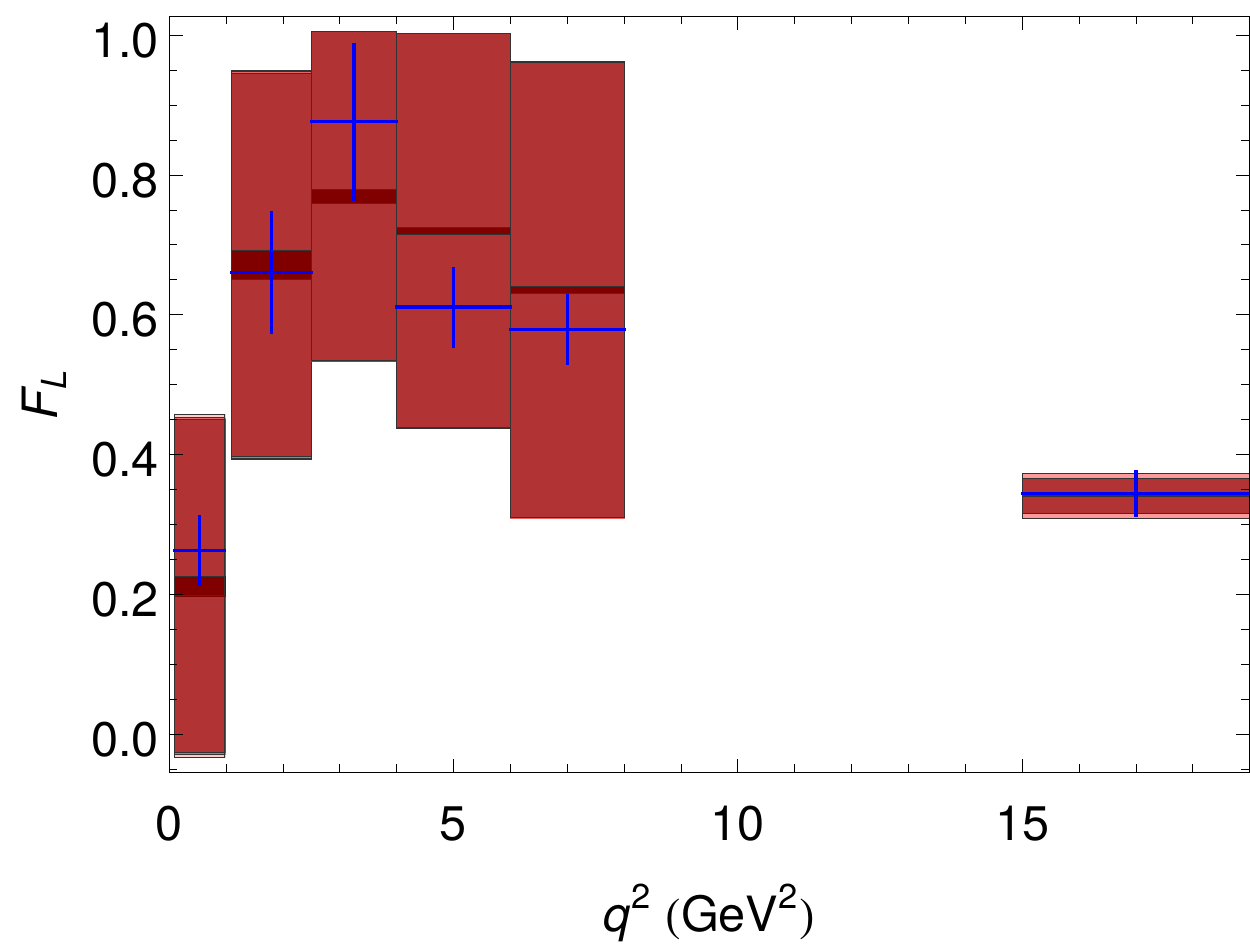}\\[4mm]
\includegraphics[width=7.5cm]{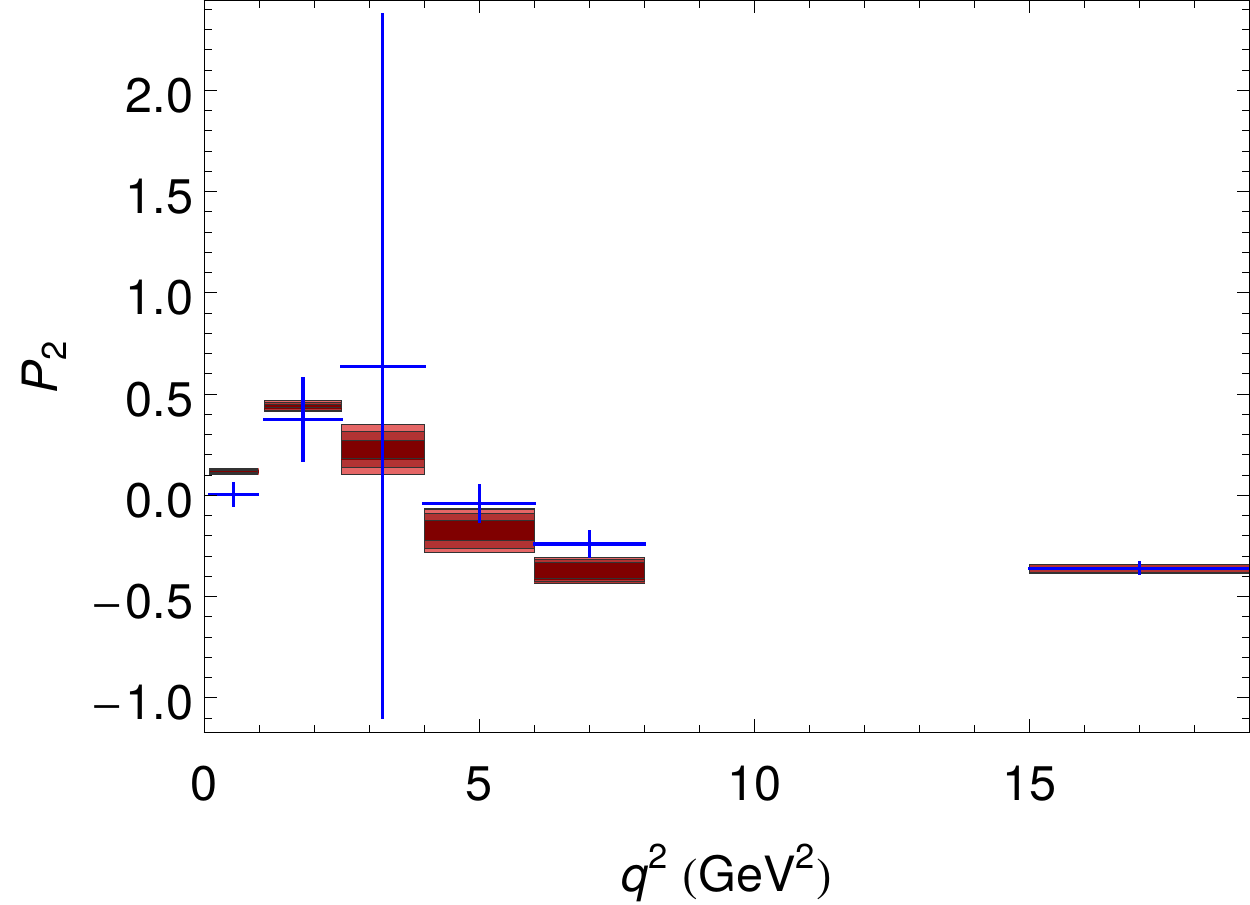}\hspace{4mm}
\includegraphics[width=7.5cm]{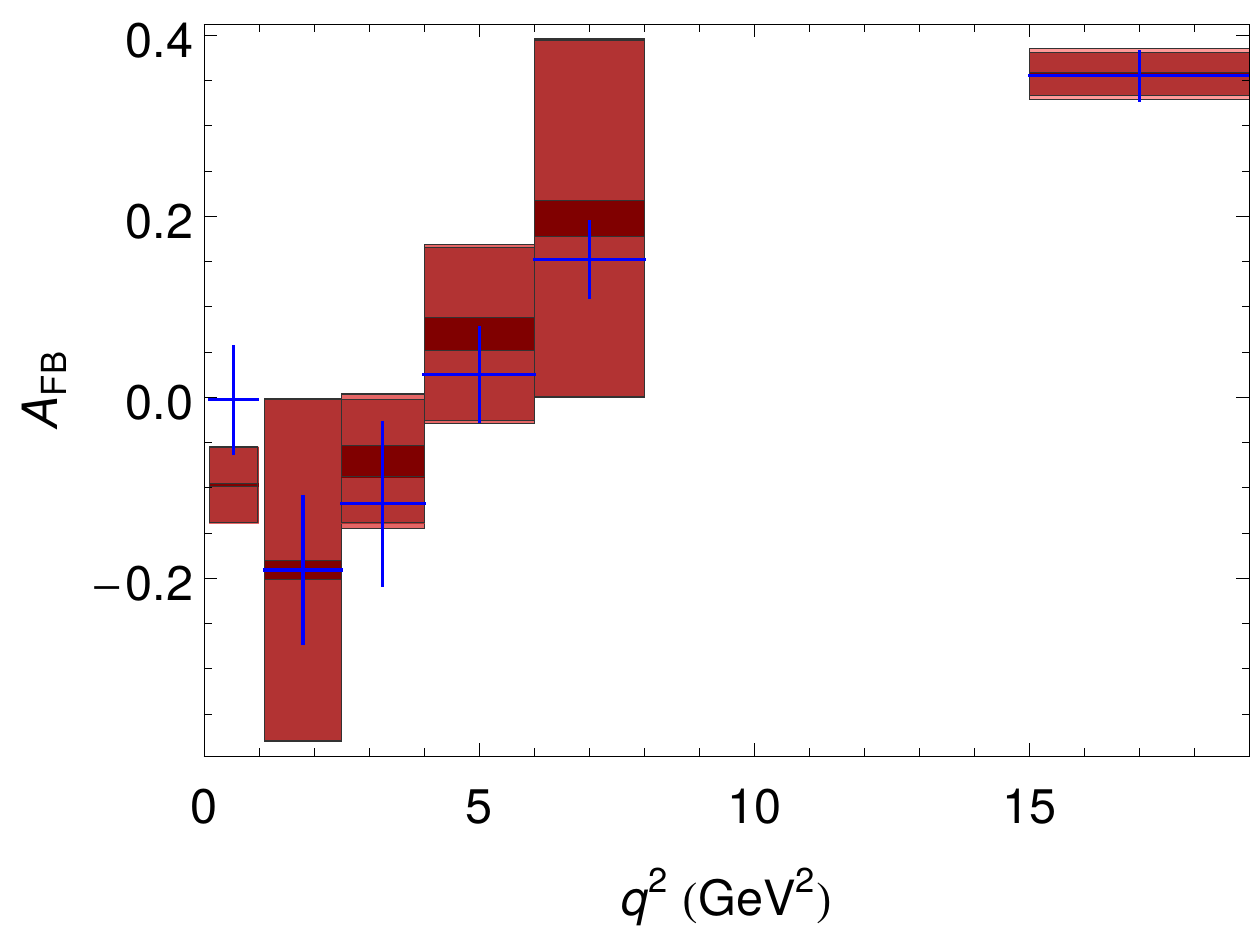}
\end{center}
\caption{\it Data (blue crosses) and SM prediction (red boxes) for $F_L$ (top), $P_2$ (bottom left), $A_{FB}$ (bottom right).  Same conventions as in Fig.~\ref{fig:P1P4}.}\label{fig:FLP2AFB}
\end{figure}

\subsubsection{\boldmath$P_2$}

The definition is~\cite{Matias:2012xw,Descotes-Genon:2013vna}
\begin{equation}
P_2=\frac{{\rm Re} (A_\|^L A_{\perp}^{L*}- A_{\perp}^R A_{\|}^{R*})}{ (|A_{\perp}|^2+|A_{\|}|^2)}\,.
\end{equation}
This observable is the optimised and clean version of the forward-backward asymmetry,
as illustrated in Fig.~\ref{fig:FLP2AFB} where the difference in the size of the uncertainties is obvious.
It was originally called $A_T^{\rm (re)}=2 P_2$ and proposed in Ref.~\cite{Becirevic:2011bp}.
$P_2$ measures a particular correlation between $A_{\rm FB}$ and $F_L$ that is independent of form factors at LO,
and combined with either $A_{\rm FB}$ or $F_L$ shows a higher NP sensitivity than the pair $\{A_{\rm FB},F_L\}$ itself. 

The observable $P_2$ contains some important pieces of information, such as the position of its zero $q_0^2$
(identical to the zero of $A_{\rm FB}$), the position of its maximum $q_1^2$, and its maximum value $P_2(q_1^2)$. 
To leading order and assuming no contribution from right-handed currents, i.e. $\C{i'}=0$, they are given by:
\begin{equation}
q_0^{2 \, \rm LO}= - 2 \frac {m_b M_B \C7^{\rm eff}}{\C9^{\rm eff}(q_0^2)}   \quad {\rm and} \quad  q_1^{2 \, \rm LO}= - 2 \frac {m_b M_B \C7^{\rm eff}}{{\rm Re}\ \C9^{\rm eff}(q_1^2) - \C{10}}\,,
\end{equation}
where for the position of the maximum we have neglected a term of ${\cal O}({\rm Im}(\C9^{\rm eff})^2)$ following
Ref.~\cite{Hofer:2015kka}. These expressions illustrate that a NP contribution to $\C9$ and $\C{7}$ would shift both
the zero and the maximum, but with a different magnitude.
Moreover, the maximum can be also shifted by a contribution to $\C{10}$.
The NLO prediction in the SM for these quantities are:
\begin{equation}
q_0^{2 \, \rm NLO}= 4.06 \pm 0.56\ {\rm GeV}^2  \quad {\rm and} \quad  q_1^{2 \, \rm NLO}= 2.03 \pm 0.26\ {\rm GeV}^2 \,,
\end{equation}
with $P_2(q_1^{2 \, \rm NLO})=0.501 \pm 0.004$.
In Refs.~\cite{Becirevic:2011bp,Hofer:2015kka}, a NP contribution to $\C{7,9,10}$ was shown to shift the position
of the maximum but not the value of its maximum that is fixed at $P_2(q_1^2)=1/2$. On the other hand, NP contributions
to the chirally flipped operators would reduce the maximum below $1/2$, even if not by a large amount.
Unfortunately, a fluctuation of the $\av{F_L}_{[2.5,4]}$  bin has induced a large experimental error in the
corresponding bin of $P_2$. This will be cured with more data and a finer binning.

Table~\ref{tab:PiNPsens} shows the sensitivity to shifts of Wilson coefficients for the two interesting [6,8] and low-recoil bins. It is clear the low sensitivity to NP of this observable at low-recoil, where the largest shift is only of +0.06. Indeed this is consistent with the perfect agreement of this observable with SM at low-recoil. Concerning the large-recoil bin, it is interesting to notice that the shifts of the Wilson coefficients pushing $\av{P_2}_{[6,8]}$ towards the data also shifts  $\av{P_2}_{ [2,5,4]}$ in the right direction (assuming that data is above the SM prediction), while all chirally flipped coefficients (positive or negative) always shift down this observable in this bin but by a relatively small amount. 

Finally, $P_2$ offers different consistency checks based on the relation~\cite{Matias:2014jua}
\begin{equation}\label{consistencyr}
P_2=\frac{1}{2} \left[P_4^\prime P_5^\prime + \sqrt{(-1+ P_1+P_4^{\prime 2})(-1-P_1+P_5^{\prime 2})} \right]\,.
\end{equation}
This relation is very useful to check the internal consistency of experimental or theoretical results for the observables\footnote{Indeed Eq.~(\ref{consistencyr}) enabled us to identify an inconsistency in the results for the $S_i$ ($P_i$) given in the first version of Ref.~\cite{Ciuchini:2015qxb} which is currently being corrected~\cite{priv}.}.

A first example is given by the bin [6,8] (or even [4,6]). Setting $P_2=-\epsilon$ (with $\epsilon>0$) one immediately obtains from the previous equation
\begin{equation} \label{order}
P_5^\prime \leq - 2 \frac{\epsilon}{P_4^\prime}\,.
\end{equation} 
Using central values for illustration and taking $\av{P_2}_{[6,8]} \sim -0.24 \equiv -\epsilon$  and $\av{P_4^\prime}_{[6,8]} \sim 1.20$, the previous equation would imply $\av{P_5^\prime}_{[6,8]} \leq -0.4$ in agreement with the data $\av{P_5^\prime}_{[6,8]} \sim -0.5$. Eq.~({\ref{order}) requires a specific ordering for $\av{P_2}_{[6,8]}$ and $\av{P_5^\prime}_{[6,8]}$, as well as in the bin [4,6] (in agreement with LHCb data). A second example comes from the zero of Eq.(\ref{order}). In Ref.~\cite{Matias:2014jua}, it was shown that the following relation should be fulfilled at the position $q_0^2$ of the zero of $P_2$ (or $A_{FB}$):
\begin{equation} \label{zero}
[P_4^{\prime 2}+P_5^{\prime 2}]_{q_0^2}=1-\eta(q_0^2) \qquad\qquad \eta(q_0^2)=[P_1^2+P_1 (P_4^{\prime 2}-P_5^{\prime 2})]_{q^2=q_0^2}\,.
\end{equation}
Since $\av{P_2}_{[4,6]}=-0.04 \pm 0.09$  is close to zero, one might expect the zero of $A_{FB}$ to lie near the center
of the bin. i.e. around 5 GeV$^2$. From $\av{P_5^{\prime}}_{[4,6]} \sim  -0.30$, $\av{P_4^{\prime}}_{[4,6]} \sim  +0.90$
and $\av{P_1}_{[4,6]} \sim  +0.18$ one finds that the l.h.s  of the first equation in Eq.~(\ref{zero}) is equal to 0.90,
while the r.h.s. is equal to 0.84, showing once again a good agreement with data.

\begin{figure}[!t]
\begin{center}
\includegraphics[width=10cm]{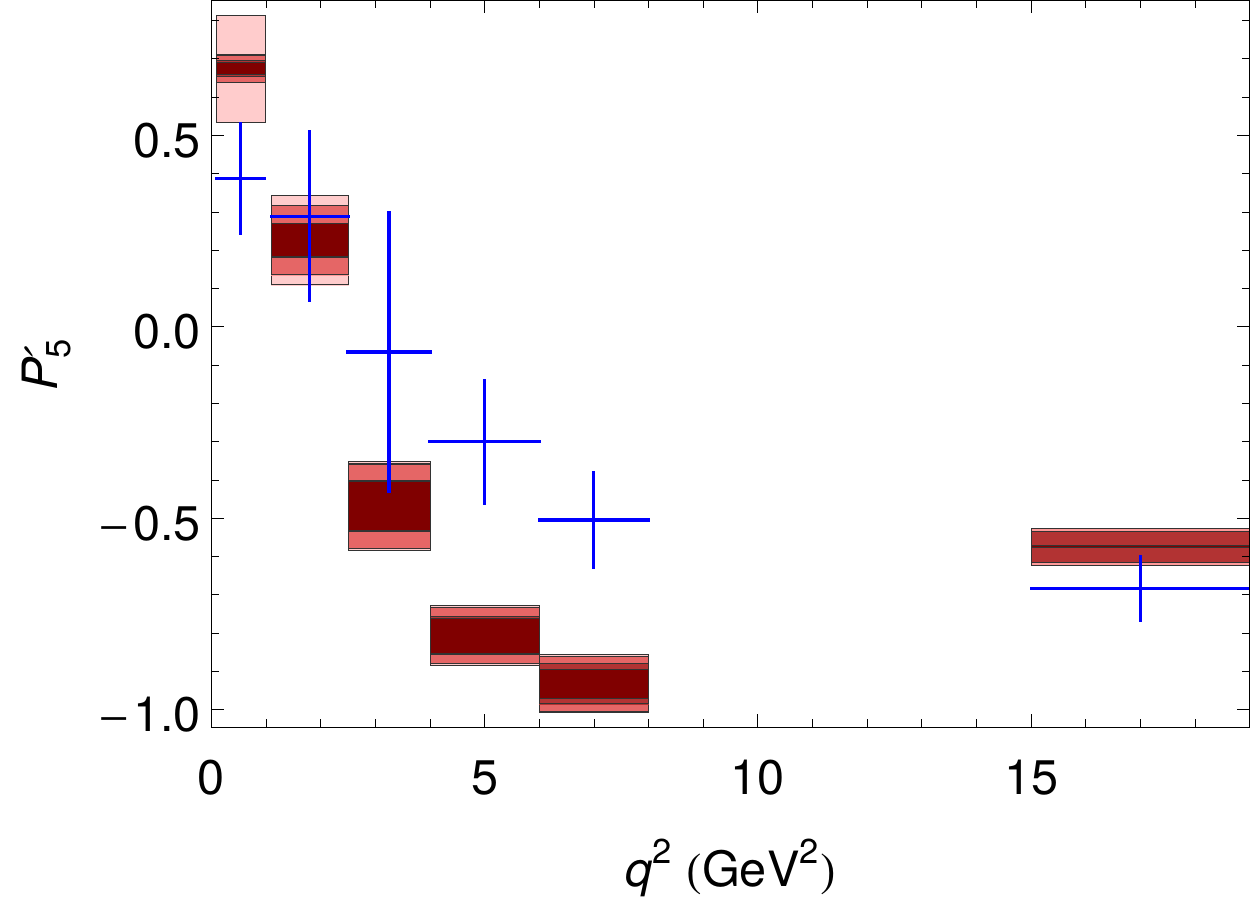}
\end{center}
\caption{\it Data (blue crosses) and SM prediction (red boxes) for $P_5^\prime$.  Same conventions as in Fig.~\ref{fig:P1P4}.}
\label{fig:P5p}
\end{figure}

\subsubsection{\boldmath$P_5^\prime$}

This observable is defined as~\cite{DescotesGenon:2012zf,Descotes-Genon:2013vna}
\begin{equation} P_5^\prime=\sqrt{2}\frac{{\rm Re} (A_0^L A_{\perp}^{L*}- A_{0}^R A_{\perp}^{R*})}{\sqrt{|A_0|^2 (|A_{\perp}|^2+|A_{\|}|^2)}} \,.
\end{equation}
One can provide an interpretation of $P_4^\prime$ and $P_5^\prime$ based on the expression in terms of the two-dimensional complex transversity vectors $n_{\perp,\|,0}$ (see Ref.~\cite{Matias:2012xw} for the definition of these vectors defined in a basis of transversity amplitudes with left- and right-handed structure for the dimuons). If we assume for simplicity that the transversity amplitudes are real, these two observables can be understood as the ``cosine'' of the relative angle between the parallel (respectively perpendicular) transversity vector and the longitudinal one 
\begin{equation} 
P_4^\prime \propto \cos \theta_{0,\|}\,, \quad P_5^\prime \propto \cos \theta_{0,\perp} \,.
\end{equation}
It is interesting to translate these expressions in the helicity basis by introducing two vectors based on the helicity $h=-1$ components of the $K^*$: $n^{(a)}_{-}=(H_{-1}^L, H_{-1}^R)$ and $n^{(b)}_{-}=(H_{-1}^L, -H_{-1}^R)$.  In the absence of right-handed currents ($H_{+1} \simeq 0$),
these observables correspond to the projection of the longitudinal helicity vector on one of the two negative helicity states, namely
\begin{equation} P_4^\prime \propto \cos \theta_{0,-1a}\,, \quad P_5^\prime \propto -\cos \theta_{0,-1b}\,. \end{equation}
Given the dominance of the left-handed part of the amplitude, this explains that $P_4^\prime$ and $P_5^\prime$ exhibit $q^2$-dependences that are almost the reflection of each other with respect to the axis $q^2=0$. Of course, this discussion is only qualitative and the details on the role of the right-handed amplitude $n^{a,b}_{-}$ are  fundamental to assess 
the sensitivity of these two observables to semileptonic coefficients.

$P_5^\prime$ exhibits the largest deviation with respect to the SM prediction in some bins, as seen in Fig.~\ref{fig:P5p},
corresponding to the so-called anomaly \cite{Descotes-Genon:2013wba}. An illustrative exercise consists in determining how
this observable can receive a large impact while keeping $P_4^\prime$ near the SM value (in agreement with data)~\footnote{
In Table~\ref{tab:PiNPsens}, one can notice the large impact of a variation of $\C9$ in $P_5^\prime$ compared to the
negligible impact on $P_4^\prime$ in the bin [6,8].
}.
A numerical analysis allows one to identify two mechanisms to enforce a suppression of $P_5^\prime$ with respect to $P_4^\prime$. The first mechanism relies on lifting the suppression of the right-handed amplitudes with respect to the left-handed amplitudes and to profit from
 the relative minus sign between the two terms in the numerator of $P_5^\prime$ versus the plus sign in $P_4^\prime$. The suppression of the right-handed amplitudes is due to the $\C9^{\rm SM} \sim - \C{10}^{\rm SM}$ cancellation, altered if the NP contribution to the Wilson coefficients does not follow the same direction~\footnote{
This can be easily seen using the large-recoil expression of the amplitudes. The numerator of $P_4^{\prime 2}$ contains a term proportional to $\C{10}^2$   that dominates and screens the partial cancellation between the $\C9$ and $\C7$  terms. There is no such $\C{10}^2$  term surviving in the numerator of $P_5^\prime$, so that the partial cancellation between $\C9$ and $\C7$ suppresses  $P_5^\prime$ with respect to  $P_4^\prime$.}.
The second mechanism is much more simple and relies on introducing a new physics contribution  that suppresses $A_{\perp}^L$ without affecting all other amplitudes.

In Table~\ref{tab:PiNPsens} we show the sensitivity to shifts of Wilson coefficients for the [6,8] and low-recoil bins. One can notice the large sensitivity of $\av{P_5^\prime}_{[6,8]}$ to a change of only  $\C9^{\rm NP}$ as compared to $\av{P_4^\prime}_{[6,8]}$ in agreement with the data. 
Similar results are found for $\av{P_5^\prime}_{[4,6]}$ albeit with a different importance. At low recoil,  $\av{P_5^\prime}_{[15,19]}$ exhibits a better sensitivity to NP than other observables in this region (though less than in the large-recoil region). This observable is already at 1~$\sigma$ consistent with SM at low-recoil, but the shifts in Wilson coefficients improving the agreement with data at large recoil go into the opposite direction at low recoil.  

\begin{figure}[!t]
\begin{center}
\includegraphics[width=7.4cm]{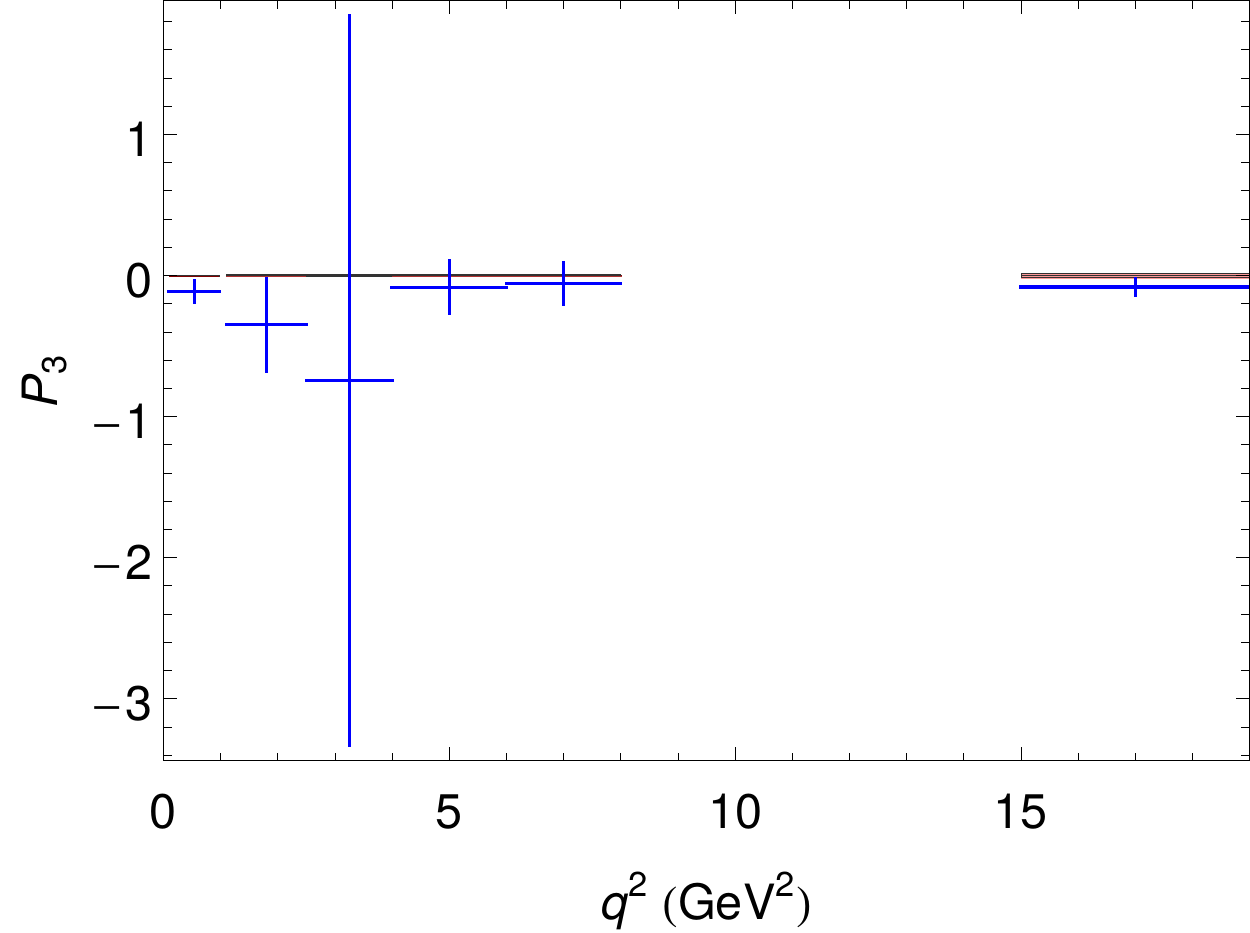}\\[4mm]
\includegraphics[width=7.4cm]{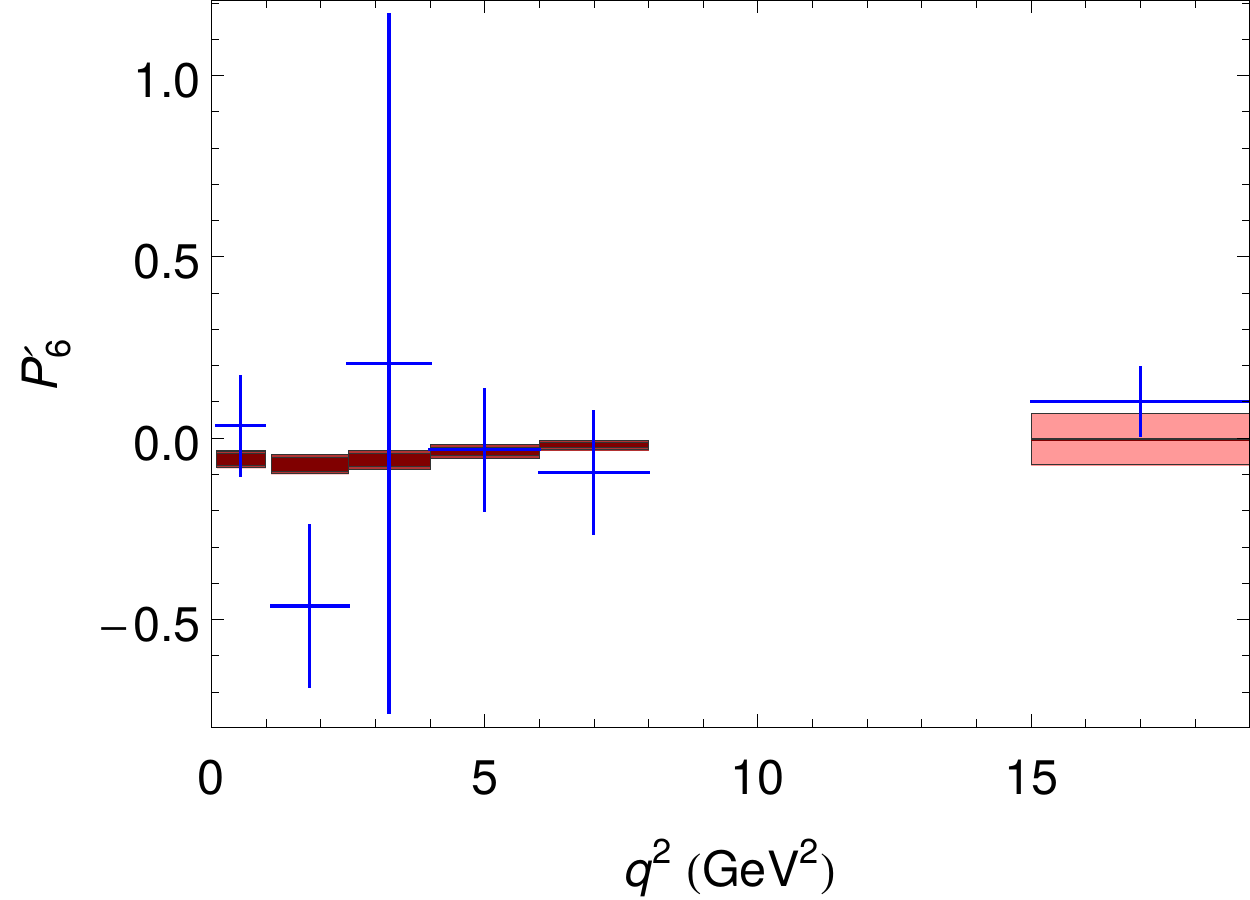}\hspace{5mm}
\includegraphics[width=7.4cm]{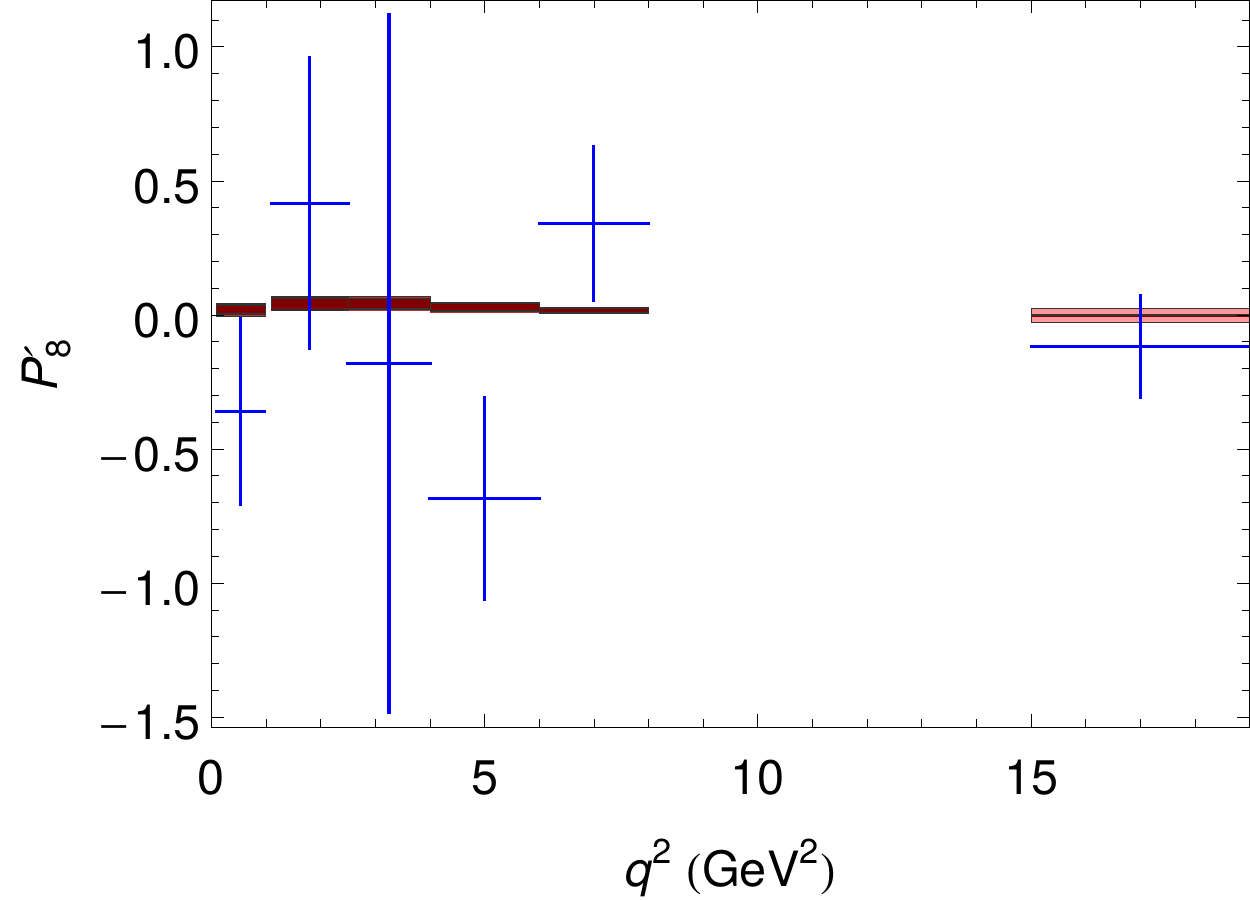}
\end{center}
\caption{\it Data (blue crosses) and SM prediction (red boxes) for $P_3$ (top), $P_6^\prime$ (bottom left), $P_8^\prime$ (bottom right).  Same conventions as in Fig.~\ref{fig:P1P4}.}
\label{fig:P3P6P8}
\end{figure}

\subsubsection{\boldmath$P_3$, $P_6^\prime$ and $P_8^\prime$}

These observables are defined as~\cite{DescotesGenon:2012zf,Descotes-Genon:2013vna}
\begin{equation}
\quad P_6^\prime=-\sqrt{2}\frac{{\rm Im} (A_0^L A_{\|}^{L*}- A_{0}^R A_{\|}^{R*})}{\sqrt{|A_0|^2 (|A_{\perp}|^2+|A_{\|}|^2)}}\quad P_8^\prime=-\sqrt{2}\frac{{\rm Im} (A_0^L A_{\perp}^{L*}+ A_{0}^R A_{\perp}^{R*})}{\sqrt{|A_0|^2 (|A_{\perp}|^2+|A_{\|}|^2)}}\,,\end{equation}
and
\begin{equation}
P_3=-\frac{{\rm Im} (A_\|^{L*} A_{\perp}^{L}+ A_{\perp}^R A_{\|}^{R*})}{ (|A_{\perp}|^2+|A_{\|}|^2)}\,.
\end{equation}

They are mainly sensitive to phases, either strong or weak, in the SM or beyond.
Present data is compatible with the SM with huge error bars, including also a local fluctuation of around 2~$\sigma$
in one bin of $P_6^\prime$ that will plausibly disappear with more data. This set of  observables also are required to fulfill bounds like 
\begin{equation} P_8^{\prime 2}-1 \leq P_1 \leq 1- P_6^{\prime 2}\,, \end{equation}
which is a natural extension of the bounds discussed in Ref.~\cite{Matias:2014jua}. Let us mention that
a more direct way to test the presence of new weak phases is the measurement of the $P_i^{\rm CP}$ observables~\cite{Descotes-Genon:2013vna}.

\subsection{Issues with specific bins}\label{sec:specificbins}

\subsubsection{The first large-recoil bin [0.1,0.98]}

The still limited statistics of LHCb data requires taking the limit of massless leptons for the determination of angular observables. The impact of this assumption is completely negligible in all bins except for the lowest bin [0.1,0.98]. Once included in the computation, the lepton mass yields a sizeable effect, pushing the SM prediction in the direction of data for $P_2$, $P_{4,5}^\prime$ and $F_L$. Indeed, the first terms of the distribution at LHCb are given by
\begin{eqnarray}
\label{lhcb}
\frac{1}{d(\Gamma + \bar\Gamma)/dq^2} \frac{d^3 (\Gamma+\bar\Gamma)}{d \Omega}=
\frac{9}{32 \pi} &\Big[& \frac{3}{4}(1-F_L^{\rm LHCb}) \sin^2\theta_K + F_L^{\rm LHCb} \cos^2\theta_K  \\
&+&\frac{1}{4}(1-F_L^{\rm LHCb}) \sin^2\theta_K \cos 2\theta_l - F_L^{\rm LHCb} \cos^2 \theta_K \cos 2\theta_l+... \Big]\nonumber
\end{eqnarray}
which is modified once lepton masses are considered~\cite{Hofer:2015kka}
\begin{eqnarray}  \frac{1}{d(\Gamma + \bar\Gamma)/dq^2} \frac{d^3 (\Gamma+\bar\Gamma)}{d \Omega}= \frac{9}{32 \pi} &\Big[& \frac{3}{4}\hat{F_T}
\sin^2\theta_K + {\hat F_L} \cos^2\theta_K  \\
&+&\frac{1}{4}F_T \sin^2\theta_K \cos 2\theta_l - F_L \cos^2 \theta_K \cos 2\theta_l+... \nonumber \Big]
\end{eqnarray}
where $\hat F_{T,L}$ and $F_{L,T}$ are detailed in Ref.~\cite{Matias:2012qz}~\footnote{Ref.~\cite{Matias:2012qz} uses $\tilde F_{L,T}$ related to $F_{L,T}=\beta^2 \tilde F_{L,T}$}.
All our observables are thus written and computed in terms of the longitudinal and transverse polarisation fractions $F_{L,T}$
\begin{equation}
F_{L}= - \frac{J_{2c}}{d(\Gamma+\bar\Gamma)/dq^2} \quad \quad F_T=4 \frac{J_{2s}}{d(\Gamma+\bar\Gamma)/dq^2}\,.
\end{equation}
However,  LHCb measures $F_L$ from the expression Eq.~(\ref{lhcb}) without lepton masses, where the dominant term
is the $\cos^2 \theta_K$ term. This means that the experimental analysis actually extracts ${\hat F_L}$, where 
\begin{equation}
{\hat F_L}= \frac{J_{1c}}{d(\Gamma + \bar\Gamma)/dq^2}\,.
\end{equation}
The difference between $F_L$ and $\hat{F}_L$ has a negligible impact in all bins except for the bin [0.1,0.98]. We have recomputed the first bin of $P_2$, $P_{4,5}^\prime$ using $\hat F_L$ instead of $F_L$  and imposing the LHCb condition $\hat F_T=1- \hat F_L$. For these observables, the central value for the SM prediction is shifted towards the data
\begin{eqnarray}
\av{F_L}_{[0.1,0.98]}=0.21 \rightarrow 0.26\,, &\qquad&
\av{P_2}_{[0.1,0.98]}=0.12 \rightarrow 0.09\,,\\
\av{P_4^\prime}_{[0.1,0.98]}=-0.49 \rightarrow -0.38\,, &\qquad&  \av{P_5^\prime}_{[0.1,0.98]}=0.68 \rightarrow 0.53\,.
\end{eqnarray}
Considering the expected accuracy during the run 2, it will be important once LHCb has enough statistics to distinguish between $F_L$ and ${\hat F}_L$. In the following, we will not attempt to correct for this effect, but instead check that the largest-recoil bin has only a minor impact in our result. 

\subsubsection{The bin [6,8]}

Some recent analyses of $B\to K^*\mu\mu$ data~\cite{Altmannshofer:2014rta,Altmannshofer:2015sma} have discarded the [6,8] bin because of the proximity of the $J/\psi$ resonance. It is obviously possible to perform analyses without this bin, as some judgement must be exerted to decide which observables are sufficiently well controlled to be included in the fit. However, we want to emphasise the role played by this bin in our analysis.
 
The smooth behaviour of $P_5^\prime$ up to bin [6,8] does not support claims of extremely large charm-loop contributions inducing a positive contribution  to $\C9$ which would affect mainly bins above 6 GeV$^2$~\cite{Lyon:2014hpa}. A direct comparison of the relative positions of $\av{P_5^\prime}_{[4,6]}$ and $\av{P_5^\prime}_{[6,8]}$ observables supports a global deviation with respect to SM predictions over a large $q^2$  range, rather than
an effect localised near the $J/\psi$ resonance that would push up $\av{P_5^\prime}_{[6,8]}$ with respect to $\av{P_5^\prime}_{[4,6]}$. Indeed, current data exhibits a pattern opposite to what was proposed in Ref.~\cite{Lyon:2014hpa} (see the plot for $P_5^\prime$  in Fig.~12 of Ref.~\cite{Lyon:2014hpa}). Of course, this cannot be considered as a proof that there are no effects coming from charm resonances, but it supports the concept of a limited impact which does not reach the size advocated in Ref.~\cite{Lyon:2014hpa}. 

On the other hand, this bin exhibits a significant discrepancy from SM expectations in $P_5^\prime$ and impacts our analysis. 
As discussed in sec.~\ref{sec:BtoKstarmumu-gen}}, we include in our predictions an estimate of the impact of charm resonances, but we also perform cross-checks concerning the role of this bin in sec.~\ref{sec:cross-checks-bin}.

\section{Other observables involved in the fit}
\label{sec:otherobs}

Here we discuss a large set of observables that we include in the fit organized in two sets, the first one involving muons and photons in the final state and the second one involving electrons.

\subsection{\boldmath$b \to s \mu\mu$ and $b \to s \gamma$ observables}
\label{sec:3.1}

This class of observables corresponds to exclusive and inclusive processes where either a real photon or a pair of muons is produced. 
It includes the decay $B\to K^*\mu\mu$ discussed at length in the previous sections, but also many other modes of interest.

\subsubsection{\boldmath$B_s\to\phi\mu\mu$}

The main difference between this mode and the decay $B\to K^*\mu\mu$ originates from the fact that $B_s\to\phi\mu\mu$ is not self-tagging, 
i.e. the final state does not contain information on whether the initial meson was a $B_s$ or a $\bar{B}_s$.
In the absence of flavour tagging, only a subset of angular observables can be easily measured at a hadron collider, some of them corresponding to CP-averaged angular coefficients ($J_{1s,1c,2s,2c,3,4,7}$) and some to CP-violating ones ($J_{5,6s,6c,8,9}$). Moreover, $B_s$-$\bar{B}_s$ mixing can interfere with 
direct decay providing additional contributions to the amplitude. This issue was addressed in detail in Ref.~\cite{Descotes-Genon:2015hea}, where it was shown that additional observables could be measured through a time-dependent analysis of the angular coefficients (in particular, promising optimised observables $Q_8$ and $Q_9$). Furthermore, the measurement 
of time-integrated angular coefficients in a hadronic environment yields $\mathcal{O}(\Delta \Gamma_s/\Gamma_s)$ corrections to the analogous $B^+\to K^{*+}\ell\ell$ expressions in terms of transversity amplitudes (related to interference between mixing and decay). 
 
One of the guidelines in our analysis is to try to test the sensitivity of the results on different choices of form factor parametrisations and thus on the specific details and assumptions of a particular form factor computation. Therefore we compare whenever possible the predictions obtained with
our default form factor parametrisation to those obtained with other choices, e.g. in the case of $B \to K\mu\mu$ and $B\to K^*\mu\mu$ results based on KMPW~\cite{Khodjamirian:2010vf} ($B$-meson LCSR) to results based on BSZ~\cite{Straub:2015ica} (light-meson LCSR). On the other hand,
for the case of $B_s \to \phi\mu\mu$, only two form factor determinations were available at low-$q^2$ (BZ~\cite{Ball:2004rg} and BSZ~\cite{Straub:2015ica}) following rather similar approaches with the latter being an update of the former one.

\begin{figure}[!t]
\begin{center}
\includegraphics[width=7cm]{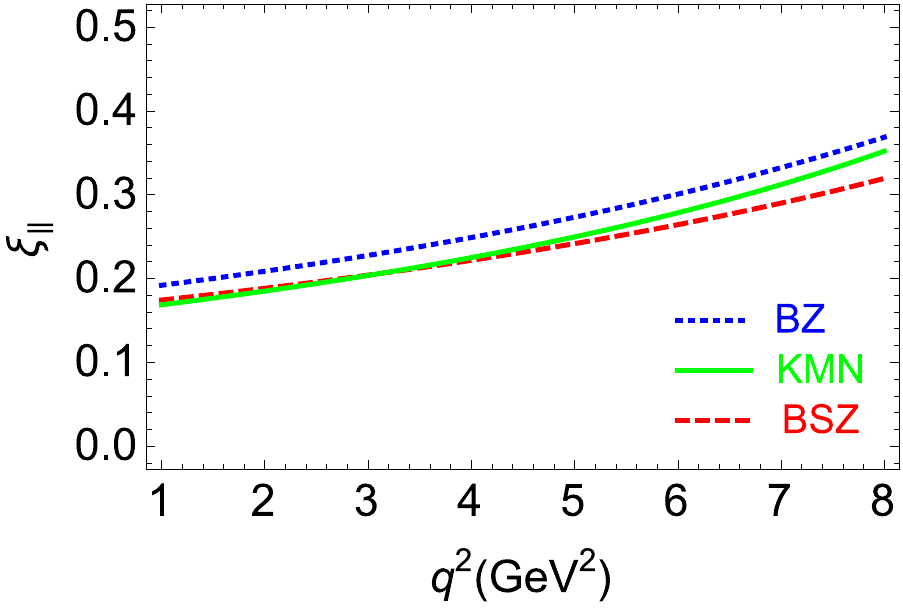}\hspace{2cm}\includegraphics[width=7cm]{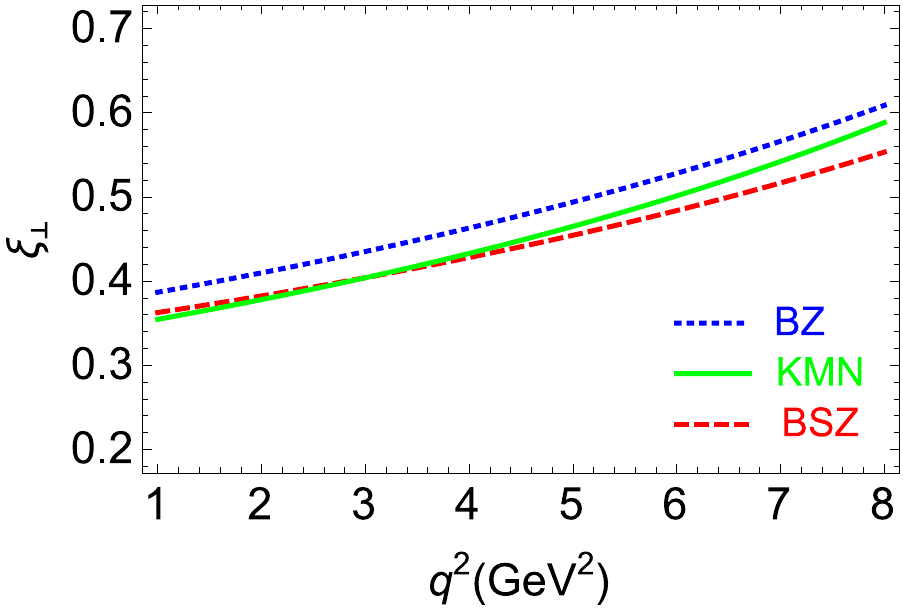}
\end{center}
\caption{\it Soft form factors for $B_s \to \phi\mu\mu$ using KMN (green, solid), BSZ (red, dashed) and BZ (blue, dotted) parametrisations: central value for $\xi_{||}$ (left) and for $\xi_\perp$ (right).}\label{fig:BsSoftformfactors}
\end{figure}

For this reason and given the importance of this mode, we implemented an alternative approach, based on the $B$-meson LCSR computation discussed in Ref.~\cite{Khodjamirian:2006st} (corresponding to the same type of method as in KMPW~\cite{Khodjamirian:2010vf}). Unfortunately, Ref.~\cite{Khodjamirian:2006st} does not provide the complete set of form factors necessary for a calculation of the $B_s\to\phi\mu\mu$ amplitudes in the full-form factor
approach, but the available subset is sufficient to construct the two soft form factors. These are extracted from the full form factors $V$, $A_1$ and $A_2$
in Ref.~\cite{Khodjamirian:2006st} using the value of decay constants, masses and hadronic inputs (we use the same threshold parameter as for $K^*$ and the Borel parameter is set to $M^2=1.0$ GeV$^2$). The results obtained for $\xi_{\perp}$ and $\xi_{\|}$ are plotted in Fig.~\ref{fig:BsSoftformfactors} where 
they are compared to the corresponding functions from BZ and BSZ. Only central values are shown, illustrating the excellent agreement between the parametrisation using Ref.~\cite{Khodjamirian:2006st} and the BSZ parametrisation up to 5 GeV$^2$, and a small deviation (below 8$\%$) in the 5 to 8 GeV$^2$ region. Considering the very good agreement with the independent computation in Ref.~\cite{Khodjamirian:2006st}, we feel confident to use 
the complete information available for the BSZ parametrisation to implement our soft form factor approach for $B_s\to\phi\mu\mu$.

We thus compute the relevant $B_s\to\phi\mu\mu$ observables with the same approach as for $B\to K^*\mu\mu$, applied to the form factors from Ref.~\cite{Straub:2015ica} as our default. The $\mathcal{O}(\Delta \Gamma_s/\Gamma_s)$ corrections to these observables are included using the expressions given in Ref.~\cite{Descotes-Genon:2015hea}, assuming all Wilson coefficients to be real.
We use a similar approach for power corrections and duality violation effects as in the case of $B\to K^*\mu\mu$, without assuming any correlation even though $SU(3)$ symmetry is expected to hold approximately. Similarly, for long-distance $c\bar{c}$ contributions, we use the same estimates for $\delta \C9^{c\bar{c}}$ as in $B\to K^*\mu\mu$~\footnote{These estimates are admittedly crude already for $B\to K^*\mu\mu$, and thus are expected to be valid at the same level of accuracy for $B_s\to\phi\mu\mu$.}, but we do not correlate the coefficients $s_i$, $a$, $b$, $c$ with those appearing for $B\to K^*\mu\mu$.

On the experimental side, LHCb has recently updated the measurement of this mode, providing the branching ratio,
its longitudinal fraction $F_L$ as well as several CP-averaged angular observables $S_{3,4,7}$ which can be recast
into optimised observables $P_1,P'_4,P'_6$ using the correlation matrix provided in Ref.~\cite{Aaij:2015esa}.
We have checked the linear propagation of errors used to obtain the optimised observables, by converting the
$B\to K^*\mu\mu$ $S_i$ observables into optimised $P_i$ observables using the same procedure and checking that
they agree very well with the values of $P_i$ quoted by the LHCb collaboration. Due to differences in the convention
in kinematics and the definition of observables, the same dictionary has to be used as in the $B\to K^*$ case
to relate our definitions of angular and optimised observables to those from LHCb articles.

\subsubsection{\boldmath$B\to K\mu\mu$}

In addition to the differential branching ratio, the angular distribution for $B\to K\mu\mu$ features two further observables, the forward-backward asymmetry $A_{FB}$ and the coefficient $F_H$~\cite{Bobeth:2007dw}. LHCb data does not suggest any deviation from SM expectations in these two quantities which are sensitive to the presence of scalar/pseudoscalar and tensor operators. Since we do not consider such NP operators, we will only examine the $B\to K\mu\mu$ branching ratio.

The theoretical description of the decay $B\to K\mu\mu$ with the scalar $K$ meson in the final state is considerably simpler 
than the one of the decay $B\to K^*\mu\mu$ with the vector $K^*$ meson, even though similar conceptual issues are involved.  
 In the large-recoil region, we apply QCD factorisation~\cite{Beneke:2000wa} to the form factors in Ref.~\cite{Khodjamirian:2010vf}, taking them as uncorrelated. The large-recoil symmetries allows us to
reduce the three form factors $f_+,f_0,f_T$ to a single soft-form factor. We use the most common scheme~\cite{Beneke:2000wa} where the soft form factor is identified with $f_+$, 
which dominates the computation of the branching ratio (contributions involving the scalar form factor $f_0$ are suppressed by the lepton mass, 
and the tensor form factor $f_T$ arises only in the presence of scalar or tensor operators). 
The dominance of the form factor $f_+$ renders correlations with the other two $f_0,f_T$ less important,
and therefore the gain of the implementation of correlations via the soft form factor approach is less significant
for $B\to K\mu\mu$ than for the vector mode (we checked that using the full form factors for $B\to K\mu\mu$ yields
indeed very similar results).
Long-distance charm-loop corrections are neglected here, as they are expected to have very little impact on branching
ratios~\cite{Khodjamirian:2010vf}. 

At low recoil, we use the lattice determination from Ref.~\cite{Bouchard:2013pna}.
In this region, the question on the size of duality-violation effects arises as in the case of $B\to K^*\ell\ell$. Again we consider a single large bin
covering this region and we implement an $\mathcal{O}(10\%)$ correction (with an arbitrary phase) to the term proportional to $\C9$ for this bin.

\subsubsection{\boldmath$B \to X_s \mu^+\mu^-$ and $B \to X_s \gamma$}

There are other important $b\to s$ penguin modes sensitive to magnetic and dimuonic operators. We consider the branching ratios ${\cal B}(B\to X_s\gamma) _{E_\gamma>1.6 {\rm GeV}}$ and ${\cal B}(B\to X_s \mu^+\mu^-)_{[1,6 ]}$.
In both cases, the SM prediction~\cite{Grinstein:1988me, Misiak:1992bc, Chetyrkin:1996vx} has gained some recent improvement, with a better control of higher QCD orders for $B \to X_s \gamma$~\cite{Misiak:2015xwa} and  the inclusion of logarithmically enhanced electromagnetic corrections for $B \to X_s \mu^+\mu^-$~\cite{Huber:2015sra}. This has induced a shift of the SM prediction, both for the central value and the uncertainty.
We update the SM predictions entering the relevant formulas for these observables in Ref.~\cite{DescotesGenon:2011yn}, but we do not modify the part 
depending on the NP coefficients $\C{i}$ (with NP being constrained to small effects, the inclusion of higher-order effects in this part can be neglected, 
considering the accuracy aimed at).

\subsubsection{\boldmath$B_s\to\mu\mu$}

The CMS and LHCb correlations have both measured the branching ratio for $B_s\to\mu\mu$, and provided an average
of the two measurements~\cite{CMS:2014xfa}. The SM theoretical prediction has been improved significantly over the
past year, including NNLO QCD corrections and NLO electroweak corrections, inducing a change in the central value
and the uncertainty~\cite{Bobeth:2013uxa}. We follow the same approach as for inclusive decays and modify the relevant formulas for these
observables in Ref.~\cite{DescotesGenon:2011yn} by updating the SM predictions, but without changing the part depending
on the NP coefficients $\C{i}$.

\subsubsection{\boldmath$B\to K^*\gamma$}

We follow the discussion in Ref.~\cite{DescotesGenon:2011yn} for $B\to K^*\gamma$ in order to constrain significantly $\C7$ (and $\C{7'}$). The observables included in our analysis are the isospin asymmetry $A_I(B\to K^* \gamma)$ and the  $B\to K^* \gamma$ time-dependent CP asymmetry $S_{K^*\gamma}$.

\subsubsection{\boldmath$\Lambda_b\to\Lambda\mu\mu$}

Another example of a $b\to s\mu\mu$ transition is the baryonic mode $\Lambda_b\to\Lambda\mu\mu$, for which the branching ratio and several angular observables have been measured by the LHCb collaboration~\cite{Aaij:2015xza}. Due to limitations of the current theoretical description of this decay (restricted to naive factorisation, with only a limited knowledge of form factors)~\cite{Wang:2008sm,Detmold:2012vy,Boer:2014kda}, we prefer not to include these results in our fits. We note, however, that the current measurement of the differential branching ratio tends to lie below its SM prediction using lattice QCD inputs~\cite{Detmold:2012vy,Aaij:2015xza}.

\subsection{\boldmath$b \to s e^+e^-$ observables}\label{sec:btoseeobs}

The analysis of $b\to s\ell\ell$ processes can be extended by considering not only muons but also electrons in the final state. As discussed in the introduction, the ratio $R_K=Br(B\to K\mu\mu)/Br(B\to Kee)$ in the [1,6]-bin shows a significant deviation from the SM expectation~\cite{Aaij:2014ora}, which can be computed to a very high accuracy since almost all hadronic effects cancel in the ratio~\cite{Hiller:2014yaa,Hiller:2014ula}. Using our setting, we find a SM value of $R_{K}=1.002\pm 0.006$ for the [1,6] bin.
The observed tension with data has triggered many interpretations in terms of various NP models
(for instance Refs.~\cite{Altmannshofer:2014cfa,Crivellin:2015era,Crivellin:2015mga,Celis:2015ara,Buras:2013dea,Buras:2014yna,Gauld:2013qja,Gauld:2013qba,Sierra:2015fma,Altmannshofer:2015mqa,Ghosh:2014awa,Glashow:2014iga,Becirevic:2015asa,Guadagnoli:2015nra,Gripaios:2015gra,Falkowski:2015zwa,Greljo:2015mma,Biswas:2014gga,Datta:2013kja}).

Instead of including directly the ratio $R_K$ together with $Br(B\to K\mu\mu)$ in the fit, we use the two branching ratios $Br(B\to K\mu\mu)$ and $Br(B\to Kee)$, keeping track of all theoretical correlations among them. Note that in this way we do not lose the information concerning the cancellation of hadronic uncertainties as it would occur in the observable $R_K$ because this cancellation is implicitly encoded in the correlations among the two branching ratios. On the experimental side, $R_K$ is significantly correlated with $Br(B\to K\mu\mu)$ (in a way not quantified yet), whereas $Br(B\to Kee)$ may only have part of the (sub-dominant) systematic uncertainties correlated with $Br(B\to K\mu\mu)$. It seems thus safer to include $Br(B\to Kee)$ in the global fit (rather $R_K$) to avoid a double counting of (correlated) deviations.

Another source of information on $b\to s ee$ is provided by $B\to K^*ee$ at very low invariant squared masses $q^2$ of the electron pair, close to the photon pole. An angular analysis~\cite{Aaij:2015dea} provides four observables $F_L, A_T^{(2)}, A_T^{re}, A_T^{im}$ (or equivalently $F_L, P_{1,2,3}$), which can be included in the fit to constrain the Wilson coefficients, in particular $\C7$ and $\C{7'}$ due to the proximity to the photon pole~\footnote{We will not provide predictions for the branching ratio $BR(B\to K^*ee)$ at very low recoil: the photon pole magnifies the uncertainty coming from the form factor $T_1(0)$, which is very large due to our choice of input for $V(0)$ with large uncertainties. Contrary to the case of angular observables, our estimate for this branching ratio at very large recoil is thus affected by large uncertainties (though in the right ball park of other estimates~\cite{Jager:2012uw}).}. Finally, we do not include information on $B\to X_s ee$, as this decay provides already little information in the muon case.

In the generic NP models, the effective Hamiltonian involves different effective $bs\ell\ell$ couplings for different lepton species ($\ell=\mu,e$), so that one should distinguish the Wilson coefficient $\C{i,\mu}$ (corresponding to $b\to s\mu\mu$ transitions) from the Wilson coefficient $\C{i\,e}$ (for $b\to see$). Hence it is not possible to include
the above data in a model-independent fit to Wilson coefficients $\C{i}$, unless an additional hypothesis concerning the value of the $\C{i}^e$ or their relationship to the $\C{i}$ is made. Therefore we will not include this set of data in our reference fit described above and in App.~\ref{app:SMpred},  but we will consider it in combined $\mu+e$ fits, assuming that NP is either absent from $\C{i\,e}$, or that it enters flavour-universally in $\C{i\,e}$ and $\C{i\,\mu}$.

\section{Global Fits to Wilson coefficients}
\label{sec:NPfits}

\subsection{General framework}\label{sec:fitgeneral}

We start with a global analysis of the data, in scenarios with potential (real)
NP contributions to the Wilson coefficients $\C{7,9,10,7',9',10'}$~\footnote{We will not consider imaginary contributions to Wilson coefficients and we do not include CP-violating observables in our fits.}. We split the SM and NP contributions at $\mu_b=$ 4.8 GeV with $\C{i}=\C{i}^{\rm SM}+\C{i}^{\rm NP}$ (with $\C{7,9,10}^{\rm SM}=-0.29, 4.07, -4.31$).

Our reference fit is obtained using
\begin{itemize}
\item the observables for $b\to s\mu\mu$ listed in App.~\ref{app:SMpred},
\item the observables for $b\to s \gamma$ discussed in Sec.~\ref{sec:3.1},
\item the form factors in Ref.~\cite{Khodjamirian:2010vf}, apart from $B_s\to\phi$ form factors~\cite{Straub:2015ica},
\item the ``improved QCD Factorisation approach'' described in Sec.~\ref{sec:BtoKstarmumu-gen}.
\end{itemize}

For our experimental inputs, we include only LHCb data for the exclusive modes considered here~\cite{Aaij:2015oid,Aaij:2014pli,Aaij:2013qta
,Aaij:2015esa,Aaij:2014ora,Aaij:2015dea,Aaij:2013hha}, as they dominate the current analysis of the anomalies and allow for a consistent inclusion of correlations. Inclusive modes and $b\to s\gamma$ inputs are taken from the HFAG review~\cite{Amhis:2014hma} and $BR(B_s\to\mu\mu)$ from the current CMS and LHCb combination~\cite{CMS:2014xfa}.
In case of asymmetric error bars, we symmetrise by taking the largest of the two uncertainties quoted, without modifying the central value.

We have to include the experimental and theoretical correlations between the different observables (and bins) for $B\to K^*\mu\mu$ and $B\to K\mu\mu$. The experimental correlations are available for $B\to K^*\mu\mu$~\cite{Aaij:2015oid}, $B_s\to\phi\mu\mu$~\cite{Aaij:2015esa} and $B\to K\mu\mu$~\cite{Aaij:2014tfa}. For $B\to K^*\mu\mu$, the correlations are given for both $S_i$ and $P_i$ observables,
whereas they are given only for $S_i$ observables for $B_s\to\phi\mu\mu$. We have performed a linear propagation of errors in the latter case in order to obtain the correlations among $P_i$ observables (we checked the validity of this procedure by reproducing the correlations among $P_i$ observables in $B\to K^*\mu\mu$ quoted in Ref.~\cite{Aaij:2015oid} starting from the information on the $S_i$ observables given in the same reference). 

For theoretical correlations, we have produced a correlation matrix by performing a propagation of error. This is achieved
by varying all input parameters following a Gaussian distribution including known correlations, and determining the
resulting distribution of the observables of interest. This is particularly necessary for the form factors: we include
correlations between parameters from the lattice QCD computation at low recoil in Ref.~\cite{Horgan:2013hoa,Horgan:2015vla}.
We treat all parameters as uncorrelated at large recoil in the case of Ref.~\cite{Khodjamirian:2010vf}, whereas we include
the available correlations when we use Ref.~\cite{Straub:2015ica}. We stress that even the uncorrelated scan of parameters
(like power corrections) induces correlations among the observables (for instance branching ratios at large recoil)
because the latter have a correlated functional dependence on these parameters. The large error bars in
Ref.~\cite{Khodjamirian:2010vf} for $B\to K^*\mu\mu$ may lead to excursions in parameter space that distort the
distribution of the $P_i$ observables and yield significant non-Gaussianities. These non-Gaussianities are avoided by
scanning over the input parameters after  scaling down all uncertainties by a global factor $\rho$, producing the
correlation matrix for the $P_i$ observables, and multiplying all its entries by $\rho^2$ .
The resulting covariance matrix is an accurate representation of the
uncertainties and correlations for the $P_i$ observables in the vicinity of the central values of the input parameters, 
as long as it is possible to propagate errors in a linearised way.  This matrix encodes all the relevant information
concerning uncertainties and correlations among observables, with all uncertainties effectively added in quadrature
(we explicitly checked that the results are independent on the exact numerical choice of the rescaling factor
$\rho$, and in practice $\rho=3$ is sufficient).
The other sets of form factors yield Gaussian distributions for the $B_s\to \phi\mu\mu$ and $B\to K\mu\mu$ observables,
because of the smaller uncertainty ranges.

Finally, we construct a single covariance matrix as the sum of the experimental ($C_{ij}^\text{exp}$) and the
theoretical one ($C_{ij}^\text{th}$), and we use it
to build the usual $\chi^2$ function corresponding to observables with correlated Gaussian distributions~\footnote
{
The theoretical correlation matrices are obtained for the observables in the context of the SM computation.
In the following, we will assume that the theory covariance matrix has only a mild dependence on the values of the
Wilson coefficients, and we will keep its SM value in the construction of our $\chi^2$ test
statistics~\cite{Altmannshofer:2014rta}. We have checked that for the scenarios considered in this paper this assumption
holds, by calculating the covariance matrix at the best-fit point and comparing the outcome of the fit with the one using
the SM covariance matrix.
}:
\eq{
\chi^2 (\C{k}) =
\sum_{i,j =1}^{N_\text{obs}} \big[O_i^{\rm exp} - O_i^{\rm th}(\C{k})\big]\,
(C_\text{exp} + C_\text{th})^{-1}_{ij}\,
\big[O_j^{\rm exp} - O_j^{\rm th}(\C{k})\big]\ .
}

Once the $\chi^2$ function is computed, it remains to exploit the information that it carries.
Following standard frequentist analysis, a first piece of information is provided by the global minimum $\chi^2_\text{min}$,
 which provides an indication of the goodness-of-fit. It can be expressed as a $p$-value 
assessing the agreement between the measurements and the scenario tested, given as the
probability for a $\chi^2$-distributed random variable with the corresponding number of degrees of freedom
(number of data points minus number of free parameters) to reach a higher value than the one obtained from the data.

If the fit is good enough, one can move on and perform the metrology of the $n$ parameters (NP in Wilson coefficients) by considering the test statistic $\Delta \chi^2 (\C{i}) \equiv \chi^2(\C{i}) - \chi^2_\text{min}$. Assuming that this quantity is distributed as a $\chi^2$ random variable with $n$ degrees of freedom,
the  $k$-sigma confidence region is obtained as
$\Delta \chi^2 (\C{i}) \leq \xi(k,n)$, where $\xi(k,n)$ is  the value at which
 the $\chi^2(n)$-cumulative distribution function 
reaches the probability $P_{k\,\sigma}$ associated to $k$ sigmas. In practice, $\xi(k,1) = \{ 1, 4, 9\}$, $\xi(k,2) = \{2.3, 6.18, 11.83\}$ and
$\xi(k,6) = \{5.89, 11.31, 18.21\}$ for $k=\{1,2,3\}$, corresponding to
$P_{k\sigma}=\{68.3, 95.4, 99.7\}\%$ defined as the probability for a Gaussian random variable to be measured within $n$ standard deviations from the mean.
In addition, the pull of the SM is the $p$-value corresponding to $\Delta \chi^2(\C{k}=0)$, i.e., the probability described above and converted in units of sigma, in the case of a $\chi^2(n)$-distributed random variable.

When we compare scenarios with different number of parameters, some care is thus needed both for the goodness-of-fit ($p$-value for $\chi^2_\text{min}$) and the metrology (pull of the SM). For instance, we note that a fit to two parameters $({\cal C}_i^{\rm NP},{\cal C}_j^{\rm NP})$
may contain the hypothesis ${\cal C}_i^{\rm NP}={\cal C}_j^{\rm NP}=0$ within the $2\,\sigma$ region, while the corresponding fit to the single parameter
${\cal C}^{\rm NP}_i$ (with ${\cal C}_j^{\rm NP}=0$ fixed) might not. In general, $p$-values and pulls tend to decrease when adding more parameters, unless the added
parameters are essential in improving the agreement with data. Having more free parameters in a fit typically reduces the significance of the SM pull and decreases the $p$-value for $\chi^2_\text{min}$  if these parameters are not relevant and do not affect the $\chi^2$ function.

\subsection{NP Fits for $b \to s \mu\mu$ and $b \to s \gamma$}

\subsubsection{One-dimensional fits to Wilson coefficients} \label{sec:fits-1D}

\begin{table}[!t]
\begin{center}
\begin{tabular}{@{}crcccr@{}}
\toprule[1.6pt] 
Coefficient & Best fit & 1$\sigma$ & 3$\sigma$ & Pull$_{\rm SM}$ & p-value (\%)\\ 
 \midrule 
 $\C7^{\rm NP}$ & $ -0.02 $ & $ [-0.04,-0.00] $ & $ [-0.07,0.03] $ &  1.2 & 17.0 \hspace{5mm}  \\[3mm] 
 $\C9^{\rm NP}$ & $ -1.09 $ & $ [-1.29,-0.87] $ & $ [-1.67,-0.39] $ &  {\bf 4.5} & 63.0 \hspace{5mm}  \\[3mm] 
 $\C{10}^{\rm NP}$ & $ 0.56 $ & $ [0.32,0.81] $ & $ [-0.12,1.36] $ &  2.5 & 25.0 \hspace{5mm}  \\[3mm] 
 $\C{7'}^{\rm NP}$ & $ 0.02 $ & $ [-0.01,0.04] $ & $ [-0.06,0.09] $ &  0.6 & 15.0 \hspace{5mm}  \\[3mm] 
 $\C{9'}^{\rm NP}$ & $ 0.46 $ & $ [0.18,0.74] $ & $ [-0.36,1.31] $ &  1.7 & 19.0 \hspace{5mm}  \\[3mm] 
 $\C{10'}^{\rm NP}$ & $ -0.25 $ & $ [-0.44,-0.06] $ & $ [-0.82,0.31] $ &  1.3 & 17.0 \hspace{5mm}  \\[3mm] 
 $\C9^{\rm NP}=\C{10}^{\rm NP}$ & $ -0.22 $ & $ [-0.40,-0.02] $ & $ [-0.74,0.50] $ &  1.1 & 16.0 \hspace{5mm}  \\[3mm] 
 $\C9^{\rm NP}=-\C{10}^{\rm NP}$ & $ -0.68 $ & $ [-0.85,-0.50] $ & $ [-1.22,-0.18] $ &  {\bf 4.2} & 56.0 \hspace{5mm}  \\[3mm] 
 $\C{9'}^{\rm NP}=\C{10'}^{\rm NP}$ & $ -0.07 $ & $ [-0.33,0.19] $ & $ [-0.86,0.68] $ &  0.3 & 14.0 \hspace{5mm}  \\[3mm] 
 $\C{9'}^{\rm NP}=-\C{10'}^{\rm NP}$ & $ 0.19 $ & $ [0.07,0.31] $ & $ [-0.17,0.55] $ &  1.6 & 18.0 \hspace{5mm}  \\[3mm] 
 $\C9^{\rm NP}=-\C{9'}^{\rm NP}$ & $ -1.06 $ & $ [-1.25,-0.86] $ & $ [-1.60,-0.40] $ &  {\bf 4.8} & 72.0 \hspace{5mm}  \\[3mm] 
 \begin{minipage}{3.5cm} \centering $\C9^{\rm NP}=-\C{10}^{\rm NP}$ \\ $=-\C{9'}^{\rm NP}=-\C{10'}^{\rm NP}$  \end{minipage} & $ -0.69 $ & $ [-0.89,-0.51] $ & $ [-1.37,-0.16] $ &  {\bf 4.1} & 53.0 \hspace{5mm}  \\[3mm] 
 \begin{minipage}{3.5cm} \centering $\C9^{\rm NP}=-\C{10}^{\rm NP}$ \\ $=\C{9'}^{\rm NP}=-\C{10'}^{\rm NP}$  \end{minipage} & $ -0.19 $ & $ [-0.30,-0.07] $ & $ [-0.55,0.15] $ &  1.7 & 19.0 \hspace{5mm}  \\[3mm] 
\bottomrule[1.6pt] 
\end{tabular}
\end{center}
\caption{\it Best-fit points, confidence intervals, pulls for the SM hypothesis and $p$-values for different one-dimensional NP scenarios.}\label{tab:1Dfits}
\end{table}

First of all, the SM itself does not yield a particularly good fit when considering all the $b \to s \mu\mu$
and $b \to s \gamma$ data, with $\chi^2_{\rm min}=110$ for $N_{\rm dof}=96$, corresponding to a p-value of 16\%.
We then include NP and start by considering 1D scenarios where only one of the Wilson coefficients is let free to
receive NP contributions. The corresponding p-values and pulls for the SM hypothesis  gathered in Table~\ref{tab:1Dfits}
show clearly that a scenario with NP in $\C9$ is the most favoured by far. A scenario with NP in $\C{10}$ and $\C{9'}$
is also preferred compared to the pure SM case, but to a lesser extent.

It is also interesting to test some scenarios where NP enters in a correlated way in two Wilson coefficients.
This occurs in particular in models preserving $SU(2)_L$ invariance in the lepton sector~\cite{Alonso:2014csa},
or models assuming a vector or axial preference for quark
couplings~\cite{Buras:2013dea,Buras:2014yna,Gauld:2013qja,Gauld:2013qba}.
From Table~\ref{tab:1Dfits}, the most favoured scenario corresponds to $\C9^{\rm NP}=-\C{9'}^{\rm NP}$, which could
for instance be generated by a $Z'$ boson with axial quark-flavour changing and vector muon couplings. 
This scenario yields a large pull due to the fact that it leads to an excellent agreement with the angular
observables at low recoil; however, it has no impact on $B\to K\ell\ell$ branching ratios, so that $R_K$ remains
unexplained.
The scenario $\C9^{\rm NP}=-\C{10}^{\rm NP}$ preserving the $SU(2)_L$ symmetry can also be considered as interesting.
One should however be careful not to overinterpret these results: any scenario allowing for NP in $\C9$ yields a large pull,
and the modification of the other Wilson coefficients might slightly improve or worsen the agreement between predictions
and measurements, but only  with limited impact.

We confirm our previous result of 2013~\cite{Descotes-Genon:2013wba} with the 3 fb$^{-1}$ dataset, namely that
$\C9$ plays a central role in the interpretation of the anomalies, and it is the main Wilson coefficient unavoidably
present in any scenario with a pull above 4 sigmas. We find that this Wilson coefficient receives typically a negative
contribution of order 25\% with respect to the SM. More details on the impact of various experimental inputs and
theoretical hypotheses can be found in App.~\ref{app:C9}.

\subsubsection{Two-dimensional fits to Wilson coefficients} \label{sec:fits2D}

\begin{table}[!t]
\begin{center}
\renewcommand{\arraystretch}{1.4}
 \setlength{\tabcolsep}{13pt}
\begin{tabular}{@{}c|cccccc@{}}
\toprule[1.6pt] 
 & $\C7^{\rm NP}$  & $\C9^{\rm NP}$ & $\C{10}^{\rm NP}$ & $\C{7'}^{\rm NP}$ & $\C{9'}^{\rm NP}$ & $\C{10'}^{\rm NP}$ \\ 
 \midrule 
 & 1.19  & 4.47 & 2.45 & 0.64 & 1.66 & 1.33 \\ 
 \midrule 
 $\C7^{\rm NP}$ &  *  &  0.07  &  0.84 & 1.18  &  1.24 & 1.21  \\ 
 $\C9^{\rm NP}$ &  4.31  &  *  &  4.04 & 4.52  &  4.61 & 4.67  \\ 
 $\C{10}^{\rm NP}$ &  2.30  &  1.54  &  * & 2.38  &  2.09 & 2.06  \\ 
 $\C{7'}^{\rm NP}$ &  0.62  &  0.92  &  0.23 & *  &  0.32 & 0.32  \\ 
 $\C{9'}^{\rm NP}$ &  1.69  &  2.00  &  1.05 & 1.56  &  * & 1.02  \\ 
 $\C{10'}^{\rm NP}$ &  1.34  &  1.89  &  0.05 & 1.20  &  0.22 & *  \\ 
\bottomrule[1.6pt] 
\end{tabular}
\end{center}
\caption{\it Pulls obtained by allowing successively NP in two Wilson coefficients: for the $\C{j}$ column,
the second row gives the pull of the SM hypothesis in the case where $\C{j}$ is let free to vary, whereas the $\C{i}$ row yields the pull of the hypothesis $\C{i}=\C{i}^{SM}$ in the scenario where $\C{i}$ and $\C{j}$ are let free to vary.}\label{tab:2Dnestedpulls}
\end{table}

\begin{table}[!t]
\begin{center}
\renewcommand{\arraystretch}{1.4}
 \setlength{\tabcolsep}{13pt}
\begin{tabular}{@{}cccr@{}}
\toprule[1.6pt] 
Coefficient & Best Fit Point & Pull$_{\rm SM}$ & p-value (\%) \\ 
 \midrule 
 $(\C7^{\rm NP},\C9^{\rm NP})$ &  $(-0.00,-1.07)$  &  {\bf 4.1} & 61.0 \hspace{5mm} \\ 
 $(\C7^{\rm NP},\C{10}^{\rm NP})$ &  $(-0.02,0.54)$  &  2.1 & 25.0 \hspace{5mm} \\ 
 $(\C7^{\rm NP},\C{7'}^{\rm NP})$ &  $(-0.02,0.01)$  &  0.8 & 15.0 \hspace{5mm} \\ 
 $(\C7^{\rm NP},\C{9'}^{\rm NP})$ &  $(-0.02,0.47)$  &  1.6 & 20.0 \hspace{5mm} \\ 
 $(\C7^{\rm NP},\C{10'}^{\rm NP})$ &  $(-0.02,-0.24)$  &  1.3 & 18.0 \hspace{5mm} \\ 
 $(\C9^{\rm NP},\C{10}^{\rm NP})$ &  $(-1.08,0.33)$  &  {\bf 4.3} & 67.0 \hspace{5mm} \\ 
 $(\C9^{\rm NP},\C{7'}^{\rm NP})$ &  $(-1.09,0.02)$  &  {\bf 4.2} & 63.0 \hspace{5mm} \\ 
 $(\C9^{\rm NP},\C{9'}^{\rm NP})$ &  $(-1.12,0.77)$  &  {\bf 4.5} & 72.0 \hspace{5mm} \\ 
 $(\C9^{\rm NP},\C{10'}^{\rm NP})$ &  $(-1.17,-0.35)$  &  {\bf 4.5} & 71.0 \hspace{5mm} \\ 
 $(\C{10}^{\rm NP},\C{7'}^{\rm NP})$ &  $(0.54,0.00)$  &  2.0 & 23.0 \hspace{5mm} \\ 
 $(\C{10}^{\rm NP},\C{9'}^{\rm NP})$ &  $(0.49,0.30)$  &  2.2 & 25.0 \hspace{5mm} \\ 
 $(\C{10}^{\rm NP},\C{10'}^{\rm NP})$ &  $(0.54,-0.01)$  &  2.0 & 23.0 \hspace{5mm} \\ 
 $(\C{7'}^{\rm NP},\C{9'}^{\rm NP})$ &  $(0.01,0.47)$  &  1.2 & 17.0 \hspace{5mm} \\ 
 $(\C{7'}^{\rm NP},\C{10'}^{\rm NP})$ &  $(0.01,-0.22)$  &  0.9 & 16.0 \hspace{5mm} \\ 
 $(\C{9'}^{\rm NP},\C{10'}^{\rm NP})$ &  $(0.38,-0.06)$  &  1.2 & 17.0 \hspace{5mm} \\ 
 $(\C{9}^{\rm NP}=-\C{9'}^{\rm NP},\C{10}^{\rm NP}=\C{10'}^{\rm NP})$ &  $(-1.15,0.34)$  &  {\bf 4.7} & 75.0 \hspace{5mm} \\ 
 $(\C{9}^{\rm NP}=-\C{9'}^{\rm NP},\C{10}^{\rm NP}=-\C{10'}^{\rm NP})$ &  $(-1.06,0.06)$  &  {\bf 4.4} & 70.0 \hspace{5mm} \\ 
 $(\C{9}^{\rm NP}=\C{9'}^{\rm NP},\C{10}^{\rm NP}=\C{10'}^{\rm NP})$ &  $(-0.64,-0.21)$  &  3.9 & 55.0 \hspace{5mm} \\ 
 $(\C{9}^{\rm NP}=-\C{10}^{\rm NP},\C{9'}^{\rm NP}=\C{10'}^{\rm NP})$ &  $(-0.72,0.29)$  &  3.8 & 53.0 \hspace{5mm} \\ 
 $(\C{9}^{\rm NP}=-\C{10}^{\rm NP},\C{9'}^{\rm NP}=-\C{10'}^{\rm NP})$ &  $(-0.66,0.03)$  &  2.0 & 23.0 \hspace{5mm} \\ 
\bottomrule[1.6pt] 
\end{tabular}
\end{center}
\caption{\it Best-fit points, pulls for the SM hypothesis and $p$-values for different two-dimensional NP scenarios.}\label{tab:2Dfits}
\end{table}

It is also interesting to proceed as in Ref.~\cite{Descotes-Genon:2013wba} and consider nested scenarios where NP is added to one Wilson coefficient after the other, starting from the SM hypothesis. In
a given scenario (where some Wilson coefficients $\C{j_1,\ldots j_N}$ receive NP and the others do not), the improvement obtained by allowing one more Wilson coefficient $\C{i}$ to receive NP contributions can be quantified by computing the pull of the $\C{i} = \C{i}^{\rm SM}$ hypothesis. This allows us to determine the NP scenarios which manage best to reproduce data.
From the results in Table~\ref{tab:2Dnestedpulls},
the most favoured scenarios correspond to $(\C9^{\rm NP},\C{9'}^{\rm NP})$
and  $(\C{9}^{\rm NP},\C{10'}^{\rm NP})$. This is supported by the actual 2D fits, with results shown in Table~\ref{tab:2Dfits}, which also indicates that $(\C{9}^{\rm NP},\C{10}^{\rm NP})$ is interesting to consider.
Other scenarios are also interesting where constraints are used to relate the various NP contributions, for instance
$\C9^{\rm NP}=-\C{9'}^{\rm NP}, \C{10}^{\rm NP}=\C{10'}^{\rm NP}$, as well as 
$\C9^{\rm NP}=-\C{9'}^{\rm NP}, \C{10}^{\rm NP}=-\C{10'}^{\rm NP}$.

\begin{figure}[!t]
\begin{center}
\begin{tabular}{cc}
\includegraphics[width=7.5cm]{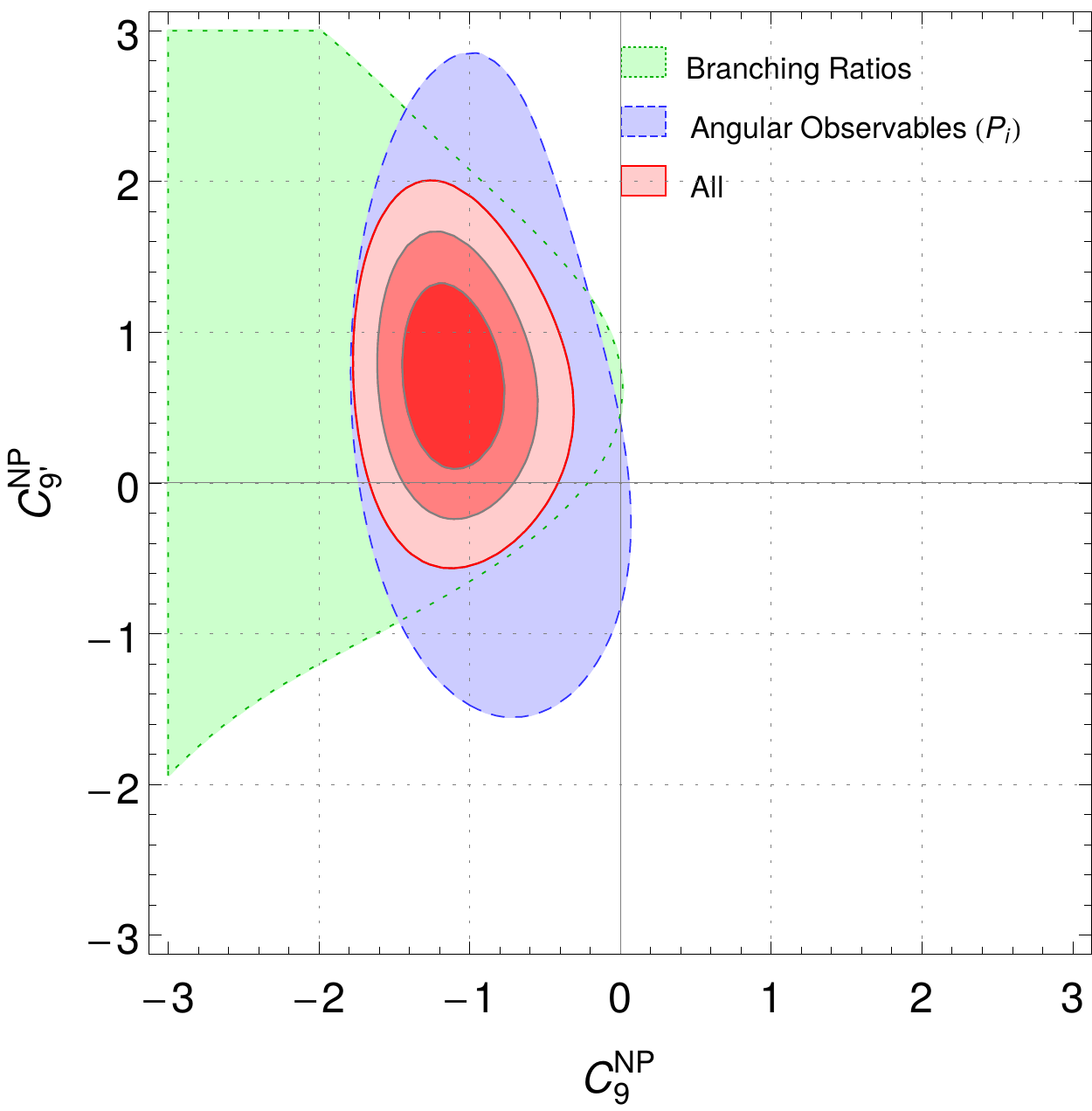} &  \includegraphics[width=7.5cm]{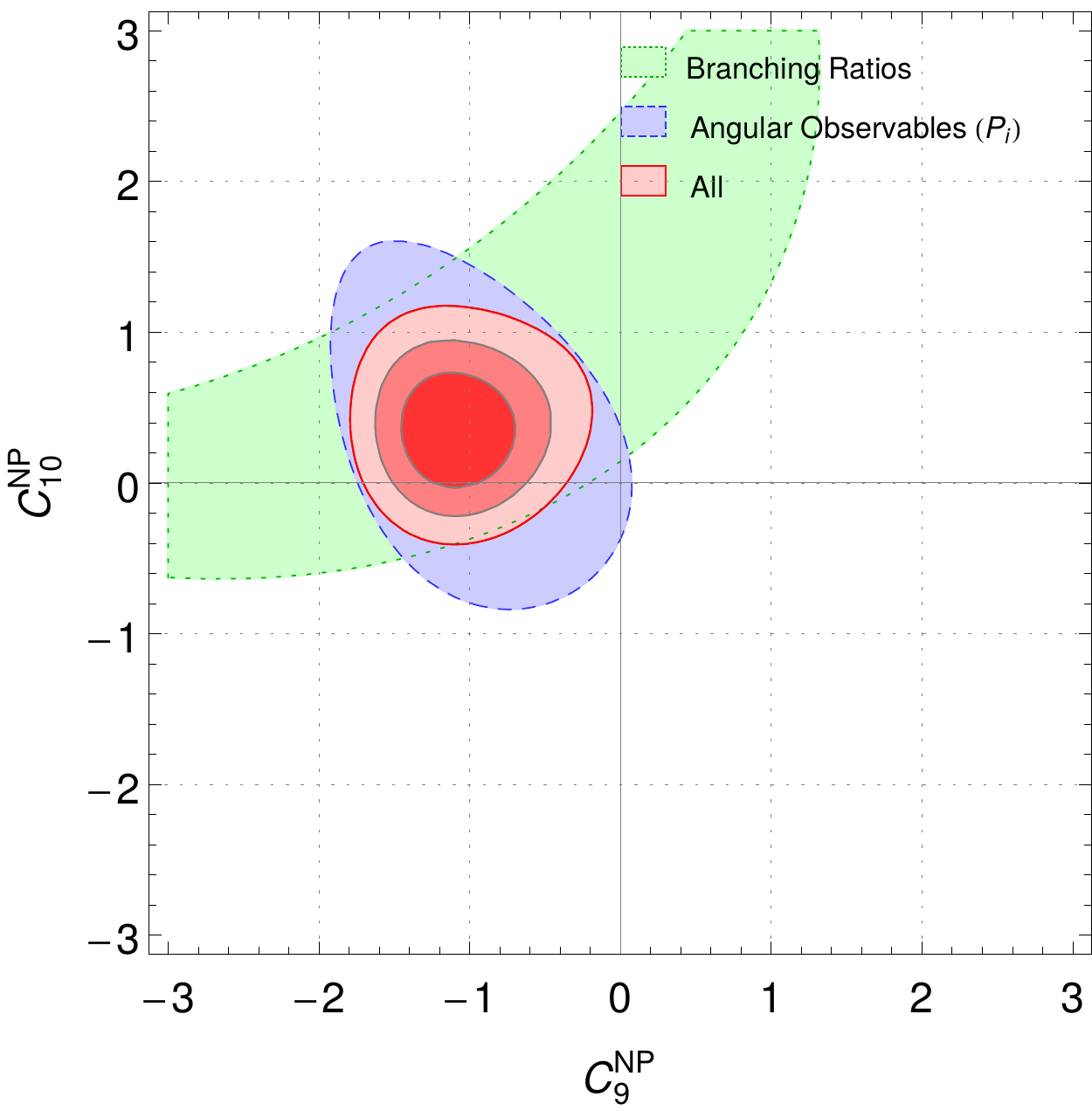}\\
\includegraphics[width=7.5cm]{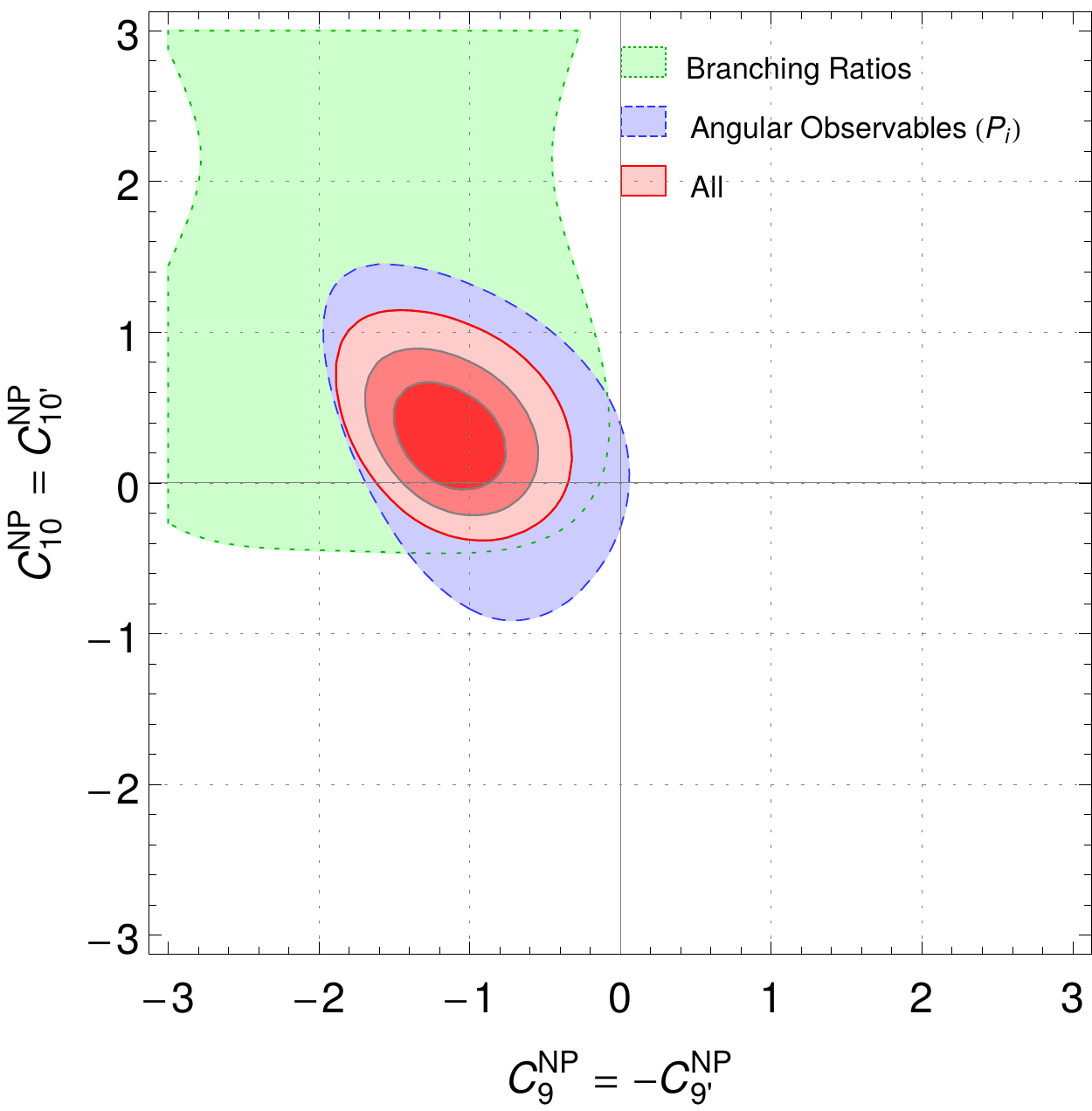} &  \includegraphics[width=7.5cm]{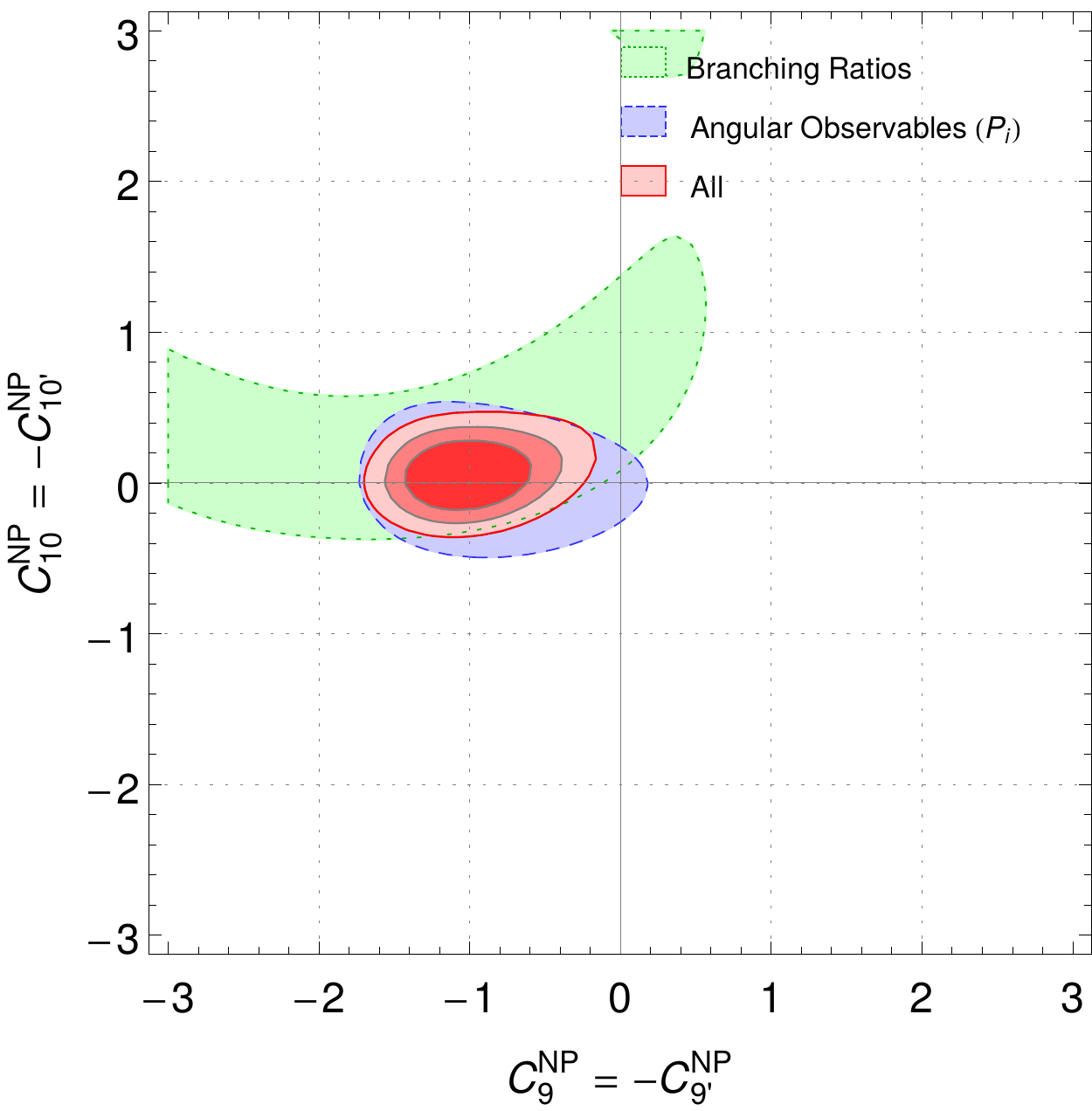}
\end{tabular}
\end{center}
\caption{\it For 4 favoured scenarios, we show the 3~$\sigma$ regions allowed by branching ratios only (dashed green), by angular observables only (long-dashed blue) and by considering both (red, with 1,2,3~$\sigma$  contours, corresponding to 68.3\%, 95.5\% and 99.7\% confidence levels). Each constraint corresponding to a subset of data includes also the inclusive and $b\to s\gamma$ data.}\label{fig:splitBRangular}
\end{figure}

\begin{figure}[!t]
\begin{center}
\begin{tabular}{cc}
\includegraphics[width=7.5cm]{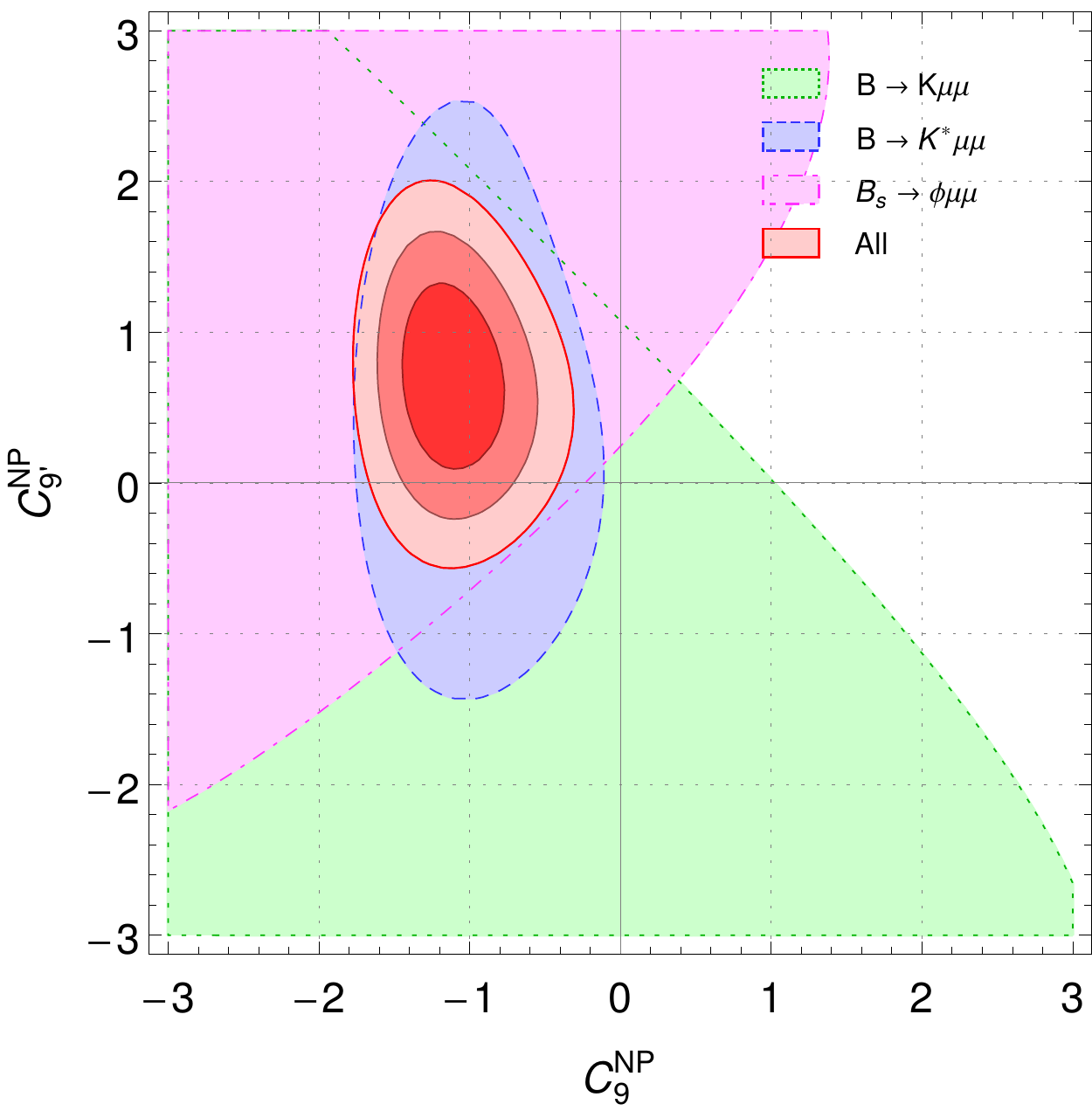} &  \includegraphics[width=7.5cm]{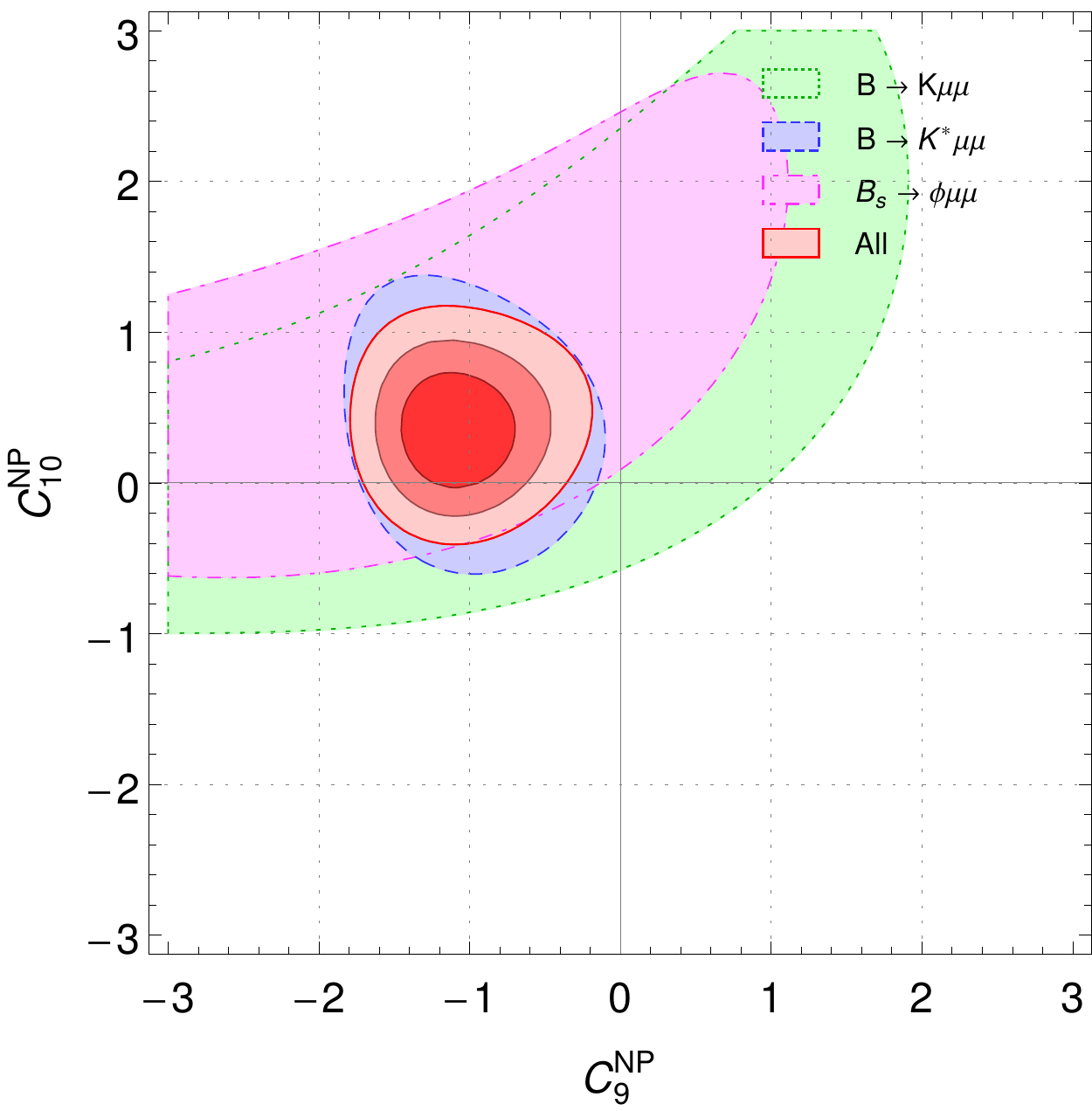}\\
\includegraphics[width=7.5cm]{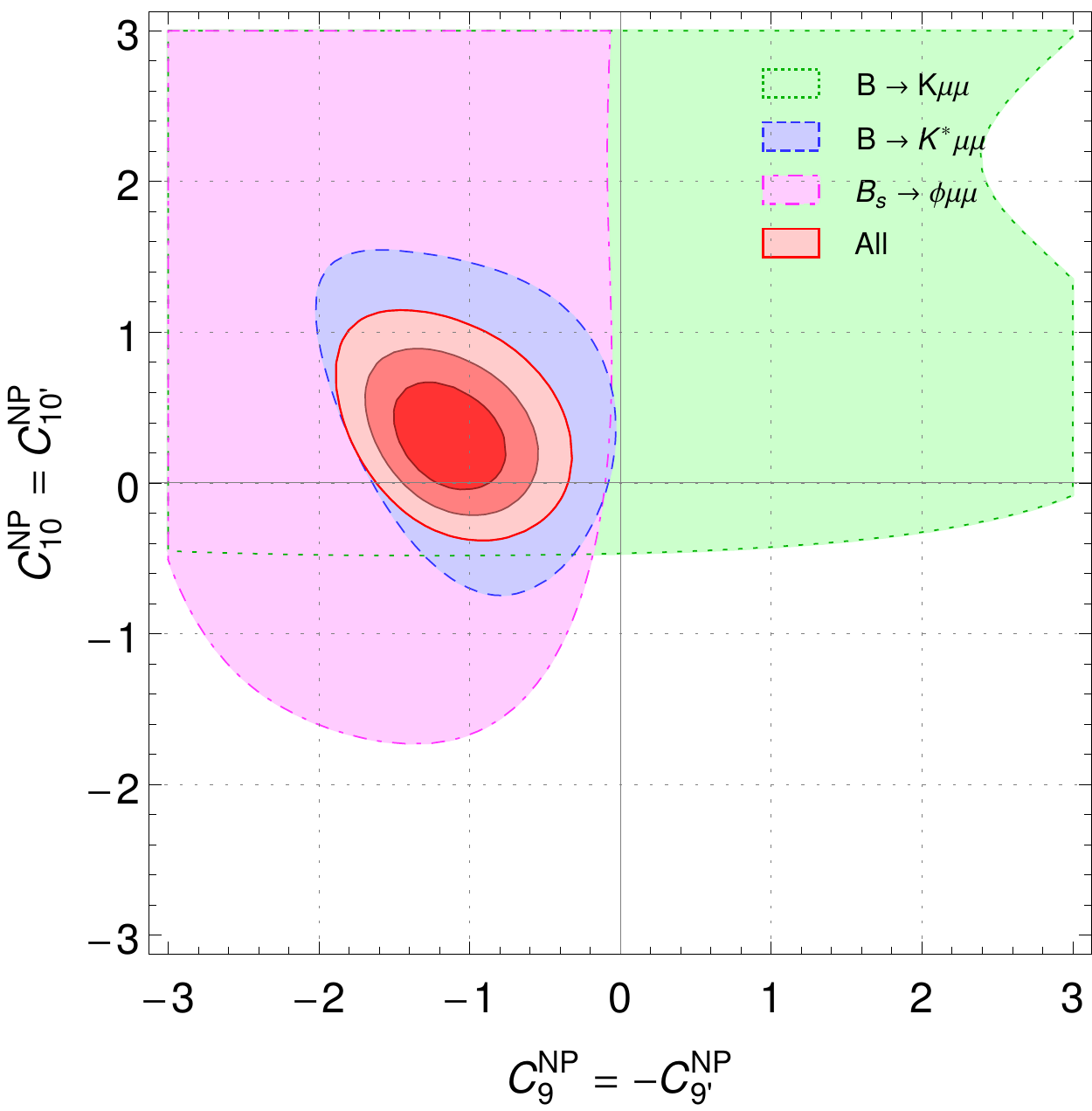} &  \includegraphics[width=7.5cm]{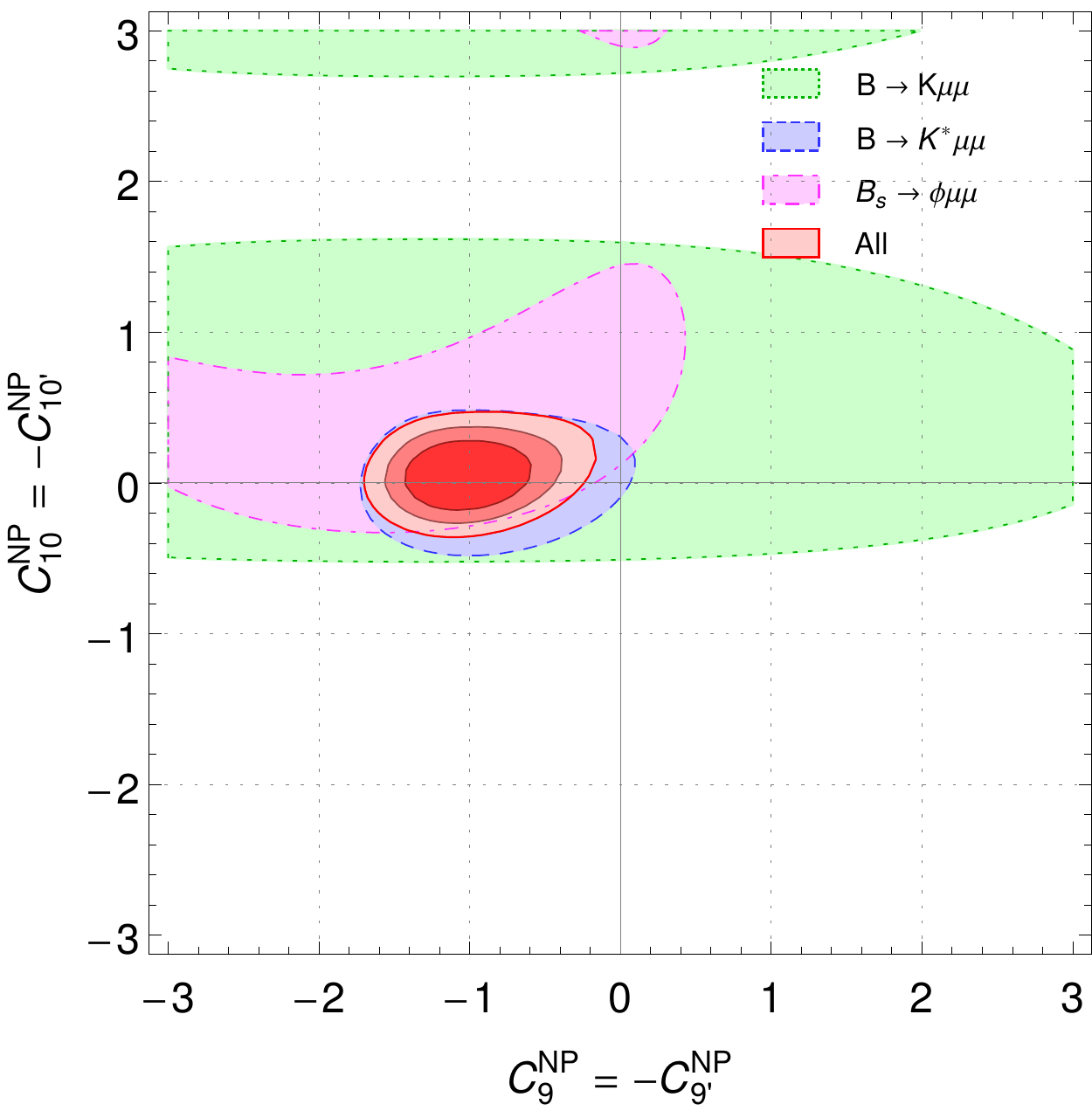}
\end{tabular}
\end{center}
\caption{\it  For 4 favoured scenarios, we show the 3~$\sigma$ regions allowed by $B\to K\mu\mu$ observables only (dashed green), by $B\to K^*\mu\mu$ observables only  (long-dashed blue), by $B_s\to \phi\mu\mu$ observables only (dot-dashed purple) and by considering all data (red, with 1,2,3~$\sigma$  contours). Same conventions for the constraints as in Fig.~\ref{fig:splitBRangular}.}\label{fig:splitbychannel}
\end{figure}

In Figs.~\ref{fig:splitBRangular} and \ref{fig:splitbychannel}, we show the 3~$\sigma$ regions corresponding to the constraints coming from branching ratios and angular observables, and from individual decay channels (respectively) for 4 favoured scenarios. Each constraint is built by considering one of the above subsets and adding the inputs from $b\to s\gamma$ and inclusive decays.  Both branching ratios and angular observables favour a negative value of $\C9$. As far as channels are concerned, the discrepancy with the Standard Model is triggered by $B\to K^*\mu\mu$ and by $B_s\to\phi\mu\mu$ (to a lesser extent).  Both scenarios 
with NP in $(\C9,\C{9'})$ or $(\C{9},\C{10})$ favour non-zero contributions for both Wilson coefficients, whereas the two scenarios $\C9^{\rm NP}=-\C{9'}^{\rm NP}, \C{10}^{\rm NP}=\C{10'}^{\rm NP}$ and $\C9^{\rm NP}=-\C{9'}^{\rm NP}, \C{10}^{\rm NP}=-\C{10'}^{\rm NP}$ favour NP in $\C9^{\rm NP}=-\C{9'}^{\rm NP}$ mainly (even though contributions to $\C{10}$ and $\C{10'}$ are allowed).

We emphasise that not all those scenarios have an interpretation in terms of a $Z^\prime$ which was first proposed by three of us in Ref.~\cite{Descotes-Genon:2013wba}, and was discussed in more detail in Refs.~\cite{Buras:2013dea,Buras:2014yna,Gauld:2013qja,Gauld:2013qba,Altmannshofer:2013foa,Altmannshofer:2014rta}. Indeed, an interpretation within a $Z^\prime$ context would reduce the subset of 2D constrained scenarios to the set of scenarios that fulfills $\C9^{\rm NP} \times \C{10^\prime}^{\rm NP}=\C{9^\prime}^{\rm NP} \times \C{10}^{\rm NP}$  (see App.~\ref{sec:Zcouplings}). 
Notice that this constraint is fulfilled by the scenarios with NP contribution only in $\C9$ or ($\C9,\C{9'})$ since both sides of the equation vanish trivially. On the other hand, if one wants to switch on NP in all four coefficients and preserve some simple pattern among them, there are  four options that may agree with a $Z^\prime$ interpretation:
\begin{itemize}
\item $(\C9^{\rm NP}=-\C{9'}^{\rm NP}, \C{10}^{\rm NP}=-\C{10'}^{\rm NP})$, with a large pull for the $b\to s\mu\mu$ reference fit, but giving $R_K=1$ by construction,
\item $(\C{9}^{\rm NP}=\C{10}^{\rm NP}, \C{9'}^{\rm NP}=\C{10'}^{\rm NP})$,  disfavoured by the data on $B_s \to \mu\mu$, which prefer a SM value for $\C{10}$, leading to a tension with the value of $\C{9}^{\rm NP}$ needed for $B\to K^*\mu\mu$
\item $(\C9^{\rm NP}=-\C{10}^{\rm NP}, \C{9'}^{\rm NP}=-\C{10'}^{\rm NP})$ and
$(\C9^{\rm NP}=\C{9'}^{\rm NP}, \C{10}^{\rm NP}=\C{10'}^{\rm NP})$ which could be interesting candidates but get
lower pulls (2.0 and 3.9~$\sigma$ respectively).
\end{itemize}
We see therefore that $Z'$ scenarios could alleviate part of the  discrepancies observed in $b\to s\mu\mu$ data, but with only one or two Wilson coefficients receiving NP contributions, corresponding to $Z'$ models with definite parity/chirality in its coupling to muons/quarks.

Another important criterion of choice among scenarios comes from considering the  main anomalies, namely, $P_5^\prime(B\to K^*\mu\mu)$, $R_K$ and $BR(B_s \to \phi\mu\mu)$,  and how they are weakened or strengthened in each scenario.
As can be seen from App.~\ref{app:SMpred},  besides the large deviations of order 2.5 to 3~$\sigma$ in different observables $P_5^\prime$, $R_K$ and ${\cal B}(B_s \to\phi\mu\mu)$ (that we called generically anomalies), there are also a large set of smaller
deviations (many of them at low recoil) that can push in different or similar directions. In App.~\ref{app:NPpred}, we illustrate how observables are affected in the presence of NP by
providing the predictions and the pulls for the observables at the best-fit point
for NP in $\C9$ only. In Table~\ref{tab:scenarcomp} we compare the best fit points for 1D and 2D scenarios involving $\C{9('),10(')}$,~\footnote{We do not consider $\C9^{\rm NP}=-\C{9^\prime}^{\rm NP}$ which has a large pull, but is not able to solve the discrepancy in $R_K$.} leading to a pull above 4.4~$\sigma$:
\begin{eqnarray}
&& {\rm I}: \C9^{\rm NP},  \qquad\qquad {\rm II}: (\C9^{\rm NP},\C{10}^{\rm NP}), \qquad\qquad {\rm III}: (\C9^{\rm NP},\C{10^\prime}^{\rm NP}),\qquad\qquad {\rm IV}: (\C9^{\rm NP},\C{9^\prime}^{\rm NP}),\nonumber\\
&& {\rm V}: (\C9^{\rm NP}=-\C{9^\prime}^{\rm NP},\C{10}^{\rm NP}=-\C{10^\prime}^{\rm NP}),\qquad {\rm VI}: (\C9^{\rm NP}=-\C{9^\prime}^{\rm NP},\C{10}^{\rm NP}=\C{10^\prime}^{\rm NP})\,.\nonumber
\end{eqnarray}

\begin{table}
\begin{center}{\small \begin{tabular}{ccccccccc}
 Impact & $ R_K$ & $\av{P_5^\prime}_{[4,6], [6,8]} $ & ${\cal B}_{B_s \to \phi\mu\mu}$ & ${\cal B}_{low\, recoil}$ \\\hline\hline
High & I,II & I,VI & VI & III,IV,VI \\
& III & II,IV & II,III,IV,V & I,II \\
& IV,VI & III,V & I & V \\
Low & V & & &
\\ \hline
\end{tabular}}
\end{center}
\caption{\it Relative impact of each scenario on the anomalies for $R_K$, $P_5^\prime$, ${\cal B}_{B_s \to \phi\mu\mu}$ and on the low-recoil bins of the different branching fractions} \label{tab:scenarcomp}\end{table}

We classify these scenarios according to how well they can fix a given anomaly or tension at their best-fit point
(reducing it below 1~$\sigma$ level awards the first position, failing to improve leads to the last position). Some scenarios are unable to improve on certain anomalies: for instance, $R_K$ which depends on the combination $(\C9^{\rm NP}+\C{9^\prime}^{\rm NP})-(\C{10}^{\rm NP}+\C{10^\prime}^{\rm NP})$ cannot be explained by a scenario of the type V. In other cases, the observables obtain contributions of opposite signs from the different NP contributions. This is the case for instance for scenarios with $\C9^{\rm NP}=-\C{9^\prime}^{\rm NP}$ where a negative $\C9^{\rm NP}$ goes in the right direction to alleviate the tension in $P_5^\prime$ whereas a positive $\C{9^\prime}^{\rm NP}$ goes in the wrong direction. However the impact of $\C{9^\prime}^{\rm NP}$ is only 25\% of $\C9^{\rm SM}$, so a small positive contribution or the contribution from other coefficients like a positive but small $\C{10^\prime}^{\rm NP}$ remains a viable possibility to explain the discrepancy in $P'_5$.
The result of this classification is that the scenario V is clearly disfavoured compared to the others which fare almost equally well, 
with a mild preference for I, II and VI. Only the scenarios II and III improve on the tiny deviation of data with respect to the SM  for $B_s \to \mu\mu$.

One can compare these results with the recent analysis in Ref.~\cite{Altmannshofer:2014rta}, which relied on a different approach (full form factor analysis, based on a different set of form factors with correlations~\cite{Straub:2015ica}, use of CP-averaged angular coefficients for the $B\to V\ell\ell$ angular analysis). We see that similar 1D scenarios are preferred with a contribution to $\C9$ alone, $\C{10}$ alone to a lesser extent, as well as $\C9^{\rm NP}=-\C{10}^{\rm NP}$. For 2D scenarios, the best-fit points for $(\C9^{\rm NP},\C{9'}^{\rm NP})$  and $(\C9^{\rm NP},\C{10}^{\rm NP})$ are also similar.

\subsubsection{Six-dimensional fits to Wilson coefficients} \label{sec:fits6D}

Even though we have seen that the anomalies can be described allowing for additional contributions in two Wilson
coefficients, it is interesting to consider the most general scenario with contributions to all six coefficients.
In this case, the best fit point is $\C7^{\rm NP} = 0.01$, $\C{7'}^{\rm NP} = 0.01$, $\C9^{\rm NP} = -1.1$,
$\C{10}^{\rm NP} = 0.5$, $\C{9'}^{\rm NP} = 1.2$, $\C{10'}^{\rm NP} = 0.3$, with the confidence intervals given
in Table~\ref{tab:6Dfits}. In agreement with the above sections, non-vanishing values for $\C9^{\rm NP}$ ($\C{9'}^{\rm NP}$)
are favoured strongly (mildly), whether the other coefficients may vanish at 1$\,\sigma$ but may also accommodate small
$\C{i}^{\rm NP}$ in their fairly large confidence intervals. The SM pull is 3.6$\,\sigma$, which is lower than the pulls
for successful 1D and 2D-scenarios, since this scenario allows for more degrees of freedom which are not all relevant to
explain the anomalies.

\begin{table}
\begin{center}
\renewcommand{\arraystretch}{1.4}
 \setlength{\tabcolsep}{13pt}
\begin{tabular}{@{}lccc@{}}
\toprule[1.2pt]
Coefficient& 1$\sigma$& 2$\sigma$& 3$\sigma$ \\ 
\midrule[1.1pt]\\[-5.5mm]
 $\C7^{\rm NP}$ & $[-0.02,0.03]$ & $[-0.04,0.04]$ & $[-0.05,0.08]$ \\ 
 $\C9^{\rm NP}$ & $[-1.4,-1.0]$ & $[-1.7,-0.7]$ & $[-2.2,-0.4]$ \\ 
 $\C{10}^{\rm NP}$ & $[-0.0,0.9]$ & $[-0.3,1.3]$ & $[-0.5,2.0]$ \\ 
 $\C{7'}^{\rm NP}$ & $[-0.02,0.03]$ & $[-0.04,0.06]$ & $[-0.06,0.07]$ \\ 
 $\C{9'}^{\rm NP}$ & $[0.3,1.8]$ & $[-0.5,2.7]$ & $[-1.3,3.7]$ \\ 
 $\C{10'}^{\rm NP}$ & $[-0.3,0.9]$ & $[-0.7,1.3]$ & $[-1.0,1.6]$ \\ 
\bottomrule[1.2pt]
\end{tabular}

\end{center}
\caption{\it Confidence intervals for Wilson coefficients in the six-dimensional NP scenario.}\label{tab:6Dfits}
\end{table}

\subsection{Fits involving $b\to see$ observables}

\begin{table}[!t]
\begin{center}
\begin{tabular}{@{}crcccr@{}}
\toprule[1.6pt] 
Coefficient & Best fit & 1$\sigma$ & 3$\sigma$ & Pull$_{\rm SM}$ & p-value (\%)\\ 
 \midrule 
 $\C7^{\rm NP}$ & $ -0.02 $ & $ [-0.04,-0.00] $ & $ [-0.07,0.03] $ &  1.2 & 17.0 \hspace{5mm}  \\[3mm] 
 $\C9^{\rm NP}$ & $ -1.11 $ & $ [-1.31,-0.90] $ & $ [-1.67,-0.46] $ &  {\bf 4.9} & 74.0 \hspace{5mm}  \\[3mm] 
 $\C{10}^{\rm NP}$ & $ 0.61 $ & $ [0.40,0.84] $ & $ [-0.01,1.34] $ &  3.0 & 32.0 \hspace{5mm}  \\[3mm] 
 $\C{7'}^{\rm NP}$ & $ 0.02 $ & $ [-0.00,0.04] $ & $ [-0.05,0.09] $ &  1.0 & 16.0 \hspace{5mm}  \\[3mm] 
 $\C{9'}^{\rm NP}$ & $ 0.15 $ & $ [-0.09,0.38] $ & $ [-0.56,0.85] $ &  0.6 & 15.0 \hspace{5mm}  \\[3mm] 
 $\C{10'}^{\rm NP}$ & $ -0.09 $ & $ [-0.26,0.08] $ & $ [-0.60,0.42] $ &  0.5 & 15.0 \hspace{5mm}  \\[3mm] 
 $\C9^{\rm NP}=\C{10}^{\rm NP}$ & $ -0.20 $ & $ [-0.38,-0.01] $ & $ [-0.70,0.47] $ &  1.0 & 16.0 \hspace{5mm}  \\[3mm] 
 $\C9^{\rm NP}=-\C{10}^{\rm NP}$ & $ -0.65 $ & $ [-0.80,-0.50] $ & $ [-1.13,-0.21] $ &  {\bf 4.6} & 66.0 \hspace{5mm}  \\[3mm] 
 $\C{9'}^{\rm NP}=\C{10'}^{\rm NP}$ & $ -0.03 $ & $ [-0.27,0.21] $ & $ [-0.76,0.68] $ &  0.1 & 15.0 \hspace{5mm}  \\[3mm] 
 $\C{9'}^{\rm NP}=-\C{10'}^{\rm NP}$ & $ 0.07 $ & $ [-0.04,0.17] $ & $ [-0.25,0.39] $ &  0.6 & 15.0 \hspace{5mm}  \\[3mm] 
 $\C9^{\rm NP}=-\C{9'}^{\rm NP}$ & $ -1.07 $ & $ [-1.25,-0.86] $ & $ [-1.60,-0.42] $ &  {\bf 4.9} & 73.0 \hspace{5mm}  \\[3mm] 
 \begin{minipage}{3.5cm} $\C9^{\rm NP}=-\C{10}^{\rm NP}$ \\ $=-\C{9'}^{\rm NP}=-\C{10'}^{\rm NP}$  \end{minipage} & $ -0.66 $ & $ [-0.84,-0.50] $ & $ [-1.25,-0.20] $ &  {\bf 4.5} & 64.0 \hspace{5mm}  \\[3mm] 
 \begin{minipage}{3.5cm} $\C9^{\rm NP}=-\C{10}^{\rm NP}$ \\ $=\C{9'}^{\rm NP}=-\C{10'}^{\rm NP}$  \end{minipage} & $ -0.22 $ & $ [-0.31,-0.13] $ & $ [-0.51,0.04] $ &  2.5 & 26.0 \hspace{5mm}  \\[3mm] 
\bottomrule[1.6pt] 
\end{tabular}
\end{center}
\caption{\it Best-fit point, confidence intervals, pulls for the SM hypothesis and $p$-value for different 1D NP scenarios,
including $b\to see$ data but assuming NP only in $b\to s\mu\mu$.}\label{tab:1Dfits-all}
\end{table}

\begin{table}[!t]
\begin{center}
\renewcommand{\arraystretch}{1.4}
 \setlength{\tabcolsep}{13pt}
\begin{tabular}{@{}cccr@{}}
\toprule[1.6pt] 
Coefficient & Best Fit Point & Pull$_{\rm SM}$ & p-value (\%) \\ 
 \midrule 
 $(C_7^{\rm NP},C_9^{\rm NP})$ &  $(-0.00,-1.10)$  &  {\bf 4.6} & 58.0 \hspace{5mm} \\ 
 $(C_7^{\rm NP},C_{10}^{\rm NP})$ &  $(-0.02,0.56)$  &  2.6 & 20.0 \hspace{5mm} \\ 
 $(C_7^{\rm NP},C_{7'}^{\rm NP})$ &  $(-0.02,0.02)$  &  1.0 & 9.4 \hspace{5mm} \\ 
 $(C_7^{\rm NP},C_{9'}^{\rm NP})$ &  $(-0.01,0.22)$  &  0.8 & 8.7 \hspace{5mm} \\ 
 $(C_7^{\rm NP},C_{10'}^{\rm NP})$ &  $(-0.02,-0.05)$  &  0.8 & 8.8 \hspace{5mm} \\ 
 $(C_9^{\rm NP},C_{10}^{\rm NP})$ &  $(-1.06,0.33)$  &  {\bf 4.8} & 65.0 \hspace{5mm} \\ 
 $(C_9^{\rm NP},C_{7'}^{\rm NP})$ &  $(-1.16,0.02)$  &  {\bf 4.7} & 62.0 \hspace{5mm} \\ 
 $(C_9^{\rm NP},C_{9'}^{\rm NP})$ &  $(-1.15,0.64)$  &  {\bf 4.9} & 67.0 \hspace{5mm} \\ 
 $(C_9^{\rm NP},C_{10'}^{\rm NP})$ &  $(-1.23,-0.29)$  &  {\bf 4.9} & 67.0 \hspace{5mm} \\ 
 $(C_{10}^{\rm NP},C_{7'}^{\rm NP})$ &  $(0.62,0.01)$  &  2.6 & 19.0 \hspace{5mm} \\ 
 $(C_{10}^{\rm NP},C_{9'}^{\rm NP})$ &  $(0.55,0.10)$  &  2.5 & 19.0 \hspace{5mm} \\ 
 $(C_{10}^{\rm NP},C_{10'}^{\rm NP})$ &  $(0.62,0.10)$  &  2.5 & 19.0 \hspace{5mm} \\ 
 $(C_{7'}^{\rm NP},C_{9'}^{\rm NP})$ &  $(0.02,0.11)$  &  0.6 & 8.2 \hspace{5mm} \\ 
 $(C_{7'}^{\rm NP},C_{10'}^{\rm NP})$ &  $(0.02,-0.09)$  &  0.5 & 8.0 \hspace{5mm} \\ 
 $(C_{9'}^{\rm NP},C_{10'}^{\rm NP})$ &  $(0.04,-0.04)$  &  0.2 & 7.4 \hspace{5mm} \\ 
 $(C_{9}^{\rm NP}=-C_{9'}^{\rm NP},C_{10}^{\rm NP}=C_{10'}^{\rm NP})$ &  $(-1.18,0.38)$  &  {\bf 5.1} & 72.0 \hspace{5mm} \\ 
 $(C_{9}^{\rm NP}=-C_{9'}^{\rm NP},C_{10}^{\rm NP}=-C_{10'}^{\rm NP})$ &  $(-1.11,0.04)$  &  {\bf 4.5} & 57.0 \hspace{5mm} \\ 
 $(C_{9}^{\rm NP}=C_{9'}^{\rm NP},C_{10}^{\rm NP}=C_{10'}^{\rm NP})$ &  $(-0.64,-0.11)$  &  {\bf 4.3} & 51.0 \hspace{5mm} \\ 
 $(C_{9}^{\rm NP}=-C_{10}^{\rm NP},C_{9'}^{\rm NP}=C_{10'}^{\rm NP})$ &  $(-0.69,0.27)$  &  {\bf 4.2} & 50.0 \hspace{5mm} \\ 
 $(C_{9}^{\rm NP}=-C_{10}^{\rm NP},C_{9'}^{\rm NP}=-C_{10'}^{\rm NP})$ &  $(-0.64,0.07)$  &  2.7 & 20.0 \hspace{5mm} \\ 
\bottomrule[1.6pt] 
\end{tabular}
\end{center}
\caption{\it Best-fit point, pulls for the SM hypothesis and $p$-value for different 2D NP scenarios, including $b\to see$ data but assuming NP only in $b\to s\mu\mu$.}\label{tab:2Dfits-all}
\end{table}

\begin{table}[!t]
{\footnotesize \begin{center}
\renewcommand{\arraystretch}{1.4}
 \setlength{\tabcolsep}{13pt}
\begin{tabular}{@{}c|ccccc}
\toprule[1.6pt] 
 & $R_K [1,6]$ & $R_{K^*} [1.1,6]$ & $R_{\phi} [1.1,6]$\\
  \midrule 
SM & $1.00\pm 0.01$   & \begin{tabular}{c} $1.00\pm 0.01$ \\ $[1.00\pm 0.01] $\end{tabular} & $1.00\pm 0.01$ \\
  \midrule $\C9^{\rm NP}=-1.11$  & $0.79\pm 0.01$ &  \begin{tabular}{c}$0.87\pm 0.08$ \\ $[0.84\pm 0.02]$\end{tabular}  & $0.84\pm 0.02$ \\
    \midrule $\C9^{\rm NP}=-\C{9'}^{\rm NP}=-1.09$  & $1.00\pm 0.01$ &  \begin{tabular}{c}$0.79\pm 0.14$ \\ $[0.74\pm 0.04]$\end{tabular}  & $0.74\pm 0.03$\\
        \midrule $\C9^{\rm NP}=-\C{10}^{\rm NP}=-0.69$  & $0.67\pm 0.01 $ &  \begin{tabular}{c}$0.71\pm 0.03$ \\ $[0.69\pm 0.01]$\end{tabular}  & $0.69\pm 0.01$\\
  \midrule $\C9^{\rm NP}=-1.15,\C{9'}^{\rm NP}=0.77$ & $0.91\pm 0.01$ &  \begin{tabular}{c}$0.80\pm 0.12$ \\ $[0.76\pm 0.03]$\end{tabular}  & $0.76 \pm 0.03 $\\
    \midrule $\C9^{\rm NP}=-1.16,\C{10}^{\rm NP}=0.35$ & $0.71\pm 0.01$&  \begin{tabular}{c}$0.78\pm 0.07$ \\ $[0.75\pm 0.02]$\end{tabular}  & $0.76\pm 0.01$ \\
  \midrule $\C9^{\rm NP}=-1.23,\C{10'}^{\rm NP}=-0.38$ & $0.87\pm 0.01$&  \begin{tabular}{c}$0.79\pm 0.11$ \\ $[0.75\pm 0.02]$\end{tabular}  & $0.76\pm 0.02$ \\
      \midrule \begin{tabular}{c}
$\C9^{\rm NP}=-\C{9'}^{\rm NP}=-1.14$\\
$\C{10}^{\rm NP}=-\C{10'}^{\rm NP}=0.04$ 
\end{tabular}\Bigg\}
& $1.00\pm 0.01$ &  \begin{tabular}{c}$0.78\pm 0.13$ \\ $[0.74\pm 0.04]$\end{tabular}  &  $0.74\pm 0.03$ \\
  \midrule \begin{tabular}{c}
$\C9^{\rm NP}=-\C{9'}^{\rm NP}=-1.17$\\
$\C{10}^{\rm NP}=\C{10'}^{\rm NP}=0.26$
\end{tabular}\Bigg\}
 & $0.88\pm 0.01$ &  \begin{tabular}{c}$0.76\pm 0.12$ \\ $[0.71\pm 0.04]$\end{tabular}  & $0.71\pm 0.03 $ \\
\bottomrule[1.6pt] 
\end{tabular}
\end{center}}
\caption{\it Predictions for $R_K$, $R_{K^*}$, $R_\phi$ at the best fit point of different scenarios of interest, assuming that NP enters only in the muon sector, and using the inputs of our reference fit, in particular the KMPW form factors in Ref.~\cite{Khodjamirian:2010vf} for $B\to K$ and $B\to K^*$, and 
Ref.~\cite{Straub:2015ica} for $B_s\to \phi$. In the case of $B\to K^*$,
we also indicate in brackets the predictions using the form factors in Ref.~\cite{Straub:2015ica}.}\label{tab:rKrKstar}
\end{table}

\subsubsection{Fits considering Lepton Flavour (non-) Universality}\label{sec:fits-LFU}

As stated in Sec.~\ref{sec:btoseeobs}, several measurements have been performed for $b\to see$ and can be included in our analysis, as long as we assume some relationship between the Wilson coefficients in the electron and muon sectors $\C{i\,e}$ and $\C{i\,\mu}$. In the following, we add to our reference fit the data described in Sec.~\ref{sec:btoseeobs}, and assume that 
NP enters $b\to see$ and $b\to s\mu\mu$ the same way (NP Lepton Flavour Universality [LFU]),
that it enters in a different way (NP LFU Violation), or even
that there is no NP in the $b\to see$ (Maximal NP LFU Violation). 

Even in the case of Maximal NP LFU Violation, adding $b\to see$ data on the fit may have an impact through the additional constraints that $b\to see$ data sets on hadronic inputs (in particular form factors). The main input here is $BR(B\to Kee)$, which has a very strong theoretical correlation with $BR(B\to K\mu\mu)$ and thus amounts to including the constraint from $R_K$. Tables~\ref{tab:1Dfits-all} and \ref{tab:2Dfits-all} (with $b\to see$ data) can be compared with Tables~\ref{tab:1Dfits} and \ref{tab:2Dfits} (without it). Since the discrepancy in $R_K$ is mainly driven by the disagreement between the SM predictions and the measurements for $BR(B\to K\mu\mu)$, it is not surprising that the 1D scenarios modifying $\C{9\,\mu}$
see their significance increase, as well as the p-value associated with the fit (apart from 
$\C9^{\rm NP}=-\C{9'}^{\rm NP}$ which remains unchanged). In particular, scenarios with  contribution to $\C9^{\rm NP}$ only and $\C9^{\rm NP}=-\C{10}^{\rm NP}$ have a large SM pull and a decent p-value. A similar situation occurs for the favoured 2D hypotheses.

It is also interesting to predict $R_K$, $R_{K^*}$ and $R_\phi$ for different scenarios in the intermediate region [1,6] or [1.1,6], assuming that NP enters only the muon sector. The results are given in Table~\ref{tab:rKrKstar}, showing some sensitivity to the scenario chosen. Varying only $\C{9}$ seems the most efficient way to get $R_K$ in agreement with the current LHCb values, with values of $R_{K^*}$ and $R_\phi$ around 0.85.
Other scenarios yield larger values of $R_K$ and smaller for $R_{K^*}$ and $R_\phi$, apart from the scenario $\C9^{\rm NP}=-\C{10}^{\rm NP}$ which leads to $R_K,R_{K^*},R_\phi$ all around 0.7.

The increase in the uncertainties for our predictions for $R_{K^*}$ and $R_\phi$ in NP scenarios comes from the fact that a part of the effects proportional to the lepton mass come from the angular coefficient $J_{1s}$ which involves $4m_\ell^2/s$ multiplied by ${\rm Re}(A_\perp^L A_\perp^{R*} + A_{||}^L A_{||}^{R*})$. This term is small in the SM where $\C9\simeq -\C{10}$ and 
thus $A_{\perp,||}^R\simeq 0$, but in presence of NP not following the same $SU(2)_L$ relationship, this contribution increases, with an uncertainty coming mainly from the form factors. We illustrate the sensitivity to the choice of form factors for $B\to K^*$ where we provide the results using the form factors of Ref.~\cite{Khodjamirian:2010vf}, compared to Ref.~\cite{Straub:2015ica} (in brackets). The larger uncertainties in the former case come mainly from the normalisation of the form factors. Moreover, one may notice that $R_{K^*}$ and $R_\phi$ are almost identical when using the form factors of
Ref.~\cite{Straub:2015ica}: these ratios are driven by the ratios $F(0)/V(0)$ with $F=A_1,A_2,T_1,T_2$ are almost identical for $B\to K^*$ and $B_s\to\phi$ in Ref.~\cite{Straub:2015ica}.

\begin{figure}[!t]
\begin{center}
\includegraphics[width=7.5cm]{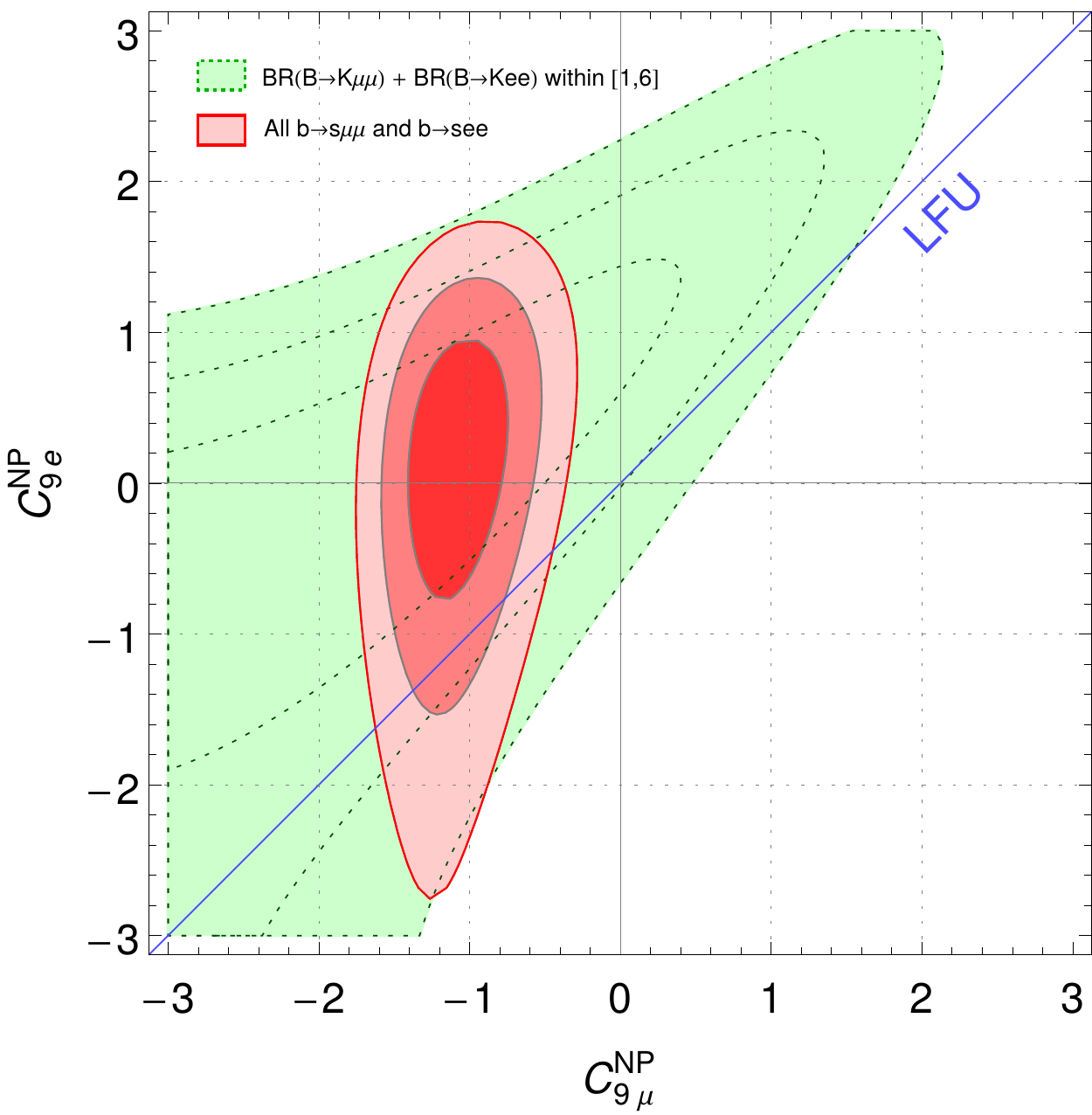}
\includegraphics[width=7.5cm]{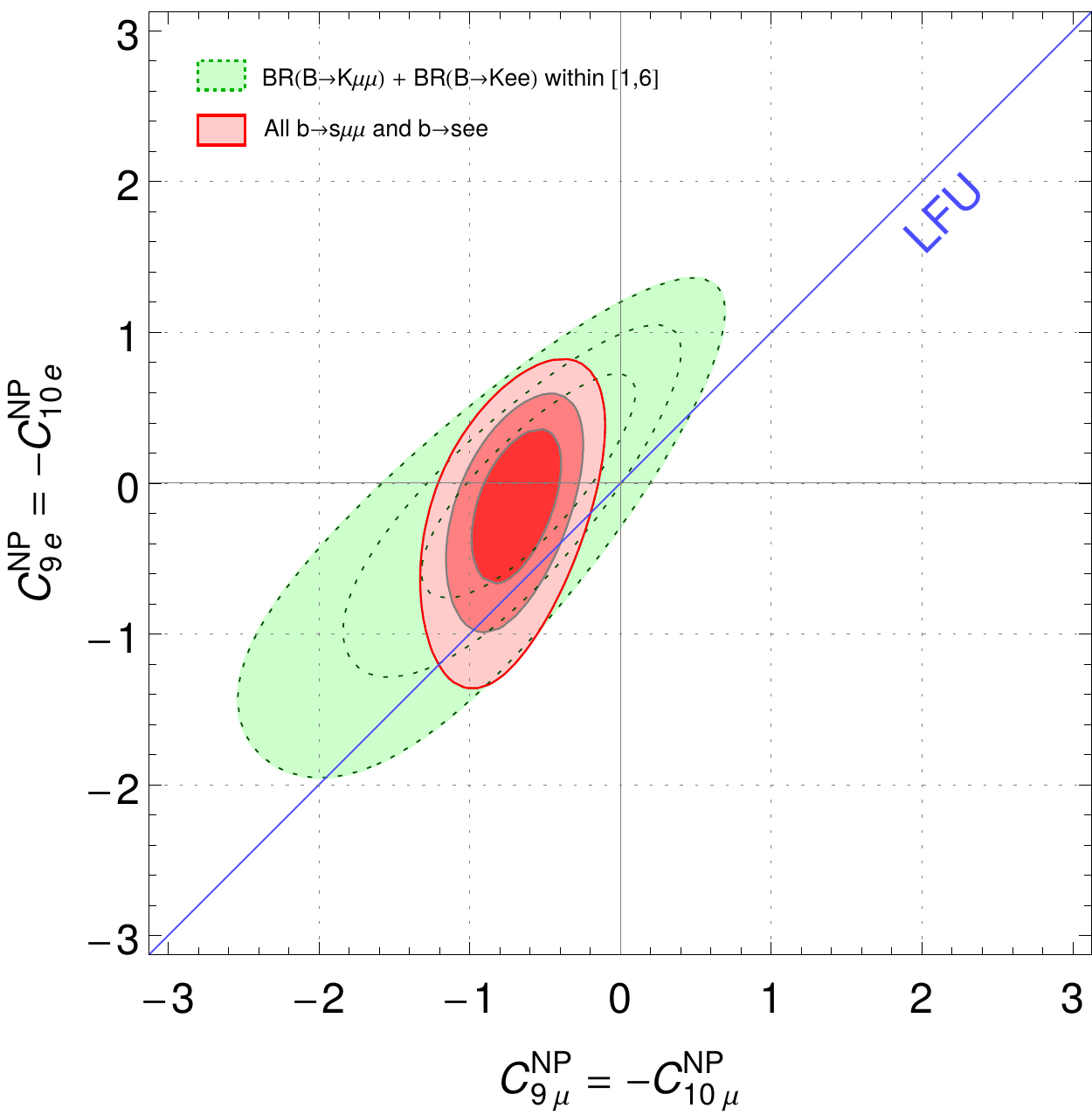}
\end{center}
\caption{\it For two scenarios where NP occurs in the two Wilson coefficients $\C{9\mu}$ and $\C{9e}$, we show the 1,2,3~$\sigma$ regions obtained  using only $BR(B^+\to K^+\mu^+\mu^-)$ and $BR(B^+\to K^+ e^+e^-)$ for bins in the [1,6] region (dashed green), and 1,2,3~$\sigma$ regions using all data from the reference fit and  $b\to s ee$ data (solid red). The two NP scenarios correspond to: $(\C{9\mu}^{\rm NP},\C{9e}^{\rm NP})$ (left) and $(\C{9\mu}^{\rm NP}=-\C{10\mu}^{\rm NP},\C{9e}^{\rm NP}=-\C{10e}^{\rm NP})$  (right). The diagonal line corresponds to the limit of Lepton Flavour Universality. Same conventions for the constraints as in Fig.~\ref{fig:splitBRangular}.}\label{fig:btoseeC9muC9e}
\end{figure}

If NP Lepton Flavour Universality Violation is assumed, NP may enter both $b\to s ee$  and $b\to s\mu\mu$ decays though potentially with different values. We show the corresponding constraints in Fig.~\ref{fig:btoseeC9muC9e} for two different
scenarios, namely $(\C{9\mu}^{\rm NP},\C{9e}^{\rm NP})$ and $(\C{9\mu}^{\rm NP}=-\C{10\mu}^{\rm NP},\C{9e}^{\rm NP}=-\C{10e}^{\rm NP})$. For each scenario, we see that there is no clear indication of a NP contribution in the electron sector, whereas one has clearly a non-vanishing contribution for the muon sector, with a deviation from the Lepton Flavour Universality line, in global agreement with Ref.~\cite{Altmannshofer:2014rta} but with a lower significance.

\subsubsection{Fits to magnetic operators at very low $q^2$}

Traditionally, the main constraints on $\C7$, $\C{7'}$ have been provided by $b\to s\gamma$ observables, both inclusive and exclusive (see e.g. Ref.~\cite{DescotesGenon:2012zf}).
Recent measurement of $b\to s ee$ observables at very low $q^2$ provides an alternative source for such constraints, as the photon pole enhances the relative importance of $\C{7,7'}$ with respect to $\C{9^{(\prime)},10^{(\prime)}}$. In order to compare the constraining power of both sets of observables separately, and to gauge the impact of including $b\to s \ell\ell$ modes in the fit, we have performed separate fits to $\C{7}^{\rm NP}$, $\C{7'}^{\rm NP}$ in two different scenarios: \emph{a)} all other Wilson coefficients have their SM value, and \emph{b)} $\C{9\mu}^{\rm NP}= -1.1$ and  the other coefficients have their SM value (a solution preferred globally by the data, as shown above).

\begin{figure}
\begin{center}
\includegraphics[width=7.6cm]{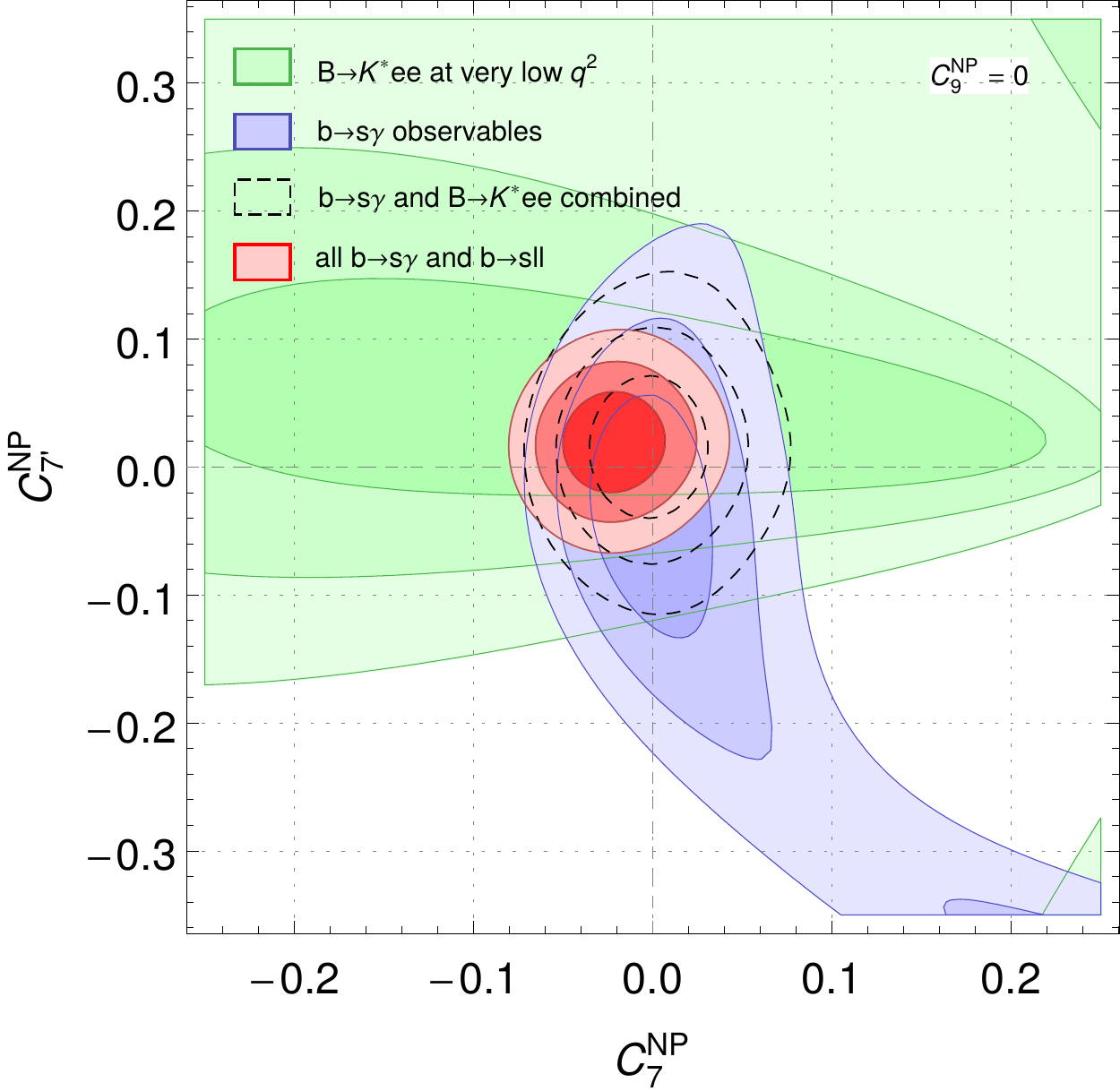} \includegraphics[width=7.6cm]{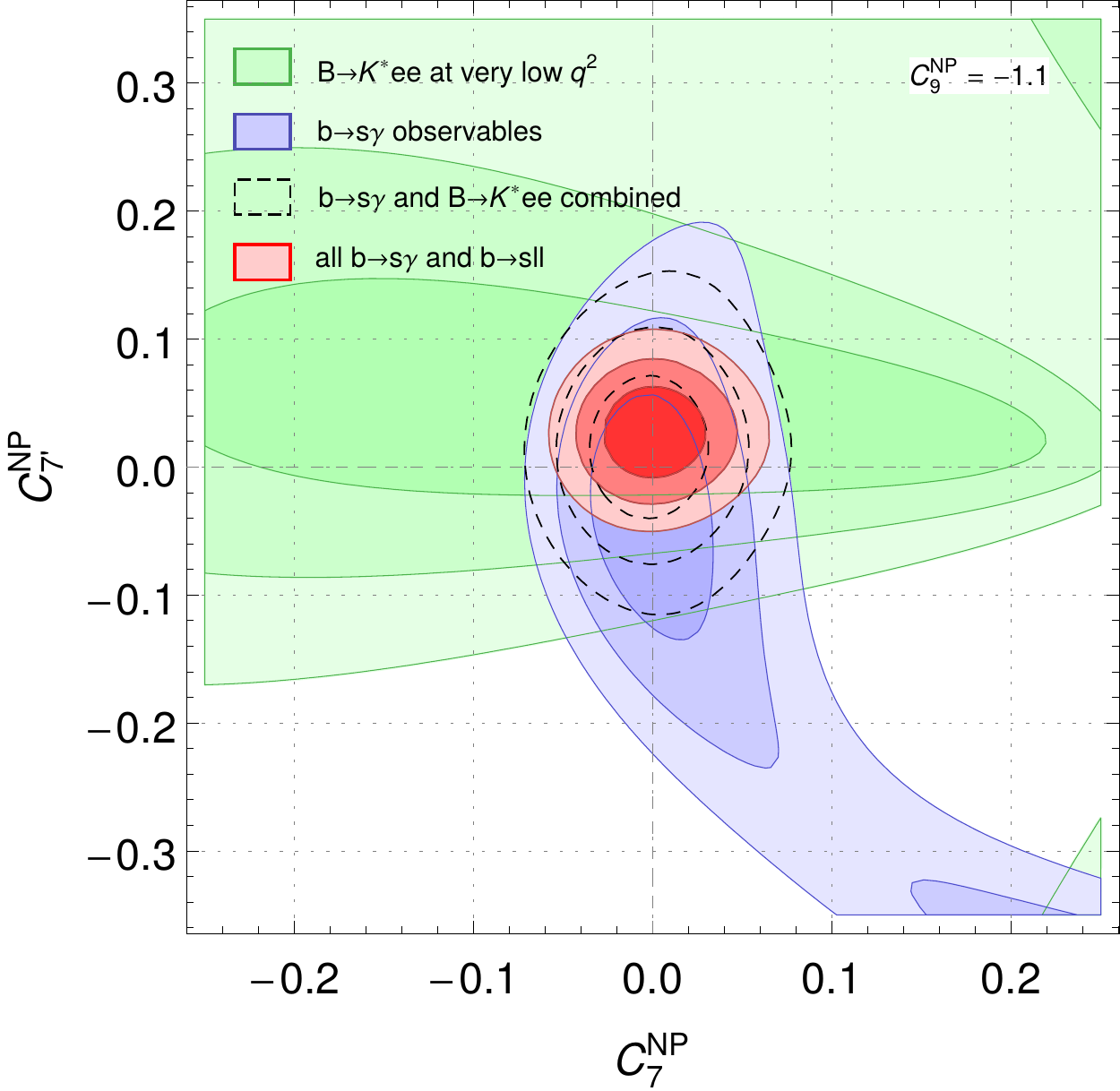}
\end{center}
\caption{\it Separate fits to $b\to s \gamma$ (blue) and $b\to see$ observables at very low $q^2$ (green). The combined fit to both sets of data is shown with dashed contours (1,2,3~$\sigma$ regions). The result of the global fit to all $b\to s \gamma$, $b\to s \ell\ell$ data is shown by the red contours (1,2,3~$\sigma$ regions). It is assumed that all the other Wilson coefficients have their SM values, except for the plot on the right, where $\C{9\mu}^\text{NP} = -1.1$.}\label{fig:C7C7p}
\end{figure}

In Fig.~\ref{fig:C7C7p} we show the resulting fits. The constraints from $b\to s ee$ observables alone (shown in green)
are milder than the $b\to s \gamma$ ones (shown in blue) but the two set of constraints are largely complementary,
leading to much tighter constraints once combined (dashed contours). As expected, all these constraints are independent
of the value of $\C{9\mu}^\text{NP}$. The result of the global fit including all observables ($b\to s \gamma$, $b\to s ee$
and $b\to s \mu\mu$) is also shown (red contours). The constraints are then (slightly) tighter, as $b\to s \mu\mu$
observables also constrain magnetic operators, with a clear dependence on $\C{9\mu}^\text{NP}$. As $\C{9\mu}^\text{NP}$ is
varied from zero to -1.1 the overall compatibility among the different sets of observables improves.

\subsection{Role of low- and large-recoil regions in the fit}\label{sec:cross-checks-bin}

\begin{figure}[!t]
\begin{center}
\begin{tabular}{cc}
\includegraphics[width=7.5cm]{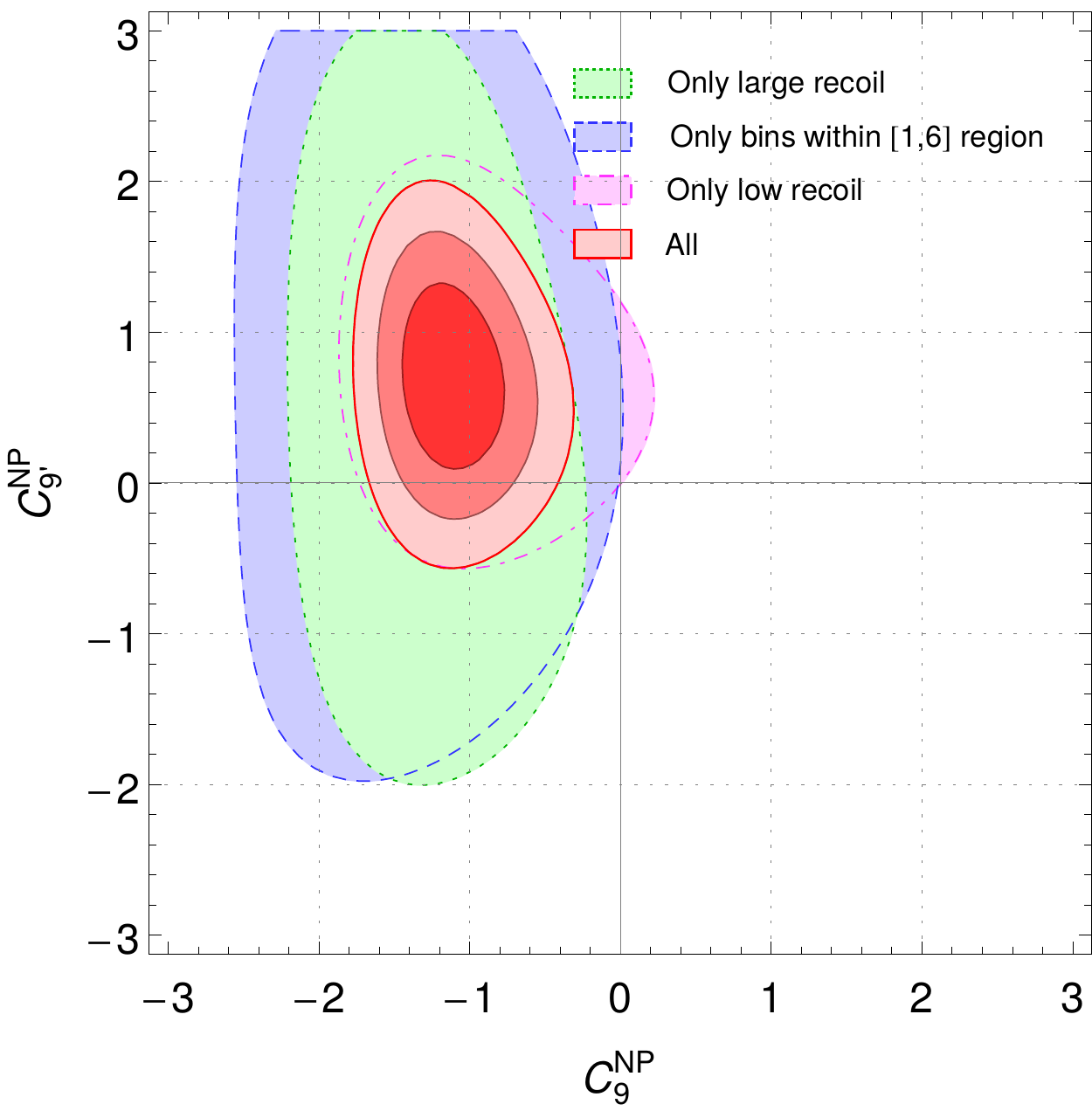} &  \includegraphics[width=7.5cm]{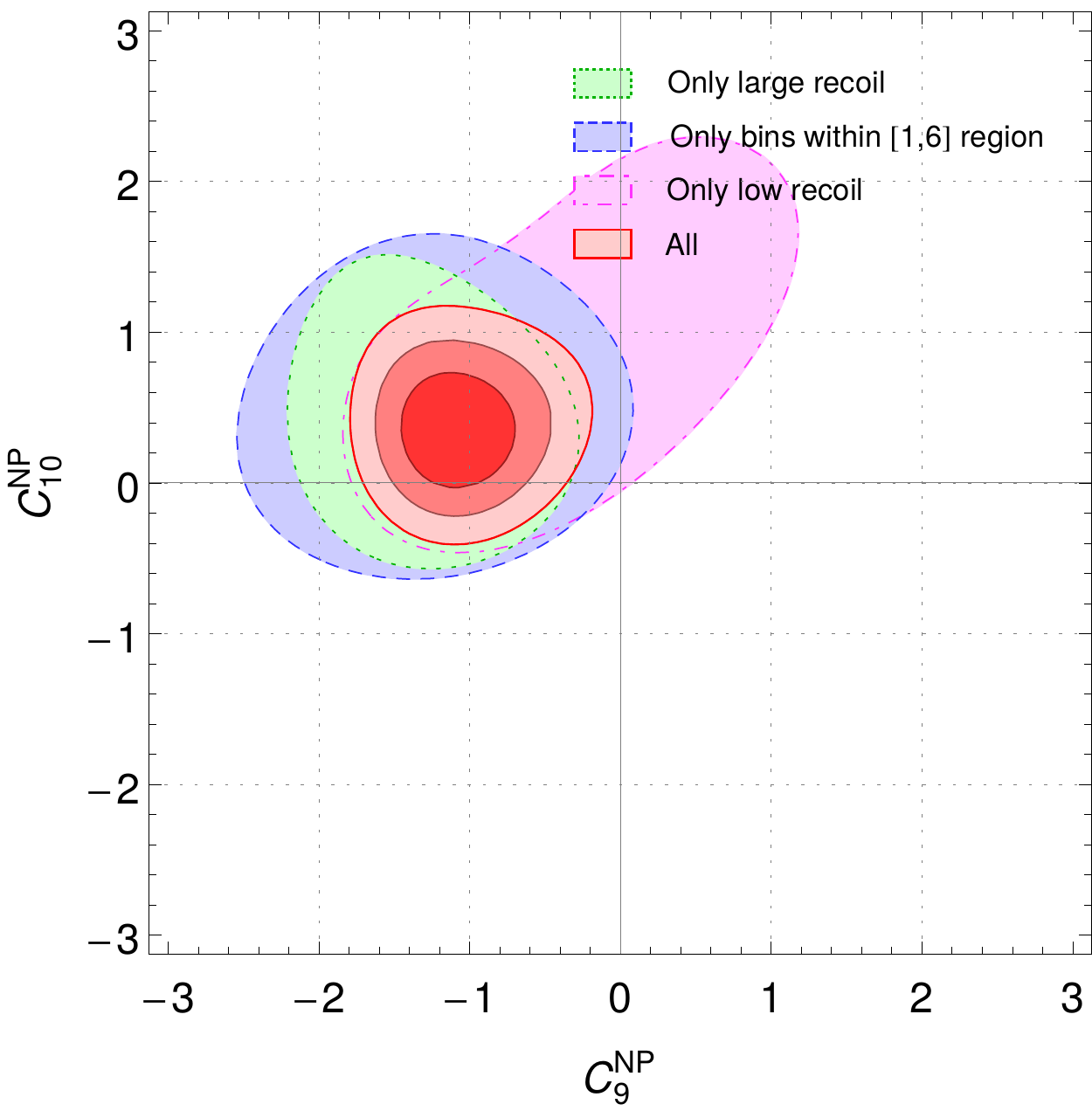}\\
\includegraphics[width=7.5cm]{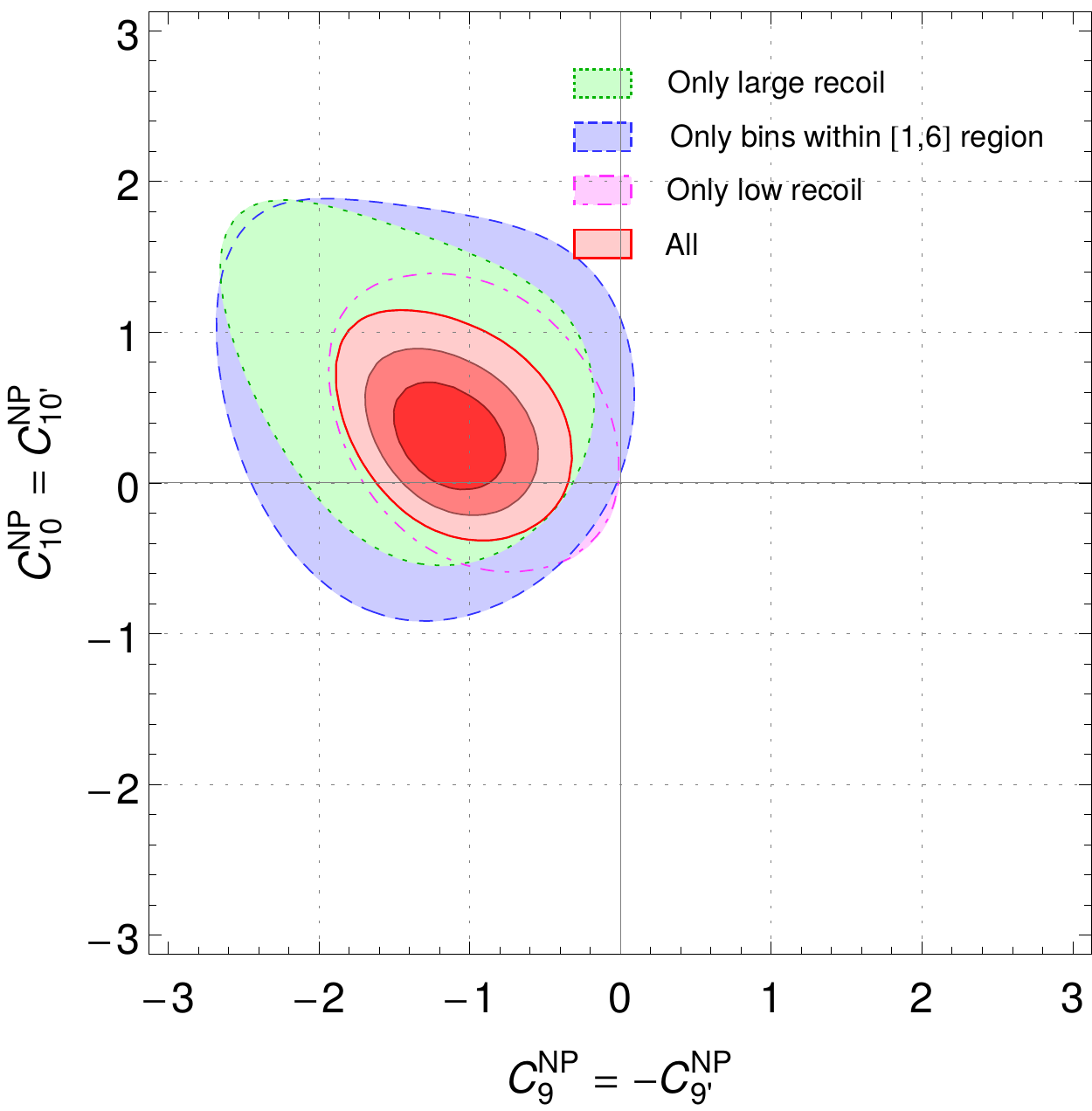} &  \includegraphics[width=7.5cm]{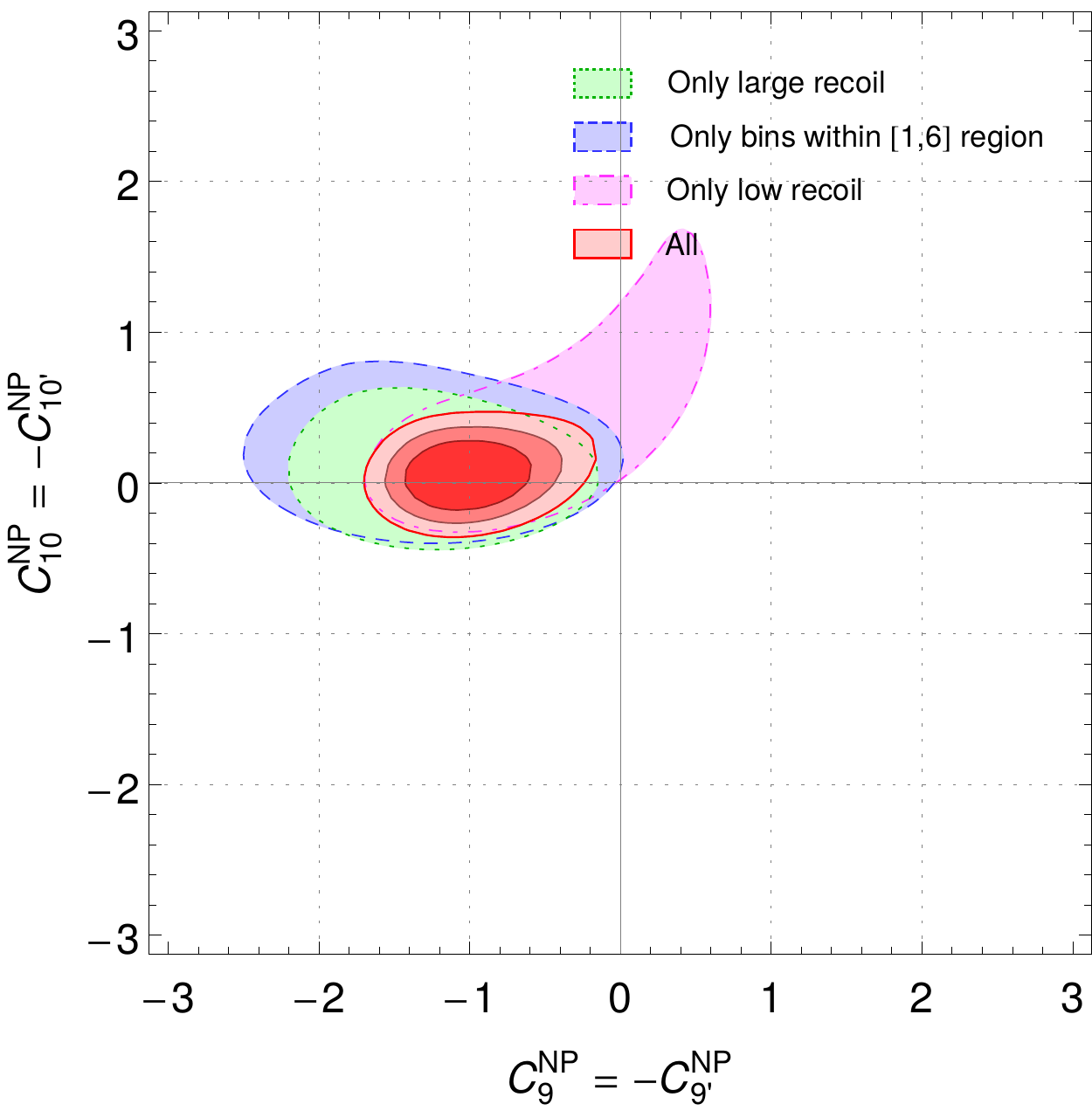}
\end{tabular}
\end{center}
\caption{\it  For 4 favoured scenarios, we show the 3~$\sigma$ regions allowed by large-recoil only (dashed green), by bins in the [1-6] range (long-dashed blue), by low recoil (dot-dashed purple) and by considering all data (red, with 1,2,3~$\sigma$  contours). Same conventions for the constraints as in Fig.~\ref{fig:splitBRangular}.}\label{fig:largevslow}
\end{figure}

\begin{figure}[!t]
\begin{center}
\includegraphics[width=7.75cm]{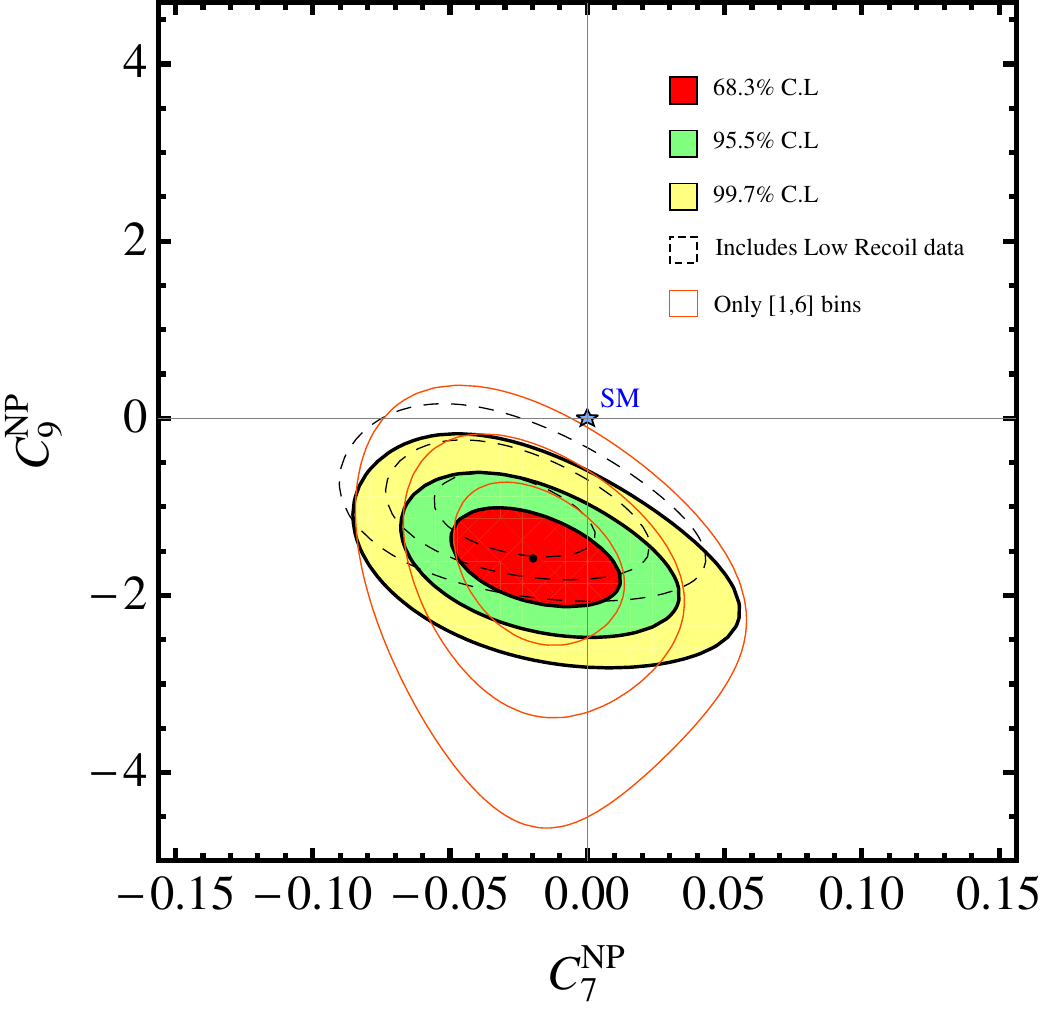}
\hspace{3mm}
\raisebox{2mm}{\includegraphics[width=7.5cm]{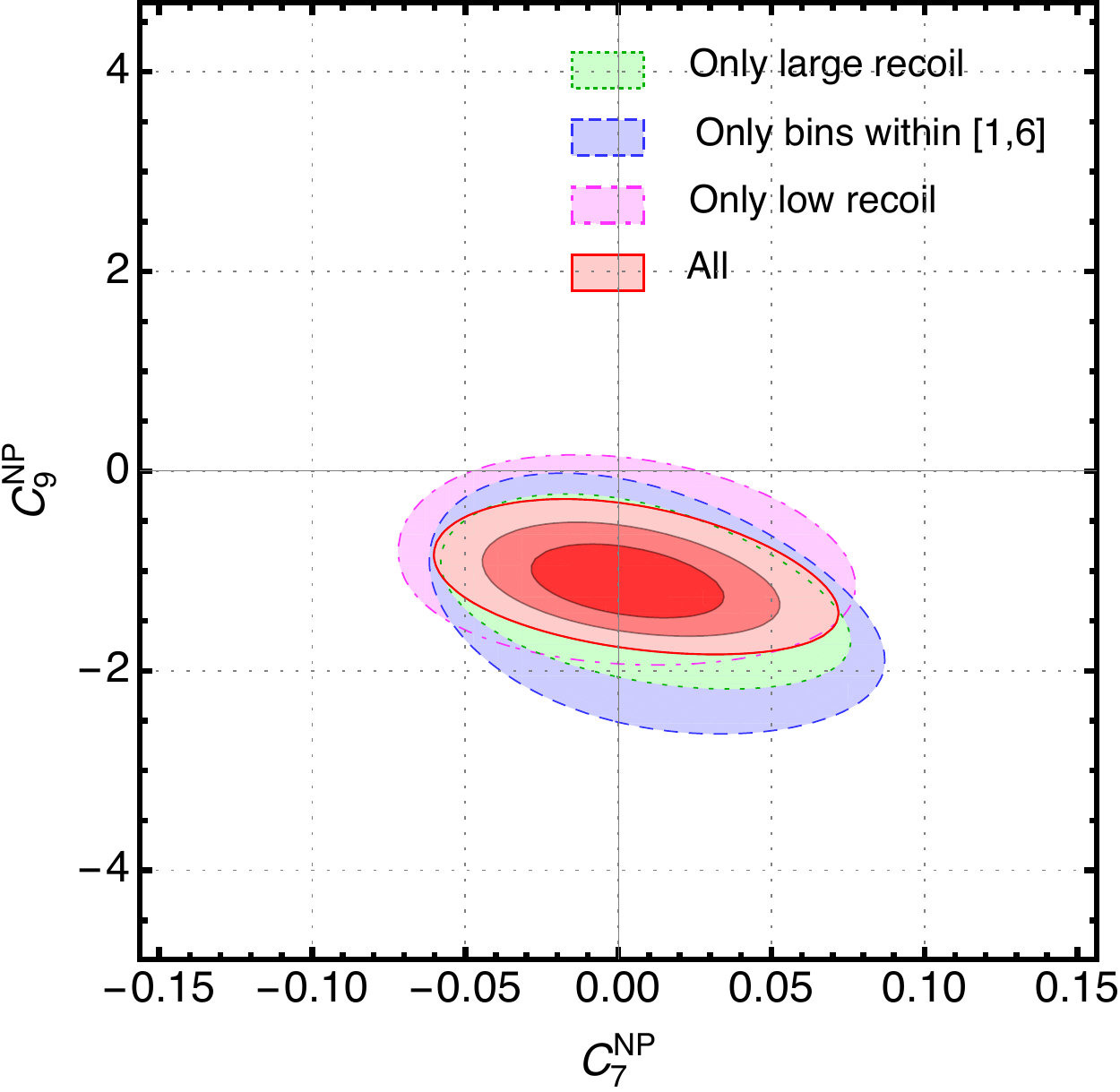}}
\end{center}
\caption{\it For the scenario where NP occurs in the two Wilson coefficients $\C{7}$ and $\C{9}$, we compare the situation from the analysis in Fig.~1 of Ref.~\cite{Descotes-Genon:2013wba} (on the left) and the current situation (on the right). On the right, 
we show the 3~$\sigma$ regions allowed by large-recoil only (dashed green), by bins in the [1-6] range (long-dashed blue), by low recoil (dot-dashed purple) and by considering all data (red, with 1,2,3~$\sigma$  contours). Same conventions for the constraints as in Fig.~\ref{fig:splitBRangular}.}\label{fig:updatedanomaly}
\end{figure}

The issues related to the first and last bins of the large-recoil region were already discussed in Sec.~\ref{sec:specificbins}. One may wonder to which extent our results depend on the inclusion of these bins, in particular the [6-8] bin where part of the discrepancies with the SM arises.  
In Sec.~\ref{sec:specificbins}, we also recalled a different issue, the size of duality-violating effects,
affecting the low-recoil bin. Even though some estimates indicate that they should not affect branching ratios significantly, we are not aware of a similar discussion for angular observables which are an important part of the reference fit. We illustrate the role played by the different bins by considering fits with only the low-recoil region, the large-recoil region, or the bins in the [1,6] GeV$^2$ range in Fig.~\ref{fig:largevslow}. It should be noticed that low recoil favours the same range of NP contributions as the large-recoil bins, but in a milder way. In addition, the [1,6] region provides similar constraints as the whole large-recoil range, implying that our results for the different NP scenarios hold even considering ranges for the dilepton invariant mass where charm contributions are expected to be less relevant.

Fig.~\ref{fig:updatedanomaly} illustrates a similar analysis for the $(\C7,\C9)$ scenario, which updates Fig.~1 in Ref.~\cite{Descotes-Genon:2013wba}. There is an overall similarity, with a best-fit point requiring almost no NP contributions to $\C7$. We stress that the right-hand plot involves a larger set of experimental measurements and a more complete understanding of the sources of theoretical uncertainties on the right. In addition, ``only [1,6] bins'' refers to observables in the single bin [1,6] only on the 2013 plot (on the left), but to those taken in any of the (smaller) bins inside the [1,6] range on the 2015 plot (on the right).

\section{Tests of SM theoretical uncertainties}
\label{sec:SMfit}

The previous studies show the robustness of the results when only part of the experimental information is included in the fit.
On the other hand, since the main discrepancies in the previous fits come from exclusive $b\to s\mu\mu$ transitions ($B\to K^*\mu\mu$, $B_s\to\phi\mu\mu$ and $B\to K\mu\mu$), one ought to consider the sources of systematics entering the SM theoretical predictions carefully, namely: form factor uncertainties, power corrections and long-distance corrections due to $c\bar{c}$ loops. We will consider these different sources of uncertainties in the following.

\begin{figure}[!t]
\begin{center}
\begin{tabular}{cc}
\includegraphics[width=7.5cm]{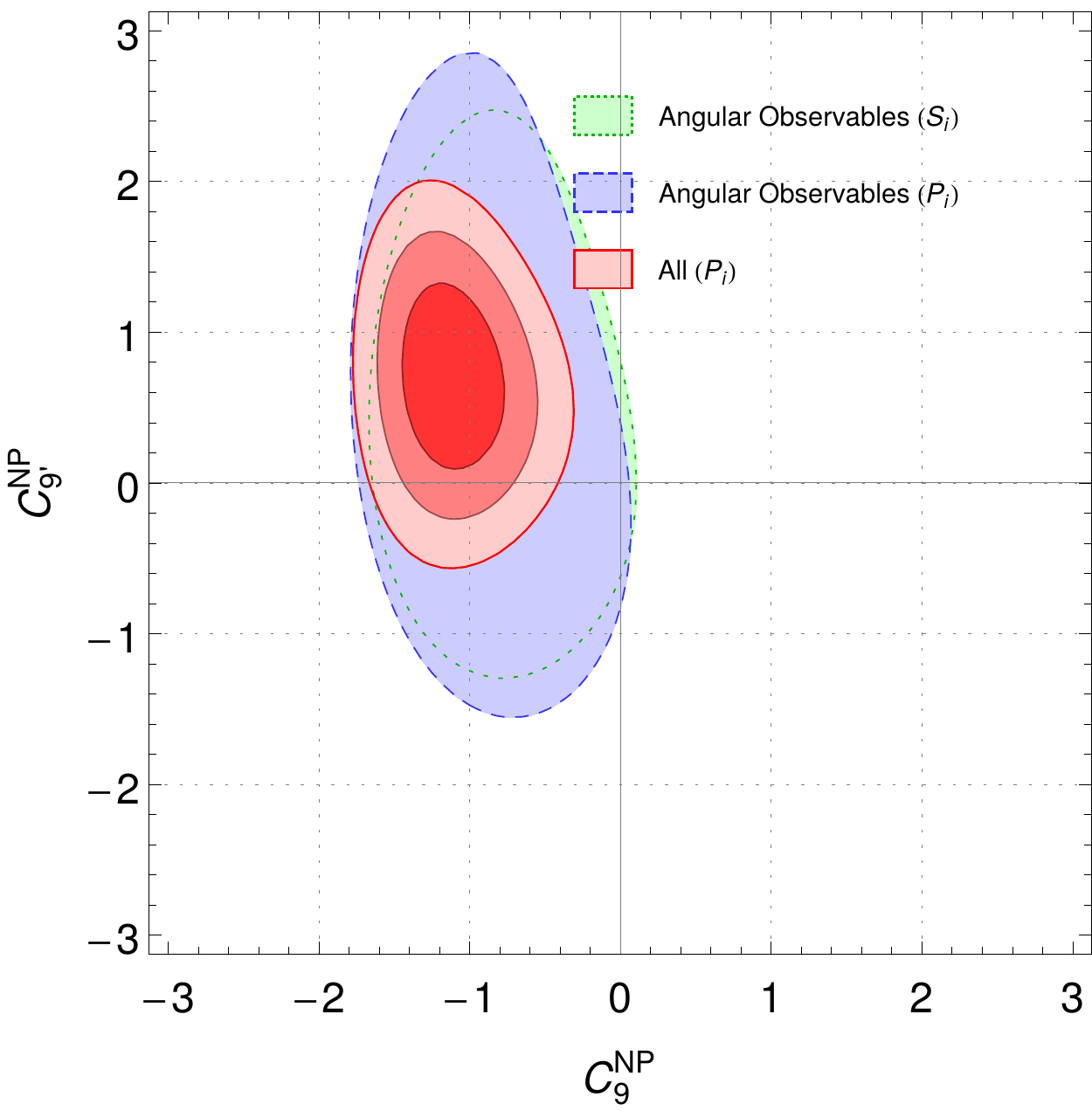} &  \includegraphics[width=7.5cm]{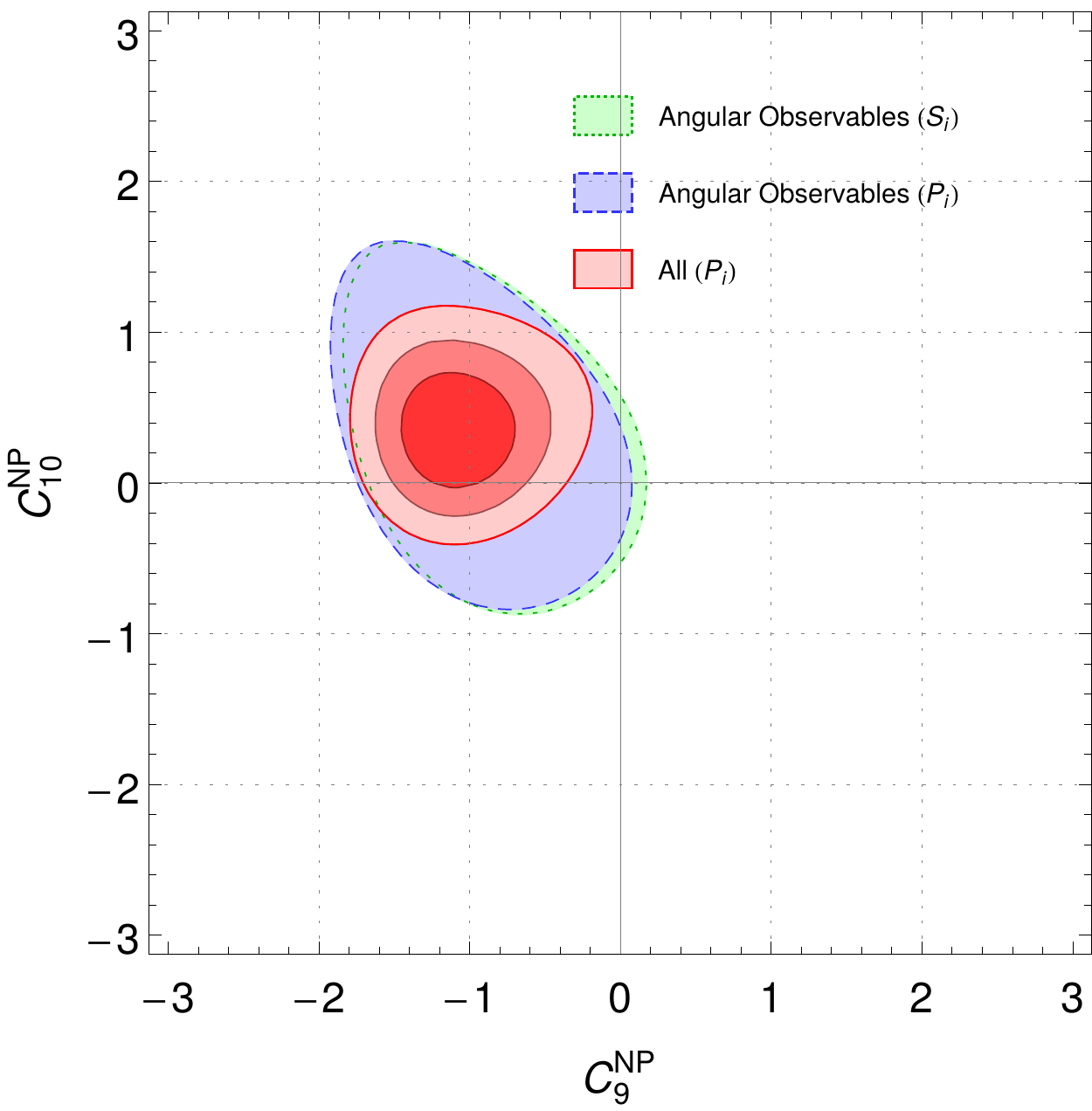}\\
\includegraphics[width=7.5cm]{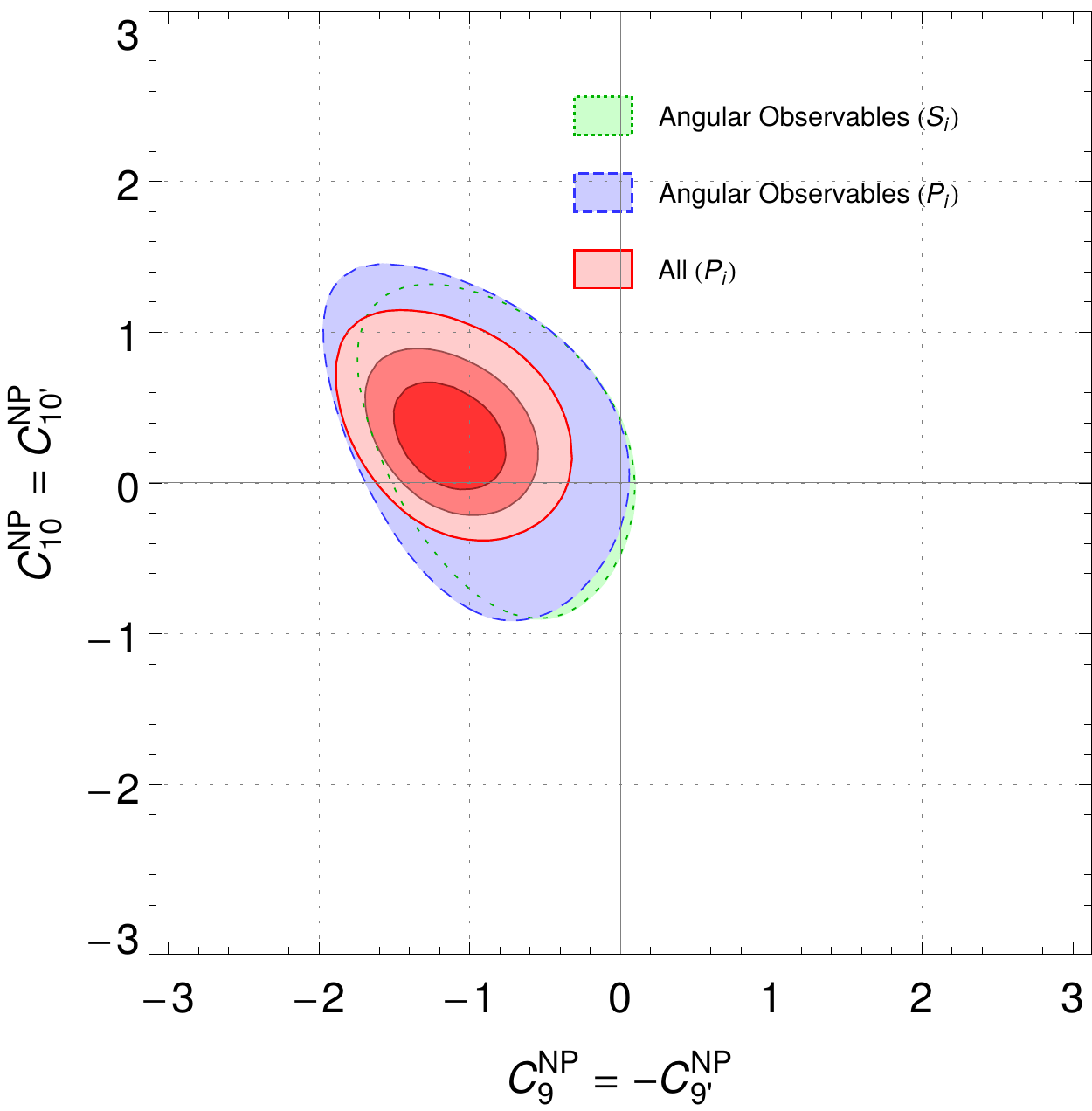} &  \includegraphics[width=7.5cm]{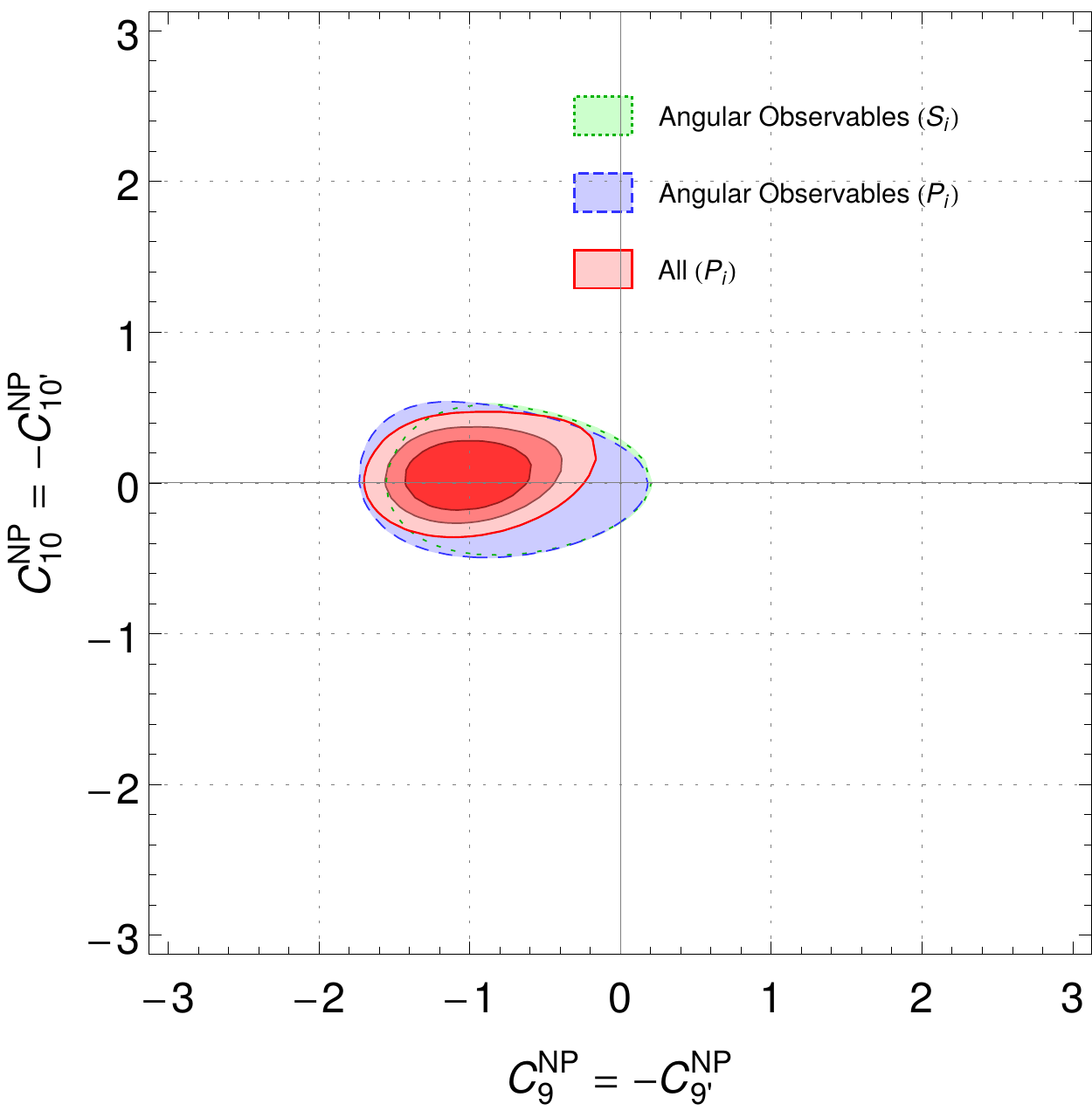}
\end{tabular}
\end{center}
\caption{\it  For 4 favoured scenarios, we show the 3~$\sigma$ regions allowed by $S_i$ angular observables for $B\to K^*\mu\mu$ and $B_s\to\phi\mu\mu$ only (dashed green), by $P_i$ angular observables for $B\to K^*\mu\mu$ and $B_s\to\phi\mu\mu$ only (long-dashed blue), and by considering all data with $P_i$ angular observables (red, with 1,2,3~$\sigma$  contours). Same conventions for the constraints as in Fig.~\ref{fig:splitBRangular}.
}\label{fig:PivsSi}
\end{figure}

\begin{figure}[!t]
\begin{center}
\begin{tabular}{cc}
\includegraphics[width=7.5cm]{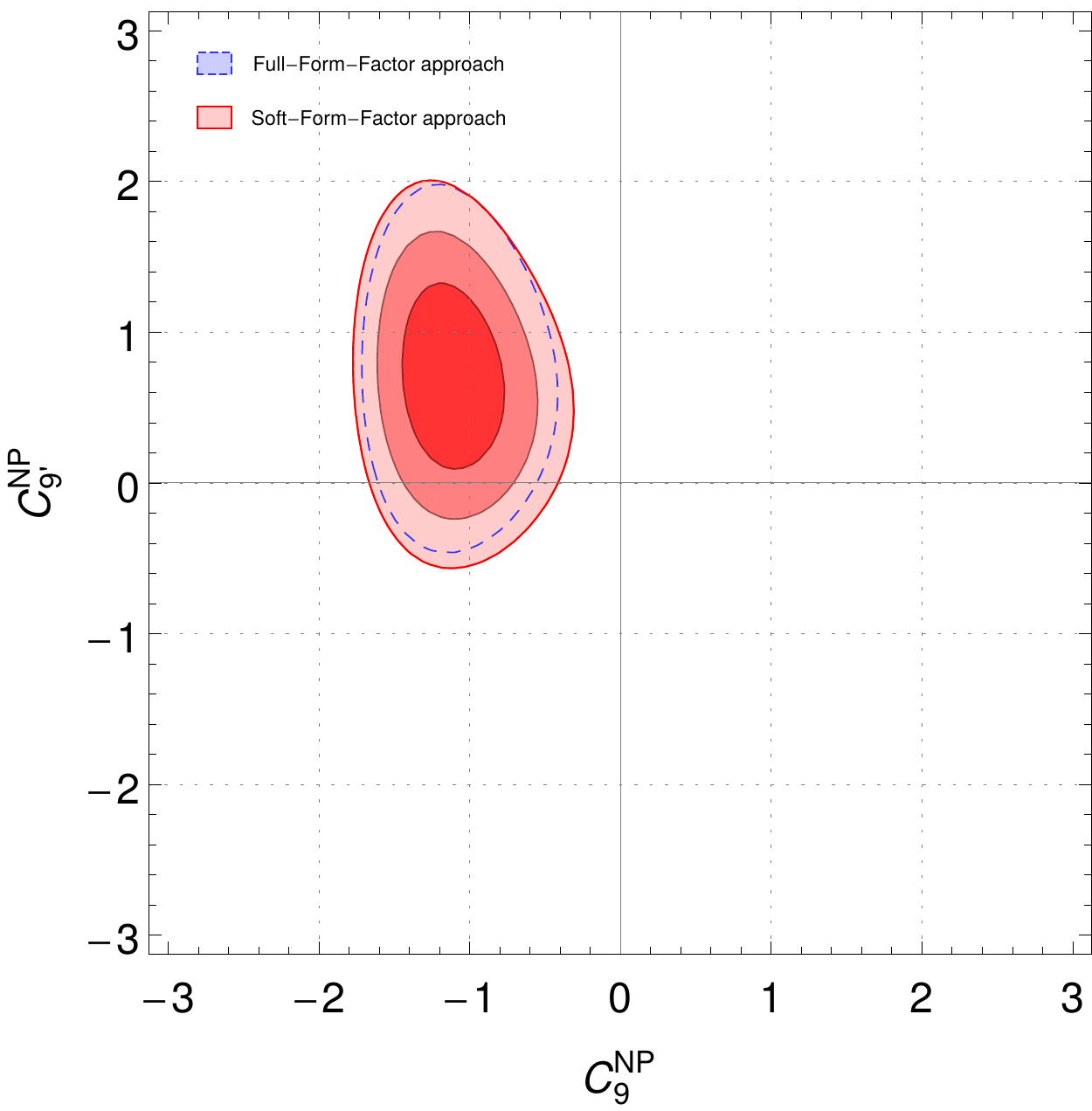} &  \includegraphics[width=7.5cm]{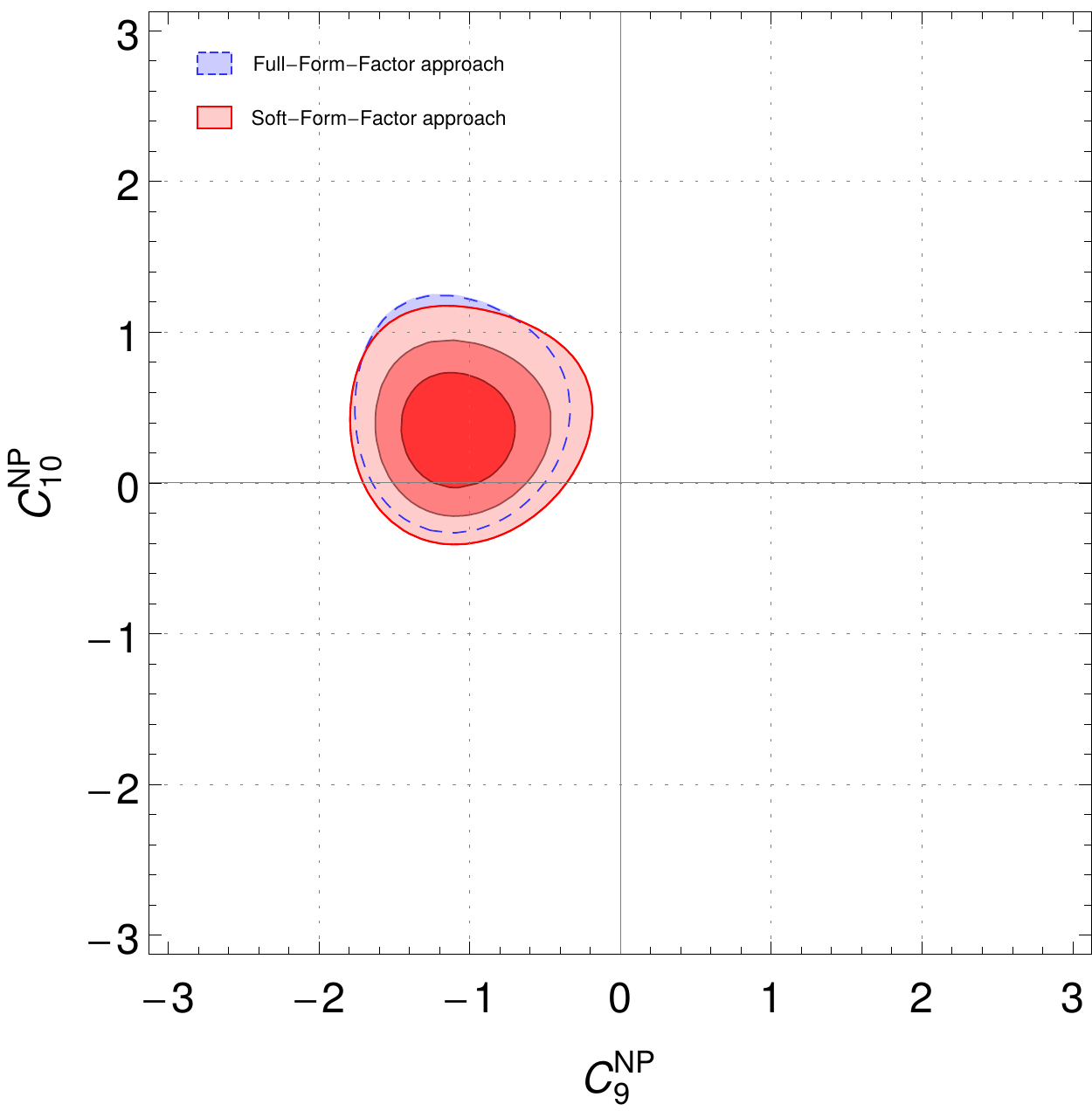}\\
\includegraphics[width=7.5cm]{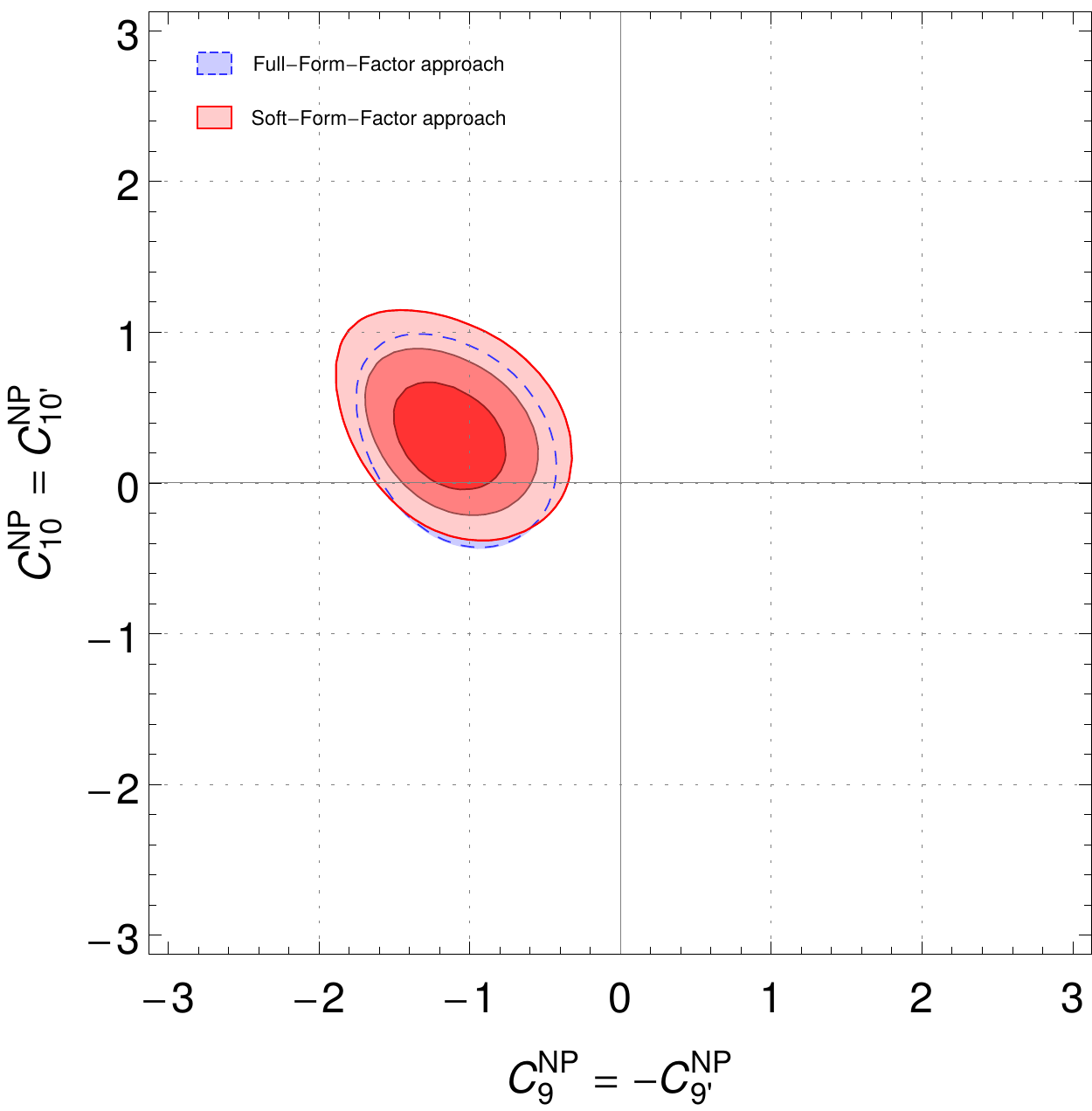} &  \includegraphics[width=7.5cm]{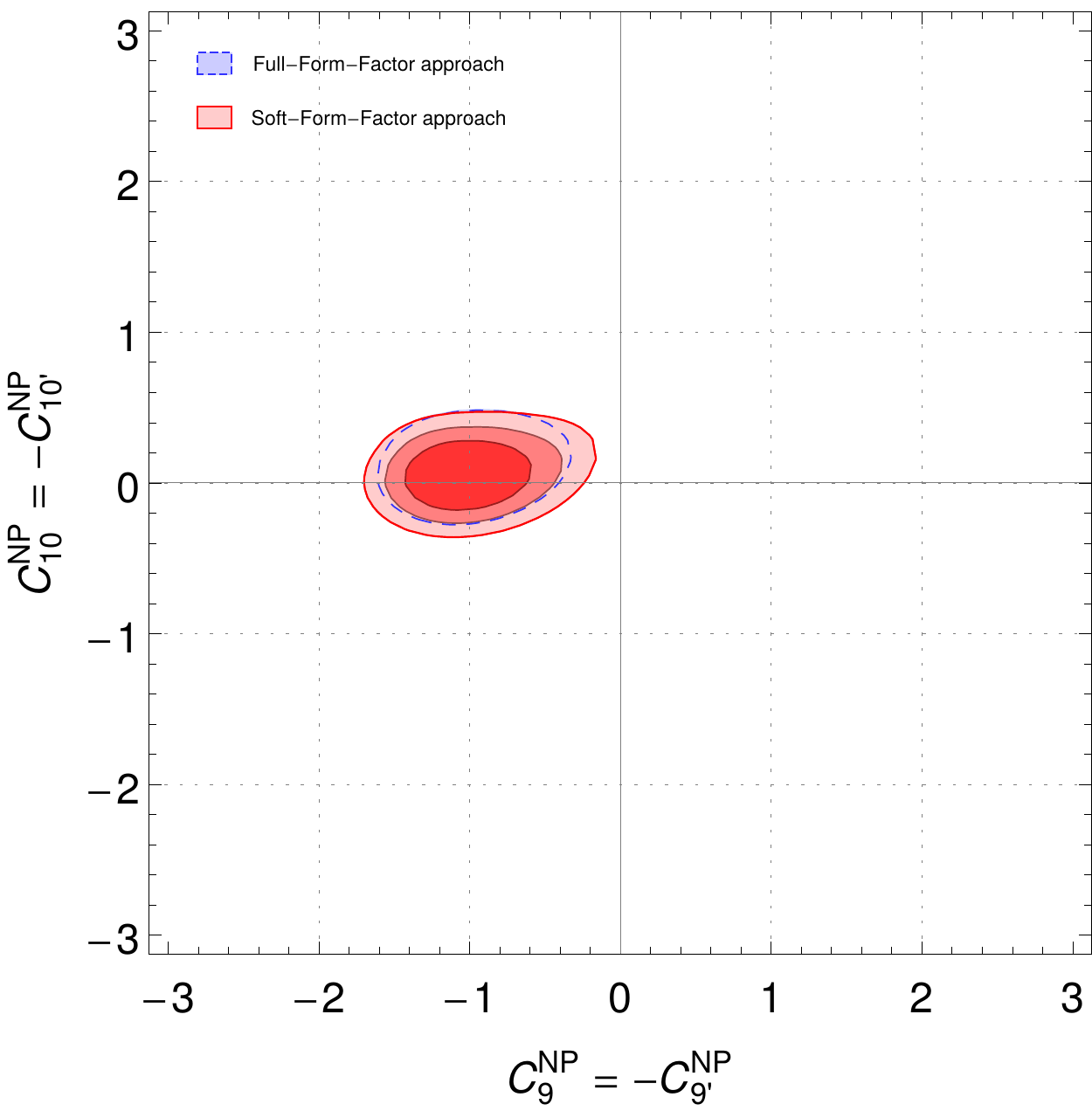}
\end{tabular}
\end{center}
\caption{\it For 4 favoured scenarios, we show the 3~$\sigma$ regions allowed using form factors in Ref.~\cite{Straub:2015ica} in the full form factor approach  (long-dashed blue) compared to our reference fit with the soft form factor approach (red, with 1,2,3~$\sigma$  contours). Same conventions for the constraints as in Fig.~\ref{fig:splitBRangular}.}\label{fig:formfactorapproach}
\end{figure}

\subsection{Role of the form factors}

Predictions for $B\to K^*\mu\mu$ observables depend on seven hadronic form factors whose calculation via non-perturbative methods like light-cone sum rules (LCSR) suffers from relatively large uncertainties (typically $\sim 20-50\%$). It is thus natural to raise the question whether an underestimation of the form factor uncertainties could be the origin of the observed anomaly \cite{Jager:2012uw}. There are two different issues to be distinguished, namely on one hand the overall size of the form factor uncertainties, and on the other hand the correlations among the errors of the different form factors.

\begin{figure}[!t]
\begin{center}
\begin{tabular}{cc}
\includegraphics[width=7.5cm]{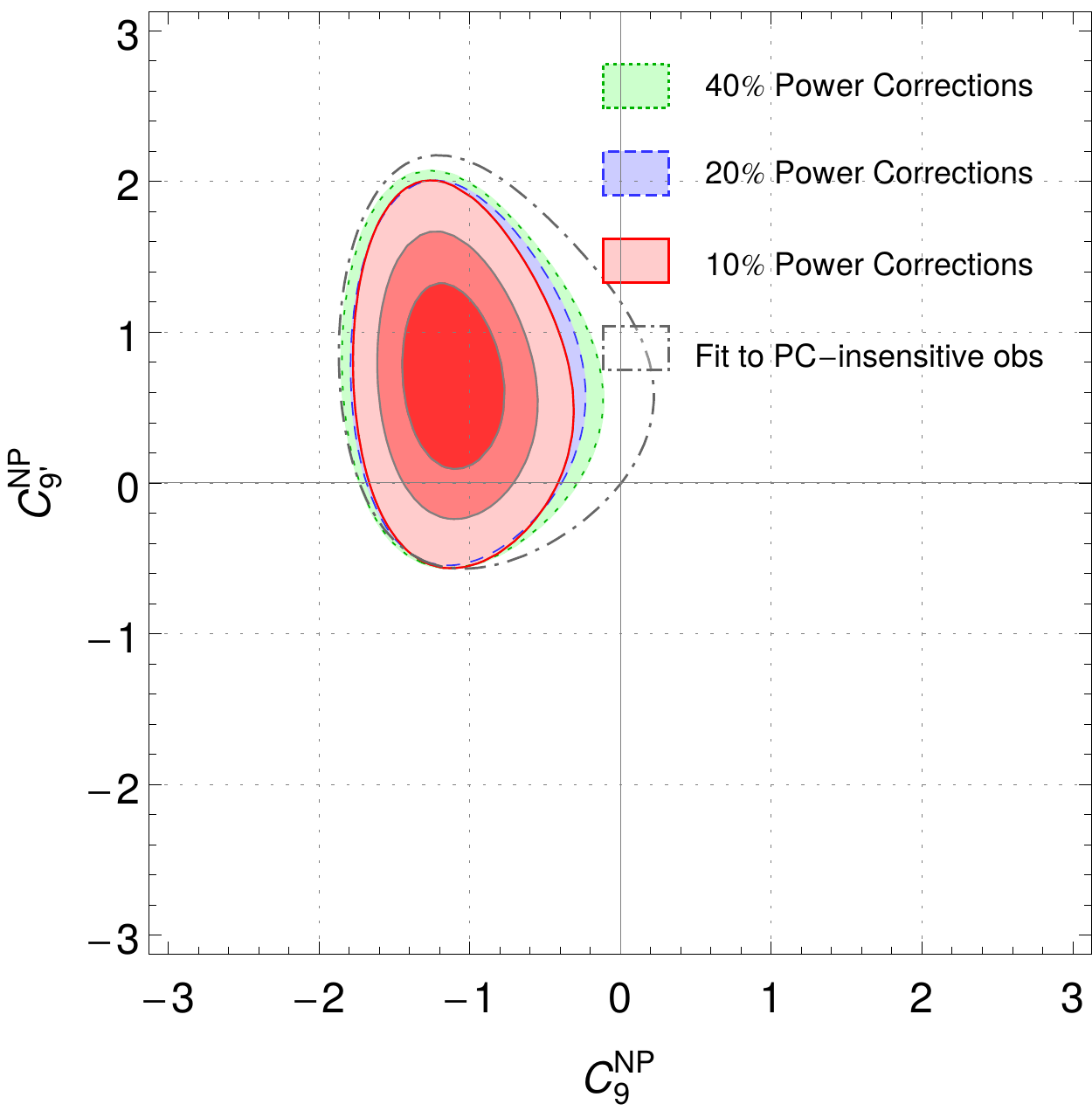} &  \includegraphics[width=7.5cm]{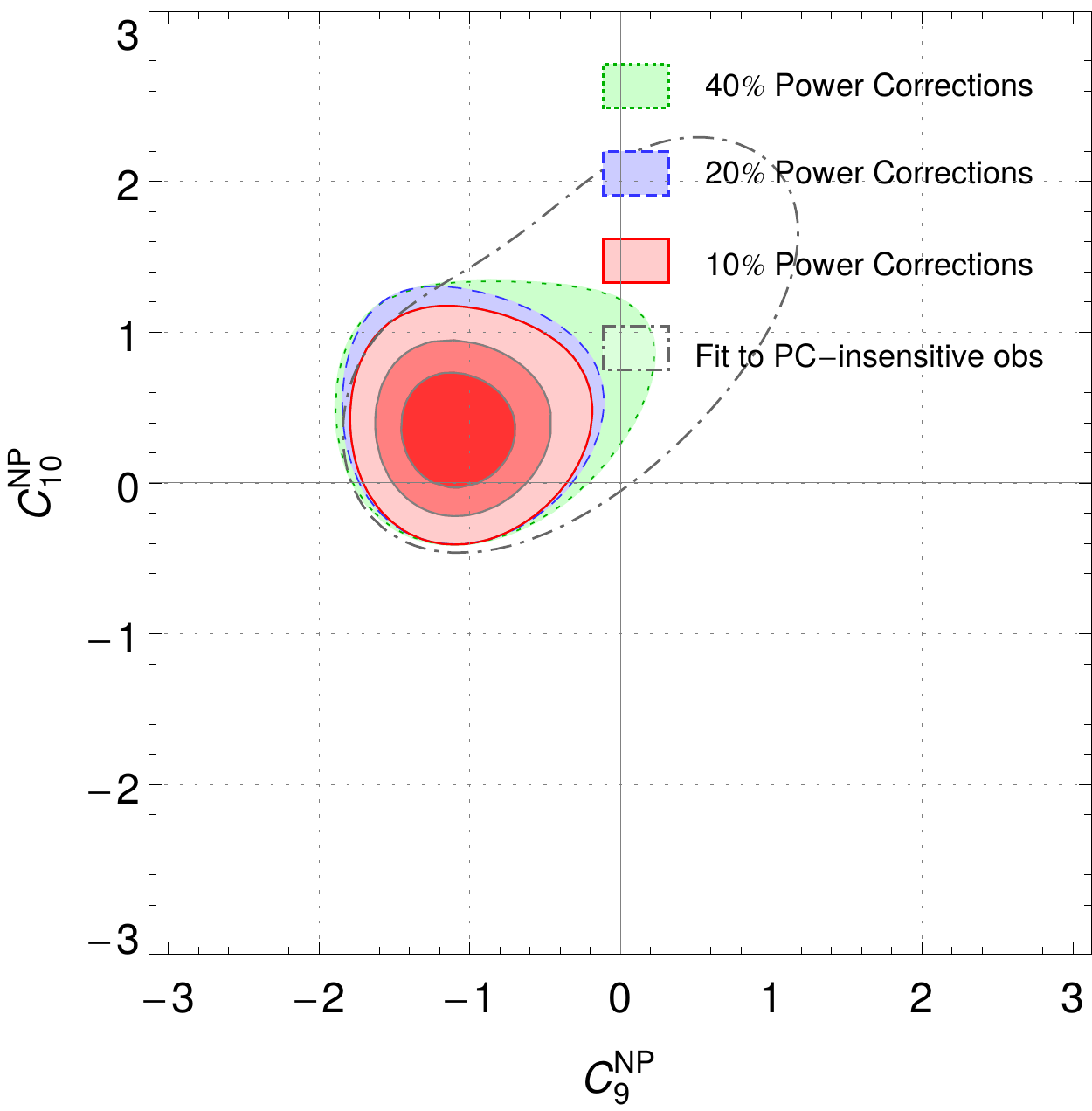}\\
\includegraphics[width=7.5cm]{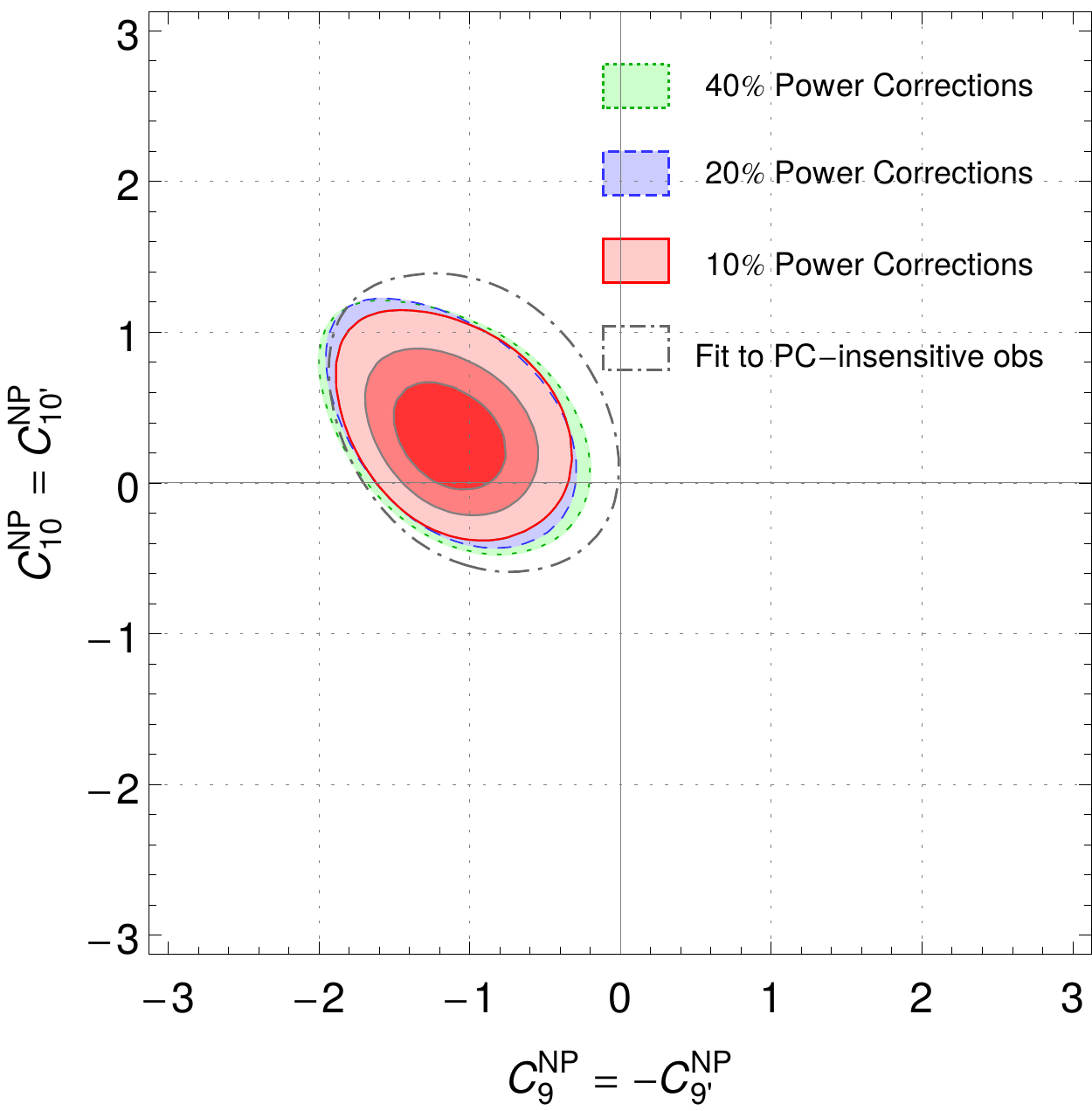} &  \includegraphics[width=7.5cm]{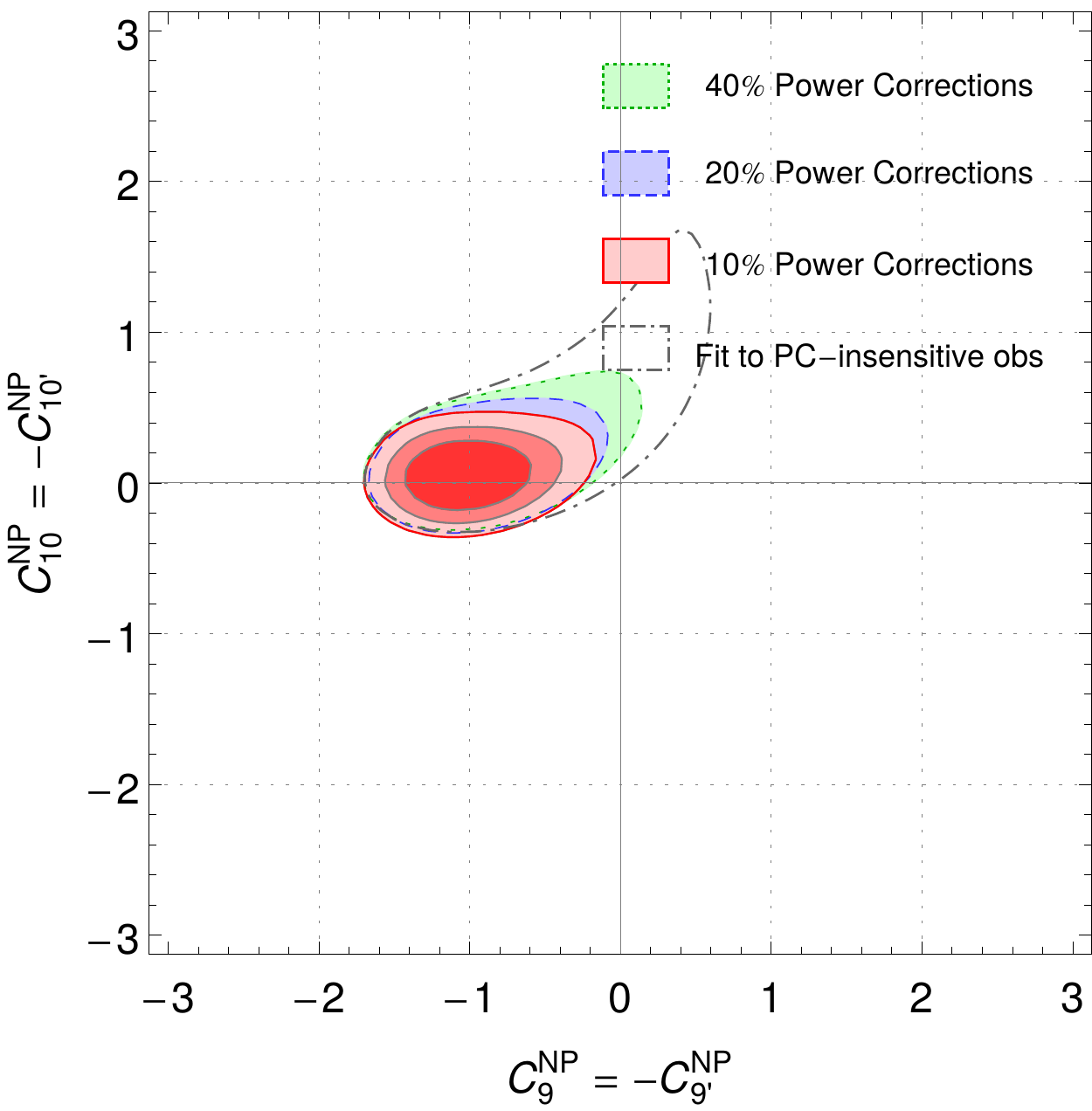}
\end{tabular}
\end{center}
\caption{\it For our 4 favoured scenarios, we show the 3~$\sigma$ regions allowed assuming 40~\% power corrections
(dashed green), 20~\% (long-dashed blue) and 10~\% (red, with 1,2,3~$\sigma$  contours).
We also show the 3~$\sigma$ region obtained from the fit to power-correction-insensitive observables (mostly low recoil),
which would correspond to the limiting fit with completely arbitrary power corrections.
Same conventions for the constraints as in Fig.~\ref{fig:splitBRangular}.}\label{fig:powercorr}
\end{figure}

\subsubsection{Overall size of uncertainties}

Let us first stress that the overall size of the form factor uncertainties has a minor impact on global fits,
and in the case of clean observables $P_i^{(\prime)}$ even on the predictions for individual observables.
The reason is that assuming a precise knowledge of the correlations among the form factors, they cancel at leading
order in the construction of the observables $P_i^{(\prime)}$ reducing the impact of their errors to a
next-to-leading-order effect $\mathcal{O}(\alpha_s,\Lambda/m_B)$. For the observables $S_i$ this effect only occurs
in a global fit where the correlation between \textit{different} observables effectively reduces the sensitivity to
the form factors, while individual $S_i$ observables display a form factor dependence at leading order.
Note that the size of the form factor errors entering our analysis is much more conservative than what is typically
assumed in other analyses~\cite{Altmannshofer:2014rta,Jager:2014rwa,Hurth:2016fbr}, as we are taking form factors from
Ref.~\cite{Khodjamirian:2010vf} where particularly large errors are assigned. In Ref.~\cite{Jager:2014rwa} the error
of the normalisation of the soft form factor $\xi_\perp(0)=0.31\pm 0.04$ is determined by considering the spread of
the central values of various different non-perturbative form factor calculations like light-cone sum
rules~\cite{Ball:2004rg,Khodjamirian:2010vf} and Dyson-Schwinger equations~\cite{Ivanov:2007cw}. This has to be compared
with our value $\xi_\perp(0)=0.31^{+0.20}_{-0.10}$ that has an error band exceeding by far the one in
Ref.~\cite{Jager:2014rwa}, implying that it covers the form factor values that would be obtained by the other
methods~\cite{Ball:2004rg,Ivanov:2007cw}.

We performed various tests on the sensitivity of our results to the choice of form factors. First,
we checked the dependence on the choice of form factors for the observables that are most sensitive to the form factors,
namely the branching ratios, in the Standard Model case. To this end we have compared our prediction for $BR(B \to K^*\mu\mu)$ using the $B$-meson LCSR determination (KMPW~\cite{Khodjamirian:2010vf}) with other predictions available in the literature based on a different form factor determination (BSZ~\cite{Straub:2015ica}). We found a good agreement at the 1~$\sigma$ level for the different bins we compared, while for the total $BR(B \to K^*\mu\mu)$ the agreement is stronger (below the 1~$\sigma$ level). In the case of $B \to K\mu\mu$ we observe a systematic difference in the branching ratio at the order of 30$\%$ compared to Ref.~\cite{Altmannshofer:2014rta}, which entirely stems from the difference between the set of form factors chosen (KMPW versus BSZ) and illustrates the sensitivity of these observables to the set of form factors considered.

To demonstrate the limited role of the size of the form factor uncertainties in a global analysis, one can trade the optimised angular observables $P_i$ for the CP-averaged angular observables $S_i$~\cite{Altmannshofer:2008dz} which are known to be more sensitive to form factor inputs~\cite{Descotes-Genon:2013vna}. The comparison presented in Fig.~\ref{fig:PivsSi} shows that the outcome of the fit is very similar in both cases, which is owed to the correlations among
the seven form factors restored via the approximate large-recoil symmetries (see below) and reducing the sensitivity to the overall size of uncertainties. We observe a systematic albeit small increase in significance of around 0.3~$\sigma$ when $P_i$ observables are used compared to  $S_i$ observables.

\subsubsection{Correlations}

The correlations among the different form factors can in principle be extracted from the corresponding calculation,
as it has been done for example in Ref.~\cite{Straub:2015ica}. On the other hand, the dominant correlations can also
be assessed from first principles relying on symmetry relations fulfilled by the form factors at low $q^2$. While this
second approach is more general and avoids any dependence on the details of a particular non-perturbative calculation,
it provides the correlations only up to symmetry-breaking corrections of the order $\Lambda/m_b$ (factorisable power
corrections). In our analysis we explicitly introduce these symmetry-breaking corrections by hand and assign to them
an error of the order of $10\%$ of the respective full form factor, corresponding to a $100\%$ error of the factorisable
power corrections. We could confirm that this assumption of $10\%$ power corrections is a realistic estimate by determining
the central values for the power corrections from a fit to a particular non-perturbative calculation: it has been done in 
Ref.~\cite{Descotes-Genon:2014uoa} for the $B\to K^*$ form factors in Refs.~\cite{Ball:2004rg} and
\cite{Khodjamirian:2010vf} and in App.~\ref{app:PowCorr} for $B_s\to\phi$ and $B\to K$ . 

We illustrate the compatibility of the two approaches at the level of the global fit analysis in Fig.~\ref{fig:formfactorapproach}. We compare the results of our reference fit, performed applying the soft-form factor approach based on the large-recoil symmetries described in Sec.~\ref{sec:BtoKstarmumu-gen} and using mainly the $B$-meson LCSR results of Ref.~\cite{Khodjamirian:2010vf}, with the full-form factor approach applied to the light-meson LCSR results of Ref.~\cite{Straub:2015ica} (including correlations, similarly to Ref.~\cite{Altmannshofer:2013foa,Altmannshofer:2014rta}). We see that both sets of results are very similar, even though 
in the soft-form factor approach we started from a set of form factors with larger uncertainties and no knowledge of correlations. This highlights the advantages of the soft-form factor approach to restore correlations among form factors. Not surprisingly, the full-form factor approach based on Ref.~\cite{Straub:2015ica} is more constraining than our soft-form factor approach based on
 the results of Ref.~\cite{Khodjamirian:2010vf}, which exhibits much larger uncertainties for the form factor parameters.

For the reasons mentioned above, our SM predictions as well as our fit results are in good agreement with Ref.~\cite{Altmannshofer:2014rta}. It is thus surprising that the authors of Ref.~\cite{Jager:2014rwa} find much larger errors from factorisable power corrections. This necessarily implies that they must implicitly have introduced a much stronger breaking of the large-recoil symmetry relations, in contradiction to expectations from dimensional arguments as well as to the explicit results for the particular LCSR calculation~\cite{Straub:2015ica}. In other words, the results in Ref.~\cite{Jager:2014rwa} taken at face value should imply that the recent LCSR estimates performed in Ref.~\cite{Straub:2015ica} are not correct.

One may wonder how big the large-recoil symmetry breaking effects (i.e. the factorisable power corrections)
should be in order to produce a similar pattern of deviations as observed in the data.
In order to study this, we performed the test of assuming twice or four times larger power corrections
(corresponding to 20\% or 40\% of the corresponding full form factors).
The results in Fig.~\ref{fig:powercorr} show that the factorisable power corrections only play a minor role in
the uncertainties and the outcome for our reference fit.
When power corrections are increased from 10\% to 40\%, the fit is more and more driven by observables with no
sensitivity to power corrections, such as low-recoil observables. Indeed, one can see that the shape of the $3\,\sigma$
regions in Fig.~\ref{fig:powercorr} evolves into the low-recoil regions shown in Fig.~\ref{fig:largevslow}
as the size of power corrections increases, which are reproduced in Fig.~\ref{fig:powercorr} to ease the comparison.

\begin{figure}[!t]
\begin{center}
\includegraphics[width=9.5cm]{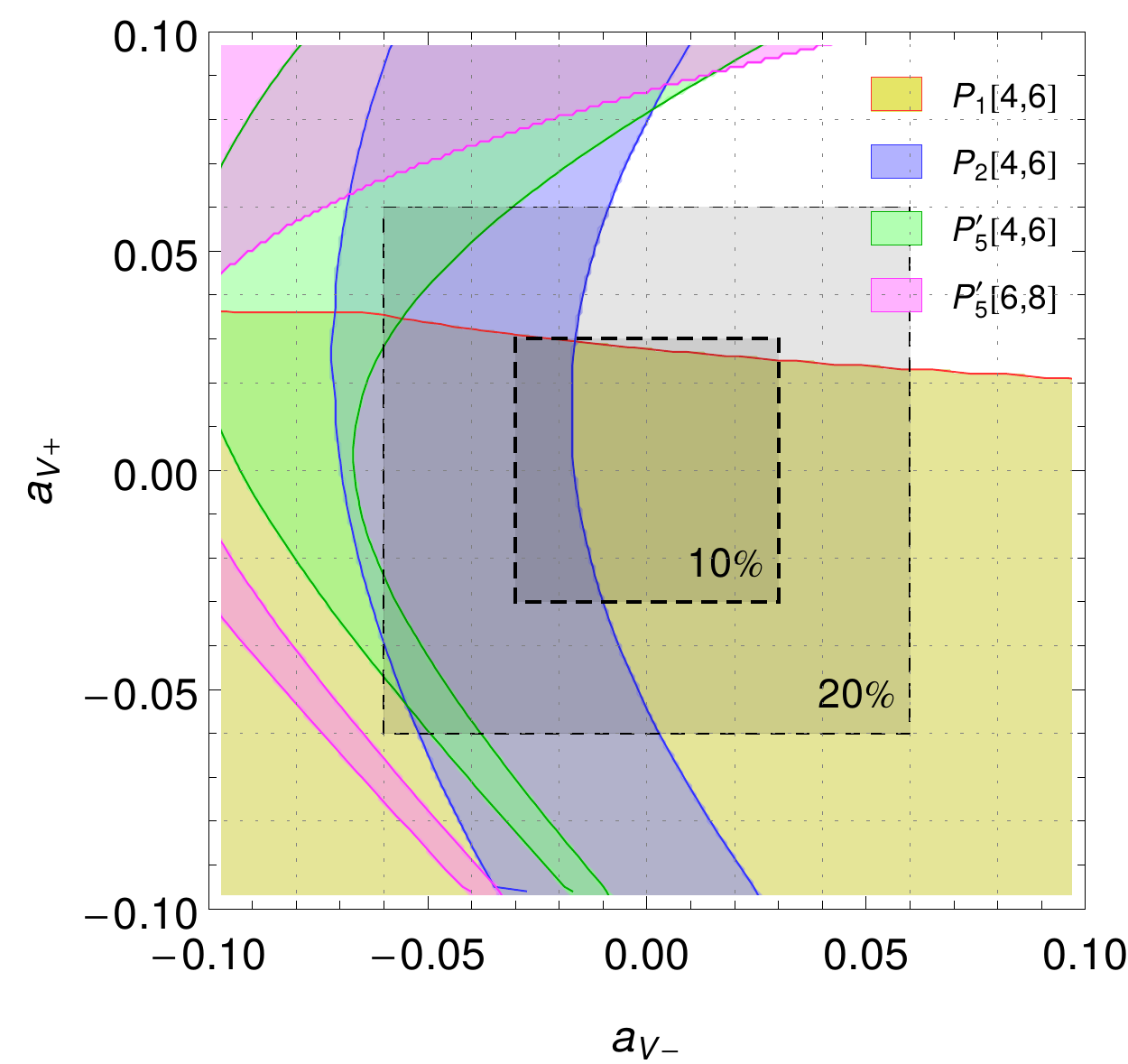}
\end{center}
\caption{\it Power corrections $a_{V-}$ and $a_{V+}$ needed to obtain agreement between SM predictions and experiment at  1~$\sigma$, considering different observables. This plot illustrates that $a_{V\pm}$ can indeed be used to obtain agreement between SM prediction and experiment in one observable, but correlations hinder a similar agreement when a larger set of observables is considered.}
\label{fig:powercorrjager}
\end{figure}

If one wants to solve the anomalies exhibited in $b\to s\mu\mu$ processes
through power corrections, it is important not to focus on one single observable, like $P_5^\prime$, alone but on the full set. Since power corrections enter many observables, trying to adjust them to fix one observable may generate a problem in another one.  The effect of correlations is illustrated in Fig.~\ref{fig:powercorrjager}, inspired by Fig.~5 of Ref.~\cite{Jager:2014rwa}. For comparison we work here in the soft-form factor scheme employed in Ref.~\cite{Jager:2014rwa} with the soft form factors defined from the full form factors $T_1$ and $A_0$. We also switch to the helicity basis used in Ref.~\cite{Jager:2014rwa} where for example the helicity form factors $V_{\pm}$ are defined in terms of the transversity form factors $V$ and $A_1$ as 
\begin{equation} V_{\pm}=\frac{1}{2} \left( \left(1+\frac{m_K^*}{m_B} \right) A_1 \mp \frac{\sqrt{\lambda}}{m_B (m_B + m_K^*)} V \right)\,. \end{equation}
For the constant terms $a_{V+}$ and $a_{V-}$ in the series of power corrections for the form factors $V_{\pm}$ this implies
\begin{equation} 
a_{V\pm}= \frac{1}{2} \left( \left(1 + \frac{m_K^*}{m_B} \right) a_1 \mp \left(1 - \frac{m_K^*}{m_B} \right) a_V \right)\,.
\end{equation}

Fig.~\ref{fig:powercorrjager} shows that power corrections can explain the data for $\av{P_5^\prime}_{[4,6]}$ within the 1~$\sigma$ range if they occur at the level of 20$\%$ for both $a_{V+}$ and $a_{V-}$ in the combination $a_{V+}-a_{V-} \propto a_V$. In a scheme where the soft form factor $\xi_\perp$  is defined from the full form factor $V$~\cite{Descotes-Genon:2014uoa}, such power corrections are absorbed into $\xi_\perp$.  Fig.~\ref{fig:powercorrjager} displays also the power corrections needed for the observables $\av{P_5^\prime}_{[6,8]}$, $\av{P_2}_{[4,6]}$ and $\av{P_1}_{[4,6]}$. Comparing the region for $\av{P_5^\prime}_{[4,6]}$ and $\av{P_1}_{[4,6]}$  one notices that a solution for $\av{P_5^\prime}_{[4,6]}$ through large power corrections moves $\av{P_1}_{[4,6]}$ away from the measured value. An explanation of all three observables within the SM in terms of power corrections requires to reach the limit of  the 20$\%$ region. Only marginal agreement is obtained then, and once $\av{P_5^\prime}_{[6,8]}$ is added to the list, no common solution is found even for power corrections beyond 20$\%$. 

This situation seems to be in contradiction with Fig.~5 of Ref.~\cite{Jager:2014rwa}. Note, however, that the $(a_{V+},a_{V-})$ profile shown there corresponds to
a scenario where all other power correction parameters have been fixed in such a way to describe best the experimental data, without specifying their presumably quite large values. In fact, already the point $a_{V+}=a_{V-}=0$ is in agreement with data nearly at the 1~$\sigma$ level, even though the power correction parameters $a_{V+}$ and $a_{V-}$ are the most relevant ones for the observable $P_5^\prime$. It is further irritating that, while it is claimed that power correction parameters are scanned only in a range of $\pm 10\%$ of the soft-form factor value, for the plot a region covering $|a_{V\pm}|\le 0.2$ has been chosen, corresponding to power corrections of order $\pm 66\%$.

\subsection{Role of long-distance charm corrections}\label{sec:charm}

Another frequent attempt to explain the $B\to K^*\mu\mu$ anomaly consists in assuming a very large charm-loop
contribution (or non-factorizable power correction). 
It is not difficult to imagine that with a sufficiently general $q^2$-dependent parametrisation one might easily
fit any data~\cite{Ciuchini:2015qxb}. However, one must check that the parametrisation itself and the resulting fit
respect all known properties of the related charm-loop correlator, as well as its behaviour at large recoil.
In the end, the assumption that the charm contribution is responsible for the anomalies leads to two kinds of predictions:
first, those arising from correlations that might survive among  observables under the most general parametrisation of the
correlator (which would give little information on $\C9$), and second, the prediction that $R_K = 1$ due to SM
lepton-flavour universality. Whatever explanation might be first assumed concerning $b\to s\mu\mu$ transitions, one
has still to invoke a NP contribution to explain $R_K^\text{LHCb}\simeq 0.75$, most plausibly
in the form of a non-SM contribution to $\C{9\mu}$. But once such NP contribution has been introduced, the other
$b\to s$ anomalies are reduced and there is no actual need for abnormally large non-perturbative hadronic effects on top
of the NP contribution.

We would like to stress that explaining some of the anomalies through such large charm contributions leads to further
difficulties. First of all, these explanations predict an enhanced effect when one gets closer to the $J/\psi$ peak.
They typically lead to an increase for the observable $P_5^\prime$ in the  [6,8] region with respect to [4,6]
(see Fig.12 of Ref.~\cite{Lyon:2014hpa} and `prediction column' of Table 6 of Ref.~\cite{Ciuchini:2015qxb}
for $P_5^\prime$). But current data does not seem to follow this behaviour: an increase in statistics in this particular
region will be very important to settle this point definitely. Another important issue comes from the comparison between
low- and large-recoil regions: the charm effects advocated in Refs.~\cite{Lyon:2014hpa,Ciuchini:2015qxb} to explain the
current data at large recoil within the SM do not provide any clue about the deviations observed in $B\to K^*\mu\mu$ and
$B_s\to\phi\mu\mu$ branching ratios at low recoil, whereas a single short-distance contribution to $\C{9}$ is able to
accommodate the deviations in both regions simultaneously. 

A confirmation of the deviation measured in $R_K$ with a higher significance, as well as the measurement of other observables exhibiting lepton-flavour universality violation
(see \emph{e.g.} Ref.~\cite{Capdevila:2016ivx}) would strongly disfavour solutions involving non-perturbative charm-loop
effects such as the ones proposed in Refs.~\cite{Lyon:2014hpa,Ciuchini:2015qxb}. Conversely, a clear evidence for
a $q^2$-dependent effect, or the need for different contributions in different transversity amplitudes in $B\to K^*\mu\mu$, would lead to prefer 
 non-perturbative QCD effects rather than New Physics. However, there is no evidence for such
a dependence on $q^2$ or transversity in the present data. A further discussion of this issue can also be found in Ref.~\cite{wip}.

\subsubsection{Increasing the size of the charm contributions}

\begin{figure}[!t]
\begin{center}
\begin{tabular}{cc}
\includegraphics[width=7.5cm]{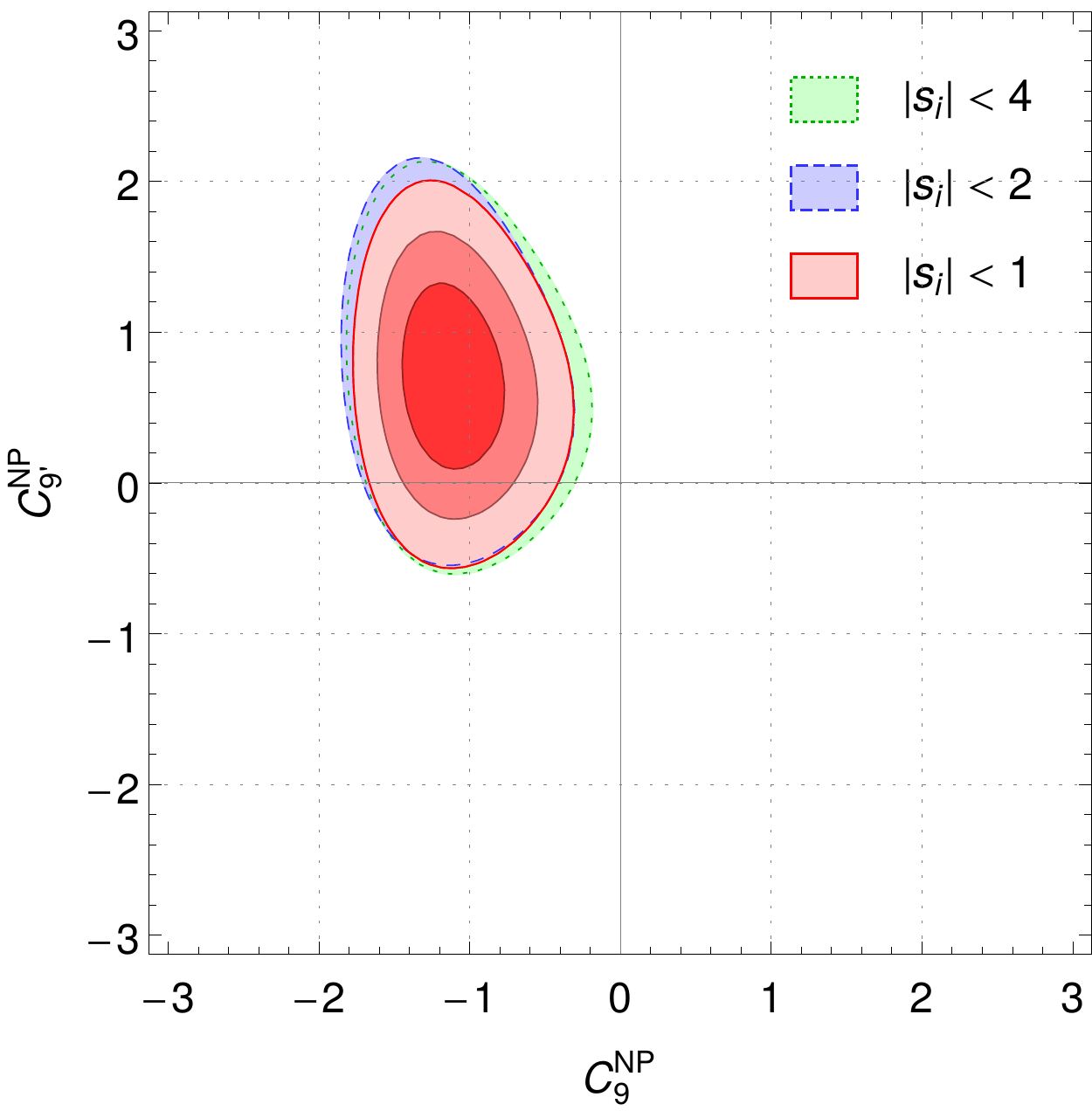} &  \includegraphics[width=7.5cm]{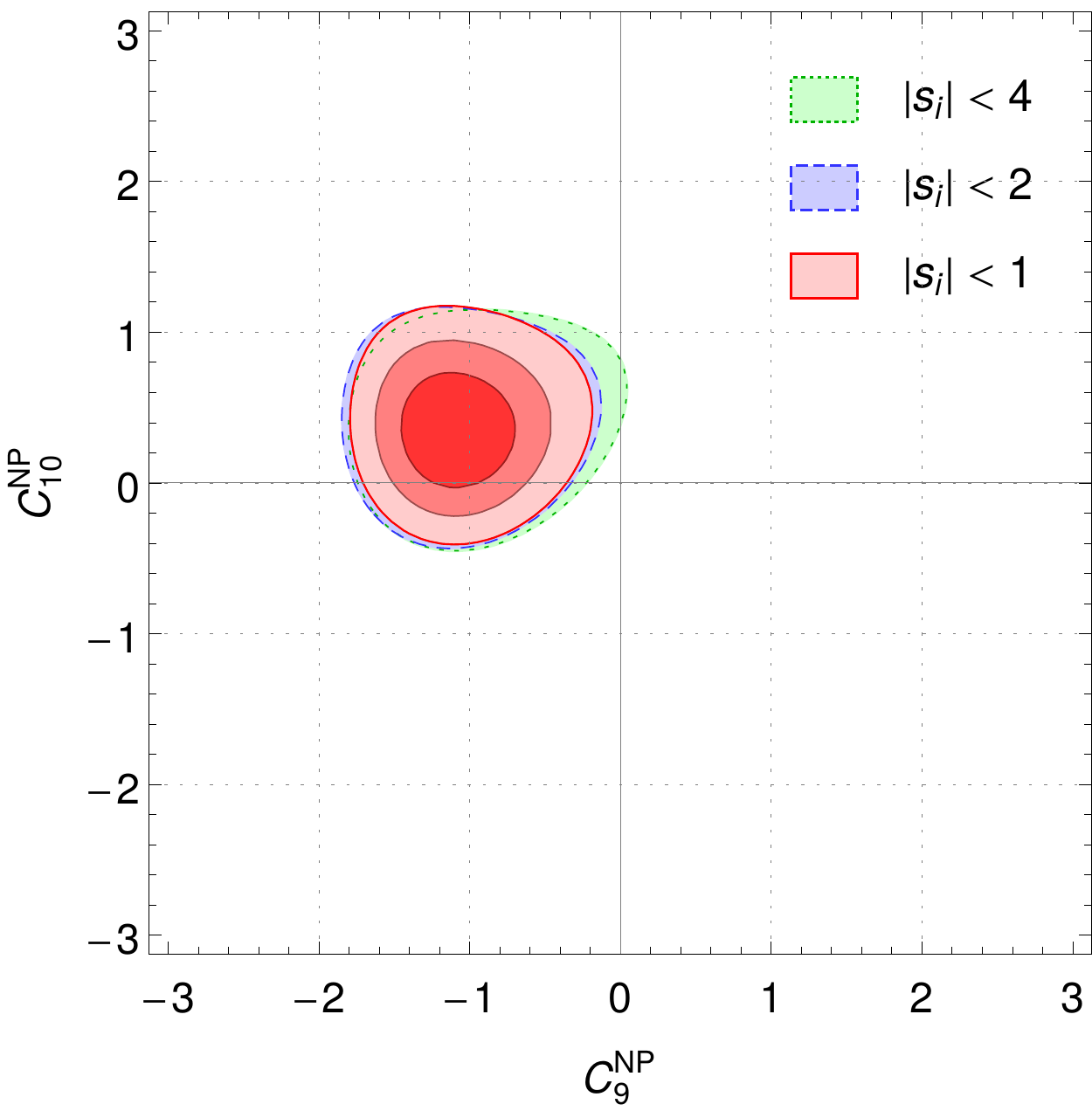}\\
\includegraphics[width=7.5cm]{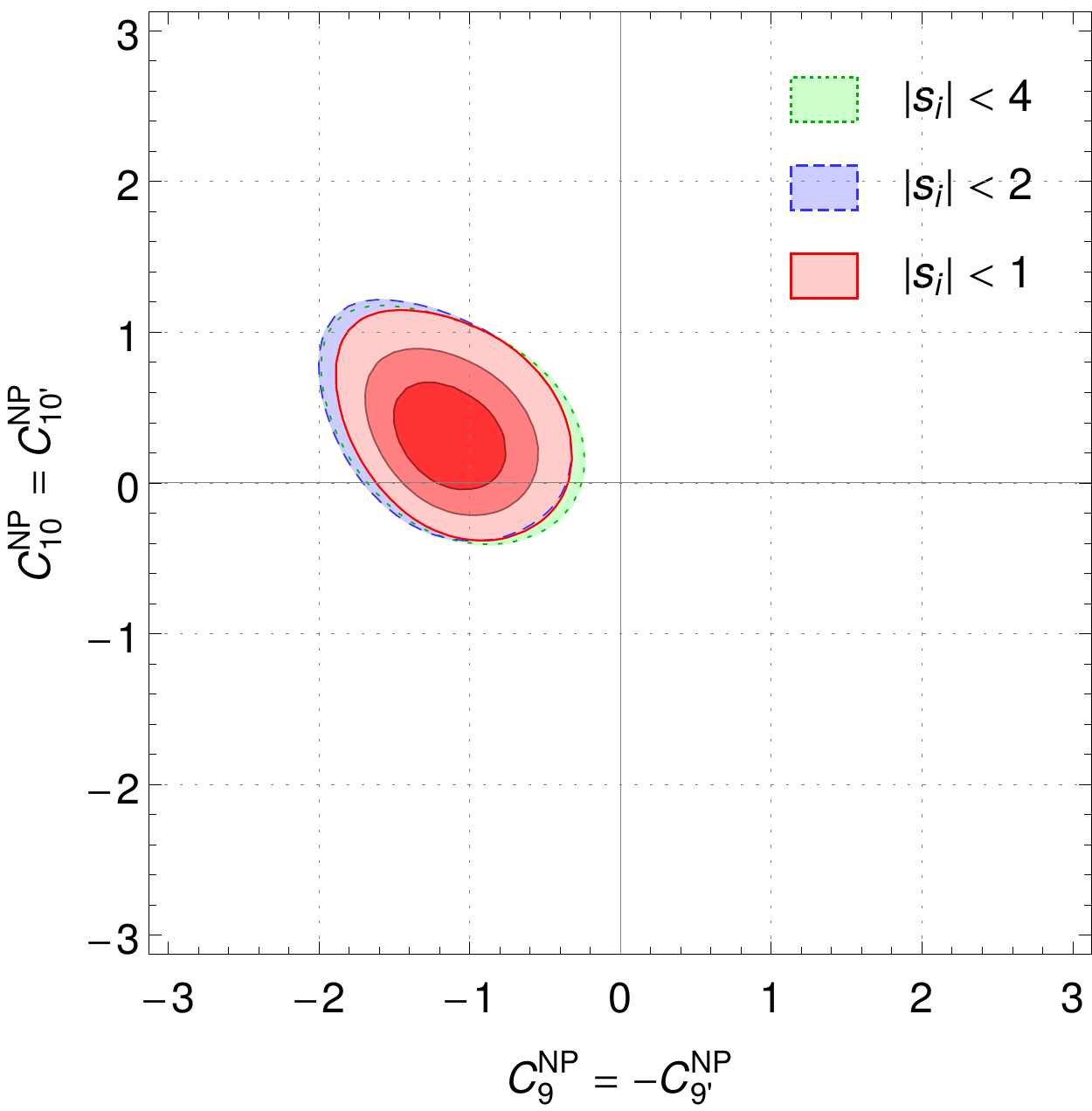} &  \includegraphics[width=7.5cm]{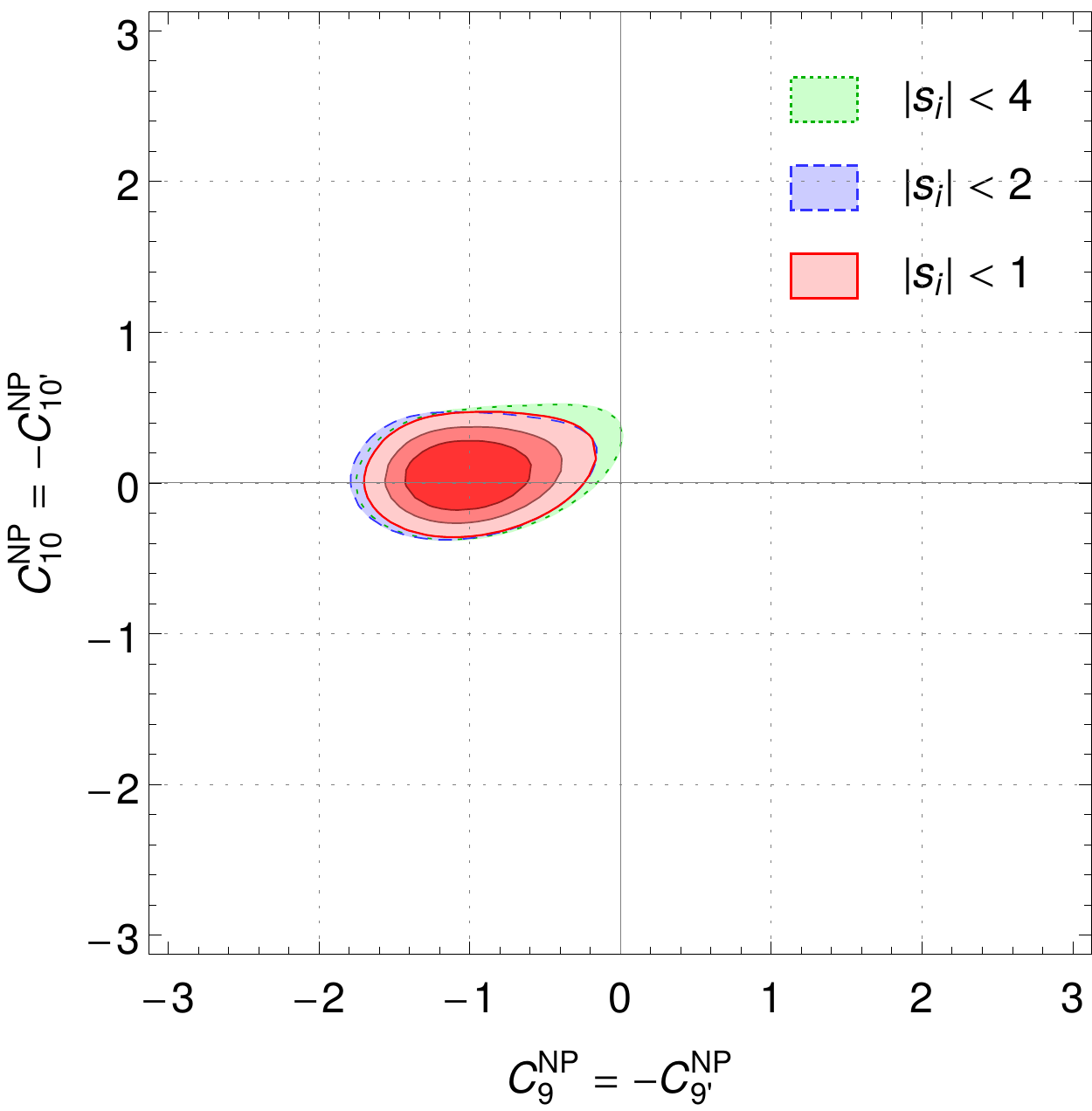}
\end{tabular}
\end{center}
\caption{\it For 4 favoured scenarios, we show the 3~$\sigma$ regions allowed assuming $s_i=0\pm 4$  (dashed green), $s_i=0\pm 2$ only (long-dashed blue) and $s_i=0\pm 1$ (red, with 1,2,3~$\sigma$  contours). Same conventions for the constraints as in Fig.~\ref{fig:splitBRangular}.}\label{fig:ccbar}
\end{figure}

Long-distance charm corrections have been subject to many recent discussions, with different estimates~\cite{Khodjamirian:2010vf,Khodjamirian:2012rm,Lyon:2014hpa}. We recalled in Sec.~\ref{sec:BtoKstarmumu-gen} that we use the work of Ref.~\cite{Khodjamirian:2010vf} as an estimate of this effect to be added on top of the perturbative contribution, but without assuming a specific sign for this contribution. In our reference fit, for each transversity amplitude of $B\to K^*\mu\mu$ and $B_s\to\phi\mu\mu$ we multiply  this contribution by $s_i=0\pm 1$ (hence six uncorrelated parameters). We present in Fig.~\ref{fig:ccbar} the corresponding results if we take contributions twice or four times larger. Increasing the size of the charm contributions reduces the significance of the deviations from Standard Model, but the discrepancy remains above 3~$\sigma$ for the various scenarios considered even if the long-distance $c\bar{c}$ contribution is multiplied by 4 compared to our reference fit.

\subsubsection{Distinguishing New Physics from charm contribution in $\C9$}

Another way of checking the robustness of our approach with respect to charm consists in determining if the fit to the
data favours an additional $q^2$-dependent contribution to $\C9$. In that case, this would be a clear indication that
some long-distance contribution has been underestimated in our analysis, as NP contributions cannot have any such
dependence. 

\begin{figure}[!t]
\begin{center}
\includegraphics[width=11.2cm]{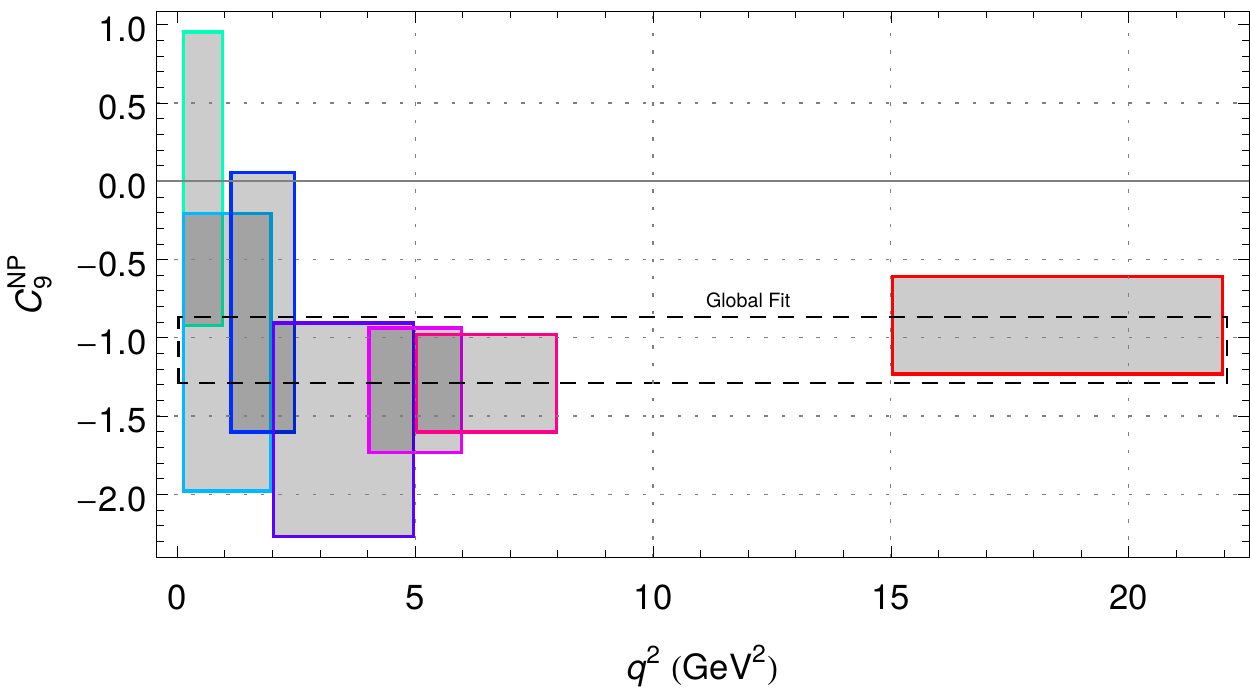}\\[3.2mm]
\includegraphics[width=11.2cm]{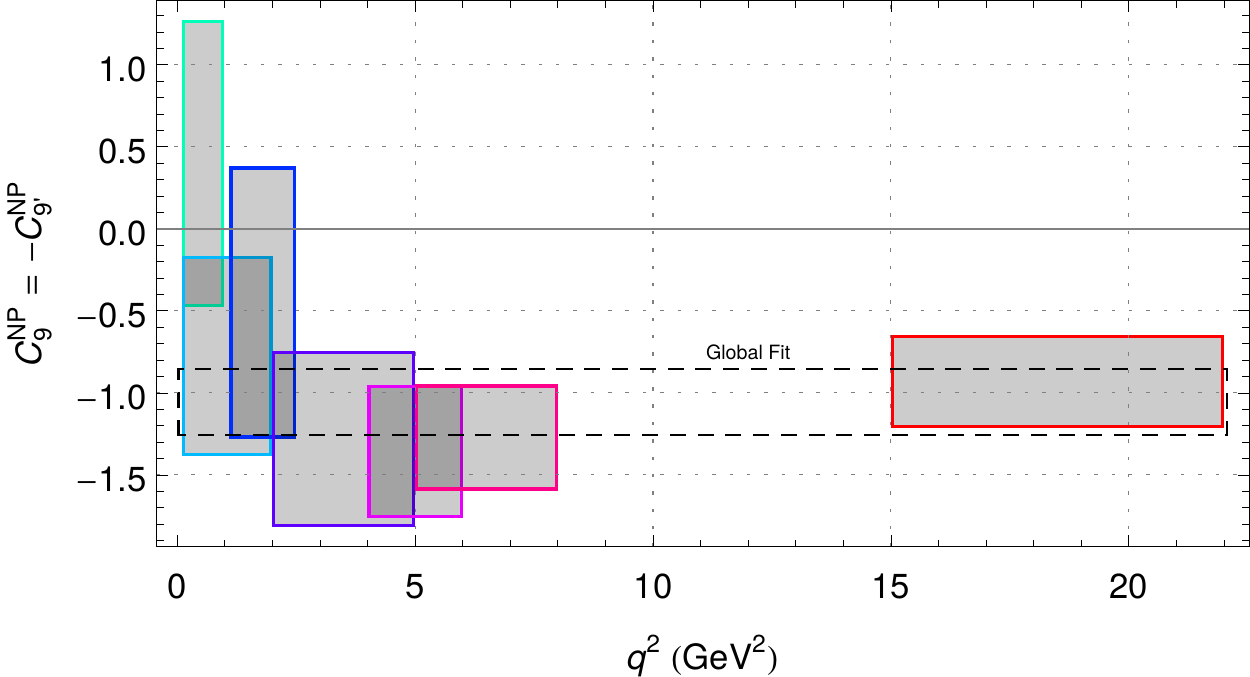}\\[3.2mm]
\includegraphics[width=11.2cm]{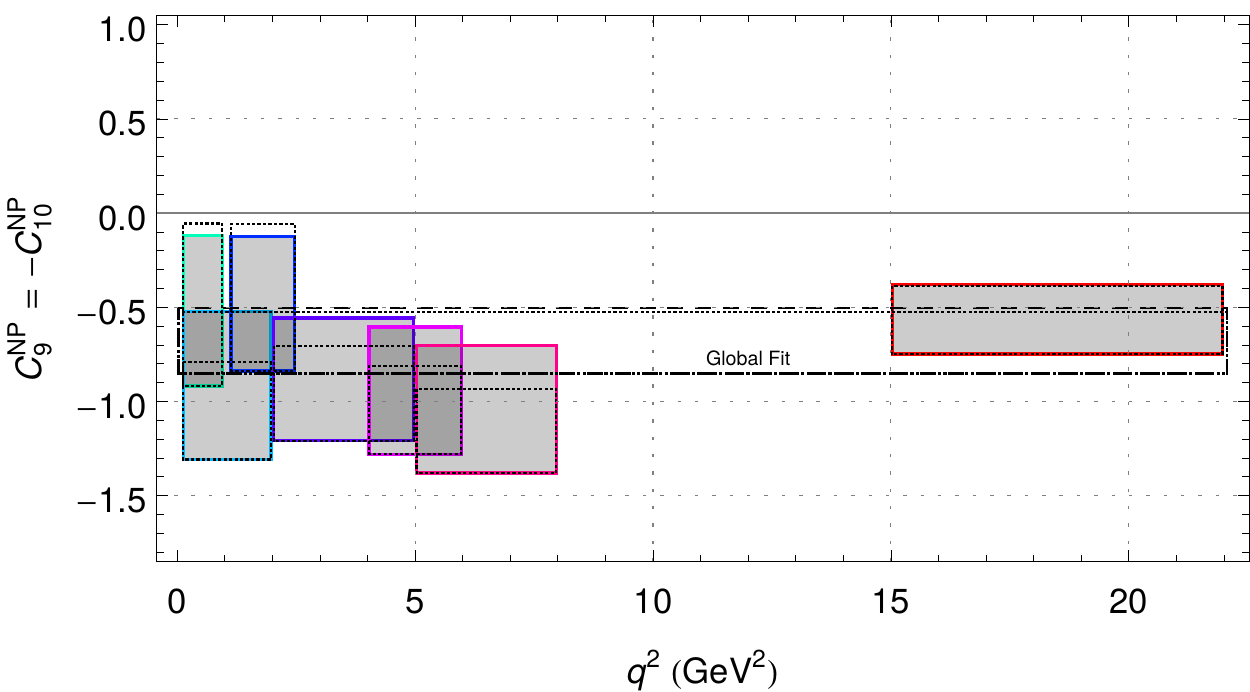}
\end{center}
\caption{\it Determination of $\C9$ from the reference fit restricted to the data available in a given $q^2$-region. We present the scenarios where NP enters $\C9$ and: all other coefficients remain SM only (top), $\C9^{\rm NP}=-\C{9'}^{\rm NP}$ (center), $\C9^{\rm NP}=-\C{10}^{\rm NP}$ (bottom). For the bottom case, the results of a fit without $B_s\to\mu\mu$ is indicated with dotted lines. See also Ref.~\cite{Altmannshofer:2015sma}.
}\label{fig:C9dep}
\end{figure}

We have performed fits to the same data as in the reference fit, but limited to particular $q^2$-ranges, in order to check the stability of the value of $\C9$ needed in different bins. 
We can perform this fit under different hypotheses: for instance, one can leave only $\C9^{\rm NP}$,
or assume that $\C9^{\rm NP}=-\C{9'}^{\rm NP}$, or that $\C9^{\rm NP}=-\C{10}^{\rm NP}$. 
An underestimated hadronic contribution from charm loop would correspond to a $q^2$-dependent contribution to $\C9$ only,  i.e., the first case. In the two other cases, the need for a $q^2$-dependent contribution  might indicate a problem of consistency in the fit that could not be understood only through a hadronic contribution.
Fig.~\ref{fig:C9dep} shows no need for a $q^2$-dependent contribution in these three situations~\footnote{In Fig.~\ref{fig:C9dep}, one should remember that the lowest bin is affected by the problems described in Sec.~\ref{sec:specificbins} and should be considered with care.}.

In the case of the last scenario $\C9^{\rm NP}=-\C{10}^{\rm NP}$ (bottom plot in Fig.~\ref{fig:C9dep}), 
the fit tries to accommodate both $B_s\to\mu\mu$ (constraining $\C{10}^{\rm NP}$ to remain small) and $B\to K^{(*)}\mu\mu$ and $B_s\to\phi\mu\mu$ observables (favouring a significant contribution to $\C{9}^{\rm NP}$). The fit exhibits thus a tension between the two types of constraints. In order to assess this effect, we have performed a fit without $B_s\to\mu\mu$. The result, indicated with dotted lines, favours lower values of $\C9^{\rm NP}=-\C{10}^{\rm NP}$, within a narrower range, without spoiling the good agreement between the global and bin-by-bin analyses.

As an alternative test, we added three $q^2$-dependent contributions to $\C9^{SM}$ of the form $\C{9,p}^{had}(s)=A_p+B_p\times s$ for $p=K,K^*,\phi$. We assumed that the same contribution entered the three transversity amplitudes identically for $B\to K^*\mu\mu$ (we assumed the same in the case of $B_s\to\phi\mu\mu$). A 6D fit to the real parameters $A_{K,K^*,\phi}$ and $B_{K,K^*,\phi}$ in the large-recoil region showed a clear preference for $A_{K^*}$ and $A_\phi$ negative and different from zero, a mild preference for $A_{K}$ negative and different from zero, whereas $B_{K,K^*,\phi}$ remained unconstrained, confirming that the current fit needs a negative contribution to $\C9$ in order to explain the data, but that it does not exhibit a preference for a $q^2$-dependent contribution.

\section{Conclusions and perspectives}
\label{sec:Conclusions}

Flavour-Changing Neutral Currents are an old favourite in the search for NP. The recent measurements performed at LHCb with
3 fb$^{-1}$ have provided a very intriguing pattern of deviations from SM predictions in $b\to s\ell\ell$ transitions.
After the discrepancy initially observed in optimised angular observables of the decay mode
$B\to K^*\mu\mu$~\cite{Descotes-Genon:2013wba}, additional tensions have arisen in the branching ratios of $B\to K\mu\mu$,
$B\to K^*\mu\mu$ and  $B_s\to\phi\mu\mu$, as well as an indication for violation of lepton flavour universality in
$B\to K\ell\ell$ (with $\ell=\mu,e$). The combined discrepancy may easily reach the 4$\,\sigma$ level.
Exploiting recent theoretical improvements concerning various sources of uncertainties (form factors, power corrections,
charm contribution), we have updated and considerably extended the analysis of Ref.~\cite{Descotes-Genon:2013wba}
performed by three of us in 2013, using the recently published LHCb results based on the full 3 fb$^{-1}$ dataset
from LHC run I.

We confirm the previous result of Ref.~\cite{Descotes-Genon:2013wba}, namely that $\C9$ plays a central role in explaining
the anomaly: a negative NP contribution to this Wilson coefficient (typically of order 25\% with respect to the SM value) is
unavoidably present in any scenario with a pull above 4~$\sigma$. Other coefficients play a secondary role but might lead
to an increase in the significance. In this sense we found several scenarios with one or two free parameters that exhibit
a pull of more than 4$\,\sigma$ compared to the SM hypothesis. One-parameter scenarios with that property are $\C9^{\rm NP}$,
$\C9^{\rm NP}=-\C{10}^{\rm NP}$ and $\C9^{\rm NP}=-\C{9^\prime}^{\rm NP}$, and two-parameter scenarios are
$(\C7^{\rm NP},\C9^{\rm NP})$, $(\C9^{\rm NP},\C{10}^{\rm NP})$, $(\C9^{\rm NP},\C{7^\prime}^{\rm NP})$,
$(\C9^{\rm NP},\C{9^\prime}^{\rm NP})$, $(\C9^{\rm NP},\C{10^\prime}^{\rm NP})$ and
$(\C9^{\rm NP}=-\C{9^\prime}^{\rm NP},\C{10}^{\rm NP}=\C{10^\prime}^{\rm NP})$,
$(\C9^{\rm NP}=-\C{9^\prime}^{\rm NP},\C{10}^{\rm NP}=-\C{10^\prime}^{\rm NP})$.
We have performed a global fit to all six Wilson coefficients simultaneously, and found that all the coefficients are
consistent with their SM values at the level of 1-2$\,\sigma$, except for $\C{9}$, in line with the results
from the more economical scenarios mentioned above.

We have also briefly discussed the situation in the context of models violating lepton-flavour universality, by allowing NP
contributions of different sizes in the electron and muon sectors. While the data requires a NP contribution in the muon
sector to explain the anomalies, it does not show preferences for a contribution in the electron sector (and thus more
generally disfavours a lepton-flavour universal NP contribution). If one restricts NP to the muon sector, some of the
above scenarios see their significance increase, with a SM pull very close (or equal) to $5\,\sigma$ in some instances.

We have performed several checks to test the robustness of our results. We have compared different possible choices for the analysis: 
QCD factorisation with soft-form factors versus the computation with full form factors, different choices for the set of LCSR form factors taken as input,
optimised observables $P_i$ versus CP-averages $S_i$, different choices for the binning.
We find a very good agreement between the various approaches. In particular, we have checked that the details of the form factor computation are not very significant for the optimised observables. 

Non-perturbative effects from power corrections and from long-distance $c\bar{c}$ contributions can only be estimated.
We have studied the effect of increasing the size of these contributions, without finding a large impact on the overall picture presented above. 
Moreover, the above-mentioned hadronic effects (in particular the $c\bar{c}$ contributions) are expected to exhibit a
$q^2$-dependence which allows them to be distinguished from a $q^2$-independent NP effect.
We have studied a possible $q^2$-dependence in a twofold way: on one hand, performing
a bin-by-bin analysis, on the other hand introducing a separate linear $q^2$-dependence in $\C{9}$ for the fit to 
$B\to K\mu\mu$, $B\to K^*\mu\mu$ and $B_s\to\phi\mu\mu$. In both cases, we found no conclusive evidence for a $q^2$-dependence.

One should notice that our results are in good agreement with those obtained in Ref.~\cite{Altmannshofer:2014rta}, even though 
the applied methods differ in many central points: different sets of form factors, a different approach to the computation (soft form factors versus full form factors), different angular observables and different estimates of hadronic uncertainties (power corrections, charm contribution).
While our method is to a large extent independent of the modelling of non-perturbative effects but has to rely on an estimation of subleading contributions
based on dimensional arguments, the analysis in Ref.~\cite{Altmannshofer:2014rta} is based on (and limited to) a particular non-perturbative
LCSR calculation. Strengths and weaknesses of the two approaches are of complementary nature, and the comparison of the obtained results
is thus a useful cross-check of the different hypotheses on which the two analyses rely.

\begin{table}
\begin{center}{\small \begin{tabular}{cccccccc}
 &  & $R_K$ &  $\av{P_5^\prime}_{[4,6], [6,8]} $ &  ${\cal B}_{B_s \to \phi\mu\mu}$  \\\hline\hline
\multirow{2}{*}{$\C9^\text{NP}$} &$+$& & &  \\
&$-$ & $\checkmark$ & $\checkmark$ & $\checkmark$ \\ \hline
\multirow{2}{*}{$\C{10}^\text{NP}$} &$+$& $\checkmark$ & & $\checkmark$ \\
&$-$ & & $\checkmark$ & \\ \hline
\multirow{2}{*}{$\C{9^\prime}^\text{NP}$} &$+$& & & $\checkmark$  \\
&$-$ & $\checkmark$ & $\checkmark$ & \\ \hline
\multirow{2}{*}{$\C{10^\prime}^\text{NP}$} &$+$& $\checkmark$ & $\checkmark$ &   \\
&$-$ &  &  & $\checkmark$ \\ \hline\\ 
\end{tabular}}
\end{center}
\caption{\it Signs of contributions to Wilson coefficients needed to explain the different anomalies observed
in $b\to s\mu\mu$ observables. A check-mark ($\checkmark$) indicates that a shift in the Wilson coefficient with this  sign moves the prediction in the right direction to solve the corresponding anomaly.
The Wilson coefficients considered here correspond to the $b\to s\mu\mu$ effective Hamiltonian (we assume that no NP enters $b\to see$).} \label{tab:anomalies}\end{table}

While the observables $S_i$ become competitive to the $P_i$ in a global fit, where their LO form factor dependence
gets cured thanks to correlations, the $P_i$ exhibit a much larger sensitivity to NP on the level of the individual observables as they are shielded to a large extent from hadronic uncertainties. Whereas for example the observable $P_5^\prime$ can be predicted in the SM with a precision of $\sim 10\%$, basically independently of the underlying form factor parametrisation, predictions for $S_5$ can develop uncertainties up to $\sim 40\%$ depending on the form factors used
as input. This feature makes the experimental measurement of the observables $P_i$ indispensable in the search for NP where it will be essential to find apart from global tensions in combined fits also some clear-cut discrepancies in individual observables.

The results we obtained from our fits are particularly encouraging as they show that at the level of the Wilson coefficients several NP scenarios provide a consistent explanation of the deviations observed in $b\to s\ell\ell$ transitions. On the other hand, the most favoured scenarios are difficult to generate in terms of simple NP models (such as a heavy $Z'$ boson, or leptoquarks). Obviously this might change over time with new experimental results. In this respect, we found it interesting to summarise in Table~\ref{tab:anomalies} how a given  NP contribution to a Wilson coefficient would affect the different anomalies. As expected, only $\C9^{\rm NP}<0$ is able to provide a consistent explanation for all of them. There is also certain preference for $\C{10}^{\rm NP}$ and $\C{10^\prime}^{\rm NP}$ to be positive in order to explain two out of three anomalies, and negative for $\C9^{\rm NP}$ and $\C{9^\prime}^{\rm NP}$. However,  whereas the best-fit point of the 1D and 2D (unconstrained) scenarios with NP in  $\C9$ and $\C{10}$ (see Tables~\ref{tab:1Dfits} and \ref{tab:2Dfits}) indeed shows a preference for a negative (respectively, positive) value in agreement with Table~\ref{tab:anomalies},
the best-fit point for $C_{9^\prime}^{\rm NP}$ and $C_{10^\prime}^{\rm NP}$ prefers a positive (respectively, negative) value in contradiction with Table~\ref{tab:anomalies}. This suggests that the chirally-flipped operators are not particularly useful to solve the anomalies but they are quite efficient (especially $\C9^\prime$)  in fixing small deviations in various bins, summing up to an overall large significance. Such a situation
arises in the scenario $\C9^{\rm NP}=-\C{9^\prime}^{\rm NP}$ which fixes neither $R_K$ nor the anomaly in $P_5^\prime$, but still manages to yield a very large pull with respect to the SM hypothesis.

In summary,  $\C9^{\rm NP}<0$ is very much favoured, providing a consistent picture for the anomalies in agreement with the results of global fits. 
A contribution $\C{10}^{\rm NP}>0$ comes in second place, while the situation with respect to NP contributions to the chirally-flipped operators is less clear. 
Obviously, this guess work is completely tributary to the current experimental situation.  Updates of these measurements, and in particular $B\to K^*\mu\mu$ observables with a finer binning, will prove particularly important to provide a more definite answer concerning the origin the anomalies observed in $b\to s\ell\ell$ transitions.

\section*{Acknowledgments}
We thank Damir Becirevic, Marcin Chrzaszcz, Andreas Crivellin, Tobias Huber, Alex Khodjamirian, Federico Mescia,
Enrique Ruiz-Arriola, Dirk Seidel, Nicola Serra and Avelino Vicente for useful discussions over the course of this work.
We also would like to thank the
organisers and participants to the workshops \emph{Rare B decays in 2015 experiment and theory} (Edinburgh) and
\emph{Novel aspects of $b\to s$ transitions: investigating new channels} (Marseille), where part of this work was discussed.
We thank Tobias Huber for sharing with us the results of Ref.~\cite{Huber:2015sra} prior to publication.
We thank Roman Zwicky for bringing Ref.~\cite{Gratrex:2015hna} to our attention while we were completing this work.
JV is funded by the DFG within research unit FOR 1873 (QFET), and acknowledges financial support from CNRS.
SDG, JM and JV acknowledge financial support from FPA2014-61478-EXP.
L.H. has been supported by FPA2011-25948 and the grant 2014 SGR 1450, and in part by the Centro de Excelencia Severo Ochoa.

\newpage


\appendix

\section{SM predictions}
\label{app:SMpred}

The prediction column corresponds to the Standard Model case.

\begin{center}\small
\begin{longtable}{@{}cccr@{}}
\toprule[1.6pt] 
$ 10^7 \times BR(B^+\to K^+\mu^+\mu^-) $ & Standard Model & Experiment & Pull \\ 
 \midrule 
 $ [0.1,0.98] $ & $ 0.31 \pm 0.09 $ & $ 0.29 \pm 0.02 $ & $ +0.2 $ \\ 
 $ [1.1,2] $ & $ 0.32 \pm 0.10 $ & $ 0.21 \pm 0.02 $ & $ +1.1 $ \\ 
 $ [2,3] $ & $ 0.35 \pm 0.11 $ & $ 0.28 \pm 0.02 $ & $ +0.6 $ \\ 
 $ [3,4] $ & $ 0.35 \pm 0.11 $ & $ 0.25 \pm 0.02 $ & $ +0.8 $ \\ 
 $ [4,5] $ & $ 0.35 \pm 0.11 $ & $ 0.22 \pm 0.02 $ & $ +1.1 $ \\ 
 $ [5,6] $ & $ 0.34 \pm 0.12 $ & $ 0.23 \pm 0.02 $ & $ +0.9 $ \\ 
 $ [6,7] $ & $ 0.34 \pm 0.12 $ & $ 0.25 \pm 0.02 $ & $ +0.8 $ \\ 
 $ [7,8] $ & $ 0.34 \pm 0.13 $ & $ 0.23 \pm 0.02 $ & $ +0.8 $ \\ 
 $ [15,22] $ & $ 0.98 \pm 0.13 $ & $ 0.85 \pm 0.05 $ & $ +0.9 $ \\ 
\midrule[1.6pt] 
$ 10^7 \times BR(B^0\to K^0\mu^+\mu^-) $ & Standard Model & Experiment & Pull \\ 
 \midrule 
 $ [0.1,2] $ & $ 0.62 \pm 0.19 $ & $ 0.23 \pm 0.11 $ & $ +1.8 $ \\ 
 $ [2,4] $ & $ 0.65 \pm 0.21 $ & $ 0.37 \pm 0.11 $ & $ +1.2 $ \\ 
 $ [4,6] $ & $ 0.64 \pm 0.22 $ & $ 0.35 \pm 0.10 $ & $ +1.2 $ \\ 
 $ [6,8] $ & $ 0.63 \pm 0.23 $ & $ 0.54 \pm 0.12 $ & $ +0.4 $ \\ 
 $ [15,19] $ & $ 0.91 \pm 0.12 $ & $ 0.67 \pm 0.12 $ & $ +1.4 $ \\ 
\midrule[1.6pt] 
$ 10^7 \times BR(B^0\to K^{*0}\mu^+\mu^-) $ & Standard Model & Experiment & Pull \\ 
 \midrule 
 $ [0.1,2] $ & $ 1.30 \pm 1.00 $ & $ 1.14 \pm 0.18 $ & $ +0.2 $ \\ 
 $ [2,4.3] $ & $ 0.85 \pm 0.59 $ & $ 0.69 \pm 0.12 $ & $ +0.3 $ \\ 
 $ [4.3,8.68] $ & $ 2.62 \pm 4.92 $ & $ 2.15 \pm 0.31 $ & $ +0.1 $ \\ 
 $ [16,19] $ & $ 1.66 \pm 0.15 $ & $ 1.23 \pm 0.20 $ & $ +1.7 $ \\ 
\midrule[1.6pt] 
$ 10^7 \times BR(B^+\to K^{*+}\mu^+\mu^-) $ & Standard Model & Experiment & Pull \\ 
 \midrule 
 $ [0.1,2] $ & $ 1.35 \pm 1.05 $ & $ 1.12 \pm 0.27 $ & $ +0.2 $ \\ 
 $ [2,4] $ & $ 0.80 \pm 0.55 $ & $ 1.12 \pm 0.32 $ & $ -0.5 $ \\ 
 $ [4,6] $ & $ 0.95 \pm 0.70 $ & $ 0.50 \pm 0.20 $ & $ +0.6 $ \\ 
 $ [6,8] $ & $ 1.17 \pm 0.92 $ & $ 0.66 \pm 0.22 $ & $ +0.5 $ \\ 
 $ [15,19] $ & $ 2.59 \pm 0.25 $ & $ 1.60 \pm 0.32 $ & $ +2.5 $ \\ 
\midrule[1.6pt] 
$ 10^7 \times BR(B_s\to \phi\mu^+\mu^-) $ & Standard Model & Experiment & Pull \\ 
 \midrule 
 $ [0.1,2.] $ & $ 1.81 \pm 0.36 $ & $ 1.11 \pm 0.16 $ & $ +1.8 $ \\ 
 $ [2.,5.] $ & $ 1.88 \pm 0.32 $ & $ 0.77 \pm 0.14 $ & $ +3.2 $ \\ 
 $ [5.,8.] $ & $ 2.25 \pm 0.41 $ & $ 0.96 \pm 0.15 $ & $ +2.9 $ \\ 
 $ [15,18.8] $ & $ 2.20 \pm 0.17 $ & $ 1.62 \pm 0.20 $ & $ +2.2 $ \\ 
\midrule[1.6pt] 
$ F_L (B\to K^*\mu^+\mu^-) $ & Standard Model & Experiment & Pull \\ 
 \midrule 
 $ [0.1,0.98] $ & $ 0.23 \pm 0.25 $ & $ 0.26 \pm 0.05 $ & $ -0.1 $ \\ 
 $ [1.1,2.5] $ & $ 0.67 \pm 0.28 $ & $ 0.66 \pm 0.09 $ & $ +0.0 $ \\ 
 $ [2.5,4] $ & $ 0.76 \pm 0.24 $ & $ 0.88 \pm 0.11 $ & $ -0.4 $ \\ 
 $ [4,6] $ & $ 0.71 \pm 0.29 $ & $ 0.61 \pm 0.06 $ & $ +0.3 $ \\ 
 $ [6,8] $ & $ 0.63 \pm 0.33 $ & $ 0.58 \pm 0.05 $ & $ +0.1 $ \\ 
 $ [15,19] $ & $ 0.34 \pm 0.03 $ & $ 0.34 \pm 0.03 $ & $ -0.1 $ \\ 
\midrule[1.6pt] 
$ P_1 (B\to K^*\mu^+\mu^-) $ & Standard Model & Experiment & Pull \\ 
 \midrule 
 $ [0.1,0.98] $ & $ 0.03 \pm 0.08 $ & $ -0.10 \pm 0.17 $ & $ +0.7 $ \\ 
 $ [1.1,2.5] $ & $ -0.00 \pm 0.06 $ & $ -0.45 \pm 0.64 $ & $ +0.7 $ \\ 
 $ [2.5,4] $ & $ 0.00 \pm 0.07 $ & $ 0.57 \pm 2.40 $ & $ -0.2 $ \\ 
 $ [4,6] $ & $ 0.02 \pm 0.12 $ & $ 0.18 \pm 0.37 $ & $ -0.4 $ \\ 
 $ [6,8] $ & $ 0.02 \pm 0.14 $ & $ -0.20 \pm 0.28 $ & $ +0.7 $ \\ 
 $ [15,19] $ & $ -0.64 \pm 0.06 $ & $ -0.50 \pm 0.11 $ & $ -1.2 $ \\ 
\midrule[1.6pt] 
$ P_2 (B\to K^*\mu^+\mu^-) $ & Standard Model & Experiment & Pull \\ 
 \midrule 
 $ [0.1,0.98] $ & $ 0.12 \pm 0.02 $ & $ 0.00 \pm 0.05 $ & $ +2.1 $ \\ 
 $ [1.1,2.5] $ & $ 0.44 \pm 0.03 $ & $ 0.37 \pm 0.20 $ & $ +0.3 $ \\ 
 $ [2.5,4] $ & $ 0.21 \pm 0.13 $ & $ 0.64 \pm 1.74 $ & $ -0.2 $ \\ 
 $ [4,6] $ & $ -0.18 \pm 0.12 $ & $ -0.04 \pm 0.09 $ & $ -1.0 $ \\ 
 $ [6,8] $ & $ -0.38 \pm 0.07 $ & $ -0.24 \pm 0.06 $ & $ -1.4 $ \\ 
 $ [15,19] $ & $ -0.36 \pm 0.02 $ & $ -0.36 \pm 0.03 $ & $ -0.0 $ \\ 
\midrule[1.6pt] 
$ P_3 (B\to K^*\mu^+\mu^-) $ & Standard Model & Experiment & Pull \\ 
 \midrule 
 $ [0.1,0.98] $ & $ -0.00 \pm 0.00 $ & $ -0.11 \pm 0.08 $ & $ +1.4 $ \\ 
 $ [1.1,2.5] $ & $ 0.00 \pm 0.01 $ & $ -0.35 \pm 0.33 $ & $ +1.1 $ \\ 
 $ [2.5,4] $ & $ 0.00 \pm 0.01 $ & $ -0.75 \pm 2.59 $ & $ +0.3 $ \\ 
 $ [4,6] $ & $ 0.00 \pm 0.01 $ & $ -0.08 \pm 0.19 $ & $ +0.5 $ \\ 
 $ [6,8] $ & $ 0.00 \pm 0.01 $ & $ -0.06 \pm 0.15 $ & $ +0.4 $ \\ 
 $ [15,19] $ & $ 0.00 \pm 0.02 $ & $ -0.08 \pm 0.06 $ & $ +1.3 $ \\ 
\midrule[1.6pt] 
$ P'_4 (B\to K^*\mu^+\mu^-) $ & Standard Model & Experiment & Pull \\ 
 \midrule 
 $ [0.1,0.98] $ & $ -0.49 \pm 0.17 $ & $ -0.37 \pm 0.32 $ & $ -0.3 $ \\ 
 $ [1.1,2.5] $ & $ -0.06 \pm 0.17 $ & $ 0.33 \pm 0.48 $ & $ -0.8 $ \\ 
 $ [2.5,4] $ & $ 0.55 \pm 0.21 $ & $ 1.43 \pm 2.61 $ & $ -0.3 $ \\ 
 $ [4,6] $ & $ 0.82 \pm 0.16 $ & $ 0.90 \pm 0.35 $ & $ -0.2 $ \\ 
 $ [6,8] $ & $ 0.93 \pm 0.12 $ & $ 1.20 \pm 0.27 $ & $ -0.9 $ \\ 
 $ [15,19] $ & $ 1.28 \pm 0.02 $ & $ 1.19 \pm 0.17 $ & $ +0.5 $ \\ 
\midrule[1.6pt] 
$ P'_5 (B\to K^*\mu^+\mu^-) $ & Standard Model & Experiment & Pull \\ 
 \midrule 
 $ [0.1,0.98] $ & $ 0.67 \pm 0.14 $ & $ 0.39 \pm 0.14 $ & $ +1.4 $ \\ 
 $ [1.1,2.5] $ & $ 0.20 \pm 0.12 $ & $ 0.29 \pm 0.22 $ & $ -0.4 $ \\ 
 $ [2.5,4] $ & $ -0.49 \pm 0.12 $ & $ -0.07 \pm 0.36 $ & $ -1.1 $ \\ 
 $ [4,6] $ & $ -0.82 \pm 0.08 $ & $ -0.30 \pm 0.16 $ & $ -2.9 $ \\ 
 $ [6,8] $ & $ -0.94 \pm 0.08 $ & $ -0.51 \pm 0.12 $ & $ -2.9 $ \\ 
 $ [15,19] $ & $ -0.57 \pm 0.05 $ & $ -0.68 \pm 0.08 $ & $ +1.2 $ \\ 
\midrule[1.6pt] 
$ P'_6 (B\to K^*\mu^+\mu^-) $ & Standard Model & Experiment & Pull \\ 
 \midrule 
 $ [0.1,0.98] $ & $ -0.06 \pm 0.02 $ & $ 0.03 \pm 0.14 $ & $ -0.7 $ \\ 
 $ [1.1,2.5] $ & $ -0.07 \pm 0.03 $ & $ -0.46 \pm 0.22 $ & $ +1.8 $ \\ 
 $ [2.5,4] $ & $ -0.06 \pm 0.03 $ & $ 0.21 \pm 0.96 $ & $ -0.3 $ \\ 
 $ [4,6] $ & $ -0.04 \pm 0.02 $ & $ -0.03 \pm 0.17 $ & $ -0.0 $ \\ 
 $ [6,8] $ & $ -0.02 \pm 0.01 $ & $ -0.10 \pm 0.17 $ & $ +0.4 $ \\ 
 $ [15,19] $ & $ -0.00 \pm 0.07 $ & $ 0.10 \pm 0.09 $ & $ -0.9 $ \\ 
\midrule[1.6pt] 
$ P'_8 (B\to K^*\mu^+\mu^-) $ & Standard Model & Experiment & Pull \\ 
 \midrule 
 $ [0.1,0.98] $ & $ 0.02 \pm 0.03 $ & $ -0.36 \pm 0.35 $ & $ +1.1 $ \\ 
 $ [1.1,2.5] $ & $ 0.04 \pm 0.03 $ & $ 0.42 \pm 0.54 $ & $ -0.7 $ \\ 
 $ [2.5,4] $ & $ 0.04 \pm 0.03 $ & $ -0.18 \pm 1.30 $ & $ +0.2 $ \\ 
 $ [4,6] $ & $ 0.03 \pm 0.02 $ & $ -0.68 \pm 0.38 $ & $ +1.9 $ \\ 
 $ [6,8] $ & $ 0.02 \pm 0.01 $ & $ 0.34 \pm 0.29 $ & $ -1.1 $ \\ 
 $ [15,19] $ & $ -0.00 \pm 0.03 $ & $ -0.12 \pm 0.19 $ & $ +0.6 $ \\ 
\midrule[1.6pt] 
$ S_3 (B\to K^*\mu^+\mu^-) $ & Standard Model & Experiment & Pull \\ 
 \midrule 
 $ [0.1,0.98] $ & $ 0.01 \pm 0.02 $ & $ -0.04 \pm 0.06 $ & $ +0.6 $ \\ 
 $ [1.1,2.5] $ & $ -0.00 \pm 0.01 $ & $ -0.08 \pm 0.10 $ & $ +0.7 $ \\ 
 $ [2.5,4] $ & $ 0.00 \pm 0.01 $ & $ 0.04 \pm 0.10 $ & $ -0.3 $ \\ 
 $ [4,6] $ & $ 0.00 \pm 0.01 $ & $ 0.04 \pm 0.07 $ & $ -0.5 $ \\ 
 $ [6,8] $ & $ 0.00 \pm 0.02 $ & $ -0.04 \pm 0.06 $ & $ +0.7 $ \\ 
 $ [15,19] $ & $ -0.21 \pm 0.02 $ & $ -0.16 \pm 0.04 $ & $ -1.1 $ \\ 
\midrule[1.6pt] 
$ S_4 (B\to K^*\mu^+\mu^-) $ & Standard Model & Experiment & Pull \\ 
 \midrule 
 $ [0.1,0.98] $ & $ -0.08 \pm 0.05 $ & $ -0.08 \pm 0.07 $ & $ -0.0 $ \\ 
 $ [1.1,2.5] $ & $ -0.01 \pm 0.03 $ & $ 0.08 \pm 0.11 $ & $ -0.8 $ \\ 
 $ [2.5,4] $ & $ 0.11 \pm 0.07 $ & $ 0.23 \pm 0.14 $ & $ -0.8 $ \\ 
 $ [4,6] $ & $ 0.18 \pm 0.08 $ & $ 0.22 \pm 0.09 $ & $ -0.3 $ \\ 
 $ [6,8] $ & $ 0.22 \pm 0.07 $ & $ 0.30 \pm 0.07 $ & $ -0.8 $ \\ 
 $ [15,19] $ & $ 0.30 \pm 0.01 $ & $ 0.28 \pm 0.04 $ & $ +0.5 $ \\ 
\midrule[1.6pt] 
$ S_5 (B\to K^*\mu^+\mu^-) $ & Standard Model & Experiment & Pull \\ 
 \midrule 
 $ [0.1,0.98] $ & $ 0.23 \pm 0.08 $ & $ 0.17 \pm 0.06 $ & $ +0.6 $ \\ 
 $ [1.1,2.5] $ & $ 0.08 \pm 0.06 $ & $ 0.14 \pm 0.10 $ & $ -0.5 $ \\ 
 $ [2.5,4] $ & $ -0.19 \pm 0.09 $ & $ -0.02 \pm 0.11 $ & $ -1.2 $ \\ 
 $ [4,6] $ & $ -0.35 \pm 0.12 $ & $ -0.15 \pm 0.08 $ & $ -1.4 $ \\ 
 $ [6,8] $ & $ -0.43 \pm 0.10 $ & $ -0.25 \pm 0.06 $ & $ -1.5 $ \\ 
 $ [15,19] $ & $ -0.27 \pm 0.03 $ & $ -0.33 \pm 0.04 $ & $ +1.2 $ \\ 
\midrule[1.6pt] 
$ A_{\rm FB} (B\to K^*\mu^+\mu^-) $ & Standard Model & Experiment & Pull \\ 
 \midrule 
 $ [0.1,0.98] $ & $ -0.10 \pm 0.04 $ & $ -0.00 \pm 0.06 $ & $ -1.3 $ \\ 
 $ [1.1,2.5] $ & $ -0.19 \pm 0.19 $ & $ -0.19 \pm 0.08 $ & $ +0.0 $ \\ 
 $ [2.5,4] $ & $ -0.07 \pm 0.07 $ & $ -0.12 \pm 0.09 $ & $ +0.5 $ \\ 
 $ [4,6] $ & $ 0.08 \pm 0.11 $ & $ 0.03 \pm 0.05 $ & $ +0.5 $ \\ 
 $ [6,8] $ & $ 0.21 \pm 0.21 $ & $ 0.15 \pm 0.04 $ & $ +0.3 $ \\ 
 $ [15,19] $ & $ 0.36 \pm 0.03 $ & $ 0.36 \pm 0.03 $ & $ +0.0 $ \\ 
\midrule[1.6pt] 
$ S_7 (B\to K^*\mu^+\mu^-) $ & Standard Model & Experiment & Pull \\ 
 \midrule 
 $ [0.1,0.98] $ & $ 0.02 \pm 0.01 $ & $ -0.02 \pm 0.06 $ & $ +0.6 $ \\ 
 $ [1.1,2.5] $ & $ 0.03 \pm 0.01 $ & $ 0.22 \pm 0.11 $ & $ -1.8 $ \\ 
 $ [2.5,4] $ & $ 0.02 \pm 0.01 $ & $ -0.07 \pm 0.12 $ & $ +0.8 $ \\ 
 $ [4,6] $ & $ 0.02 \pm 0.01 $ & $ 0.02 \pm 0.08 $ & $ -0.0 $ \\ 
 $ [6,8] $ & $ 0.01 \pm 0.00 $ & $ 0.05 \pm 0.07 $ & $ -0.6 $ \\ 
 $ [15,19] $ & $ 0.00 \pm 0.03 $ & $ -0.05 \pm 0.04 $ & $ +0.9 $ \\ 
\midrule[1.6pt] 
$ S_8 (B\to K^*\mu^+\mu^-) $ & Standard Model & Experiment & Pull \\ 
 \midrule 
 $ [0.1,0.98] $ & $ -0.00 \pm 0.01 $ & $ 0.08 \pm 0.08 $ & $ -1.1 $ \\ 
 $ [1.1,2.5] $ & $ -0.01 \pm 0.00 $ & $ -0.10 \pm 0.12 $ & $ +0.7 $ \\ 
 $ [2.5,4] $ & $ -0.01 \pm 0.00 $ & $ 0.03 \pm 0.13 $ & $ -0.3 $ \\ 
 $ [4,6] $ & $ -0.01 \pm 0.00 $ & $ 0.17 \pm 0.10 $ & $ -1.8 $ \\ 
 $ [6,8] $ & $ -0.00 \pm 0.00 $ & $ -0.09 \pm 0.07 $ & $ +1.1 $ \\ 
 $ [15,19] $ & $ 0.00 \pm 0.01 $ & $ 0.03 \pm 0.05 $ & $ -0.6 $ \\ 
\midrule[1.6pt] 
$ S_9 (B\to K^*\mu^+\mu^-) $ & Standard Model & Experiment & Pull \\ 
 \midrule 
 $ [0.1,0.98] $ & $ 0.00 \pm 0.00 $ & $ 0.08 \pm 0.06 $ & $ -1.4 $ \\ 
 $ [1.1,2.5] $ & $ -0.00 \pm 0.00 $ & $ 0.12 \pm 0.10 $ & $ -1.2 $ \\ 
 $ [2.5,4] $ & $ -0.00 \pm 0.00 $ & $ 0.09 \pm 0.13 $ & $ -0.7 $ \\ 
 $ [4,6] $ & $ -0.00 \pm 0.00 $ & $ 0.03 \pm 0.07 $ & $ -0.4 $ \\ 
 $ [6,8] $ & $ -0.00 \pm 0.00 $ & $ 0.02 \pm 0.06 $ & $ -0.4 $ \\ 
 $ [15,19] $ & $ -0.00 \pm 0.01 $ & $ 0.05 \pm 0.04 $ & $ -1.3 $ \\ 
\midrule[1.6pt] 
$ P_1 (B_s\to \phi\mu^+\mu^-) $ & Standard Model & Experiment & Pull \\ 
 \midrule 
 $ [0.1,2.] $ & $ 0.11 \pm 0.07 $ & $ -0.13 \pm 0.33 $ & $ +0.7 $ \\ 
 $ [2.,5.] $ & $ -0.10 \pm 0.10 $ & $ -0.38 \pm 1.47 $ & $ +0.2 $ \\ 
 $ [5.,8.] $ & $ -0.20 \pm 0.10 $ & $ -0.44 \pm 1.27 $ & $ +0.2 $ \\ 
 $ [15,18.8] $ & $ -0.69 \pm 0.03 $ & $ -0.25 \pm 0.34 $ & $ -1.3 $ \\ 
\midrule[1.6pt] 
$ P'_4 (B_s\to \phi\mu^+\mu^-) $ & Standard Model & Experiment & Pull \\ 
 \midrule 
 $ [0.1,2.] $ & $ -0.28 \pm 0.14 $ & $ -1.35 \pm 1.46 $ & $ +0.7 $ \\ 
 $ [2.,5.] $ & $ 0.81 \pm 0.11 $ & $ 2.02 \pm 1.84 $ & $ -0.7 $ \\ 
 $ [5.,8.] $ & $ 1.06 \pm 0.06 $ & $ 0.40 \pm 0.72 $ & $ +0.9 $ \\ 
 $ [15,18.8] $ & $ 1.30 \pm 0.01 $ & $ 0.62 \pm 0.49 $ & $ +1.4 $ \\ 
\midrule[1.6pt] 
$ P'_6 (B_s\to \phi\mu^+\mu^-) $ & Standard Model & Experiment & Pull \\ 
 \midrule 
 $ [0.1,2.] $ & $ -0.06 \pm 0.02 $ & $ 0.10 \pm 0.30 $ & $ -0.5 $ \\ 
 $ [2.,5.] $ & $ -0.05 \pm 0.02 $ & $ -0.06 \pm 0.49 $ & $ +0.0 $ \\ 
 $ [5.,8.] $ & $ -0.02 \pm 0.01 $ & $ 0.08 \pm 0.40 $ & $ -0.2 $ \\ 
 $ [15,18.8] $ & $ -0.00 \pm 0.07 $ & $ 0.29 \pm 0.24 $ & $ -1.1 $ \\ 
\midrule[1.6pt] 
$ F_L (B_s\to \phi\mu^+\mu^-) $ & Standard Model & Experiment & Pull \\ 
 \midrule 
 $ [0.1,2.] $ & $ 0.45 \pm 0.08 $ & $ 0.20 \pm 0.09 $ & $ +2.2 $ \\ 
 $ [2.,5.] $ & $ 0.79 \pm 0.03 $ & $ 0.68 \pm 0.15 $ & $ +0.6 $ \\ 
 $ [5.,8.] $ & $ 0.65 \pm 0.05 $ & $ 0.54 \pm 0.10 $ & $ +1.0 $ \\ 
 $ [15,18.8] $ & $ 0.36 \pm 0.02 $ & $ 0.29 \pm 0.07 $ & $ +0.9 $ \\ 
\midrule[1.6pt] 
$ S_3 (B_s\to \phi\mu^+\mu^-) $ & Standard Model & Experiment & Pull \\ 
 \midrule 
 $ [0.1,2.] $ & $ 0.02 \pm 0.01 $ & $ -0.05 \pm 0.13 $ & $ +0.5 $ \\ 
 $ [2.,5.] $ & $ -0.01 \pm 0.01 $ & $ -0.06 \pm 0.21 $ & $ +0.3 $ \\ 
 $ [5.,8.] $ & $ -0.03 \pm 0.02 $ & $ -0.10 \pm 0.25 $ & $ +0.3 $ \\ 
 $ [15,18.8] $ & $ -0.22 \pm 0.01 $ & $ -0.09 \pm 0.12 $ & $ -1.1 $ \\ 
\midrule[1.6pt] 
$ S_4 (B_s\to \phi\mu^+\mu^-) $ & Standard Model & Experiment & Pull \\ 
 \midrule 
 $ [0.1,2.] $ & $ -0.06 \pm 0.03 $ & $ -0.27 \pm 0.23 $ & $ +0.8 $ \\ 
 $ [2.,5.] $ & $ 0.16 \pm 0.03 $ & $ 0.47 \pm 0.37 $ & $ -0.7 $ \\ 
 $ [5.,8.] $ & $ 0.25 \pm 0.02 $ & $ 0.10 \pm 0.17 $ & $ +1.0 $ \\ 
 $ [15,18.8] $ & $ 0.31 \pm 0.00 $ & $ 0.14 \pm 0.11 $ & $ +1.5 $ \\ 
\midrule[1.6pt] 
$ S_7 (B_s\to \phi\mu^+\mu^-) $ & Standard Model & Experiment & Pull \\ 
 \midrule 
 $ [0.1,2.] $ & $ 0.03 \pm 0.01 $ & $ -0.04 \pm 0.12 $ & $ +0.6 $ \\ 
 $ [2.,5.] $ & $ 0.02 \pm 0.01 $ & $ 0.03 \pm 0.21 $ & $ -0.0 $ \\ 
 $ [5.,8.] $ & $ 0.01 \pm 0.00 $ & $ -0.04 \pm 0.18 $ & $ +0.3 $ \\ 
 $ [15,18.8] $ & $ 0.00 \pm 0.03 $ & $ -0.13 \pm 0.11 $ & $ +1.1 $ \\ 
\midrule[1.6pt] 
$ 10^7 \times BR(B^+\to K^+ e^+ e^-) $ & Standard Model & Experiment & Pull \\ 
 \midrule 
 $ [1.,6.] $ & $ 1.62 \pm 0.52 $ & $ 1.56 \pm 0.18 $ & $ +0.1 $ \\ 
\midrule[1.6pt] 
$ B^0\to K^{*0}e^+e^- $ & Standard Model & Experiment & Pull \\ 
 \midrule 
 $ F_L[0.0020,1.120] $ & $ 0.11 \pm 0.17 $ & $ 0.16 \pm 0.07 $ & $ -0.2 $ \\ 
 $ P_1[0.0020,1.120] $ & $ 0.03 \pm 0.08 $ & $ -0.23 \pm 0.24 $ & $ +1.1 $ \\ 
 $ P_2[0.0020,1.120] $ & $ 0.03 \pm 0.00 $ & $ 0.05 \pm 0.09 $ & $ -0.2 $ \\ 
 $ P_3[0.0020,1.120] $ & $ -0.00 \pm 0.00 $ & $ -0.07 \pm 0.11 $ & $ +0.6 $ \\ 
\bottomrule[1.6pt] 
\end{longtable}
\end{center}

\newpage

\section{Predictions at the best-fit point for NP in $\C9$ only}

\label{app:NPpred}

The prediction column corresponds to  to the best-fit point $\C9^{\rm NP}=-1.10$.

\begin{center}\small
\begin{longtable}{@{}crrr@{}}
\toprule[1.6pt] 
$ 10^7 \times BR(B^+\to K^+\mu^+\mu^-) $ & Prediction & Experiment & Pull \\ 
 \midrule 
 $ [0.1,0.98] $ & $ 0.24 \pm 0.07 $ & $ 0.29 \pm 0.02 $ & $ -0.6 $ \\ 
 $ [1.1,2] $ & $ 0.25 \pm 0.08 $ & $ 0.21 \pm 0.02 $ & $ +0.5 $ \\ 
 $ [2,3] $ & $ 0.28 \pm 0.09 $ & $ 0.28 \pm 0.02 $ & $ -0.1 $ \\ 
 $ [3,4] $ & $ 0.27 \pm 0.09 $ & $ 0.25 \pm 0.02 $ & $ +0.2 $ \\ 
 $ [4,5] $ & $ 0.27 \pm 0.09 $ & $ 0.22 \pm 0.02 $ & $ +0.6 $ \\ 
 $ [5,6] $ & $ 0.27 \pm 0.09 $ & $ 0.23 \pm 0.02 $ & $ +0.4 $ \\ 
 $ [6,7] $ & $ 0.27 \pm 0.09 $ & $ 0.25 \pm 0.02 $ & $ +0.2 $ \\ 
 $ [7,8] $ & $ 0.27 \pm 0.10 $ & $ 0.23 \pm 0.02 $ & $ +0.4 $ \\ 
 $ [15,22] $ & $ 0.77 \pm 0.10 $ & $ 0.85 \pm 0.05 $ & $ -0.7 $ \\ 
\midrule[1.6pt] 
$ 10^7 \times BR(B^0\to K^0\mu^+\mu^-) $ &  Prediction & Experiment & Pull \\ 
 \midrule 
 $ [0.1,2] $ & $ 0.49 \pm 0.15 $ & $ 0.23 \pm 0.11 $ & $ +1.4 $ \\ 
 $ [2,4] $ & $ 0.51 \pm 0.16 $ & $ 0.37 \pm 0.11 $ & $ +0.7 $ \\ 
 $ [4,6] $ & $ 0.50 \pm 0.17 $ & $ 0.35 \pm 0.10 $ & $ +0.8 $ \\ 
 $ [6,8] $ & $ 0.49 \pm 0.18 $ & $ 0.54 \pm 0.12 $ & $ -0.2 $ \\ 
 $ [15,19] $ & $ 0.71 \pm 0.09 $ & $ 0.67 \pm 0.12 $ & $ +0.3 $ \\ 
\midrule[1.6pt] 
$ 10^7 \times BR(B^0\to K^{*0}\mu^+\mu^-) $ &  Prediction & Experiment & Pull \\ 
 \midrule 
 $ [0.1,2] $ & $ 1.25 \pm 1.06 $ & $ 1.14 \pm 0.18 $ & $ +0.1 $ \\ 
 $ [2,4.3] $ & $ 0.75 \pm 0.51 $ & $ 0.69 \pm 0.12 $ & $ +0.1 $ \\ 
 $ [4.3,8.68] $ & $ 2.10 \pm 2.35 $ & $ 2.15 \pm 0.31 $ & $ -0.0 $ \\ 
 $ [16,19] $ & $ 1.31 \pm 0.11 $ & $ 1.23 \pm 0.20 $ & $ +0.4 $ \\ 
\midrule[1.6pt] 
$ BR(B^+\to K^{*+}\mu^+\mu^-) $ &  Prediction & Experiment & Pull \\ 
 \midrule 
 $ [0.1,2] $ & $ 1.29 \pm 1.11 $ & $ 1.12 \pm 0.27 $ & $ +0.1 $ \\ 
 $ [2,4] $ & $ 0.70 \pm 0.47 $ & $ 1.12 \pm 0.32 $ & $ -0.7 $ \\ 
 $ [4,6] $ & $ 0.79 \pm 0.58 $ & $ 0.50 \pm 0.20 $ & $ +0.5 $ \\ 
 $ [6,8] $ & $ 0.95 \pm 0.76 $ & $ 0.66 \pm 0.22 $ & $ +0.4 $ \\ 
 $ [15,19] $ & $ 2.05 \pm 0.18 $ & $ 1.60 \pm 0.32 $ & $ +1.2 $ \\ 
\midrule[1.6pt] 
$ 10^7 \times BR(B_s\to \phi\mu^+\mu^-) $ &  Prediction & Experiment & Pull \\ 
 \midrule 
 $ [0.1,2.] $ & $ 1.71 \pm 0.34 $ & $ 1.11 \pm 0.16 $ & $ +1.6 $ \\ 
 $ [2.,5.] $ & $ 1.58 \pm 0.25 $ & $ 0.77 \pm 0.14 $ & $ +2.8 $ \\ 
 $ [5.,8.] $ & $ 1.81 \pm 0.32 $ & $ 0.96 \pm 0.15 $ & $ +2.4 $ \\ 
 $ [15,18.8] $ & $ 1.74 \pm 0.13 $ & $ 1.62 \pm 0.20 $ & $ +0.5 $ \\ 
\midrule[1.6pt] 
$ F_L (B\to K^*\mu^+\mu^-) $ &  Prediction & Experiment & Pull \\ 
 \midrule 
 $ [0.1,0.98] $ & $ 0.19 \pm 0.22 $ & $ 0.26 \pm 0.05 $ & $ -0.3 $ \\ 
 $ [1.1,2.5] $ & $ 0.58 \pm 0.32 $ & $ 0.66 \pm 0.09 $ & $ -0.2 $ \\ 
 $ [2.5,4] $ & $ 0.70 \pm 0.28 $ & $ 0.88 \pm 0.11 $ & $ -0.6 $ \\ 
 $ [4,6] $ & $ 0.67 \pm 0.31 $ & $ 0.61 \pm 0.06 $ & $ +0.2 $ \\ 
 $ [6,8] $ & $ 0.61 \pm 0.33 $ & $ 0.58 \pm 0.05 $ & $ +0.1 $ \\ 
 $ [15,19] $ & $ 0.34 \pm 0.03 $ & $ 0.34 \pm 0.03 $ & $ -0.1 $ \\ 
\midrule[1.6pt] 
$ P_1 (B\to K^*\mu^+\mu^-) $ &  Prediction & Experiment & Pull \\ 
 \midrule 
 $ [0.1,0.98] $ & $ 0.03 \pm 0.07 $ & $ -0.10 \pm 0.17 $ & $ +0.7 $ \\ 
 $ [1.1,2.5] $ & $ -0.00 \pm 0.05 $ & $ -0.45 \pm 0.64 $ & $ +0.7 $ \\ 
 $ [2.5,4] $ & $ -0.01 \pm 0.05 $ & $ 0.57 \pm 2.40 $ & $ -0.2 $ \\ 
 $ [4,6] $ & $ 0.00 \pm 0.09 $ & $ 0.18 \pm 0.37 $ & $ -0.5 $ \\ 
 $ [6,8] $ & $ 0.00 \pm 0.12 $ & $ -0.20 \pm 0.28 $ & $ +0.7 $ \\ 
 $ [15,19] $ & $ -0.64 \pm 0.05 $ & $ -0.50 \pm 0.11 $ & $ -1.2 $ \\ 
\midrule[1.6pt] 
$ P_2 (B\to K^*\mu^+\mu^-) $ &  Prediction & Experiment & Pull \\ 
 \midrule 
 $ [0.1,0.98] $ & $ 0.11 \pm 0.02 $ & $ 0.00 \pm 0.05 $ & $ +2.0 $ \\ 
 $ [1.1,2.5] $ & $ 0.43 \pm 0.03 $ & $ 0.37 \pm 0.20 $ & $ +0.3 $ \\ 
 $ [2.5,4] $ & $ 0.38 \pm 0.07 $ & $ 0.64 \pm 1.74 $ & $ -0.1 $ \\ 
 $ [4,6] $ & $ 0.06 \pm 0.12 $ & $ -0.04 \pm 0.09 $ & $ +0.7 $ \\ 
 $ [6,8] $ & $ -0.19 \pm 0.10 $ & $ -0.24 \pm 0.06 $ & $ +0.4 $ \\ 
 $ [15,19] $ & $ -0.31 \pm 0.02 $ & $ -0.36 \pm 0.03 $ & $ +1.4 $ \\ 
\midrule[1.6pt] 
$ P_3 (B\to K^*\mu^+\mu^-) $ &  Prediction & Experiment & Pull \\ 
 \midrule 
 $ [0.1,0.98] $ & $ -0.00 \pm 0.00 $ & $ -0.11 \pm 0.08 $ & $ +1.4 $ \\ 
 $ [1.1,2.5] $ & $ 0.00 \pm 0.00 $ & $ -0.35 \pm 0.33 $ & $ +1.1 $ \\ 
 $ [2.5,4] $ & $ 0.00 \pm 0.00 $ & $ -0.75 \pm 2.59 $ & $ +0.3 $ \\ 
 $ [4,6] $ & $ 0.00 \pm 0.00 $ & $ -0.08 \pm 0.19 $ & $ +0.5 $ \\ 
 $ [6,8] $ & $ 0.00 \pm 0.00 $ & $ -0.06 \pm 0.15 $ & $ +0.4 $ \\ 
 $ [15,19] $ & $ 0.00 \pm 0.02 $ & $ -0.08 \pm 0.06 $ & $ +1.3 $ \\ 
\midrule[1.6pt] 
$ P'_4 (B\to K^*\mu^+\mu^-) $ &  Prediction & Experiment & Pull \\ 
 \midrule 
 $ [0.1,0.98] $ & $ -0.36 \pm 0.21 $ & $ -0.37 \pm 0.32 $ & $ +0.0 $ \\ 
 $ [1.1,2.5] $ & $ 0.02 \pm 0.15 $ & $ 0.33 \pm 0.48 $ & $ -0.6 $ \\ 
 $ [2.5,4] $ & $ 0.51 \pm 0.18 $ & $ 1.43 \pm 2.61 $ & $ -0.4 $ \\ 
 $ [4,6] $ & $ 0.78 \pm 0.15 $ & $ 0.90 \pm 0.35 $ & $ -0.3 $ \\ 
 $ [6,8] $ & $ 0.91 \pm 0.11 $ & $ 1.20 \pm 0.27 $ & $ -1.0 $ \\ 
 $ [15,19] $ & $ 1.28 \pm 0.02 $ & $ 1.19 \pm 0.17 $ & $ +0.5 $ \\ 
\midrule[1.6pt] 
$ P'_5 (B\to K^*\mu^+\mu^-) $ &  Prediction & Experiment & Pull \\ 
 \midrule 
 $ [0.1,0.98] $ & $ 0.80 \pm 0.14 $ & $ 0.39 \pm 0.14 $ & $ +2.0 $ \\ 
 $ [1.1,2.5] $ & $ 0.43 \pm 0.12 $ & $ 0.29 \pm 0.22 $ & $ +0.6 $ \\ 
 $ [2.5,4] $ & $ -0.12 \pm 0.13 $ & $ -0.07 \pm 0.36 $ & $ -0.1 $ \\ 
 $ [4,6] $ & $ -0.50 \pm 0.11 $ & $ -0.30 \pm 0.16 $ & $ -1.0 $ \\ 
 $ [6,8] $ & $ -0.73 \pm 0.12 $ & $ -0.51 \pm 0.12 $ & $ -1.3 $ \\ 
 $ [15,19] $ & $ -0.50 \pm 0.05 $ & $ -0.68 \pm 0.08 $ & $ +1.9 $ \\ 
\midrule[1.6pt] 
$ P'_6 (B\to K^*\mu^+\mu^-) $ &  Prediction & Experiment & Pull \\ 
 \midrule 
 $ [0.1,0.98] $ & $ -0.06 \pm 0.03 $ & $ 0.03 \pm 0.14 $ & $ -0.7 $ \\ 
 $ [1.1,2.5] $ & $ -0.07 \pm 0.03 $ & $ -0.46 \pm 0.22 $ & $ +1.7 $ \\ 
 $ [2.5,4] $ & $ -0.06 \pm 0.03 $ & $ 0.21 \pm 0.96 $ & $ -0.3 $ \\ 
 $ [4,6] $ & $ -0.04 \pm 0.02 $ & $ -0.03 \pm 0.17 $ & $ -0.1 $ \\ 
 $ [6,8] $ & $ -0.02 \pm 0.02 $ & $ -0.10 \pm 0.17 $ & $ +0.4 $ \\ 
 $ [15,19] $ & $ -0.00 \pm 0.09 $ & $ 0.10 \pm 0.09 $ & $ -0.8 $ \\ 
\midrule[1.6pt] 
$ P'_8 (B\to K^*\mu^+\mu^-) $ &  Prediction & Experiment & Pull \\ 
 \midrule 
 $ [0.1,0.98] $ & $ 0.01 \pm 0.02 $ & $ -0.36 \pm 0.35 $ & $ +1.1 $ \\ 
 $ [1.1,2.5] $ & $ 0.03 \pm 0.02 $ & $ 0.42 \pm 0.54 $ & $ -0.7 $ \\ 
 $ [2.5,4] $ & $ 0.03 \pm 0.02 $ & $ -0.18 \pm 1.30 $ & $ +0.2 $ \\ 
 $ [4,6] $ & $ 0.03 \pm 0.02 $ & $ -0.68 \pm 0.38 $ & $ +1.9 $ \\ 
 $ [6,8] $ & $ 0.02 \pm 0.01 $ & $ 0.34 \pm 0.29 $ & $ -1.1 $ \\ 
 $ [15,19] $ & $ -0.00 \pm 0.03 $ & $ -0.12 \pm 0.19 $ & $ +0.6 $ \\ 
\midrule[1.6pt] 
$ S_3 (B\to K^*\mu^+\mu^-) $ &  Prediction & Experiment & Pull \\ 
 \midrule 
 $ [0.1,0.98] $ & $ 0.01 \pm 0.02 $ & $ -0.04 \pm 0.06 $ & $ +0.6 $ \\ 
 $ [1.1,2.5] $ & $ -0.00 \pm 0.01 $ & $ -0.08 \pm 0.10 $ & $ +0.7 $ \\ 
 $ [2.5,4] $ & $ -0.00 \pm 0.01 $ & $ 0.04 \pm 0.10 $ & $ -0.4 $ \\ 
 $ [4,6] $ & $ -0.00 \pm 0.01 $ & $ 0.04 \pm 0.07 $ & $ -0.5 $ \\ 
 $ [6,8] $ & $ -0.00 \pm 0.02 $ & $ -0.04 \pm 0.06 $ & $ +0.7 $ \\ 
 $ [15,19] $ & $ -0.21 \pm 0.02 $ & $ -0.16 \pm 0.04 $ & $ -1.2 $ \\ 
\midrule[1.6pt] 
$ S_4 (B\to K^*\mu^+\mu^-) $ &  Prediction & Experiment & Pull \\ 
 \midrule 
 $ [0.1,0.98] $ & $ -0.06 \pm 0.05 $ & $ -0.08 \pm 0.07 $ & $ +0.3 $ \\ 
 $ [1.1,2.5] $ & $ 0.00 \pm 0.03 $ & $ 0.08 \pm 0.11 $ & $ -0.6 $ \\ 
 $ [2.5,4] $ & $ 0.11 \pm 0.06 $ & $ 0.23 \pm 0.14 $ & $ -0.8 $ \\ 
 $ [4,6] $ & $ 0.17 \pm 0.08 $ & $ 0.22 \pm 0.09 $ & $ -0.4 $ \\ 
 $ [6,8] $ & $ 0.21 \pm 0.07 $ & $ 0.30 \pm 0.07 $ & $ -0.9 $ \\ 
 $ [15,19] $ & $ 0.30 \pm 0.01 $ & $ 0.28 \pm 0.04 $ & $ +0.4 $ \\ 
\midrule[1.6pt] 
$ S_5 (B\to K^*\mu^+\mu^-) $ &  Prediction & Experiment & Pull \\ 
 \midrule 
 $ [0.1,0.98] $ & $ 0.26 \pm 0.10 $ & $ 0.17 \pm 0.06 $ & $ +0.7 $ \\ 
 $ [1.1,2.5] $ & $ 0.20 \pm 0.07 $ & $ 0.14 \pm 0.10 $ & $ +0.5 $ \\ 
 $ [2.5,4] $ & $ -0.05 \pm 0.06 $ & $ -0.02 \pm 0.11 $ & $ -0.2 $ \\ 
 $ [4,6] $ & $ -0.22 \pm 0.08 $ & $ -0.15 \pm 0.08 $ & $ -0.7 $ \\ 
 $ [6,8] $ & $ -0.34 \pm 0.08 $ & $ -0.25 \pm 0.06 $ & $ -0.8 $ \\ 
 $ [15,19] $ & $ -0.24 \pm 0.02 $ & $ -0.33 \pm 0.04 $ & $ +1.9 $ \\ 
\midrule[1.6pt] 
$ A_{\rm FB} (B\to K^*\mu^+\mu^-) $ &  Prediction & Experiment & Pull \\ 
 \midrule 
 $ [0.1,0.98] $ & $ -0.10 \pm 0.04 $ & $ -0.00 \pm 0.06 $ & $ -1.4 $ \\ 
 $ [1.1,2.5] $ & $ -0.24 \pm 0.21 $ & $ -0.19 \pm 0.08 $ & $ -0.2 $ \\ 
 $ [2.5,4] $ & $ -0.16 \pm 0.15 $ & $ -0.12 \pm 0.09 $ & $ -0.2 $ \\ 
 $ [4,6] $ & $ -0.03 \pm 0.04 $ & $ 0.03 \pm 0.05 $ & $ -0.7 $ \\ 
 $ [6,8] $ & $ 0.11 \pm 0.13 $ & $ 0.15 \pm 0.04 $ & $ -0.3 $ \\ 
 $ [15,19] $ & $ 0.31 \pm 0.03 $ & $ 0.36 \pm 0.03 $ & $ -1.2 $ \\ 
\midrule[1.6pt] 
$ S_7 (B\to K^*\mu^+\mu^-) $ &  Prediction & Experiment & Pull \\ 
 \midrule 
 $ [0.1,0.98] $ & $ 0.02 \pm 0.01 $ & $ -0.02 \pm 0.06 $ & $ +0.6 $ \\ 
 $ [1.1,2.5] $ & $ 0.03 \pm 0.01 $ & $ 0.22 \pm 0.11 $ & $ -1.8 $ \\ 
 $ [2.5,4] $ & $ 0.03 \pm 0.01 $ & $ -0.07 \pm 0.12 $ & $ +0.8 $ \\ 
 $ [4,6] $ & $ 0.02 \pm 0.01 $ & $ 0.02 \pm 0.08 $ & $ +0.0 $ \\ 
 $ [6,8] $ & $ 0.01 \pm 0.01 $ & $ 0.05 \pm 0.07 $ & $ -0.5 $ \\ 
 $ [15,19] $ & $ 0.00 \pm 0.04 $ & $ -0.05 \pm 0.04 $ & $ +0.8 $ \\ 
\midrule[1.6pt] 
$ S_8 (B\to K^*\mu^+\mu^-) $ &  Prediction & Experiment & Pull \\ 
 \midrule 
 $ [0.1,0.98] $ & $ -0.00 \pm 0.00 $ & $ 0.08 \pm 0.08 $ & $ -1.0 $ \\ 
 $ [1.1,2.5] $ & $ -0.01 \pm 0.00 $ & $ -0.10 \pm 0.12 $ & $ +0.8 $ \\ 
 $ [2.5,4] $ & $ -0.01 \pm 0.00 $ & $ 0.03 \pm 0.13 $ & $ -0.3 $ \\ 
 $ [4,6] $ & $ -0.01 \pm 0.00 $ & $ 0.17 \pm 0.10 $ & $ -1.8 $ \\ 
 $ [6,8] $ & $ -0.00 \pm 0.00 $ & $ -0.09 \pm 0.07 $ & $ +1.1 $ \\ 
 $ [15,19] $ & $ 0.00 \pm 0.01 $ & $ 0.03 \pm 0.05 $ & $ -0.6 $ \\ 
\midrule[1.6pt] 
$ S_9 (B\to K^*\mu^+\mu^-) $ &  Prediction & Experiment & Pull \\ 
 \midrule 
 $ [0.1,0.98] $ & $ 0.00 \pm 0.00 $ & $ 0.08 \pm 0.06 $ & $ -1.4 $ \\ 
 $ [1.1,2.5] $ & $ -0.00 \pm 0.00 $ & $ 0.12 \pm 0.10 $ & $ -1.2 $ \\ 
 $ [2.5,4] $ & $ -0.00 \pm 0.00 $ & $ 0.09 \pm 0.13 $ & $ -0.7 $ \\ 
 $ [4,6] $ & $ -0.00 \pm 0.00 $ & $ 0.03 \pm 0.07 $ & $ -0.4 $ \\ 
 $ [6,8] $ & $ -0.00 \pm 0.00 $ & $ 0.02 \pm 0.06 $ & $ -0.4 $ \\ 
 $ [15,19] $ & $ -0.00 \pm 0.01 $ & $ 0.05 \pm 0.04 $ & $ -1.3 $ \\ 
\midrule[1.6pt] 
$ P_1 (B_s\to \phi\mu^+\mu^-) $ &  Prediction & Experiment & Pull \\ 
 \midrule 
 $ [0.1,2.] $ & $ 0.10 \pm 0.08 $ & $ -0.13 \pm 0.33 $ & $ +0.7 $ \\ 
 $ [2.,5.] $ & $ -0.06 \pm 0.07 $ & $ -0.38 \pm 1.47 $ & $ +0.2 $ \\ 
 $ [5.,8.] $ & $ -0.18 \pm 0.09 $ & $ -0.44 \pm 1.27 $ & $ +0.2 $ \\ 
 $ [15,18.8] $ & $ -0.69 \pm 0.03 $ & $ -0.25 \pm 0.34 $ & $ -1.3 $ \\ 
\midrule[1.6pt] 
$ P'_4 (B_s\to \phi\mu^+\mu^-) $ &  Prediction & Experiment & Pull \\ 
 \midrule 
 $ [0.1,2.] $ & $ -0.18 \pm 0.17 $ & $ -1.35 \pm 1.46 $ & $ +0.8 $ \\ 
 $ [2.,5.] $ & $ 0.74 \pm 0.10 $ & $ 2.02 \pm 1.84 $ & $ -0.7 $ \\ 
 $ [5.,8.] $ & $ 1.04 \pm 0.06 $ & $ 0.40 \pm 0.72 $ & $ +0.9 $ \\ 
 $ [15,18.8] $ & $ 1.30 \pm 0.01 $ & $ 0.62 \pm 0.49 $ & $ +1.4 $ \\ 
\midrule[1.6pt] 
$ P'_6 (B_s\to \phi\mu^+\mu^-) $ &  Prediction & Experiment & Pull \\ 
 \midrule 
 $ [0.1,2.] $ & $ -0.07 \pm 0.02 $ & $ 0.10 \pm 0.30 $ & $ -0.6 $ \\ 
 $ [2.,5.] $ & $ -0.06 \pm 0.02 $ & $ -0.06 \pm 0.49 $ & $ +0.0 $ \\ 
 $ [5.,8.] $ & $ -0.02 \pm 0.01 $ & $ 0.08 \pm 0.40 $ & $ -0.3 $ \\ 
 $ [15,18.8] $ & $ -0.00 \pm 0.09 $ & $ 0.29 \pm 0.24 $ & $ -1.1 $ \\ 
\midrule[1.6pt] 
$ F_L (B_s\to \phi\mu^+\mu^-) $ &  Prediction & Experiment & Pull \\ 
 \midrule 
 $ [0.1,2.] $ & $ 0.38 \pm 0.08 $ & $ 0.20 \pm 0.09 $ & $ +1.6 $ \\ 
 $ [2.,5.] $ & $ 0.74 \pm 0.04 $ & $ 0.68 \pm 0.15 $ & $ +0.4 $ \\ 
 $ [5.,8.] $ & $ 0.63 \pm 0.05 $ & $ 0.54 \pm 0.10 $ & $ +0.8 $ \\ 
 $ [15,18.8] $ & $ 0.35 \pm 0.02 $ & $ 0.29 \pm 0.07 $ & $ +0.9 $ \\ 
\midrule[1.6pt] 
$ S_3 (B_s\to \phi\mu^+\mu^-) $ &  Prediction & Experiment & Pull \\ 
 \midrule 
 $ [0.1,2.] $ & $ 0.02 \pm 0.02 $ & $ -0.05 \pm 0.13 $ & $ +0.6 $ \\ 
 $ [2.,5.] $ & $ -0.01 \pm 0.01 $ & $ -0.06 \pm 0.21 $ & $ +0.3 $ \\ 
 $ [5.,8.] $ & $ -0.03 \pm 0.02 $ & $ -0.10 \pm 0.25 $ & $ +0.3 $ \\ 
 $ [15,18.8] $ & $ -0.22 \pm 0.01 $ & $ -0.09 \pm 0.12 $ & $ -1.1 $ \\ 
\midrule[1.6pt] 
$ S_4 (B_s\to \phi\mu^+\mu^-) $ &  Prediction & Experiment & Pull \\ 
 \midrule 
 $ [0.1,2.] $ & $ -0.04 \pm 0.04 $ & $ -0.27 \pm 0.23 $ & $ +0.8 $ \\ 
 $ [2.,5.] $ & $ 0.16 \pm 0.03 $ & $ 0.47 \pm 0.37 $ & $ -0.7 $ \\ 
 $ [5.,8.] $ & $ 0.25 \pm 0.02 $ & $ 0.10 \pm 0.17 $ & $ +1.0 $ \\ 
 $ [15,18.8] $ & $ 0.31 \pm 0.00 $ & $ 0.14 \pm 0.11 $ & $ +1.5 $ \\ 
\midrule[1.6pt] 
$ S_7 (B_s\to \phi\mu^+\mu^-) $ &  Prediction & Experiment & Pull \\ 
 \midrule 
 $ [0.1,2.] $ & $ 0.03 \pm 0.01 $ & $ -0.04 \pm 0.12 $ & $ +0.6 $ \\ 
 $ [2.,5.] $ & $ 0.03 \pm 0.01 $ & $ 0.03 \pm 0.21 $ & $ -0.0 $ \\ 
 $ [5.,8.] $ & $ 0.01 \pm 0.00 $ & $ -0.04 \pm 0.18 $ & $ +0.3 $ \\ 
 $ [15,18.8] $ & $ 0.00 \pm 0.04 $ & $ -0.13 \pm 0.11 $ & $ +1.1 $ \\ 
\midrule[1.6pt] 
$ 10^7 \times BR(B^+\to K^+ e^+ e^-) $ &  Prediction & Experiment & Pull \\ 
 \midrule 
 $ [1.,6.] $ & $ 1.26 \pm 0.40 $ & $ 1.56 \pm 0.18 $ & $ -0.7 $ \\ 
\midrule[1.6pt] 
$ B^0\to K^{*0}e^+e^- $ &  Prediction & Experiment & Pull \\ 
 \midrule 
 $ F_L[0.0020,1.120] $ & $ 0.09 \pm 0.13 $ & $ 0.16 \pm 0.07 $ & $ -0.5 $ \\ 
 $ P_1[0.0020,1.120] $ & $ 0.03 \pm 0.08 $ & $ -0.23 \pm 0.24 $ & $ +1.1 $ \\ 
 $ P_2[0.0020,1.120] $ & $ 0.03 \pm 0.00 $ & $ 0.05 \pm 0.09 $ & $ -0.2 $ \\ 
 $ P_3[0.0020,1.120] $ & $ -0.00 \pm 0.00 $ & $ -0.07 \pm 0.11 $ & $ +0.6 $ \\ 
\bottomrule[1.6pt] 
\end{longtable}
\end{center}

\newpage

\section{Confidence regions for selected 2D New Physics scenarios}\label{app:ImpFit}

\begin{figure}[!h]
\begin{center}
\includegraphics[width=7cm]{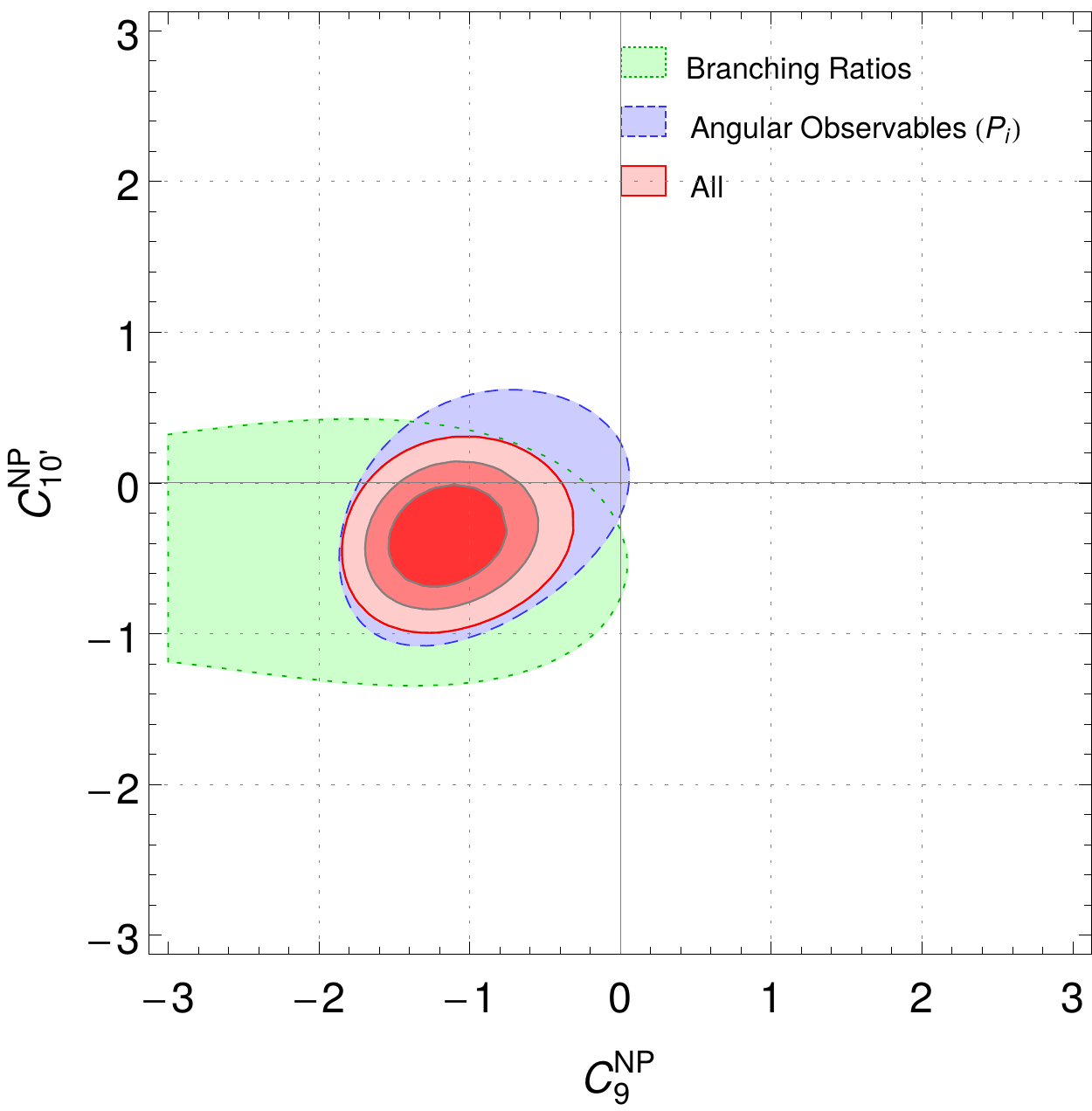}\qquad  
\includegraphics[width=7cm]{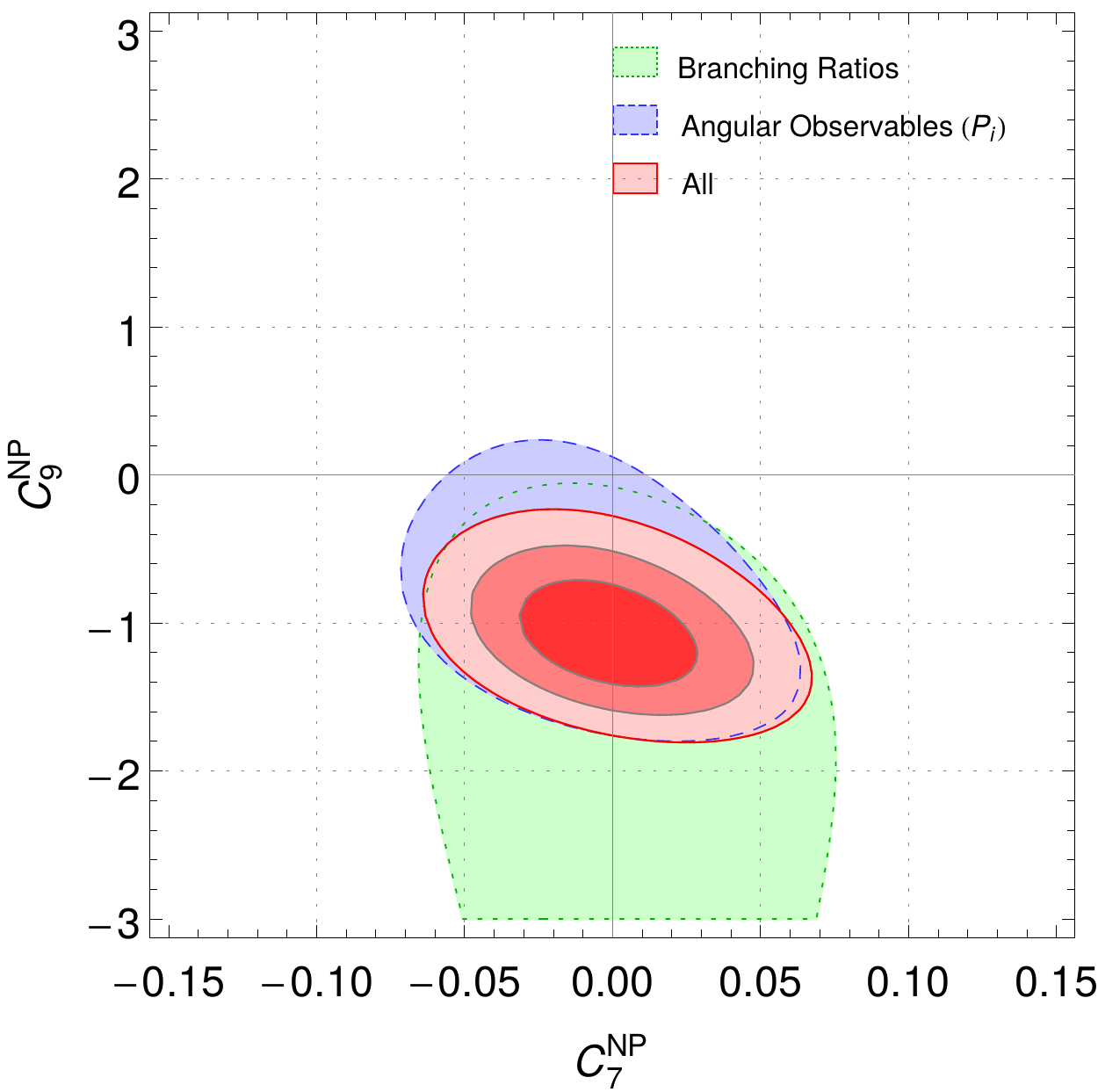}

\includegraphics[width=7cm]{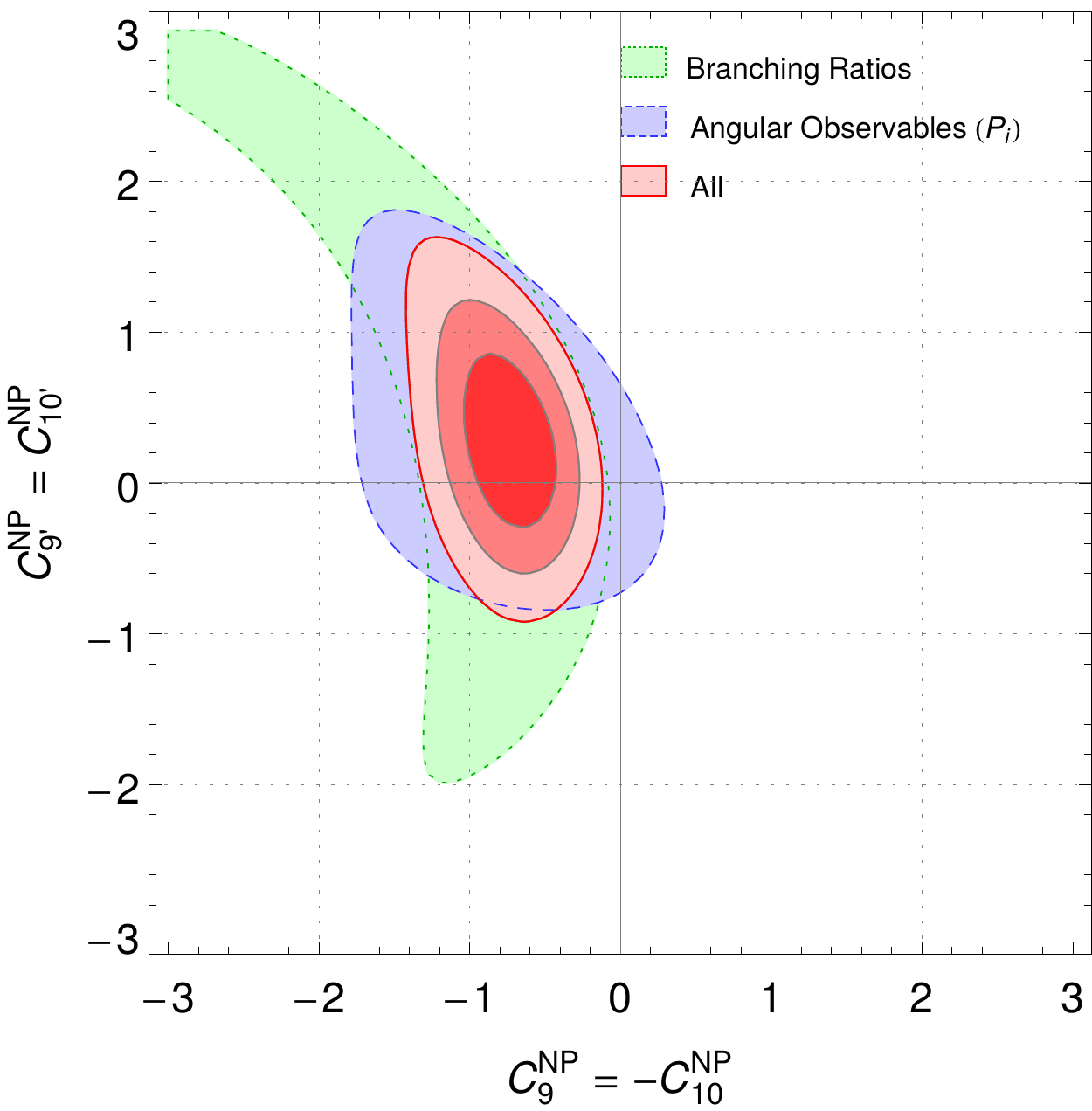}\qquad
\includegraphics[width=7cm]{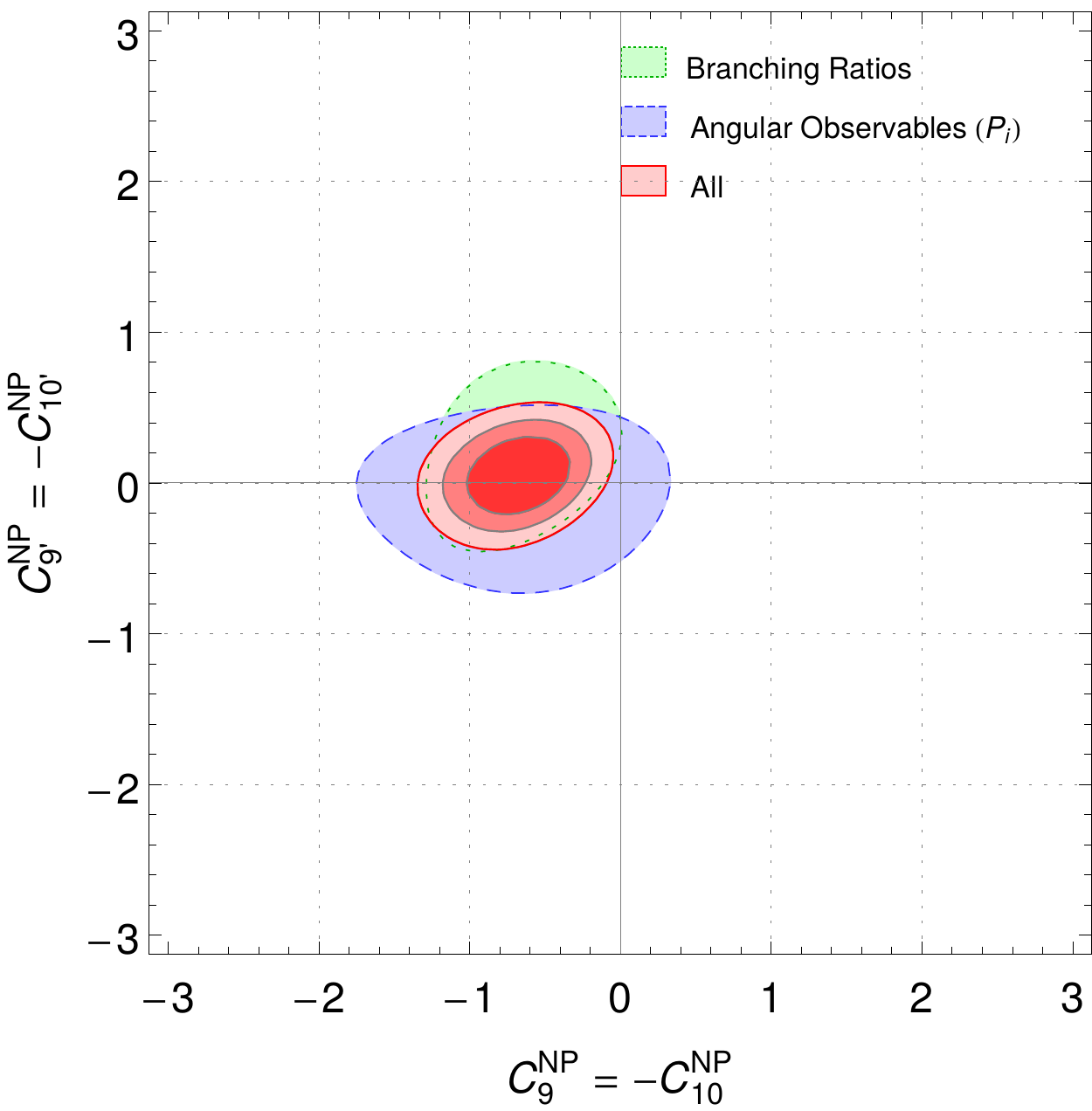} 
\end{center}
\caption{\it For the scenarios $(\C9,\C{10'})$ (upper left), $(\C7,\C9)$ (upper right)
$(\C9^{\rm NP}=-\C{10}^{\rm NP},\C{9'}^{\rm NP}=\C{10'}^{\rm NP})$ (lower left),
$(\C9^{\rm NP}=-\C{10}^{\rm NP},\C{9'}^{\rm NP}=-\C{10'}^{\rm NP})$ (lower right), we show the 3~$\sigma$ regions allowed by branching ratios only (dashed green), by angular observables only (long-dashed blue) and by considering both (red, with 1,2,3~$\sigma$  contours). Same conventions for the constraints as in Fig.~\ref{fig:splitBRangular}.}\label{fig:otherscenarios}
\end{figure}

In Fig.~\ref{fig:otherscenarios}, we provide the confidence regions of interest for two-dimensional scenarios less favoured from the point of view of the fit, but which might be of interest for model building, namely contributions to $(\C9^{\rm NP},\C{10'}^{\rm NP})$, $(\C7^{\rm NP},\C9^{\rm NP})$, $(\C9^{\rm NP}=-\C{10}^{\rm NP},\C{9'}^{\rm NP}=\C{10'}^{\rm NP})$  and
$(\C9^{\rm NP}=-\C{10}^{\rm NP},\C{9'}^{\rm NP}=-\C{10'}^{\rm NP})$.

\clearpage

\section{Impact of the fit inputs on NP in $\C{9\mu}$ only}
\label{app:C9}

\begin{center}
{\footnotesize \renewcommand\arraystretch{1}
\begin{tabular}{@{}lccccr@{}}
\toprule[1.6pt] 
Fit & $\C{9\ \rm Best fit}^{\rm NP}$ & 1$\sigma$ & Pull$_{\rm SM}$ &  $N_{\rm dof}$   & p-value (\%)\\ 
 \midrule 
All $b\to s\mu\mu$ in SM & -- & -- & -- &  96   & 16.0 \hspace{5mm}  \\[3mm] 
 All $b\to s\mu\mu$ & $ -1.09 $ & $ [-1.29,-0.87] $ &  4.5 & $ 95 $ & 63.0 \hspace{5mm}  \\[3mm] 
 All $b\to s\ell\ell$, $\ell = e,\mu$ & $ -1.11 $ & $ [-1.31,-0.90] $ &  4.9 & $ 101 $ & 74.0 \hspace{5mm}  \\[3mm] 
 All $b\to s\mu\mu$ excluding [5,8] region & $ -0.99 $ & $ [-1.23,-0.75] $ &  3.8 & $ 77 $ & 37.0 \hspace{5mm}  \\[3mm] 
 Only $b\to s\mu\mu$ BRs & $ -1.58 $ & $ [-2.22,-1.07] $ &  3.7 & $ 31 $ & 43.0 \hspace{5mm}  \\[3mm] 
 Only $b\to s\mu\mu$ $P_i$'s & $ -1.01 $ & $ [-1.25,-0.73] $ &  3.1 & $ 68 $ & 75.0 \hspace{5mm}  \\[3mm] 
 Only $b\to s\mu\mu$ $S_i$'s & $ -0.95 $ & $ [-1.19,-0.68] $ &  2.9 & $ 68 $ & 96.0 \hspace{5mm}  \\[3mm] 
 Only $B\to K\mu\mu$ & $ -0.85 $ & $ [-1.67,-0.20] $ &  1.4 & $ 18 $ & 20.0 \hspace{5mm}  \\[3mm] 
 Only $B\to K^*\mu\mu$ & $ -1.05 $ & $ [-1.27,-0.80] $ &  3.7 & $ 61 $ & 74.0 \hspace{5mm}  \\[3mm] 
 Only $B_s\to \phi\mu\mu$ & $ -1.98 $ & $ [-2.84,-1.29] $ &  3.5 & $ 24 $ & 94.0 \hspace{5mm}  \\[3mm] 
 Only $b\to s\mu\mu$ at large recoil & $ -1.30 $ & $ [-1.57,-1.02] $ &  4.0 & $ 78 $ & 61.0 \hspace{5mm}  \\[3mm] 
 Only $b\to s\mu\mu$ at low recoil & $ -0.93 $ & $ [-1.23,-0.61] $ &  2.8 & $ 21 $ & 75.0 \hspace{5mm}  \\[3mm] 
 Only $b\to s\mu\mu$ within [1,6] & $ -1.30 $ & $ [-1.66,-0.93] $ &  3.4 & $ 43 $ & 73.0 \hspace{5mm}  \\[3mm] 
 Only $BR(B\to K\ell\ell)_{[1,6]}$, $\ell = e,\mu$ & $ -1.55 $ & $ [-2.73,-0.81] $ &  2.4 & $ 10 $ & 76.0 \hspace{5mm}  \\[3mm] 
 All $b\to s\mu\mu$, 20\% PCs & $ -1.10 $ & $ [-1.31,-0.87] $ &  4.3 & $ 95 $ & 69.0 \hspace{5mm}  \\[3mm] 
 All $b\to s\mu\mu$, 40\% PCs & $ -1.08 $ & $ [-1.32,-0.82] $ &  3.8 & $ 95 $ & 73.0 \hspace{5mm}  \\[3mm] 
 All $b\to s\mu\mu$, charm$\times 2$ & $ -1.12 $ & $ [-1.33,-0.89] $ &  4.4 & $ 95 $ & 73.0 \hspace{5mm}  \\[3mm] 
 All $b\to s\mu\mu$, charm$\times 4$ & $ -1.06 $ & $ [-1.29,-0.82] $ &  4.0 & $ 95 $ & 81.0 \hspace{5mm}  \\[3mm] 
 Only $b\to s\mu\mu$ within [0.1,6] & $ -1.21 $ & $ [-1.57,-0.84] $ &  3.1 & $ 60 $ & 30.0 \hspace{5mm}  \\[3mm] 
 Only $b\to s\mu\mu$ within [0.1,0.98] & $ +0.08 $ & $ [-0.92,0.95] $ &  0.1 & $ 13 $ & 33.0 \hspace{5mm}  \\[3mm] 
 Only $b\to s\mu\mu$ within [0.1,2] & $ -1.03 $ & $ [-1.98,-0.20] $ &  1.3 & $ 22 $ & 4.6 \hspace{5mm}  \\[3mm] 
 Only $b\to s\mu\mu$ within [1.1,2.5] & $ -0.74 $ & $ [-1.60,0.06] $ &  0.9 & $ 13 $ & 85.0 \hspace{5mm}  \\[3mm] 
 Only $b\to s\mu\mu$ within [2,5] & $ -1.56 $ & $ [-2.27,-0.91] $ &  2.5 & $ 23 $ & 95.0 \hspace{5mm}  \\[3mm] 
 Only $b\to s\mu\mu$ within [4,6] & $ -1.34 $ & $ [-1.73,-0.94] $ &  3.1 & $ 16 $ & 93.0 \hspace{5mm}  \\[3mm] 
 Only $b\to s\mu\mu$ within [5,8] & $ -1.30 $ & $ [-1.60,-0.98] $ &  3.5 & $ 22 $ & 96.0 \hspace{5mm}  \\[3mm] 
 All $b\to s\mu\mu$ excluding large-recoil $B_s\to\phi\mu\mu$ & $ -1.04 $ & $ [-1.26,-0.81] $ &  4.0 & $ 80 $ & 55.0 \hspace{5mm}  \\[3mm] 
 All $b\to s\ell\ell$, $\ell = e,\mu$ excl. large-recoil $B_s\to\phi\mu\mu$ & $ -1.06 $ & $ [-1.26,-0.84] $ &  4.5 & $ 86 $ & 35.0 \hspace{5mm}  \\[3mm] 
\bottomrule[1.6pt] 
\end{tabular}}
\end{center}

\clearpage

\section{Power corrections to $B_s\to\phi$, $B\to K$ form factors}
\label{app:PowCorr}

The hadronic form factors $F$ for $B\to M$ decays (where $M$ denotes a light vector or pseudo-scalar meson) can be decomposed into a 
soft part $F^{\textrm{soft}}$, an $\alpha_s$-correction $\Delta F^{\alpha_s}$ and a factorisable power correction $\Delta F^\Lambda$:
\begin{equation}
  F(q^2)\,=\,F^{\textrm{soft}}(q^2)\,+\,\Delta F^{\alpha_s}(q^2)\,+\,\Delta F^\Lambda(q^2).
  \label{eq:PC1}
\end{equation}
The soft component $F^{\textrm{soft}}$ is a linear combination of two soft form factors $\xi_\perp$ and $\xi_\parallel$ for $M$  vector, and proportional to a single soft form factor $\xi_P$ for $M$ pseudoscalar. 
The decomposition Eq.~(\ref{eq:PC1}) is not unique: depending on the exact definition of the soft form factors $\xi_i$, a part of $\Delta F^{\alpha_s}$ and 
$\Delta F^\Lambda$ can be reabsorbed into the soft contribution $F^{\textrm{soft}}$. This introduces a scheme dependence for $\Delta F^{\alpha_s}$ and
$\Delta F^\Lambda$ which has been discussed in detail for the $B\to K^*$ form factors in Ref.~\cite{Descotes-Genon:2014uoa}.

While QCD corrections $\Delta F^{\alpha_s}$ can be calculated employing QCD factorisation, the power corrections
$\Delta F^{\Lambda}$ cannot be computed directly and in general, they must be estimated on dimensional grounds.
However, one can perform explicit computations of the full form factors $F(q^2)$ (say, from light-cone sum rules) in order to extract $\Delta F^{\Lambda}(q^2)$  through a fit. In the case of the $B\to K^*$ form factors, this determination has been performed in Ref.~\cite{Descotes-Genon:2014uoa}. The parameters $\hat{a}_F,\hat{b}_F,\hat{c}_F$ arising in the parametrisation 
\begin{equation}\label{eq:PCparaJager}
   \Delta F^{\Lambda}(q^2) = \hat{a}_F\,+\,\hat{b}_F\,\frac{q^2}{m_B^2}\,+\,\hat{c}_F\,\frac{q^4}{m_B^4}\,,
\end{equation}
can be found in Tables 1 and 2 of that paper for two different choices of scheme for $\xi_\perp$ and 
$\xi_\parallel$ (considering either the LCSR calculation in Ref.~\cite{Khodjamirian:2010vf}  or that in
Ref.~\cite{Ball:2004rg} as inputs). In Table~\ref{tab:fitPC},
we give the corresponding results for $B_s\to\phi$ and $B\to K$ form factors using LCSR input 
from Refs.~\cite{Straub:2015ica} and \cite{Khodjamirian:2010vf}, respectively. We follow
 scheme 1 in Ref.~\cite{Descotes-Genon:2014uoa} and define the soft form factors as
\begin{eqnarray}
  \xi_{\perp}(q^2)&\equiv&\frac{m_B}{m_B+m_{K^*}}V(q^2),\nonumber\\
  \xi_{\parallel}(q^2)&\equiv&\frac{m_B+m_{K^*}}{2E}A_1(q^2)\,-\,\frac{m_B-m_{K^*}}{m_B}A_2(q^2),
\end{eqnarray}
for $B_s\to\phi$, and as
\begin{equation}
   \xi_P(q^2) = f_+(q^2),
\end{equation}
for $B\to K$. We further quantify the relative size of power corrections for the various form factors
though the ratio
\begin{equation}
   r(q^2)\,=\,\left|\frac{\hat{a}_F+\hat{b}_F\frac{q^2}{m_B^2}+\hat{c}_F\frac{q^4}{m_B^4}}{F(q^2)}\right|
\end{equation}
at $q^2=0, 4, 8$\,GeV$^2$.

From dimensional arguments one expects $r(q^2)=\mathcal{O}(\Lambda/m_B)\lesssim 10\%$. The results in Table~\ref{tab:fitPC} show that the 
LCSR form factors from Refs.~\cite{Straub:2015ica} and \cite{Khodjamirian:2010vf} indeed comply with this expectation, except for the form factor
$A_2$ where larger power corrections occur. In our SM predictions as well as in the NP fits, we use the results from Table~\ref{tab:fitPC}
as central values for the parameters $a_F,b_F,c_F$, to which we assign error ranges of the order of $10\%\times F$. Comparing with $r(q^2)$, we see that this
corresponds to the assumption of $\mathcal{O}(100\%)$ uncertainties for the coefficients $a_F,b_F,c_F$. Since our error estimate is based only on dimensional arguments,
it is independent of the detail of the particular LCSR calculation. On the other hand, taking into account correlations among the LCSR form factors,
it is also possible to determine the uncertainties of $a_F,b_F,c_F$ from a particular set of LCSR input, which will be detailed in an upcoming publication~\cite{wip}.

\begin{table}
\footnotesize
\centering
\begin{tabular}{@{}l|rrr|ccc@{}}
\toprule[1.6pt] 
$B_s\to\phi$ & $\hat{a}_F\qquad$ & $\hat{b}_F\qquad$ & $\hat{c}_F\qquad$ &
$r(0\,\rm{GeV}^2)$ & $r(4\,\rm{GeV}^2)$ & $r(8\,\rm{GeV}^2)$ \\ 
\hline 
$A_0$& $0.000\pm 0.000$ & $0.047\pm 0.057$ & $0.192\pm 0.120$ & $0.000$ & $0.020$ & $0.041$ \\ 
$A_1$& $0.028\pm 0.032$ & $0.053\pm 0.018$ & $0.115\pm 0.027$ & $0.090$ & $0.110$ & $0.135$ \\  
$A_2$& $0.042\pm 0.026$ & $0.103\pm 0.028$ & $0.431\pm 0.050$ & $0.162$ & $0.209$ & $0.274$ \\ 
$T_1$& $-0.024\pm 0.033$ & $-0.016\pm 0.045$ & $-0.015\pm 0.100$ & $0.072$ & $0.063$ & $0.054$ \\  
$T_2$& $-0.023\pm 0.033$ & $0.016\pm 0.016$ & $0.174\pm 0.035$ & $0.071$ & $0.049$ & $0.013$ \\ 
$T_3$& $-0.015\pm 0.019$ & $-0.007\pm 0.018$ & $0.133\pm 0.031$ & $0.080$ & $0.061$ & $0.026$ \\ 
\midrule[1.6pt] 
$B\to K$ & $\hat{a}_F\qquad$ & $\hat{b}_F\qquad$ & $\hat{c}_F\qquad$ &
$r(0\,\rm{GeV}^2)$ & $r(4\,\rm{GeV}^2)$ & $r(8\,\rm{GeV}^2)$ \\
\hline
$f_0$& $0.000\pm 0.000$ & $0.159\pm 0.034$ & $0.065\pm 0.015$ & $0.000$ & $0.062$ & $0.113$ \\ 
$f_T$& $0.053\pm 0.034$ & $0.065\pm 0.046$ & $0.113\pm 0.079$ & $0.136$ & $0.133$ & $0.130$ \\ 
\bottomrule[1.6pt]
\end{tabular}
\caption{Fit results for the power-correction parameters to the $B_s\to\phi$ and $B\to K$ form factors, choosing a scheme with the soft form factors $(\xi_{\perp},\xi_{\parallel})$ defined from $V$ and the difference of $A_1$ and $A_2$ in the case of $B_s\to\phi$, and with $\xi_P$ defined from $f_+$ in the case of
$B\to K$. The corresponding LCSR input has been taken from Ref.~\cite{Straub:2015ica} for $B_s\to\phi$ and from Ref.~\cite{Khodjamirian:2010vf}
for $B\to K$. Furthermore, the relative size $r(q^2)$ with which the power corrections contribute to the full form factors is shown for $q^2=0,4,8\,\rm{GeV}^2$.}
\label{tab:fitPC}
\end{table}

\newpage

\section{$Z'$ couplings}
\label{sec:Zcouplings}

In Ref.~\cite{Descotes-Genon:2013wba}, we proposed to explain the deviation in $B\to K^*\mu\mu$ using a $Z^\prime$ gauge
boson contributing to
\begin{equation}
{\cal O}_9=e^2/(16\pi^2) \, (\bar s\gamma_\mu P_Lb)(\bar \ell \gamma^{\mu} \ell)\,
\end{equation}
with specific couplings, as a possible explanation of the anomaly in $P_5^\prime$.
This possibility was embedded in several models~\cite{Altmannshofer:2014cfa,Crivellin:2015era,Crivellin:2015mga,Celis:2015ara,Buras:2012jb,Buras:2013dea,Buras:2014yna,Gauld:2013qja,Sierra:2015fma,Altmannshofer:2015mqa,Boucenna:2016wpr}.
With the notation of Ref.~\cite{Buras:2012jb},
\begin{eqnarray}
{\cal L}^{q}&=&\left( {\bar s} \gamma_{\nu} P_L b \Delta_L^{sb}+ {\bar s} \gamma_{\nu} P_R b \Delta_R^{sb} +h.c. \right) Z^{\prime\nu} \\
{\cal L}^{lep}&=&\left( {\bar \mu} \gamma_{\nu} P_L \mu \Delta_L^{\mu{\bar \mu}} +{\bar \mu} \gamma_{\nu} P_R \mu \Delta_R^{\mu{\bar \mu}}+...\right) Z^{\prime\nu}
\end{eqnarray}
the Wilson coefficients of the semileptonic operators receive the following contributions:
\begin{eqnarray} \C{ \{9,10\}}^{\rm NP}= -\frac{1}{ s_W^2 g_{SM}^2} \frac{1}{M_{Z^{\prime}}^2} \frac{\Delta_L^{sb} \Delta^{\mu\mu}_{ \{V,A\}}}{\lambda_{ts}}\,, \quad
\C{ \{9^{\prime},10^{\prime}\}}^{\rm NP}= -\frac{1}{s_W^2 g_{SM}^2} \frac{1}{M_{Z^{\prime}}^2} \frac{\Delta_R^{sb} \Delta^{\mu\mu}_{ \{V,A\}}}{\lambda_{ts}} \,,\end{eqnarray}
with the vector and axial couplings to muons defined in terms of the couplings of the Lagrangian by $\Delta^{\mu\mu}_{ V,A}=\Delta^{\mu\mu}_R { \pm} \Delta^{\mu\mu}_L$.

These couplings obey the relationship
\begin{equation}
\C9^{\rm NP}\times \C{10^\prime}^{\rm NP}=\C{10}^{\rm NP}\times \C{9^\prime}^{\rm NP}\,.
\end{equation}
A $Z^\prime$ model can therefore belong to the following categories:
\begin{itemize}
\item NP only in the following pairs, with a priori arbitrary contributions,
\begin{equation}
(\C9,\C{10})\,,\qquad (\C9,\C{9^\prime})\,,\qquad (\C{10},\C{10^\prime})\,,\qquad (\C{9^\prime},\C{10^\prime})\,,\end{equation}
each case corresponding to the vanishing of some of the couplings $\Delta_{L,R}^{sb},\Delta_{V,A}^{\mu\mu}$. These models have a definite chirality for quark-flavour changing coupling currents and/or a definite parity for the couplings to muons.
\item NP enters all four semileptonic coefficients with the following relationships 
\begin{equation}
\frac{\C9^{\rm NP}}{\C{10}^{\rm NP}}=\frac{\C{9^\prime}^{\rm NP}}{\C{10^\prime}^{\rm NP}}=\frac{\Delta_V^{\mu\mu}}{\Delta_A^{\mu\mu}}\,,\qquad 
\frac{\C9^{\rm NP}}{\C{9^\prime}^{\rm NP}}=\frac{\C{10}^{\rm NP}}{\C{10^\prime}^{\rm NP}}=\frac{\Delta_L^{sb}}{\Delta_R^{sb}}\,.
\end{equation}
\end{itemize}

\newpage


\end{document}